\documentclass[jou,floatsintext, hidelinks, table]{apa7}

\usepackage{lipsum}

\usepackage[american]{babel}

\usepackage{csquotes}
\usepackage[fleqn]{amsmath}

\newcommand{\appropto}{\mathrel{\vcenter{
  \offinterlineskip\halign{\hfil$##$\cr
    \propto\cr\noalign{\kern2pt}\sim\cr\noalign{\kern-2pt}}}}}

\usepackage{algorithm}
\usepackage{algpseudocode}
\usepackage{multirow}
\newcolumntype{?}{!{\vrule width 1.5pt}}
\usepackage{hyperref}
\hypersetup{
    colorlinks=true,
    linkcolor=.,
    filecolor=.,
    citecolor=.,
    urlcolor=blue
    }

\algnewcommand\algorithmicinput{\textbf{Input:}}
\algnewcommand\INPUT{\item[\algorithmicinput]}

\algnewcommand\algorithmicinit{\textbf{Initialize:}}
\algnewcommand\INIT{\item[\algorithmicinit]}

\usepackage[style=apa,sortcites=true,sorting=nyt,backend=biber]{biblatex}
\DeclareLanguageMapping{american}{american-apa}
\title{The Mixed-Sparse-Smooth-Model Toolbox (MSSM): Efficient Estimation and Selection of Large Multi-Level Statistical Models}
\shorttitle{Mixed Sparse Smooth Models (MSSM)}

\author{Joshua Krause, Jelmer P. Borst, Jacolien van Rij}
\affiliation{{Bernoulli Institute for Mathematics, Computer Science, and Artificial Intelligence}}

\leftheader{Joshua Krause, Jelmer P. Borst, Jacolien van Rij}

\abstract{Additive smooth models, such as Generalized additive models (GAMs) of location, scale, and shape (GAMLSS), are a popular choice for modeling experimental data. However, software available to fit such models is usually not tailored specifically to the estimation of mixed models. As a result, estimation can slow down dramatically as the number of random effects in the model increases. Additionally, users often have to provide a substantial amount of problem-specific information in case they are interested in more general non-standard smooth models, such as higher-order derivatives of the likelihood function. In contrast, here we combined and extended recently proposed strategies to reduce memory requirements and matrix infill into a theoretical framework that supports efficient estimation of general mixed sparse smooth models, including GAMs \& GAMLSS, based only on the Gradient and Hessian of the log-likelihood. To make non-standard smooth models more accessible, we developed an approximate estimation algorithm (the L-qEFS update) based on limited-memory quasi-Newton methods. This enables estimation of any general (mixed) smooth model based only on the log-likelihood function. In addition, we considered the problem of model selection for general mixed sparse smooth models. To facilitate practical application we provide an accessible Python implementation of the theoretical framework, algorithms, and model selection strategies presented here: the Mixed-Sparse-Smooth-Model (MSSM) toolbox. MSSM supports estimation and selection of massive additive multi-level models that are impossible to estimate with alternative software, for example of trial level EEG data. Additionally, when the L-qEFS update is used for estimation, implementing a new non-standard smooth model in MSSM is straightforward. Results from multiple simulation studies and real data examples are presented, showing that the framework implemented in MSSM is both efficient and robust to numerical instabilities.

}


\authornote{
   \addORCIDlink{Joshua Krause}{0000-0002-3165-2584}
   \addORCIDlink{Jelmer P. Borst}{0000-0002-4493-8223}
   \addORCIDlink{Jacolien van Rij}{0000-0001-7445-5647}

  Correspondence concerning this article should be addressed to Joshua Krause, Bernoulli Institute for Mathematics, Computer Science and Artificial Intelligence, University of Groningen, Nijenborgh 9, 9747 AG Groningen, Netherlands.  E-mail: j.krause@rug.nl}

\begin{document}
\maketitle

\section{Introduction}\label{sec:Introduction}
Generalized additive models \parencite[GAMs;][]{hastie_generalized_1986,hastie_generalized_1990,wood_generalized_2017-2} are a popular class of regression models, because they allow one to model the mean of response variables as an additive combination of potentially non-linear functions of covariates (e.g., time or trials, or experimental manipulations such as word frequency, dot motion coherence, etc.). Theoretically, this flexibility makes them an excellent choice for modeling the many different signals that are of interest to researchers working with experimental data. In the cognitive sciences, for example, researchers commonly work with reaction times (RTs), accuracy, the electroencephalogram (EEG), pupil dilation, gaze fixation patterns, and blood-oxygen-level-dependent responses in different cortical regions (BOLD) -- which can all be modeled with GAMs.

\begin{figure*}[!t]
    \caption{Smooth Function Overview}
    \includegraphics[width=\textwidth]{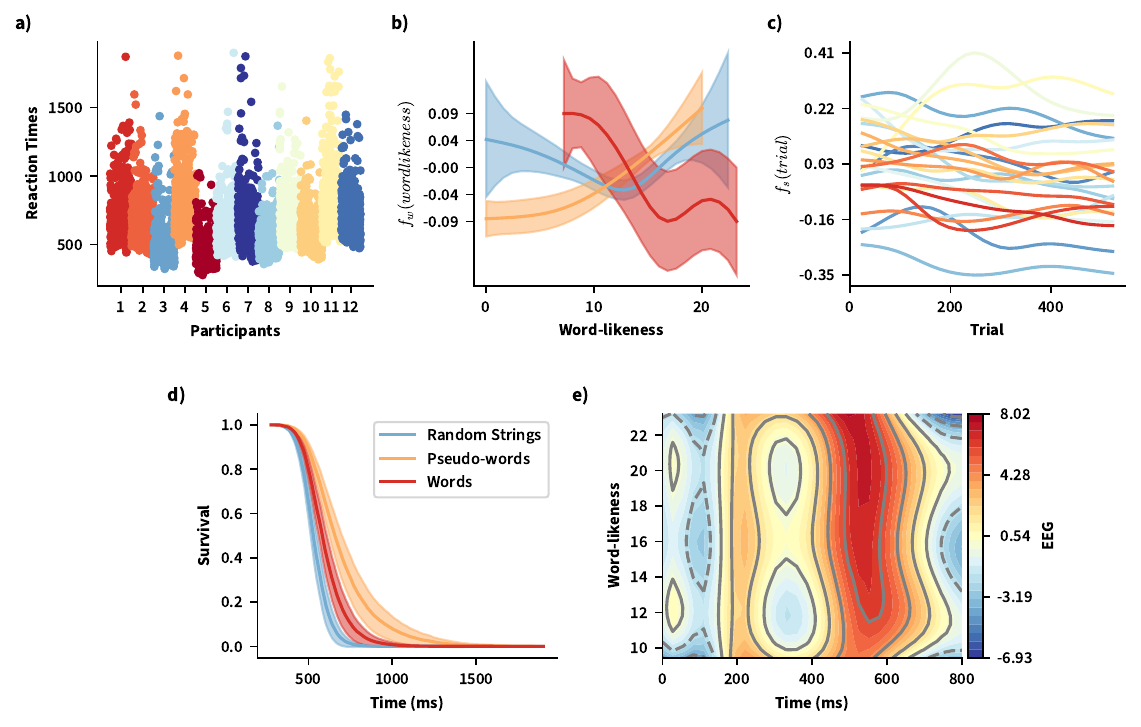}
    \label{fig:smooth_overview}
    {\small
        \textit{Note.} Figure \ref{fig:smooth_overview} depicts a subset of the RT data collected by \Textcite{krause_word_2024} as well as selected smooth function estimates obtained from models of their RT and EEG data. Panel a shows RT values for a subset of 12 subjects who completed a lexical decision experiment involving three types of stimuli (words, pseudo-words, and random strings). Observations belonging to a specific subject were plotted in the same color. Panel b shows estimated smooth functions of the ``word-likeness'' covariate utilized by \Textcite{krause_word_2024} taken from a Gamma model of the RTs and approximate 95\% credible intervals. Panel c shows subject-specific ``random smooth functions'' of the time spent in the experiment (i.e., over trials) taken from the same model. Panel d shows the estimated survival functions and approximate 95\% credible intervals for the three word types at their average word-likeness rating obtained from a Cox-proportional survival model of the RTs. Panel e shows the estimated mean EEG amplitude at a single electrode and only for words as a function of the time-course of a lexical decision and word-likeness of the stimuli.
        }
\end{figure*}

Another advantage is that \emph{random} effects can readily be included in GAMs, allowing to address the kind of dependencies common in experimental data, which is almost always collected from different participants who contribute multiple observations \parencite[e.g.,][]{baayen_analyzing_2010,baayen_cave_2017}. As illustrated in Figure \ref{fig:smooth_overview}a, dependencies in reaction time (RT) data might arise because some participants consistently respond faster than others (compare for example the observations of participants 4 and 5 in Figure \ref{fig:smooth_overview}a). Evidently, a model only including \emph{fixed} effects of the experimentally manipulated variables, which by definition are assumed not to differ between different participants, cannot account for these dependencies between observations. A GAM of RT data collected in a lexical decision experiment (i.e., judging whether a presented string of characters corresponds to a word or not), might for example include a \emph{fixed} smooth function of the word-likeness of the presented stimuli in the model of the mean. Such a model could account for a smooth non-linear effect of word-likeness on the expected RTs (see Figure \ref{fig:smooth_overview}b for a visualization). However, without random effects it would lack a mechanism to explain dependencies in the data resulting from subject to subject differences in their average RTs. Importantly, failing to address such dependencies is well known to result in biased inference \parencite[e.g., confidence intervals that become too narrow][]{baayen_cave_2017,wood_generalized_2017-2}, which can lead to wrong conclusions about the importance of different predictors or the strength of different effects.

To address these dependencies, a multi-level or mixed model is needed with \emph{random} effects to model factor-level-specific deviations in the mean of the response variables \parencite[e.g.,][]{wood_generalized_2017-2}\footnote{Or a known function of the mean.}. A simple \emph{random intercept} model of subject-specific RT deviations can be obtained for example, by assuming that they reflect independent realizations from a zero-centered normal distribution. The variance of the latter then informs about the range of deviations that could be expected in additional replications of the experiment - each of which would involve different subjects and thus result in a different set of subject-specific deviations.

This approach generalizes to accommodate more complex random effect models: in GAMs the factor-level-specific deviations in the mean can take on the form of ``random smooth functions'' of experimental covariates \parencite[e.g., ][]{wood_generalized_2017-2}. Considering again the RT data example, it will often be the case that the mean RT changes over the time-course of an experiment (i.e., over trials), and does so differently for different participants \parencite[e.g.,][]{baayen_analyzing_2010,baayen_cave_2017}. Some subjects will tire more easily than others and their responses might slow down towards the end of the experiment. Others might get faster over time because they figure out a response strategy that works particularly well for them. A random intercept model as described in the previous paragraph would not be able to account for these fluctuations since the subject-specific deviation from the mean is assumed to be constant. Instead, it would be prudent to adjust the mean of the response variables by a subject-specific \emph{random} smooth function of trials. Figure \ref{fig:smooth_overview}c shows how the latter would look in a model of this RT data; the upcoming Background section introduces these random smooth terms more formally \parencite[see also;][]{wood_generalized_2017-2}.

Considerable effort has been dedicated to enable stable estimation of multi-level or mixed GAMs (i.e., GAMMs) and related models, and several software packages have been developed to facilitate this, including mgcv \parencite[][]{wood_fast_2011,wood_smoothing_2016,wood_generalized_2017-2}, INLA \parencite[]{rue_approximate_2009}, and VGAM \parencite[]{yee_vector_2015}. Among those, mgcv is the most general and robust. Implemented in R \parencite[]{r_core_team_r_2024}, it offers stable estimation, automatic regularization, and computation of model-selection criteria \parencite[e.g., AIC;][]{akaike_information_1992} for the broad class of ``general or generic smooth models'' \parencite[]{wood_smoothing_2016}. The latter includes any statistical model with a regular likelihood function, the parameters of which are represented as an additive combination of (smooth) functions of experimental covariates \parencite[see][for discussions and an introduction to the specific regularity conditions]{wood_smoothing_2016,wood_core_2015}. This includes GAMMs, which come with additional assumptions that facilitate estimation (see the Background section), in which only a single parameter of the likelihood (i.e., the mean of the response variables) is represented as an additive model \parencite[e.g.,][]{wood_smoothing_2016}.  

In the context of modeling experimental data, estimation is usually considered only a means to an end; instead, inference about the effect of experimental manipulations is of primary interest. Access to well-behaved model-selection criteria is therefore extremely important. Additionally, support for general smooth models is particularly relevant to researchers working with complex statistical models with a likelihood function that does not take on the convenient form of a GAMM, potentially due to weaker independence assumptions. An example are Markov models, which have been used to estimate the stages of cognitive processing required to complete experimental tasks \parencite[e.g.,][]{anderson_discovery_2016,krause_word_2024,weindel_trial-by-trial_2024}. Another example are survival models of RT data, allowing researchers to investigate how the probability of responding changes over the time-course of an experimental trial (see Figure \ref{fig:smooth_overview}d for a visualization of such ``survival functions'' and also \Textcite{klein_survival_2003} for an introduction). These and related models are becoming increasingly popular in the cognitive sciences \parencite[e.g.,][]{hendrix_word_2020}. Often it would be appropriate to represent some or all of the parameters of such models as an additive combination of (smooth) functions of experimental covariates, just like the mean in a regular GAMM. The resulting models would fall into the aforementioned class of general smooth models \parencite[]{wood_smoothing_2016}. Despite their ``non-standard'' likelihood these models can thus still be estimated, regularized, and subjected to selection using the theoretical framework by \Textcite{wood_smoothing_2016} implemented in mgcv.

To enable fitting generic smooth models and selecting between such models, mgcv principally requires access to derivatives of the model's log-likelihood with respect to model coefficients up to the fourth order\footnote{For GAMMs an alternative strategy exists that only requires higher-order derivatives of the model's link function with respect to the mean (see the Background section for more details) rather than the log-likelihood \parencite[e.g.,][]{wood_fast_2011}.} \parencite[e.g.,][]{wood_smoothing_2016}. In practice, this comes with a computational cost even when working with conventional GAMMs: As discussed more extensively in the Background section, because mgcv's fitting algorithms involve higher-order derivatives of the log-likelihood, they cannot generally exploit the sparsity induced by random effects and random smooth terms to speed up model fitting. This in turn means that estimation starts to slow down noticeably once separate (random) smooths are to be estimated for more than a handful of levels of a factor variable. Already the estimation of simple models of, for example, 5000 RT values collected from 20 participants (i.e. 250 observations per participant) might take 5-10 times longer when the sparsity induced by random effects is not exploited, as shown in Sections \ref{sec:AM} and \ref{sec:GAM} of this paper. While the time-to-fit required by mgcv will often be acceptable for such simple cases\footnote{For these simple cases, the time-to-fit will usually stay below a minute. Less time will be required when using the ``bam'' function for ``big additive models'' as discussed by \Textcite{wood_generalized_2015,wood_generalized_2017-1}.}, it will quickly become prohibitive for more complex models. An example are additive models of trial-level time-course data such as EEG or pupil dilation time-course recordings. A GAMM-based approach to the analysis of such data has the advantage that it can readily be investigated how the effect of continuous predictor variables changes smoothly over time. For example, Figure \ref{fig:smooth_overview}e shows an estimate of the \emph{non-linear interaction effect} of word-likeness and time on the expected EEG amplitude measured by a particular electrode of interest. However, because such models usually require series-specific random smooths of time to address the dependencies present in the data \parencite[cf.][]{van_rij_analyzing_2019}, fitting them might conceivably take half a year when model sparsity is not exploited -- or approximately 10,000 times as long as it would take when it is (see Section \ref{sec:DataExamples}).

Researchers wanting to work with general models featuring a non-standard likelihood will additionally have to pay the full price for the high theoretical cost of this framework: not only will estimation of these models quickly slow down as the number of random effects increases, but researchers will also have to derive and code the model-specific higher-order derivatives (up to fourth order) of the log-likelihood function. While obtaining these derivatives can be partially automated, this requires familiarity with (symbolic) computer algebra systems (e.g., Maxima \parencite{maxima_maxima_2023}, or the Deriv package \parencite{clausen_deriv_2024} in R -- see \Textcite{wood_core_2015} for an introduction). Additionally, even when these quantities can be obtained or are available, they still have to be translated into numerically stable and efficient computer code which can be just as challenging\footnote{For an example, we encourage readers to consult the Online Supplementary materials from \Textcite{wood_smoothing_2016} -- which provide examples for multiple models and show how to obtain the derivatives required for estimation, regularization, and selection. Computations of the Gradient and Hessian are often fairly straightforward still, but the derivatives of the latter with respect to the log regularization parameters (denoted as $\rho$) generally become quite complex. Many of these higher-order derivatives also suffer from numerical instabilities and thus have to be implemented very carefully.}. Alternatively, automatic differentiation could be performed directly based on the implementation of the log-likelihood in computer code \parencite[e.g.,][]{wood_core_2015,wood_inference_2020}. However, most automatic differentiation libraries rely on specific data-structures or cannot handle fairly conventional coding practices (e.g., list assignment) which still means that in practice researchers will first have to familiarize themselves with quite complex computer software before they can even get started with the task of modeling. For many researchers and other end-users of estimation software, any number of these requirements might be perceived as an insurmountable obstacle, increasing the risk that they might instead fall back to simpler, potentially miss-specified or otherwise inappropriate, models of their data.

Importantly, the log-likelihood function itself, as well as lower order derivatives (up to second order, i.e., Gradient and Hessian) are usually less complex. As a consequence, they are generally easier to derive. They also require less work to be translated, often resulting only in a few lines of efficient and stable computer code. Importantly, at least the objectives of estimation and automatic regularization of general smooth models can be achieved (approximately) based on the Gradient and Hessian matrix alone \parencite[e.g.,][]{wood_generalized_2017}. For GAMMs and many members of the related class of Generalized Additive Models of Location, Scale, and Shape \parencite[GAMMLSS;][]{rigby_generalized_2005} this reduction in complexity also enables exploiting sparsity of the Hessian to speed up estimation of multi-level models including many random effects (see Sections \ref{sec:AM}-\ref{sec:GAMMLSS} in this paper and \citeauthor{wood_generalized_2017}, \citeyear{wood_generalized_2017}).

Unfortunately, for more generic models still, the Hessian matrix itself will often either be completely dense (i.e., feature no or only a few zeroes) or have a sparsity structure that might not be known in advance. This makes efficiently accumulating the matrix in sparse matrix storage impossible or at least difficult. An example are Cox' proportional Hazard models introduced in the next section \parencite[see also][]{cox_regression_1972,klein_survival_2003,wood_smoothing_2016}. To ensure that estimation remains memory-efficient for generic multi-level smooth models and to further lower the burden placed on researchers wanting to estimate these models would require developing an algorithm that enables estimation and automatic regularization based on the Gradient alone, without the need to accumulate the Hessian matrix explicitly.

While omitting higher-order derivatives facilitates the estimation and regularization of smooth models, it makes quantifying the uncertainty in the estimated amount of regularization more difficult. This is problematic, because model selection criteria like the conditional variant of the Akaike Information Criterion \parencite[cAIC;][]{akaike_information_1992,greven_behaviour_2010} have to be corrected for this source of uncertainty in order to be well-behaved \parencite[e.g.,][]{greven_behaviour_2010,saefken_unifying_2014,greven_comment_2016,wood_smoothing_2016}. \Textcite{wood_smoothing_2016} outlined how to quantify and correct for this uncertainty, even for the most generic smooth models. However, their strategy again requires access to higher order derivatives of the log-likelihood, which implies that the same issues complicating the estimation \& regularization of multi-level and generic smooth models continue to apply here: quantifying the uncertainty in the estimated amount of regularization again becomes computationally expensive for multi-level models and places a high burden on researchers who will principally have to provide these higher-order derivatives. Because of the resulting inconvenience or in case higher-order derivatives are not readily available, researchers might opt not to correct for this source of uncertainty instead. However, it is well known that this can result in a drastically higher false positive rate \parencite[e.g.,][]{greven_behaviour_2010,greven_comment_2016,wood_smoothing_2016}. It thus remains an important open problem to investigate how to enable approximate quantification of the uncertainty in the estimated degree of regularization, in the absence of higher-order derivatives and for sparse multi-level models.

To summarize the problems outlined above, we conclude that working with multi-level smooth models quickly becomes time-consuming, because estimating them is computationally demanding and, for more generic models, because considerable theoretical work is necessary upfront to obtain the necessary third and fourth order derivatives of the log-likelihood to be optimized. Omitting higher-order derivatives goes a long way to address these problems, because this reduces the theoretical work necessary upfront and enables efficient exploitation of model sparsity for GAMMs and more generic models with a sparse Hessian. However, exploiting sparsity is not really possible in case the Hessian is dense or has a sparsity structure unknown to the researcher. Additionally, omitting higher-order derivatives means that we can no longer quantify uncertainty in the estimated amount of regularization for conventional smooth models (e.g., GAMMS) and more generic models alike \parencite[e.g.,][]{wood_smoothing_2016,wood_inference_2020}.

\subsection{Contributions}

In an effort to address these problems, we derive a theoretical framework optimized for fast estimation and automatic regularization of mixed or multi-level generic smooth models, that still enables selection between generic models with a ``non-standard'' likelihood -- even when the model-specific higher-order derivatives of the log-likelihood function are unavailable. To facilitate practical application, we provide a Python implementation of the resulting framework: the Mixed-Sparse-Smooth-Model (MSSM) toolbox (available on GitHub at \url{https://github.com/JoKra1/mssm} and distributed via the Python Package Index). The toolbox supports massive mixed additive models and as such can be used to estimate models of trial-level EEG data, often in less than an hour. We provide a more detailed description of such a model in Section \ref{sec:DataExamples}. The MSSM toolbox is not limited to GAMMs, but offers support for multi-level generic smooth models as defined by \Textcite{wood_smoothing_2016}.

Before we describe the theoretical framework implemented in MSSM in detail, we conduct a more formal review of GAMMs, general smooth models, and the challenges surrounding their estimation, regularization, and selection -- which we present in the upcoming Background section. In three separate sections we then outline how the sparsity of multi-level models can be exploited for Gaussian additive mixed models (Section \ref{sec:AM}), GAMMS (Section \ref{sec:GAM}), GAMMLSS and generic smooth models of likelihood functions with a sparse Hessian (Section \ref{sec:GAMMLSS}), to enable efficient estimation and automatic regularization of these models. Many of the algorithms and techniques outlined in these sections have previously been described in the literature. Our main contribution here is the combination of these algorithms into a unified framework, applicable to the broad class of models described above, and the implementation of this framework in the MSSM toolbox.

We then develop a novel algorithm based on limited-memory quasi-Newton methods \parencite[e.g.,][]{nocedal_numerical_2006} that enables approximate estimation and automatic regularization for any (mixed) general smooth model (Section \ref{sec:lqefs}). The algorithm only requires researchers to provide the log-likelihood function and (optionally) ways to compute its Gradient. Because this update does not require the Hessian matrix, it remains memory efficient even for likelihoods with a Hessian that is dense or has an unknown sparsity structure. Finally, in Section \ref{sec:Uncertainty}, we derive multiple strategies, based on the methods proposed by \Textcite{wood_smoothing_2016} and \Textcite{greven_comment_2016}, that can be used to approximately quantify and correct the cAIC for uncertainty in the estimated amount of regularization.

In Section \ref{sec:Results} we put the entire framework, and our implementation in the MSSM toolbox, to the test. We compare it to existing alternatives on a range of different performance metrics (e.g., mean square error, time to fit, model selection outcomes) across multiple simulation studies. The results reveal that the methods discussed here are both stable and efficient, typically producing estimates that are virtually indistinguishable from those produced by alternative methods -- often in a fraction of the time. Finally, in Section \ref{sec:DataExamples}, we present examples of large survival and EEG time-course models, which further demonstrate the usefulness of mixed sparse smooth models and the estimation framework presented here for the cognitive sciences in particular.

\section{Background}\label{sec:Background}

\subsection{Additive Smooth Models}

We start with a formal definition of general additive smooth models. Subsequently, we show how specific models like strictly Additive (i.e., Gaussian) Models (AMs), Generalized Additive Models (GAMs), and Generalized Additive Models of Location, Scale, and Shape (GAMLSS) and their mixed variants fit into this definition. 

Following \Textcite{wood_smoothing_2016}, the log-likelihood function of a generic or general smooth model can be defined as $\mathcal{L} = log(p(\mathbf{y}|\boldsymbol{\beta},\phi))$, where $p(\mathbf{y}|\boldsymbol{\beta},\phi)$ denotes the probability of the collected observations available in the $N$ dimensional vector $\mathbf{y}$ given a set of model coefficients $\boldsymbol{\beta}$ and optional scale parameter $\phi$ which for many models can simply be omitted. The dependence of the log-likelihood function on the vector of coefficients, which is required to be smooth and regular \parencite[e.g.,][]{wood_smoothing_2016}, is usually indirect. That is, the log-likelihood will typically be a function of a small set of parameters (e.g., the mean of a normal distribution), and the coefficients $\boldsymbol{\beta}$ are used to parameterize additive models of these parameters or a known ``link'' function of them. The resulting additive models become additive smooth models when some of the coefficients are used to represent smooth functions $f_1,f_2,...,f_{N_J}$. To achieve this, the latter are parameterized as weighted sums of known basis functions $b$ (e.g., piece-wise polynomials, such as B-spline basis functions; \citeauthor{eilers_flexible_1996}, \citeyear{eilers_flexible_1996} but see also \citeauthor{wood_generalized_2017-2}, \citeyear{wood_generalized_2017-2} for alternatives), so that

\begin{equation}\label{eq:smooth_term}
f_j(x_{j,1},x_{j,2},...) = \sum_k b_{j,k}(x_{j,1},x_{j,2},...)\beta_{j,k}.
\end{equation}

\noindent
Here, $x_{j,1},x_{j,2},...$ are the covariates of which $f_j$ (and each $b_{j,k}$) is a function and $\beta_{j,k}$ denotes element $k$ in the subset of coefficients $\boldsymbol{\beta}_{j} \subseteq \boldsymbol{\beta}$ used to weigh the basis functions of $f_j$. Examples of smooth functions represented like this are visualized in Figure \ref{fig:smooth_overview}.

Generalized additive models correspond to a single parameter case under this broad definition, but also come with additional assumptions that drastically simplify estimation \parencite[e.g.,][]{hastie_generalized_1986,hastie_generalized_1990,wood_generalized_2017-2}. Specifically, each of the $N$ collected observations $y_i \in \mathbf{y}$ is assumed to be a realization of a corresponding response variable $Y_i \sim \mathcal{E}(\mu_i,\phi)$ which are all assumed to be independent from one another. Conventionally, $\mathcal{E}$ is taken to be a distribution belonging to the exponential family \parencite[e.g., Gaussian $\mathcal{N}$, Binomial $\mathcal{B}$, Gamma $\Gamma$, etc.;][]{hastie_generalized_1986,hastie_generalized_1990,wood_generalized_2017-2} and all response variables are assumed to share the same scale parameter $\phi$. Because of this independence assumption, computation of the log-likelihood $\mathcal{L}$ for a GAM simplifies to a sum over $N$ log-densities, so that $\mathcal{L} =\sum_i^N log(p_{Y_i}(y_i | \mu_i, \phi))$, where $p_{Y_i}$ denotes the probability or density function of $Y_i$ and hence of distribution $\mathcal{E}$.

In a GAM, the only parameter of the log-likelihood that is represented as an additive model -- and hence parameterized by the coefficients $\boldsymbol{\beta}$ -- is the mean $\mu_i$ of the response variables, or a known function $g(\mu_i)$ of it, so that

\begin{equation}\label{eq:gam}
Y_i  \sim \mathcal{E}(\mu_i,\phi)\ \text{where}\ g(\mu_i) = \mathbf{X}_i\boldsymbol{\beta}.
\end{equation}

\noindent
$\mathbf{X}_i$, denoting the $i$-th row of the \emph{model matrix} $\mathbf{X}$, is used in Equation (\ref{eq:gam}) to represent the specific model of the mean. What distinguishes GAMs from other linear models, and makes them attractive for modeling experimental data, is that the columns of $\mathbf{X}$ can be filled with basis functions as shown in Equation (\ref{eq:smooth_term}), to allow for a broad range of flexible models of $g(\mu_i)$. The function $g(\mu_i)$ can for example be modeled as a, potentially non-linear, function of covariate $x$: $g(\mu_i)= \alpha + f(x_i)$ with $\alpha$ representing an offset term\footnote{For this example model $\mathbf{X}_i = [1,b_1(x_i),b_2(x_i),...,b_k(x_i)]$ and $\boldsymbol{\beta} = [\alpha,\beta_1, \beta_2,...,\beta_k]^\top$.} -- necessary because the $f$ terms are usually subjected to sum-to-zero constraints \parencite[to ensure identifiability, see][for details]{wood_generalized_2017-2}. 

This additive structure can even be relaxed further, with a model like $g(\mu_i) = \alpha + f(x_{1,i},x_{2,i},...,,x_{j,i})$ allowing for two or more covariates to interact \emph{non-linearly} in the effect they have on the mean (see Figure \ref{fig:smooth_overview}e). Terms like $f(x_{1,i},x_{2,i},...,,x_{j,i})$ can be represented as \emph{tensor smooths} \parencite[][]{wood_lowrank_2006,wood_straightforward_2013}. The latter can conveniently and efficiently be constructed from the corresponding marginal smooths $f_1, f_2, ..., f_j$ \parencite[see][for details]{wood_lowrank_2006}. Another popular extension is to estimate separate versions of some or all of the $f$ per level of a categorical predictor variable, resulting in a model of the form $g(\mu_i) = \alpha_{s(i)} + f_{s(i)}(x_i)$ with $s(i)$ denoting the level of categorical predictor variable $s$ at observation $i$ \parencite[e.g.,][]{wood_generalized_2017-2}. This is particularly useful when modeling experimental time-series data, where it is often reasonable to believe that the signal of interest will not just generally vary smoothly over time but will do so differently under different experimental conditions. An example is provided in Figure \ref{fig:smooth_overview}b, which visualizes how the reaction times in a lexical decision task are modulated by word-likeness, but differently for words, pseudo-words (non-existing words following the rules of the language) and random letter strings.

Related to this last example is the type of mixed or multi-level model (e.g., GAMMs) discussed briefly in the Introduction. In such a model, smooth terms are utilized as classical \emph{random} effects, for example to capture level-specific deviations from some \emph{fixed} effect: $g(\mu_i) = \alpha + f(z_i) + \tilde{f}_{s(i)}(z_i)$. To enable this, terms like $\tilde{f}_{s}$ are not subjected to identifiability constraints, but set up in a way that the approximated functions shrink towards zero \parencite[i.e., in mgcv; see Appendix \ref{sec:AppendixSmoothCon} and also][]{marra_practical_2011}. The resulting terms are thus comparable to regular, albeit slightly more complex, random effects and, in this example, capture the potentially non-linear deviation from $f$ that is specific to each level of $s$ \parencite[see][for early discussions on the relationship between random effects and smooth terms]{kimeldorf_correspondence_1970}. More generally speaking, $\tilde{f}_s$ terms adjust $g(\boldsymbol{\mu})$ by a level-specific \emph{random} smooth function of $z$ so that it would also be possible to replace $f(z)$ with smooth function(s) of other covariate(s) (see Section \ref{sec:challenges_mixed} for more details). Figure \ref{fig:smooth_overview}c shows an example of such $\tilde{f}_s$ terms.

One limitation of GAMMs is that the variance of the response variables $Y_i$ is assumed to be described by a relatively simple function of the mean $\mu$ \parencite[e.g.,][]{faraway_extending_2016,wood_generalized_2017-2,rigby_generalized_2005}. Specifically, each member $\mathcal{E}$ of the exponential family assumes that the variance of the $Y_i$ is given by $V_{\mathcal{E}}(\mu_i)\phi$, where $V_{\mathcal{E}}$ corresponds to the family's ``variance function'', capturing the aforementioned relationship between the mean and variance of $Y_i$ \parencite[e.g.,][]{wood_generalized_2017-2}. For the Gaussian model, $V_{\mathcal{E}}(\mu_i)=1$ while the Gamma distribution for example assumes that $V_{\mathcal{E}}(\mu_i) = \mu_i^2$ -- a fact that is commonly exploited in models of reaction time data where it is quite reasonable to expect greater variability for longer RTs \parencite[][]{lo_transform_2015}. 

Often it will however be more reasonable to expect the variance in the response variables to also change as a function of some predictor variables. In a random dot motion task for example, decreasing the coherence of the moving dots is likely to result in longer RTs on average but also in greater variance in the responses. If this change in variance cannot be accounted for by a (variance) function of the mean, the obvious solution would be to represent the scale parameter $\phi$, or a known function of it, as an additive model of dot motion coherence and other predictors as well. All of the specific smooth setups discussed earlier, captured under the general definition provided in Equation (\ref{eq:smooth_term}), can readily be used for this purpose. The resulting model would be a member of the GAMMs of Location, Scale, and Shape \parencite[GAMMLSS;][]{rigby_generalized_2005} class, which includes models of the form

\begin{equation}\label{eq:gamlss}
\begin{split}
&Y_i \sim \mathcal{F}(\mu_i,\phi_,\ldots,\tau_i)\ \text{where}\ \ g_\mu(\mu_i) = \mathbf{X}^\mu_i\boldsymbol{\beta}_\mu,\\&g_\phi(\phi_i) = \mathbf{X}^\phi_i\boldsymbol{\beta}_\phi,~g_\tau(\tau_i) = \mathbf{X}^\tau_i\boldsymbol{\beta}_\tau....
\end{split}
\end{equation}

\noindent
Here, $\mathcal{F}$ has replaced $\mathcal{E}$ to indicate that this broader case also includes distributions outside of the exponential family, which might include additional parameters (e.g., $\tau$) or not have a scale parameter \parencite[e.g., a Tweedie distribution with variable power parameter or Multinomial models;][]{wood_generalized_2017,wood_smoothing_2016}. Additive models can now be specified for each of these additional parameters, so that each parameter comes with a specific model matrix $\mathbf{X}$ (distinguished by super-scripts in Equation (\ref{eq:gamlss})) and a specific link function $g$. Note, that model matrices associated with different parameters (e.g., $\mathbf{X}^\mu$ and $\mathbf{X}^\phi$) can, but do not have to, include different (functions) of continuous and categorical predictor variables so that different parameters (or functions of these, e.g., $g_\mu(\mu)$ and $g_\phi(\phi)$) can depend on different predictors. Additionally, different parameters generally have separate coefficient vectors (e.g., $\boldsymbol{\beta}_\mu$ and $\boldsymbol{\beta}_\phi$). In consequence, estimated functions of predictor variables generally differ between parameters, even if the parameters depend on the same (functions of) predictor variables. It is however also possible for some coefficients to be shared between different parameters \parencite[e.g.,][]{wood_smoothing_2016}. For example, while $g_\mu(\mu)$ and $g_\phi(\phi)$ would both depend on a function of coherence in the random dot motion example outlined above, it would be possible to estimate parameter-specific functions $f^\mu$ and $f^\phi$ or a single shared function $f$ of coherence. In addition, the model of $g_\mu(\mu)$ might include subject-specific random smooths of coherence, while the model of $g_\phi(\phi)$ might only include a random intercept per subject. Finally note, that Equation (\ref{eq:gamlss}) as a special case includes the definition of GAMMs provided in Equation (\ref{eq:gam}), when the model of the scale parameter includes a single intercept\footnote{As we will discuss, this will not generally result in equivalent estimates for $\phi$, since the latter is typically chosen to maximize a restricted version of the log-likelihood for GAMMs \parencite[e.g.,][]{wood_fast_2011,wood_generalized_2017-2}.}. 

The log-likelihood of a GAMMLSS remains a simple sum over independent log-densities, which is convenient but not actually required by the broad definition of a general smooth model. An example of a more general smooth model is the Cox proportional hazard model, which is a specific type of survival model \parencite[][]{cox_regression_1972}. The proportional hazard model is instructive, because the corresponding log-likelihood function is clearly ``non-standard'', in the sense that it does not correspond to a sum of independent log-densities as would be the case for GAMMs or GAMMLSS \parencite[see][]{klein_survival_2003}. Furthermore, survival models, like the proportional hazard model, are very useful models of reaction time data, collected for example from binary choice tasks \parencite[e.g.,][]{hendrix_word_2020}. Survival analyses typically focus on the so-called survival function $S(t)$ of time \parencite[e.g.,][]{klein_survival_2003}. In a typical binary choice experiment such as the lexical decision task, the latter refers to the probability of still providing a correct response after time-point $t$, when no response has been given so far and $t=0$ was the moment of stimulus onset \parencite[e.g.,][]{hendrix_word_2020}.

When including smooth functions in a survival model it becomes possible to investigate how $S(t)$ changes as a function of continuous and categorical predictor variables alike. For estimation this principally means that $S(t)$ needs to be parameterized by the coefficients $\boldsymbol{\beta}$. However, the Cox proportional hazard model instead parametrizes $h(t)$, which is usually called the hazard function and relates to the survival function $S(t) = exp({-H(t)})$ through the \emph{cumulative} hazard function $H(t) = \int_0^t h(u)du$ \parencite[see][]{klein_survival_2003}. Specifically, the additive model is specified for the log-hazard function given covariates $log(h(t|\mathbf{X}_i))=log(h_0(t))+ \mathbf{X}_i\boldsymbol{\beta}$, where $h_0(t)$ denotes the so-called ``baseline hazard''. Note that this model is rather restrictive, since it assumes that the effect of predictor variables remains stable throughout the time-course considered (e.g., the time spent to reach a lexical decision): only the ``baseline log-hazard'' $log(h_0(t))$ changes over time, with $\mathbf{X}_i\boldsymbol{\beta}$ acting as a constant additive correction factor \parencite[][]{klein_survival_2003}. In Section \ref{sec:DataExamples} we consider a more plausible model of the hazard function, which permits for time-varying effects of the covariates \parencite[see also][]{bender_generalized_2018}.

As mentioned, the proportional hazard model is nevertheless instructive, because of the corresponding log-likelihood function\footnote{This is actually a \emph{partial} log-likelihood which does however take on the role of a regular log-likelihood (i.e., $\mathcal{L}$) for estimation \parencite[see][for a discussion]{klein_survival_2003}.}, which \Textcite{wood_smoothing_2016} state as \parencite[see also][Section 8.3]{hastie_generalized_1990} 

\begin{equation}\label{eq:prpHazLLK}
\begin{split}
\mathcal{L}_h = \sum_l^{N_t} \left(\sum_{i:~t_i=\tilde{t}_l}\delta_i\mathbf{X}_i\boldsymbol{\beta} - log\left(\sum_{i:~t_i\leq\tilde{t}_l}exp({\mathbf{X}_i\boldsymbol{\beta}})\right)r_l\right).
\end{split}
\end{equation}

\noindent
Here, $\tilde{\mathbf{t}}$ contains the $N_t$ \emph{unique} observed response times, ordered so that $\tilde{t}_1 > ... >\tilde{t}_l ... > \tilde{t}_{N_t}$ and $\mathbf{t}$ contains the $N \geq N_t$ \emph{recorded} response times, again ordered\footnote{Essentially, the data has to be ordered so that RTs are non-decreasing. I.e., the first RT in $\mathbf{t}$ corresponds to a trial on which the longest RT was observed. The second element will either hold a RT from a trial on which the same RT was observed (since RTs do not have to be unique) or the RT from a trial on which the next shorter RT was observed \parencite[see][]{hastie_generalized_1990,wood_smoothing_2016}.} so that $t_1 \geq ... \geq t_i ... \geq t_{N}$, some of which might be set to a pre-specified upper time-limit in case no response was observed beforehand (i.e., the response was ``censored'') in which case $\delta_i$ will be set to zero so that $r_l = \sum_{i:~t_i = \tilde{t}_l}\delta_i$ corresponds to the number of trials on which a response was observed at (unique) time-point $\tilde{t}_l$ \parencite[see][]{hastie_generalized_1990,wood_smoothing_2016}. While the log-likelihood in Equation (\ref{eq:prpHazLLK}) is clearly different from the log-likelihood of a simple GAMM, it is nevertheless captured under the broad definition of general smooth models and can thus be estimated via the methods proposed by \Textcite{wood_smoothing_2016} and described in this paper.

The next sub-section outlines the main challenge faced when estimating any kind of smooth model (risk of overfitting), including GAMMs and GAMMLSS, and how this problem can be tackled efficiently.

\subsection{Estimating \& Regularizing Smooth Models}

An integral part of each smooth model are the $f$ terms. When expressed as a sum of $k$ weighted basis functions, as shown in Equation (\ref{eq:smooth_term}), it becomes evident that their complexity, and hence the complexity of the model to be estimated, depends on the number chosen for $k$: too few basis functions and the model might miss out on non-linear effects of covariates, too many and the risk of overfitting to the observed data will increase quickly \parencite[e.g.,][]{wood_generalized_2017-2,wood_inference_2020}. Fortunately, much research has been dedicated to the effective regularization of these terms, ensuring that unnecessary flexibility available to a smooth term $f$ can be suppressed efficiently by means of regularization parameters (conventionally denoted as $\boldsymbol{\lambda}$), if the data only supports fewer than the specified number of basis functions \parencite[see][for an overview of recent developments]{wood_inference_2020}. Here we follow the general approach taken by \Textcite{wood_smoothing_2016}, since this allows estimating the vector of regularization parameters $\boldsymbol{\lambda}$ together with the coefficients $\boldsymbol{\beta}$, while also quantifying uncertainty in these estimates, for any generic smooth model.

To enable efficient regularization, \Textcite{wood_smoothing_2016} suggested to estimate the coefficients $\boldsymbol{\beta}$ of a smooth model by maximizing the \emph{penalized} log-likelihood function

\begin{equation}\label{eq:pen_llk}
\mathcal{L}_{\boldsymbol{\lambda}} = \mathcal{L} - \frac{1}{2\phi}\sum_j \boldsymbol{\beta}^\top\mathbf{S}^j_{\lambda}\boldsymbol{\beta}.
\end{equation}

\noindent
That is, the log-likelihood function $\mathcal{L}$ of the model is \emph{penalized} by the $\sum_j \boldsymbol{\beta}^\top\mathbf{S}^j_{\lambda}\boldsymbol{\beta}$ sum, which is optionally scaled by $\phi$. This sum corresponds to the \emph{smoothness penalty} associated with the model \parencite[see also][]{kimeldorf_correspondence_1970,wahba_comparison_1985,eilers_flexible_1996,wood_generalized_2017-2}. The $\mathbf{S}^j_{\lambda}$ denote zero-embedded versions of the $k_j*k_j$ ``parameterized'' penalty matrix $\mathcal{S}^j_{\lambda}$ associated with a particular smooth term $f_j$, so that $\boldsymbol{\beta}^\top\mathbf{S}^j_{\lambda}\boldsymbol{\beta} = \boldsymbol{\beta}_j^\top\mathcal{S}^j_{\lambda}\boldsymbol{\beta}_j$. The penalty matrix of a specific smooth term $f_j$ is ``parameterized'' by a subset of regularization parameters $\boldsymbol{\lambda}_j\subset\boldsymbol{\lambda}$. For univariate smooth terms (i.e., of a single covariate) for example, $\mathcal{S}^j_{\lambda}$ is often parameterized by a single regularization parameter and set to $\lambda_j\mathcal{S}^j$, where $\mathcal{S}^j$ is a single $k_j*k_j$ penalty matrix of fixed structure. However, smooth terms can also be associated with more complex penalties \parencite[e.g., tensor smooths][]{wood_lowrank_2006}, so that in general $\mathcal{S}^j_{\lambda} = \sum_{l} \lambda_{lj}\mathcal{S}^{lj}$. Additionally, the same subset of regularization parameters might be used to penalize different smooth terms \parencite[see Appendix \ref{sec:AppendixSmoothCon} of this manuscript and][for a more detailed discussion]{wood_generalized_2017-2}. To avoid notational clutter we will simply use $\mathcal{S}^r$ ($\mathbf{S}^r$) to refer to the ``non-parameterized'' (embedded) penalty matrix associated with an individual parameter $\lambda_r \in \boldsymbol{\lambda}$.

From a regularization perspective, the structure of the $\mathcal{S}^j$ then needs to be chosen such that $\boldsymbol{\beta}_j^\top\mathcal{S}^j_{\lambda}\boldsymbol{\beta}_j$ captures the complexity \parencite[or ``roughness''][]{kimeldorf_correspondence_1970} of the resulting estimate for $f_j$ given a set of coefficients $\boldsymbol{\beta}_j$ \parencite[e.g.,][]{kimeldorf_correspondence_1970,wahba_comparison_1985,wood_smoothing_2016}. How the structure of the $\mathcal{S}^j$ needs to be chosen to achieve this will vary\footnote{Unifying approaches to the construction of univariate and tensor smooths have been discussed by \Textcite{wood_smoothing_2016} and \Textcite{wood_straightforward_2013} respectively \parencite[see also][]{wood_generalized_2017-2}. The approach for univariate smooths, and how ``random smooths'' follow from it, is reviewed in Appendix \ref{sec:AppendixSmoothCon}.} depending on the specific type of smooth and the choice of basis function used to parameterize $f_j$ as shown in Equation (\ref{eq:smooth_term}) \parencite[see for example][]{eilers_flexible_1996}. Once suitable $\mathcal{S}^j$ have been constructed however, the elements in $\boldsymbol{\lambda}$ can fulfill the same role for any smooth term: they act as weights, controlling how strongly the flexibility available to a specific $f_j$ is suppressed \parencite[see also Section 5.4.2 in][]{wood_generalized_2017-2}.

This can conveniently be formalized by introducing the additional distributional assumption that $\boldsymbol{\beta} \sim \mathcal{N}(\mathbf{0},\mathbf{S}_{\lambda}^{-}\phi)$, where $\mathbf{S}_{\lambda}^{-}$ denotes any generalized inverse of $\mathbf{S}_{\lambda} = \sum_j \mathbf{S}^j_{\lambda}$ \parencite[][]{kimeldorf_correspondence_1970,wahba_comparison_1985,wood_fast_2011,wood_smoothing_2016,wood_inference_2020}. Expressed like this, $\mathbf{S}_{\lambda}^{-}$ takes on the role of the ($\phi$-scaled) covariance matrix belonging to an (improper) Bayesian prior \parencite[e.g.,][]{kimeldorf_correspondence_1970,wood_inference_2020} which permits for complex functions in the model only if $\boldsymbol{\lambda}$ contains elements of small magnitude. Note, that the prior is ``improper'', necessitating the \emph{generalized} inverse, because the Kernel of $\mathbf{S}_{\lambda}$ will usually be non-trivial \parencite[e.g.,][]{wood_smoothing_2016,wood_generalized_2017-2}. For once, $\mathbf{S}_{\lambda}$ will usually contain a zero block on the diagonal reflecting any un-penalized ``parametric'' (e.g., linear or offset such as $\alpha$) terms present in the models of some or all parameters (e.g., $\mu$). Similarly, it will often be desirable to avoid penalizing the case of any $f_j$ taking on the form of a simple (e.g., linear or quadratic) function. For this to be possible formally, any constellation $\boldsymbol{\beta}_j$ that would result in a sufficiently simple function has to lie in the Kernel of $\mathcal{S}^j_{\lambda}$, and thus $\mathbf{S}_{\lambda}$ so that $\boldsymbol{\beta}_j^\top\mathcal{S}^j_{\lambda}\boldsymbol{\beta}_j=0$, independent of the value of the elements in $\boldsymbol{\lambda}_j$. Consequently, $\mathbf{S}_{\lambda}$ will \emph{generally} be a positive semi-definite matrix \parencite[e.g.,][]{wood_smoothing_2016,wood_generalized_2017-2}.

An immediate benefit of the Bayesian prior assumption outlined above is that identically and independently distributed (i.i.d) Gaussian random effects (e.g., random intercept or slope terms) can simply be treated as another type of smooth function, with a suitable identity matrix replacing $\mathcal{S}^j$, and are thus available at no additional theoretical cost in the general smooth model framework \parencite[see][]{kimeldorf_correspondence_1970}. Additionally, the prior assumption also provides a means to estimate $\boldsymbol{\lambda}$, and thus the necessary amount of regularization, \emph{together} with $\boldsymbol{\beta}$ \parencite[e.g.,][]{wood_smoothing_2016}. Specifically, consider the product of the likelihood function and said prior density $p(\mathbf{y},\boldsymbol{\beta}|\boldsymbol{\lambda}) =p(\mathbf{y}|\boldsymbol{\beta},\boldsymbol{\lambda}) p(\boldsymbol{\beta}|\boldsymbol{\lambda})$, where the optional scale parameter $\phi$ has been absorbed into $\boldsymbol{\lambda}$ and the notation for the likelihood function $exp(\mathcal{L}) = p(\mathbf{y}|\boldsymbol{\beta},\boldsymbol{\lambda}) $ has been updated accordingly \parencite[e.g.,][]{wood_fast_2011,wood_smoothing_2016}. The result $p(\mathbf{y},\boldsymbol{\beta}|\boldsymbol{\lambda})$ is called the ``joint likelihood'' of $\mathbf{y}$ and $\boldsymbol{\beta}$, for a given set of regularization parameters $\boldsymbol{\lambda}$ \parencite[e.g.,][]{wood_fast_2011}. Application of Bayes rule then yields the \emph{Bayesian marginal likelihood} $p(\mathbf{y}|\boldsymbol{\lambda}) = \int p(\mathbf{y},\boldsymbol{\beta}|\boldsymbol{\lambda})d\boldsymbol{\beta}$, which is often referred to as the ``restricted marginal likelihood'' (REML) instead \parencite[see][]{wahba_comparison_1985,wood_fast_2011,wood_smoothing_2016,wood_inference_2020}.

Note that the latter, which is no longer a function of $\boldsymbol{\beta}$, can be maximized to find an estimate for $\boldsymbol{\lambda}$, which includes any optional scale parameter $\phi$ of the model \parencite[e.g.,][]{wood_fast_2011,wood_smoothing_2016,wood_generalized_2017}. Unfortunately, an exact analytic solution to the integral required to compute $p(\mathbf{y}|\boldsymbol{\lambda})$ is only available for strictly additive models \parencite[i.e., $\mathcal{E} = \mathcal{N}$ and $g(\mu)=\mu$;][]{wood_fast_2011}. In any other case the solution can however be approximated by means of a Laplace approximation \parencite[e.g.,][]{wood_fast_2011,wood_core_2015,wood_smoothing_2016}, which results in the ``Laplace-approximate'' estimate of the Bayesian marginal likelihood/REML

\begin{equation}\label{eq:laplace_bml}
p_L(\mathbf{y}|\boldsymbol{\lambda}) = p(\mathbf{y},\hat{\boldsymbol{\beta}}|\boldsymbol{\lambda})\frac{\sqrt{2\pi}^{N_p}}{|\mathcal{H}|^{0.5}},
\end{equation}

\noindent
that can again be optimized for $\boldsymbol{\lambda}$ instead. Here, $\hat{\boldsymbol{\beta}}$ denotes the set of coefficients that maximize $p(\mathbf{y},\boldsymbol{\beta}|\boldsymbol{\lambda})$. When considering the density of the normal prior, denoted here by $p(\boldsymbol{\beta}|\boldsymbol{\lambda})$, it becomes evident that $exp({\mathcal{L}_{\lambda}}) \propto  p(\mathbf{y}|\boldsymbol{\beta},\boldsymbol{\lambda}) p(\boldsymbol{\beta}|\boldsymbol{\lambda})$, that is the penalized likelihood $exp({\mathcal{L}_{\lambda}})$ is proportional to the joint likelihood \parencite[e.g.,][]{wood_fast_2011}. As a consequence, $\hat{\boldsymbol{\beta}}$ in Equation (\ref{eq:laplace_bml}) is simply the estimate that maximizes $\mathcal{L}_{\lambda}$ -- the penalized log-likelihood of Equation (\ref{eq:pen_llk}) -- for any given $\boldsymbol{\lambda}$ \parencite[e.g.,][]{wood_generalized_2017-2}. Furthermore, $N_p$ here corresponds to the number of coefficients to be estimated for the model, and $|\mathcal{H}_{|\hat{\boldsymbol{\beta}}}|$ denotes the determinant of the negative Hessian of the penalized log-likelihood $\mathcal{H} = -\frac{\partial^2 \mathcal{L}_{\lambda}}{\partial \boldsymbol{\beta} \partial \boldsymbol{\beta}^\top}\Big\rvert_{\hat{\boldsymbol{\beta}}}$ \parencite[e.g.,][]{wood_fast_2011,wood_smoothing_2016}. In practice, the log of the Laplace-approximate REML,

\begin{equation}\label{eq:laplace_reml}
\mathcal{V}(\boldsymbol{\lambda}) = \mathcal{L}_\lambda(\hat{\boldsymbol{\beta}}) + \frac{1}{2}log|\mathbf{S}_\lambda/\phi|_+ - \frac{1}{2}log|\mathcal{H}| + c,
\end{equation}

\noindent
is optimized instead, where $c$ holds any terms present in Equation (\ref{eq:laplace_bml}) that are constant with respect to $\boldsymbol{\lambda}$ and can thus be omitted during estimation \parencite[e.g.,][]{wood_fast_2011,wood_smoothing_2016}. Hence, we will refer to $\mathcal{V}(\boldsymbol{\lambda})$ as the \emph{REML criterion} \parencite[i.e, the log of the restricted marginal likelihood;][]{wood_fast_2011}. Note, that $log|\mathbf{S}_{\lambda}/\phi|_+$ in Equation (\ref{eq:laplace_reml}) denotes the log of a generalized determinant in contrast to  $log|\mathcal{H}|$, which is a ``proper'' log-determinant. The generalization is again necessary to deal with the aforementioned non-trivial Kernel of $\mathbf{S}_{\lambda}$, resulting from the rank deficiency (of some) of the $\mathcal{S}^r$ \parencite[e.g.,][]{wood_fast_2011}.

Crucially, optimizing Equation (\ref{eq:laplace_reml}) for $\boldsymbol{\lambda}$ automatically yields the final coefficient estimate $\hat{\boldsymbol{\beta}}$ in addition to an estimate $\hat{\boldsymbol{\lambda}}$ at convergence. \Textcite{wood_smoothing_2016} propose to optimize Equation (\ref{eq:laplace_reml}) via Newton's method \parencite[see][for an introduction]{wood_core_2015}, which requires the first order $\frac{\partial \mathcal{V}}{\partial \lambda_r}$ and second order $\frac{\partial^2 \mathcal{V}}{\partial \lambda_r \partial \lambda_r}$ partial derivatives of Equation (\ref{eq:laplace_reml}) with respect to the elements $\lambda_l,\lambda_j \in \boldsymbol{\lambda}$. Since $\mathcal{L}_\lambda(\hat{\boldsymbol{\beta}})$ depends on $\lambda_r$ both directly, through the penalty, and indirectly, through the estimate for $\boldsymbol{\beta}$, evaluating $\frac{\partial \mathcal{V}}{\partial \lambda_r}$ would in principle require computation of $\frac{\partial \mathcal{L}_\lambda}{\partial \lambda_r} + \frac{\partial \mathcal{L}_\lambda}{\partial \boldsymbol{\beta}}\frac{d \boldsymbol{\beta}}{d \lambda_r}$ \parencite[e.g.,][]{wood_generalized_2017-2}. However, because $\frac{\partial \mathcal{L}_\lambda}{\partial \boldsymbol{\beta}}\Big\rvert_{\hat{\boldsymbol{\beta}}}=\mathbf{0}$ it follows from Equation \parencite[\ref{eq:laplace_reml}; see also][]{wood_smoothing_2016,wood_generalized_2017}, that

\begin{equation}\label{eq:laplace_reml_grad}
\frac{\partial \mathcal{V}}{\partial \lambda_r}=-\frac{\hat{\boldsymbol{\beta}}^\top\mathbf{S}^r\hat{\boldsymbol{\beta}}}{2\phi} + \frac{1}{2}\frac{\partial log|\mathbf{\mathbf{S}_\lambda/\phi}|_+}{\partial \lambda_r} - \frac{1}{2}\frac{\partial log|\mathcal{H}|}{\partial \lambda_r}.
\end{equation}

\noindent
Since $\mathcal{H} = \mathbf{H} + \mathbf{S}_\lambda/\phi$ \parencite[e.g.,][]{wood_smoothing_2016}, application of the standard result, stated for example in Appendix B of \Textcite{wood_fast_2011}, that $\frac{\partial log|\mathbf{B}|}{\partial a} = tr\left(\mathbf{B}^{-1}\frac{\partial\mathbf{B}}{\partial a}\right)$ yields $\frac{\partial log|\mathbf{S}_\lambda/\phi|_+}{\partial \lambda_r}=tr\left(\mathbf{S}^{-}_\lambda \frac{\partial \mathbf{S}_\lambda}{\partial \lambda_r}\right)=tr\left(\mathbf{S}^{-}_\lambda \mathbf{S}^r\right)$ and $\frac{\partial log|\mathcal{H}|}{\partial \lambda_r}=tr\left(\mathcal{H}^{-1}\left[\frac{\partial \mathbf{H}}{\partial \lambda_r} + \mathbf{S}^r\right]\right)$, where $\phi$ has been omitted in the final trace term, since it anyway cancels out for GAMMs (see Section \ref{sec:AM}). As pointed out by \Textcite{wood_smoothing_2016}, evaluation of $\frac{\partial \mathbf{H}}{\partial \lambda_r}$ for all $\lambda_r \in \boldsymbol{\lambda}$ by means of implicit differentiation would generally require the additional third-order derivatives of the log-likelihood $\mathcal{L}$ with respect to the coefficients. Similarly, evaluating $\frac{\partial^2 \mathcal{V}}{\partial \lambda_l \partial \lambda_j}$ requires $\frac{\partial^2 \mathbf{H}}{\partial \lambda_l \partial \lambda_l}$ for all $\lambda_l,\lambda_j \in \boldsymbol{\lambda}$ \parencite[see][or Appendix \ref{sec:AppendixRemlDeriv} in this paper]{wood_smoothing_2016}, which could again be evaluated by means of implicit differentiation, but this would require the additional fourth-order derivatives as well.

As mentioned in the Introduction, these higher-order derivatives can become quite complex and difficult to compute efficiently and, in some cases, might simply not be known \parencite[e.g.,][]{wood_generalized_2017}. Driven by this problem, \Textcite{wood_generalized_2017} presented the Extended Fellner Schall (EFS) update for $\boldsymbol{\lambda}$ that (approximately) optimizes the Laplace-approximate criterion defined in Equation (\ref{eq:laplace_reml}) and does not require these higher-order terms because it simply assumes that $\forall~ \lambda_r \in \boldsymbol{\lambda}~ \frac{\partial \mathbf{H}}{\partial \lambda_r}=\mathbf{0}$ holds in general. They motivate the update by considering that the sign of the derivative defined in Equation (\ref{eq:laplace_reml_grad}) (and thus the direction an update to $\lambda_r$ should take to increase $\mathcal{V}$) is determined by $\left[\frac{\partial log|\mathbf{S}_{\lambda}|_+}{\partial \lambda_r} - \frac{\partial log|\mathcal{H}|}{\partial \lambda_r} \right]\phi - \boldsymbol{\hat{\beta}}^\top\mathbf{S}^r\boldsymbol{\hat{\beta}}$. Based on this insight, and after substituting the results stated above for the derivatives of the (generalized) log-determinants, the authors arrive at

\begin{equation}\label{eq:efs}
    \Delta_{\lambda_r} = \left(\lambda_{r} \frac{tr\left(\mathbf{S}^{-}_\lambda \mathbf{S}^r\right) - tr\left(\mathcal{H}^{-1} \mathbf{S}^r\right)}{\boldsymbol{\hat{\beta}}^\top\mathbf{S}^r\boldsymbol{\hat{\beta}}} \phi \right) - \lambda_{r},
\end{equation}

\noindent
which is the additive\footnote{The original \emph{multiplicative} EFS update by \Textcite{wood_generalized_2017} is simply the term in parentheses of Equation (\ref{eq:efs}).} EFS update for a specific regularization parameter $\lambda_r$. For strictly additive models and some canonical GAMs \parencite[e.g.,][]{hastie_generalized_1986,hastie_generalized_1990,faraway_extending_2016}, iterating Equation (\ref{eq:efs}) will result in an estimate $\hat{\boldsymbol{\lambda}}$ that maximizes Equation (\ref{eq:laplace_reml}) when step-length control is performed for each update \parencite{wood_generalized_2017}. Section \ref{sec:AM} covers how this can be achieved in practice \parencite[see also][]{wood_generalized_2017,wood_generalized_2017-2}.

As mentioned, the update in Equation (\ref{eq:efs}) is particularly attractive for more generic smooth models, because computing it avoids higher-order derivatives of the log-likelihood. There are however some caveats. For once, the update will only be approximate for these models whenever $\mathbf{H}$ depends on at least one regularization parameter $\lambda_r\in\boldsymbol{\lambda}$ \parencite[e.g.,][]{wood_generalized_2017}. In that case, $\frac{\partial \mathbf{H}}{\partial \lambda_r} \neq \mathbf{0}$ and $tr\left(\mathcal{H}^{-1} \mathbf{S}^r\right)$ will not correspond to the exact result for the required derivative. As a consequence, the estimate $\hat{\boldsymbol{\lambda}}$ obtained by iterating Equation (\ref{eq:efs}) will not necessarily correspond to the set of regularization parameters maximizing Equation (\ref{eq:laplace_reml}) \parencite[e.g.,][]{wood_generalized_2017}. However, \Textcite{wood_generalized_2017} point to asymptotic results, which provide some justification for assuming that $\forall~ \lambda_r \in \boldsymbol{\lambda}~ \frac{\partial \mathbf{H}}{\partial \lambda_r}=\mathbf{0}$ holds in general, and in practice the update performs well. We revisit this issue and related challenges in Sections \ref{sec:GAM}, \ref{sec:GAMMLSS}, and  \ref{sec:Uncertainty}.

Additionally, as discussed in detail by \Textcite{wood_generalized_2017}, the update only results in valid updates for $\boldsymbol{\lambda}$ if $\mathbf{H}$ is at least positive semi-definite, which might not generally be the case for more generic models \parencite[e.g.,][]{wood_generalized_2017}. In those situations, $\mathbf{H}$ will either have to be modified, to ensure that the smallest eigenvalue is larger than or equal to zero, or replaced, for example with the negative expected Hessian of the log-likelihood $\mathbb{E}\{\mathbf{H}\}$ \parencite[e.g.,][]{wood_generalized_2017}. We revisit this issue in Sections \ref{sec:GAMMLSS} and \ref{sec:lqefs}.

Another advantage of the EFS update that applies to all models considered in this paper, is that the computations required by Equation (\ref{eq:efs}) can be set up to efficiently to exploit sparsity of $\mathbf{H}$ \parencite[e.g.,][]{wood_generalized_2017}. This becomes particularly important when wanting to estimate larger multi-level or mixed models, as discussed in the following section. 

\subsubsection{Challenges faced when Estimating Mixed Sparse Smooth Models}\label{sec:challenges_mixed}

We will rely on the special case of GAMMs to illustrate the specific problems that arise when estimating multi-level or mixed smooth models -- and how they can be alleviated in case the negative Hessian of the log-likelihood $\mathbf{H}$ has a sparsity structure that can be exploited (which, as discussed below, will generally be the case for multi-level GAMMs). For this specific example, we will work with the following model $g(\mu_i) = \alpha + f(x_i) + f_{s(i)}(z_i)$, which involves separate smooth terms $f_{s}$ of covariate $z$ per level of a factor variable $s$. $s$ can reflect an experimentally manipulated factor variable (e.g., difficulty conditions, control groups, etc.) but could also reflect a factor variable with levels that we would prefer to treat as random (e.g., individual subjects).
Note, that in the latter case $x$ and $z$ can, but do not have to, refer to different covariates. This covers the case of the $f_{s}$ terms taking on the role of level-specific random smooths (i.e., denoted as $\tilde{f}$ in the previous section). For example, in a model of RTs, $x$ might be a continuous measure of the word-likeness of stimuli and $z$ might reflect the time-course of an experiment (e.g., trials). As discussed in the previous section, $f_{s}$ (i.e., $\tilde{f}_{s}$) terms could then be added to adjust the mean of the RT variables by a subject-specific random smooth function of $z$ (e.g., trials). Similarly, if there is reason to believe that there is a shared trend in the mean RT over trials $z$, for example because the experiment is designed to show practice or habituation effects, $f$ could be used to estimate the shape of this shared trend when setting $x=z$. The $f_{s}$ (i.e., $\tilde{f}_{s}$) terms would then correspond to level-specific random difference curves of $x=z$, to be added to the model of the mean. 

Now, consider how the model matrix $\mathbf{X}$ parameterizing any of these model variants changes with every additional level of factor $s$. Assuming, without loss of generality, that the collected observations are sorted by levels of factor $s$, we can write $\mathbf{X}=$

$$
\begin{bmatrix}
\mathbf{1} & \mathbf{X}^f_{\mathbf{1}} & \mathbf{X}^{f_s}_{\mathbf{1}} & \mathbf{0} & \mathbf{0} & \ldots & \mathbf{0}\\
\vdots & \mathbf{X}^f_{\mathbf{2}} & \mathbf{0} & \mathbf{X}^{f_s}_{\mathbf{2}} & \mathbf{0} & \ldots & \mathbf{0}\\
\vdots & \mathbf{X}^f_{\mathbf{3}} & \mathbf{0} & \ddots & \ddots & \ddots & \mathbf{0}\\
\vdots & \vdots & \vdots & \ddots & \ddots & \ddots & \mathbf{0}\\
\mathbf{1} & \mathbf{X}^f_{\mathbf{N_s}} & \mathbf{0} &\ldots & \mathbf{0} & \mathbf{0} & \mathbf{X}^{f_s}_{\mathbf{N_s}}
\end{bmatrix}.
$$

\noindent
Bold sub-scripts here indicate to which level of factor variable $s$ a specific sub-block of $\mathbf{X}$ belongs. Super-scripts on the other hand indicate whether a subset of columns is associated with $f$ or $f_s$. For example, each column of $\mathbf{X}_{\mathbf{1}}^{f}$ corresponds to a basis function of $f$ evaluated for all values $x_{i:s(i)=1}$ -- i.e., for all values of covariate $x$ recorded at the first level of factor $s$. Note, that $f$ takes up only $k_f$ columns of $\mathbf{X}$: those containing row-blocks $\mathbf{X}^f_{\mathbf{1}},...,\mathbf{X}^f_{\mathbf{N_s}}$. In contrast, all of the $f_{s}$ together take up $N_s*k_{f_s}$ columns, where $N_s$ denotes the number of levels of $s$. In other words, the overall model matrix $\mathbf{X}$ grows vertically by $k_{f_s}$ columns for every level of factor variable $s$. This problem persists, although to a lesser extend, for other random effects (e.g., random intercepts and slopes) as well, because they will also result in vertical additions to $\mathbf{X}$.

Similarly, for every $f_s$ term a zero-embedded version of the corresponding penalty matrix would have to be stored. In principle, these can be combined into a single matrix, since the term-specific penalty matrices do not overlap, but the dimensions of the resulting matrix, and also any other zero-embedded penalties for remaining terms, would still grow by $k_{f_s}$ columns and rows for every level of factor variable $s$. Note that these storage requirements would remain the same, even if the $f_s$ would all share the same smoothness penalties (as would be the case if the $f_s$ were to be set up as random smooths; see Appendix \ref{sec:AppendixSmoothCon}).

Use of the Laplace approximation to derive the REML criterion in Equation (\ref{eq:laplace_reml}) requires that, for all but strictly additive Gaussian models, $N_p << N$ \parencite[e.g.,][]{wood_generalized_2015,wood_smoothing_2016} so that the resulting memory requirements in practice are not actually prohibitive. However, strictly additive models are popular for time-series modeling in particular \parencite[e.g.,][]{van_rij_analyzing_2019}. Additionally, the computations required during estimation of multi-level models as considered here start to take noticeably more time long before the $N_p << N$ assumption becomes questionable -- in particular those involving the negative Hessian $\mathbf{H}$ (i.e., Equations \ref{eq:laplace_reml}, \ref{eq:laplace_reml_grad}, and \ref{eq:efs}).

However, under the EFS update, and as long as any sparsity of $\mathbf{X}$ is reflected in sparsity of $\mathbf{H}$, utilizing sparse matrix storage combined with sparsity enhancing transformations \parencite[e.g.,][]{golub_matrix_2013,scott_algorithms_2023}, can substantially lower both the memory footprint and the time required to complete the aforementioned computations \parencite[e.g.,][]{wood_generalized_2017}. This contrasts with the Newton update to $\boldsymbol{\lambda}$ \parencite[e.g.,][]{wood_fast_2011,wood_smoothing_2016} which would generally involve computing by-products that are inherently dense. At best this would simply nullify any advantage gained by utilizing sparse matrices in the first place but because of the additional overhead associated with sparse matrix algorithms, this would more likely result in longer computation times compared to what could be achieved with dense matrices and algorithms optimized for them \parencite[e.g.,][]{wood_generalized_2017}.

For GAMMs, $\mathbf{H}$ will be equal to $\mathbf{X}^\top\mathbf{W}\mathbf{X}/\phi$ \parencite[e.g.,][]{wood_fast_2011}, where $\mathbf{W}$ is diagonal. $\mathbf{H}$ will thus generally be sparse for multi-level variants of these models and Sections \ref{sec:AM} and \ref{sec:GAM} review the necessary computations to exploit this sparsity, ensuring efficient estimation of $\boldsymbol{\lambda}$ and $\boldsymbol{\beta}$. Similar savings will be possible for many GAMMLSS models (see Section \ref{sec:GAMMLSS}). However, as discussed in the Introduction, providing a unified account for more generic models is complicated by the fact that $\mathbf{H}$ might not be sparse at all or have an unknown sparsity structure. Attempting to use sparse matrix algorithms when $\mathbf{H}$ is actually dense would again slow computations down unnecessarily due to the associated overhead. Similarly, relying on sparse matrix storage for $\mathbf{H}$ would not actually result in lower memory requirements, which could become problematic for generic multi-level models. Fortunately, the limited-memory quasi-Newton update to $\boldsymbol{\lambda}$ developed in Section \ref{sec:lqefs} remains memory efficient for these models as well and requires no prior knowledge about the structure of $\mathbf{H}$ \parencite[see][for an introduction to limited-memory quasi-Newton methods]{nocedal_numerical_2006}. For now, we return to the final concept to be reviewed in this Background section: the problem of selecting between different smooth models.

\subsection{Quantifying Uncertainty in $\hat{\boldsymbol{\lambda}}$ and Selecting Between Smooth Models}

In applied work involving smooth models, researchers are generally more interested in the effect of some experimental manipulations, with parameter estimation taking on a secondary role. A researcher might for example want to check whether a function of one or more covariates can be expected to be zero everywhere over the range of the covariates, signaling that the latter are unlikely to have any effect on the observed signal of interest. For a model like $g(\mu_i) = \alpha + f(z) + f(x_i)$ it could for example be of interest to test whether the smooth function of $x$ contributes strongly to the model fit or can be removed. Other times it might be of interest to evaluate whether there is greater support for an additive combination of separate smooth functions (e.g., $f(x) + f(t)$) or for their non-linear interaction (e.g., $f(t,x)$).

Answers to these questions are commonly sought via model comparisons of two nested models: one including the smooth term of interest and one excluding it. A popular criterion used to compare such models, which conveniently generalizes to all the models discussed in this paper \parencite[e.g.,][]{wood_smoothing_2016}, is the conditional version of the Akaike Information Criterion \parencite[cAIC;][]{akaike_information_1992,greven_behaviour_2010}. Intuitively, the cAIC taxes increases in log-likelihood based on the increase in model complexity necessary to achieve the better fit: a model that requires many additional parameters to slightly increase the log-likelihood, will thus be ranked less favorable compared to a much simpler model that achieves only a marginally worse fit \parencite[see][for an introduction]{burnham_model_2004}. Computation of the cAIC for generic smooth models proceeds as shown in Equation (\ref{eq:aic_intro}) \parencite[e.g.,][]{hastie_generalized_1990,wood_smoothing_2016}: 

\begin{equation}\label{eq:aic_intro}
cAIC = -2\mathcal{L}({\hat{\beta}}) + 2tr(\mathbf{V}\mathbf{H})
\end{equation}

\noindent
Here $\mathcal{L}({\hat{\beta}})$ again denotes the \emph{un-penalized} log-likelihood of the model, evaluated at the final coefficient estimate $\hat{\boldsymbol{\beta}}$ obtained from the routine used to optimize $\mathcal{V}$ for $\boldsymbol{\lambda}$ \parencite[e.g., via a Newton or the EFS update;][]{wood_fast_2011,wood_smoothing_2016,wood_generalized_2017}. The trace $\tau =tr(\mathbf{V}\mathbf{H})$ provides the ``effective degrees of freedom'' of the model \parencite[][]{wood_smoothing_2016} that acts as the required measure of model complexity for smooth models \parencite[see][for an interpretation in terms of coefficient shrinkage]{wood_generalized_2017-2}. In this trace term, $\mathbf{V}=\mathcal{H}^{-1}$ denotes the inverse of the negative Hessian of the penalized log-likelihood. Importantly, as discussed also by \Textcite{wood_smoothing_2016}, under the Laplace approximation used to obtain the result in Equation (\ref{eq:laplace_reml}), $\mathbf{V}$ takes on the role of the covariance matrix of the normal approximation\footnote{Note that, just like Equation (\ref{eq:laplace_reml}) itself, the posterior result shown in Equation (\ref{eq:laplace_reml}) is actually not approximate but exact for strictly additive models.} to the \emph{conditional Bayesian posterior} of $\boldsymbol{\beta}$ given $\boldsymbol{\lambda}$ (and $\mathbf{y}$), denoted as

\begin{equation}\label{eq:posteriorBeta}
\boldsymbol{\beta} | \boldsymbol{\lambda}, \mathbf{y} \sim \mathcal{N}(\hat{\boldsymbol{\beta}},\mathbf{V}).
\end{equation}

\noindent
This result highlights a problem with computing the cAIC as shown in Equation (\ref{eq:aic_intro}): use of $\mathbf{V}$, the covariance matrix of the (approximate) posterior for $\boldsymbol{\beta}$ \emph{conditional} on $\boldsymbol{\lambda}$ implies that the final estimate $\hat{\boldsymbol{\lambda}}$ is actually treated as a set of parameters known prior to estimation -- i.e., it is treated as being free of uncertainty \parencite[e.g.,][]{wood_smoothing_2016}. This is well known to result in bias towards more complex models\footnote{A conditional generalized likelihood test (GLRT) would be affected similarly, while marginal versions of both AIC and GLRT tend to be too conservative, favoring the simpler model too often \parencite[e.g.,][]{greven_behaviour_2010,wood_smoothing_2016,wood_generalized_2017-2}.} in nested comparisons, especially when these comparisons involve a random effect \parencite[e.g.,][]{greven_behaviour_2010,saefken_unifying_2014,wood_smoothing_2016}.

\Textcite{wood_smoothing_2016} proposed a simple two-step correction procedure, designed to eliminate the conditioning and thus the neglect of uncertainty in the estimate $\hat{\boldsymbol{\lambda}}$. Specifically, they devised two additive correction terms, $\mathbf{V}^J$ and $\mathbf{V}^L$, that transform the conditional covariance matrix $\mathbf{V}$ into the ``marginal'' covariance matrix $\tilde{\mathbf{V}}$, so that $\boldsymbol{\beta} | \mathbf{y} \sim \mathcal{N}(\hat{\boldsymbol{\beta}},\tilde{\mathbf{V}})$ holds approximately \parencite[see also][]{greven_comment_2016}. In simulation studies the performance of the ``uncertainty-corrected'' cAIC, computed as shown in Equation (\ref{eq:aic_intro}) but with $\tau'=r(\tilde{\mathbf{V}}\mathbf{H})$ as a measure of model complexity, was improved markedly.

While other corrections have been proposed to address this issue \parencite[e.g.,][]{greven_behaviour_2010,saefken_unifying_2014}, the two-step procedure by \Textcite{wood_smoothing_2016} again has the advantage that it applies to more general models as well, not just to GAMMs. In addition, access to an approximate result for the covariance matrix $\tilde{\mathbf{V}}$ of the normal approximation to the marginal posterior $\boldsymbol{\beta}|\mathbf{y}$ is useful, not just for the specific problem of selecting between models but for many other post-estimation tasks as well \parencite[e.g., to compute credible intervals or to sample from the normal approximation to the marginal posterior $\boldsymbol{\beta}|\mathbf{y}$;][]{wood_generalized_2017-2,greven_comment_2016}.

\begin{table*}[!t]
    \centering
    \caption{Section Overview}
    \renewcommand{\arraystretch}{1.25}
    \begin{tabular}{p{.075\textwidth}|p{.25\textwidth}|p{.37\textwidth}|p{.2\textwidth}}
         \toprule
         \centering \textbf{Section} & \centering \textbf{Topic} &  \centering \textbf{Important Concepts} &  \centering \textbf{MSSM Implementation} \tabularnewline
         \midrule
         \centering 3.1 & AMM estimation when $\mathbf{X}$ is sparse & Cholesky \& QR decomposition with sparsity-enhancing pivoting, efficient computation of $tr\left(\mathbf{S}^{-}_\lambda \mathbf{S}^r\right)$ and $tr\left(\mathcal{H}^{-1}\mathbf{S}^r\right)$ required for the EFS update, step-length control for $\hat{\boldsymbol{\lambda}}$ without the need to compute $log|\mathbf{S}_{\lambda}|_+$ & The \texttt{GAMM} class can be used to estimate massive additive mixed models\\
         \centering 3.2 & GAMM estimation when $\mathbf{X}$ is sparse & Fisher scoring \& the PQL/POI approach to GAMM estimation, discussion of the approximate nature of the EFS update for models for which $\frac{\partial \mathbf{W}}{\partial \lambda_l} \neq \mathbf{0}$ for some or all $\lambda_r\in\boldsymbol{\lambda}$ & The \texttt{GAMM} class also supports other members of the Exponential family beyond the Gaussian\\ [1ex]
         \centering 3.3 & General mixed smooth model estimation when $\mathbf{H}$ is sparse  & Using Newton's method to estimate $\boldsymbol{\beta}$, identifying miss-specified models with unidentifiable elements in $\boldsymbol{\beta}$, asymptotic justification for assuming that $\forall~\lambda_r \in \boldsymbol{\lambda}~\frac{\partial \mathbf{H}}{\partial \lambda_r} = \mathbf{0}$  & The \texttt{GAMMLSS} and \texttt{GSMM} classes can be used to efficiently estimate more general smooth models\\[0.5ex]
         \midrule
         \centering 4 & General (mixed) smooth model estimation when $\mathbf{H}$ is not sparse or not known  & Quasi-Newton estimation of $\boldsymbol{\beta}$, limited-memory representations for quasi-Newton approximations to $\mathbf{H}$ (and $\mathbf{V}$ and $\mathcal{H}$), how to use the latter as part of a limited-memory quasi EFS (L-qEFS) update & The \texttt{GSMM} class only requires code for the log-likelihood function to efficiently estimate a general smooth model\\[0.5ex]
         \midrule
         \centering 5 & Approximate estimation of uncertainty in $\hat{\boldsymbol{\lambda}}$ and how to correct for it & Approximate computation of $\mathbf{V}^\rho$ and $\tilde{\mathbf{V}}$ -- estimates of the covariance matrices for the normal approximations to the marginal posteriors $\boldsymbol{\rho}|\mathbf{y}$ and $\boldsymbol{\beta}|\mathbf{y}$ -- for models estimated via the EFS update, computing an uncertainty-corrected cAIC (i.e., $\tau'$) given EFS estimates of $\mathbf{V}^\rho$ and $\tilde{\mathbf{V}}$, alternative Monte Carlo (MC) estimation of $\tau'$  & The \texttt{compare} and \texttt{utils} modules contain functions to facilitate uncertainty correction, model selection, and posterior sampling for any general smooth model\\
         \bottomrule
    \end{tabular}
    \flushleft
    {\small
    $~$
    
    \textit{Note.} Table \ref{tab:sec_overview} provides an overview of the topics covered in Sections \ref{sec:Methods}-\ref{sec:Uncertainty} of this paper and lists key concepts addressed within them as well as the relevant classes and modules of the MSSM toolbox.
    }
    \label{tab:sec_overview}
\end{table*}

Unfortunately, computing the two correction terms $\mathbf{V}^J$ and $\mathbf{V}^L$ again requires computation of $\frac{\partial \mathcal{V}}{\partial \lambda_l \partial \lambda_j}$ for all $\lambda_l,\lambda_j \in \boldsymbol{\lambda}$ and thus generally access to third and fourth order derivatives of the log-likelihood \parencite[e.g.,][]{wood_smoothing_2016}. In situations where these derivatives are available and for models with a modest number of coefficients, it might be tempting to simply compute $\frac{\partial \mathcal{V}}{\partial \lambda_l \partial \lambda_j}$ for all $\lambda_l,\lambda_j \in \boldsymbol{\lambda}$ once, based on the final estimate $\hat{\boldsymbol{\lambda}}$ obtained efficiently by means of the aforementioned EFS update, which would again enable computation of the correction terms proposed by \Textcite{wood_smoothing_2016}. However, because the resulting estimate $\hat{\boldsymbol{\lambda}}$ does not necessarily maximize Equation (\ref{eq:laplace_reml}) for more generic smooth models, the asymptotic result justifying the correction by \Textcite{wood_smoothing_2016} is not guaranteed to apply when relying on the EFS update either. At least for more generic models the strategy by \Textcite{wood_smoothing_2016} can thus not be used directly to quantify uncertainty in $\hat{\boldsymbol{\lambda}}$, when the EFS update is used to estimate the latter \parencite[e.g.,][]{wood_inference_2020}. Apart from these theoretical issues, computing $\mathbf{V}^L$ also quickly becomes prohibitively expensive for large variants of the multi-level models of interest here (see also Wood et al., 2016). We revisit these issues in more detail, and discuss possible solutions, in Section \ref{sec:Uncertainty}.

This concludes our review of the topics of estimation, regularization, and selection for additive smooth models. In summary, optimizing the Bayesian marginal likelihood $p(\mathbf{y}|\boldsymbol{\lambda})$ theoretically provides a means to obtain estimates $\hat{\boldsymbol{\beta}}$ and $\hat{\boldsymbol{\lambda}}$. In practice, the log of the Laplace approximate Bayesian marginal likelihood $\mathcal{V}(\boldsymbol{\lambda})$ is optimized instead \parencite[i.e., the REML criterion;][]{wood_fast_2011,wood_smoothing_2016}. Using Newton's method for this task quickly becomes expensive for multi-level smooth models involving many random effects \parencite[e.g.,][]{wood_generalized_2017}. Instead the EFS update by \Textcite{wood_generalized_2017} can be utilized, which benefits from sparsity of the model matrices and negative Hessian of the log-likelihood $\mathbf{H}$ and thus remains efficient for these kind of models. However, under the EFS update it becomes more difficult to quantify uncertainty in the final estimate $\hat{\boldsymbol{\lambda}}$ \parencite[e.g.,][]{wood_inference_2020}. This is problematic, because failing to account for this source of uncertainty when selecting between models (for example via the cAIC) can result in a high rate of false positives \parencite[e.g.,][]{greven_behaviour_2010,wood_smoothing_2016}.

In the upcoming Sections \ref{sec:AM}, \ref{sec:GAM}, \ref{sec:GAMMLSS} we discuss in detail how the EFS update by \Textcite{wood_generalized_2017} can be used in combination with sparse matrix algorithms to (approximately) optimize $\mathcal{V}$ for AMMs, GAMMs, GAMMLSS and generic multi-level smooth models for which $\mathbf{H}$ is sparse, respectively. We also discuss how to efficiently optimize the penalized log-likelihood $\mathcal{L}_\lambda$ for these different models, when relying on sparse matrix algorithms. As stated previously, many of the individual algorithms and techniques reviewed in these sections have previously been discussed in the literature. Apart from the review, our main contribution is their combination into a unified framework applicable to such a broad range of models and the software implementation of this framework in the MSSM toolbox. At the end of each section we briefly state which classes and functions in MSSM implement the computations described in each section. For a more detailed overview we refer interested readers to the documentation publicly available at \url{https://jokra1.github.io/mssm}. 

Efficient estimation of the aforementioned models is in large parts contingent on the negative Hessian $\mathbf{H}$ being sparse. If this is not the case, estimation of multi-level models will again slow down and quickly require more memory as the number of random effects increases -- even when relying on the EFS update by \Textcite{wood_generalized_2017}. To address this problem, we developed a novel algorithm to approximately optimize $\mathcal{V}$, based on limited-memory quasi-Newton methods \parencite[Section \ref{sec:lqefs}, see also][]{byrd_representations_1994,nocedal_updating_1980,nocedal_numerical_2006}. This algorithm, which we refer to as the L-qEFS update, is a modification of the EFS update by \Textcite{wood_generalized_2017} that relies exclusively on the Gradient of the log-likelihood to compute an update to $\boldsymbol{\lambda}$. Because the L-qEFS update does not require the Hessian matrix, estimation and automatic regularization of any (mixed) general smooth model remains memory-efficient, independent of the sparsity structure of $\mathbf{H}$.

Combined, these sections achieve two of the three goals, outlined initially in the Introduction, for which this section provided the necessary context: efficient estimation and automatic regularization of generic (multi-level) smooth models in the absence of higher-order derivatives of the log-likelihood. Section \ref{sec:Uncertainty} addresses the remaining goal: enabling (approximate) model selection for generic smooth models estimated using the methods outlined in this paper. To this end, the section describes multiple strategies, based on the methods proposed by \Textcite{wood_smoothing_2016} and \Textcite{greven_comment_2016}, to quantify the uncertainty in the final estimate $\hat{\boldsymbol{\lambda}}$ and shows how the results can be used to approximately correct the cAIC.

Table \ref{tab:sec_overview} provides an overview of the topics covered and includes a list of important concepts to be reviewed or utilized in these sections.

\section{Theoretical Framework}\label{sec:Methods}

This section outlines the core theoretical framework implemented in MSSM, enabling efficient estimation and automatic regularization for generic multi-level smooth models in the absence of higher-order derivatives of the log-likelihood, under the condition that the negative Hessian of the log-likelihood $\mathbf{H}$ is sparse (Section \ref{sec:lqefs} extends the framework to models for which this will not be the case). As mentioned in the last section, we will first review estimation of strictly additive Gaussian models (i.e., assuming that $Y_i \sim \mathcal{N}(\mu_i,\sigma)$). Fortunately, many of the equations introduced along the way and most of the strategies used can readily be extended to accommodate more general models later.

\subsection{Estimating Mixed Sparse Additive Models}\label{sec:AM}

As discussed already briefly in the Background section, GAMM estimation mainly focuses on obtaining two quantities: an estimate for the vector of coefficients, $\boldsymbol{\beta}$, and an estimate for the vector of regularization parameters, $\boldsymbol{\lambda}$. Since the REML criterion optimized for the latter (i.e., Equation (\ref{eq:laplace_reml})) anyway requires $\hat{\boldsymbol{\beta}}$, we first consider the issue of computing this estimate \emph{given} $\boldsymbol{\lambda}$. As mentioned in the Background section, $\hat{\boldsymbol{\beta}}$ can be obtained by simply maximizing (\ref{eq:pen_llk}), which for an additive model only requires solving the penalized ``normal equations'' $\left(\mathbf{X}^\top\mathbf{X} + \mathbf{S}_{\lambda}\right)\boldsymbol{\beta}=\mathbf{X}^\top\mathbf{y}$ for $\boldsymbol{\beta}$ \parencite[e.g.,][]{wood_generalized_2017-2,golub_matrix_2013}.

For sparse models, as discussed by \Textcite{wood_generalized_2017}, this can be achieved most efficiently by means of a pivoted Cholesky decomposition, formed so that $\mathbf{L}\mathbf{L}^\top = \mathbf{P}\mathbf{X}^\top\mathbf{X} + \mathbf{S}_{\lambda}\mathbf{P^\top}$. $\mathbf{P}$ here is a sparsity-enhancing transformation matrix, designed to reorder the rows and columns of $\mathbf{X}^\top\mathbf{X} + \mathbf{S}_{\lambda}$ so that $\mathbf{L}$ is less dense than the Cholesky obtained from decomposing $\mathbf{X}^\top\mathbf{X} + \mathbf{S}_{\lambda}$ directly \parencite[e.g.,][]{golub_matrix_2013,wood_generalized_2017,scott_algorithms_2023}. This approach is highly efficient: solving $\mathbf{L}\mathbf{v} = \mathbf{P}\mathbf{X}^\top\mathbf{y}$ for $\mathbf{v}$ by means of forward substitution followed by solving $\mathbf{L}^\top\hat{\boldsymbol{\beta}}_P = \mathbf{v}$ for $\hat{\boldsymbol{\beta}}_P$ yields the estimated coefficients $\boldsymbol{\hat{\beta}} = \mathbf{P}^\top\hat{\boldsymbol{\beta}}_P$ in mere seconds, even if $\boldsymbol{\hat{\beta}}$ contains tens of thousands of elements \parencite[e.g.,][]{golub_matrix_2013,wood_generalized_2017}.

This efficiency comes at a cost, however. Specifically, $\mathbf{P}$ is obtained in a pre-processing step and not at all designed to handle potential rank deficiencies of $\mathbf{X}^\top\mathbf{X} + \mathbf{S}_{\lambda}$. This distinguishes it from pivoting strategies that are employed to stabilize the computation of the decomposition \parencite[e.g.,][]{golub_matrix_2013,scott_algorithms_2023}. \Textcite{wood_generalized_2017-1} for example employed stability-oriented pivoting in their estimation algorithm for GAMMs of many observations (i.e., $N$ is very large and $N_p << N$). Relying on stability-oriented pivoting comes with the additional advantage that persistent rank deficiencies of $\mathbf{X}^\top\mathbf{X} + \mathbf{S}_{\lambda}$ can not only be detected but also accommodated gracefully during model fitting \parencite[see][for details]{wood_fast_2011,wood_smoothing_2016,wood_generalized_2017-2}. Furthermore, relying on stability-oriented pivoting generally leads to greater numerical stability, a benefit which is lost when $\mathbf{P}$ is chosen in a pre-processing step to optimize for sparsity of $\mathbf{L}$ \parencite[see][for a more detailed discussion]{golub_matrix_2013}.

While the sparsity-oriented pivoting strategy by default does not permit graceful accommodation of rank-deficiencies during model fitting, the decomposition will be stable even if no stability-oriented pivoting is employed if $\mathbf{X}^\top\mathbf{X} + \mathbf{S}_{\lambda}$ is far from singular \parencite[][]{golub_matrix_2013,hansen_detection_1987}. Additionally, the Cholesky will still fail if the decomposable matrix (e.g., $\mathbf{X}^\top\mathbf{X} + \mathbf{S}_{\lambda}$) is clearly singular, which will at least indicate to the end-user of estimation software that something about their model is wrong. Usually, these problems arise when some smooth terms in a given model remain unidentifiable despite absorbing identifiability constraints and applying regularization\footnote{This could be addressed either by dropping the problematic terms from the model or by placing additional penalties on the Kernel of smooth terms \parencite[see][for an efficient strategy]{marra_practical_2011}.}. The greater risk is then, that the decomposable matrix ends up being \emph{nearly singular}, in which case the decomposition might not fail but result in an estimated $\hat{\boldsymbol{\beta}}$ that is far off from the more exact estimate recovered under stability-oriented pivoting \parencite[][]{golub_matrix_2013,hansen_detection_1987}.

Fortunately, efficient algorithms are available \parencite[e.g.,][]{cline_estimate_1979} to compute an estimate of the ``condition number'' for $\mathbf{X}^\top\mathbf{X} + \mathbf{S}_{\lambda}$ from $\mathbf{L}$, which can be used to check whether, at convergence, $\mathbf{X}^\top\mathbf{X} + \mathbf{S}_{\lambda}$ was close to being singular \parencite[see][for discussions]{golub_matrix_2013,hansen_detection_1987}. In that case, or generally if efficiency is not of utmost importance, it will be prudent to instead rely on an alternative strategy to estimate $\hat{\boldsymbol{\beta}}$ and $\mathbf{L}$. Specifically, we can follow approximately the approach proposed by \Textcite{wood_fast_2011} and instead form a sparsity-preserving QR-decomposition $\begin{bmatrix}
\mathbf{X}\\
\mathbf{E}_\lambda^\top\\
\end{bmatrix}\mathbf{P}^\top=\mathbf{Q}\mathbf{R}$ in a first step, where $\mathbf{E}_{\lambda}\mathbf{E}_{\lambda}^\top=\mathbf{S}_\lambda$. Note, that $\mathbf{R}^\top\mathbf{R}=\mathbf{P}\mathbf{X}^\top\mathbf{X} + \mathbf{S}_{\lambda}\mathbf{P}^\top$ so that $\mathbf{R}^\top$ can be used as a replacement for $\mathbf{L}$, which does not require forming $\mathbf{X}^\top\mathbf{X} + \mathbf{S}_{\lambda}$.

Because the columns of $\begin{bmatrix}
\mathbf{X}\\
\mathbf{E}_\lambda^\top\\
\end{bmatrix}$ are again pivoted to maximize sparsity of $\mathbf{R}$, and $\mathbf{Q}$ does not have to be formed explicitly \parencite[e.g.,][]{wood_generalized_2017-2}, the QR strategy still remains quite memory-efficient. In addition, since the QR approach does not require formation of $\mathbf{X}^\top\mathbf{X} + \mathbf{S}_{\lambda}$, the estimate for $\hat{\boldsymbol{\beta}}$ will generally be more stable, especially in situations where decomposition of $\mathbf{X}^\top\mathbf{X} + \mathbf{S}_{\lambda}$ via a Cholesky would be ill-advised \parencite[e.g.,]{golub_matrix_2013}. Naturally, this improved stability comes at the cost of the greater complexity of the QR-decomposition. However, for all but very large multi-level models, the cost will generally still be justifiable - especially when considering that under the QR approach Heath's method \parencite[][]{heath_extensions_1982} can be used to get an indication of any terms that are clearly not identifiable. This again allows for the option to gracefully accommodate the resulting rank deficiency by excluding these terms from further fitting iterations \parencite[e.g.,][]{wood_fast_2011,wood_generalized_2017-2}. While Heath's method can result in less accurate estimates of rank deficiency compared to the strategies proposed for example by \Textcite{wood_fast_2011} and \Textcite{wood_smoothing_2016}, in practice it generally performs well \parencite[see][]{davis_algorithm_2011}.

Given $\hat{\boldsymbol{\beta}}$, the EFS update (see Equation (\ref{eq:efs})) can be computed to update $\boldsymbol{\lambda}$ \parencite[e.g.,][]{wood_generalized_2017}. This requires, $\mathbf{H}$, which for a Gaussian additive model is equal to $\mathbf{X}^\top\mathbf{X}/\phi$. $\phi$ (i.e., $\sigma^2$ for the Normal model), which is often not known, is also required for the EFS update itself. However, as mentioned briefly in the Background section, it can readily be replaced with the estimate $\hat{\phi}$ based on the REML criterion (for conciseness the calculations are omitted here, they can be found for example in the paper by \Textcite{wood_generalized_2017}). At this point, all that remains necessary is to efficiently compute the two required traces $tr\left(\mathbf{S}^{-}_\lambda \mathbf{S}^r\right)$ and $tr\left(\left[\mathbf{X}^\top\mathbf{X}/\phi + \mathbf{S}_{\lambda}/\phi\right]^{-1}\mathbf{S}^r/\phi\right)=tr\left(\left[\mathbf{X}^\top\mathbf{X} + \mathbf{S}_{\lambda}\right]^{-1}\mathbf{S}^r\right)$, where, as mentioned in the Background section, $\phi$ cancels in the final trace. The next two sub-sections discuss how this can be achieved efficiently.

\subsubsection{Computing $tr\left(\mathbf{S}^{-}_\lambda \mathbf{S}^r\right)$}

Computing $tr\left(\mathbf{S}^{-}_\lambda \mathbf{S}^r\right)$ relies on the aforementioned generalized inverse $\mathbf{S}^{-}_\lambda$ of the overall penalty matrix $\mathbf{S}_\lambda$. However, as pointed out previously, actually computing $\mathbf{S}^{-}_\lambda $ or even the trace term itself is often not necessary \parencite[e.g.,][]{wood_fast_2011,wood_generalized_2017-1,wood_generalized_2017-2,wood_generalized_2017}. To see this, note that for smooth terms $f_j$ with a single penalty, 

$$tr(\mathbf{S}_{\boldsymbol{\lambda}}^{-}\mathbf{S}^r)=tr([\mathcal{S}^r\lambda_r]^{-}\mathcal{S}^r)=tr([\mathcal{S}^r]^-\mathcal{S}^r)/\lambda_r$$

\noindent
which, if $\mathcal{S}^r$ would be of (full) rank $k_j$, would further simplify to $tr([\mathcal{S}^r]^{-1}\mathcal{S}^r)/\lambda_r=k_j/\lambda_r$. \Textcite{wood_generalized_2017-2} shows that, for single-penalty terms, it is always possible, using for example the Demmler \& Reinsch \parencite{demmler_oscillation_1975} transformation reviewed in Appendix \ref{sec:AppendixSmoothCon}, to transform the rank-deficient matrix $\mathcal{S}^r$ into a full-rank matrix of reduced dimensions $rank(\mathcal{S}^r)*rank(\mathcal{S}^r)$, to which the aforementioned simplified computations could be applied. Therefore, for single-penalty terms, but also for terms with multiple penalties that can be separated into individual full-rank sub-blocks $tr\left(\mathbf{S}^{-}_\lambda \mathbf{S}^r\right) = rank(\mathcal{S}^r)/\lambda_r$ \parencite{wood_fast_2011}. In addition, a similar re-parameterization can be applied to terms with multiple penalties, again allowing for the pseudo-inverse to be replaced by a more efficiently computable inverse \parencite[i.e., via a Cholesky decomposition as discussed in the next sub-section; see also][]{wood_fast_2011,wood_generalized_2017-1,wood_generalized_2017-2}. 

\subsubsection{Computing $tr\left(\left[\mathbf{X}^\top\mathbf{X} + \mathbf{S}_{\lambda}\right]^{-1}\mathbf{S}^r\right)$}

The final trace term, $tr\left(\left[\mathbf{X}^\top\mathbf{X} + \mathbf{S}_{\lambda}\right]^{-1}\mathbf{S}^r\right)$, is the most difficult term to compute efficiently, because of the inverse in the trace operator \parencite[]{wood_generalized_2017}. In principle, because $\left[\mathbf{X}^\top\mathbf{X} + \mathbf{S}_{\lambda}\right]^{-1}=\mathbf{P}^\top\mathbf{L}^{-\top}\mathbf{L}^{-1}\mathbf{P}$, the inverse is readily available. However, the resulting product would no longer benefit from the enhanced sparsity of $\mathbf{L}$ \parencite[e.g.,][]{wood_generalized_2017}. Instead, \Textcite{wood_generalized_2017} thus suggest to first compute $\mathbf{B}^r = \mathbf{L}^{-1}\mathbf{P}\mathbf{D}^r$, where $\mathbf{D}^r[\mathbf{D}^r]^\top = \mathbf{S}_r$. Since $tr(\mathbf{P}^\top\mathbf{L}^{-\top}\mathbf{L}^{-1}\mathbf{P}\mathbf{D}^r[\mathbf{D}^r]^\top)=$

$$tr([\mathbf{D}^r]^\top\mathbf{P}^\top\mathbf{L}^{-\top}\mathbf{L}^{-1}\mathbf{P}\mathbf{D}^r)=tr([\mathbf{B}^r]^\top\mathbf{B}^r),$$

\noindent
due to the cyclic property of the trace operator, summing up the squared (non-zero) elements of $\mathbf{B}^r$ then produces the same result obtained when computing the required trace $tr\left(\left[\mathbf{X}^\top\mathbf{X} + \mathbf{S}_{\lambda}\right]^{-1}\mathbf{S}^r\right)$ directly, but benefits far more from sparsity of $\mathbf{L}$ \parencite[e.g.,][]{wood_generalized_2017}.

In principle, obtaining $\mathbf{B}^r$ is straightforward after direct computation of $\mathbf{L}^{-1}$, which would require solving $\mathbf{L}\mathbf{L}^{-1} = \mathbf{I}$ and hence obtaining solutions to $N_p$ (the number of elements in $\boldsymbol{\beta}$) linear systems by means of forward substitution\footnote{If multiple CPU cores are available, the time it takes to form $\mathbf{L}^{-1}$ can be reduced drastically by solving multiple of these systems in parallel}. This approach was taken for example by \Textcite{wood_generalized_2017-1}. For models including only single-penalty smooth terms it would be slightly more efficient\footnote{Because $\mathbf{D}^r$ has only $rank(\mathcal{S}^r)$ non-zero columns. Thus, only $rank(\mathcal{S}^r)$ systems would need solving to obtain $\mathbf{B}^r$ from $\mathbf{L}\mathbf{B}^r = \mathbf{P}\mathbf{D}^r$.} to simply solve $\mathbf{L}\mathbf{B}^r = \mathbf{P}\mathbf{D}^r$ for $\mathbf{B}^r$. Naturally, the advantage disappears if all the $\mathcal{S}^r$ are of full rank. Similarly, this second strategy would actually require finding \emph{more} solutions, in case the model includes terms for which $\mathcal{S}^j_{\lambda}$ represents a sum over multiple penalty matrices. The first strategy, to form $\mathbf{L}^{-1}$ explicitly, will thus likely be faster in most situations. However, because already $\mathbf{L}^{-1}$ will often be less sparse than $\mathbf{L}$, the second strategy might still be preferred in case the available memory is limited or for models for which $N_p$ is very large.

Once the two trace terms have been evaluated for every regularization parameter $\lambda_r$, the EFS update shown in Equation (\ref{eq:efs}) can be used to compute the updates $\Delta_{\lambda_r}$ for every $\lambda_r \in\boldsymbol{\lambda}$, followed by setting $\boldsymbol{\lambda} = \boldsymbol{\lambda} + \boldsymbol{\Delta}_\lambda$, where the latter term denotes the vector holding all updates. Since $\mathbf{H}$ is generally independent of $\boldsymbol{\lambda}$ for Gaussian additive models as discussed in this section, the EFS update is exact and will eventually produce an estimate for $\boldsymbol{\lambda}$ that maximizes the REML criterion $\mathcal{V}$ \parencite[]{wood_generalized_2017}. However, as would also be the case for the Newton update, care needs to be taken to ensure that $\mathcal{V}(\boldsymbol{\lambda} + \boldsymbol{\Delta}_{\lambda}) > \mathcal{V}(\boldsymbol{\lambda})$, i.e., that the criterion in Equation (\ref{eq:laplace_reml}) is actually increased by the EFS update applied to $\boldsymbol{\lambda}$ \parencite[e.g.,][]{wood_core_2015,wood_generalized_2017-2,wood_generalized_2017}. \Textcite{wood_generalized_2017} recommend to control the step length by simply halving each element in $\boldsymbol{\Delta}_\lambda$ until the aforementioned inequality is achieved. However, in the context of the EFS update it is more efficient to rely on the step-length control strategy discussed by \Textcite{wood_generalized_2017-1}, which is based on terms anyway required to evaluate Equation (\ref{eq:efs}).

The strategy by \Textcite{wood_generalized_2017-1} relies exclusively on Equation (\ref{eq:laplace_reml_grad}), the partial derivatives of $\mathcal{V}$ with respect to the regularization parameters, to determine whether $\boldsymbol{\Delta}_\lambda$ needs to be corrected. Specifically, they point out that as long as convergence of the entire estimation routine \parencite[e.g., change in $\mathcal{L}_\lambda$ or alternatively the \emph{penalized deviance};][]{wood_generalized_2017-1} is consistently monitored at every iteration, \emph{before} applying $\boldsymbol{\Delta}_\lambda$, a correction would be necessary only if $\boldsymbol{\nabla}^\top_{\boldsymbol{\lambda} + \boldsymbol{\Delta}_\lambda}\boldsymbol{\Delta}_\lambda < 0$, where $\boldsymbol{\nabla}^\top_{\boldsymbol{\lambda} + \boldsymbol{\Delta}_\lambda}$ denotes the Gradient of $\mathcal{V}$ evaluated at $\boldsymbol{\lambda} + \boldsymbol{\Delta}_\lambda$ (i.e., the vector containing the partial derivatives obtained from Equation (\ref{eq:laplace_reml_grad})). Intuitively, this dot product will be negative if there is a decisive mismatch between the sign (direction) of the REML Gradient and the sign of the EFS update across lambda terms, which would imply that applying the current update would go past the optimum of $\mathcal{V}$ \parencite[e.g.,][]{wood_core_2015}. 

Importantly, the terms involved in computing $\boldsymbol{\nabla}_{\boldsymbol{\lambda} + \boldsymbol{\Delta}_\lambda}$ are exactly what is required to compute the next EFS update (i.e., starting from $\boldsymbol{\lambda} + \boldsymbol{\Delta}_\lambda$). Therefore, as long as no correction to $\boldsymbol{\Delta}_\lambda$ is required, this strategy adds minimal computational overhead. Conversely, for every attempted correction to $\boldsymbol{\Delta}_\lambda$ all terms required for Equations (\ref{eq:laplace_reml_grad}) and (\ref{eq:efs}) will have to be re-computed to test whether the halving was sufficient \parencite[e.g.,][]{wood_generalized_2017-1}. If repeated corrections were necessary during fitting, estimation could thus be prolonged quite drastically for more complex models. Fortunately, as shown by \citeauthor{wood_generalized_2017} (see Theorem 3, omitted here; \citeyear{wood_generalized_2017}), the EFS updates do not exceed the step-size achieved under a Newton proposal and are thus, in general, also not more likely to overshoot the $\mathcal{V}$ optimum than alternative methods.

\begin{algorithm}
\caption{\textit{Fitting algorithm for sparse AMMs}}\label{alg:sparse_gamm}
\begin{algorithmic}[1]
\INPUT Vector of observations $\mathbf{y}$, model matrix $\mathbf{X}$, individual penalty matrices $\mathbf{S}^r$ where $r \in \{1,2,...,N_\lambda\}$
\INIT $\boldsymbol{\lambda} \gets \boldsymbol{1}$, $\mathbf{S}_{\lambda} = \sum_{r=1,\ldots,N_\lambda} \lambda_r\mathbf{S}^r$
\State Obtain $\boldsymbol{\hat{\beta}}$ by solving the penalized normal equations. This provides $\mathbf{P}$ and $\mathbf{L}$
\State If necessary compute $\hat{\phi}$, compute derivatives $\frac{\partial log|\mathbf{S}_{\lambda}|_+}{\partial \lambda_r}$ and $\frac{\partial log|\mathbf{X}^\top\mathbf{X} + \mathbf{S}_{\lambda}|}{\partial \lambda_r}$ for every $\lambda_r \in \boldsymbol{\lambda}$
\State Compute first $\boldsymbol{\Delta}_{\boldsymbol{\lambda}}$ as shown in Equation (\ref{eq:efs})
\For{$i = 1, 2, ...$}
    \If{i > 1}
        \State Exit the loop if converged
    \EndIf
    \State Conditionally apply the proposed $\boldsymbol{\Delta}_{\boldsymbol{\lambda}}$ to $\boldsymbol{\lambda}$
    \State Re-form $\mathbf{S}_{\lambda}$ and re-compute steps 1 to 2
    \State Compute Gradient of REML criterion $\boldsymbol{\nabla}_{\boldsymbol{\lambda} + \boldsymbol{\Delta}_\lambda}$\linebreak\hspace*{.9em} as shown in Equation (\ref{eq:laplace_reml_grad}) under the conditional\linebreak\hspace*{1.2em} smoothing penalties
    \If{$\boldsymbol{\nabla}_{\boldsymbol{\lambda} + \boldsymbol{\Delta}_{\boldsymbol{\lambda}}}^\top \boldsymbol{\Delta}_{\boldsymbol{\lambda}} < 0$}
        \State Undo the update applied to $\boldsymbol{\lambda}$ and divide $\boldsymbol{\Delta}_{\boldsymbol{\lambda}}$ by 2
        \State Repeat steps 8-11
    \EndIf
    \State Compute new $\boldsymbol{\Delta}_{\boldsymbol{\lambda}}$ as shown in Equation (\ref{eq:efs})
\EndFor
\end{algorithmic}
\end{algorithm}

The steps discussed in this section enable efficient estimation of mixed sparse additive models and are summarized in Algorithm (\ref{alg:sparse_gamm}). Subsequent sections will generalize this basic algorithm to ensure that it handles the estimation of more general smooth models as well, starting with the estimation of GAMMS in the next section. Before considering these extensions, we briefly describe how the steps in this section are implemented in MSSM.

\subsubsection{Implementation in MSSM}

The \texttt{GAMM} class in MSSM can be used to estimate additive models as discussed in this section. By default MSSM relies on the sparse QR decomposition approach to estimate $\boldsymbol{\beta}$, but the sparse Cholesky approach is supported as well. To compute the sparse decompositions MSSM interfaces with the C++ header library Eigen \parencite{guennebaud_eigen_2010}. When relying on the QR decomposition to estimate $\boldsymbol{\beta}$, MSSM relies on Heath's method \parencite{heath_extensions_1982} to identify coefficients that cannot be estimated and sets those to zero. When relying on the Cholesky approach, MSSM by default estimates the condition number of the final state of $\mathcal{H} = \mathbf{X}^\top\mathbf{X} + \mathbf{S}_\lambda$ and prompts the user with a warning in case the estimate suggests that estimation should be repeated based on the QR decomposition approach \parencite[e.g.,][]{golub_matrix_2013}. Useful properties of the estimated model, such as $\hat{\boldsymbol{\beta}}$, $\hat{\boldsymbol{\lambda}}$, the un-pivoted version of $\mathbf{L}^{-1}$, and the effective degrees of freedom $\tau=tr(\mathbf{V}\mathbf{X}^\top\mathbf{X})$ are accessible after fitting through instance variables. In addition, the \texttt{GAMM} class implements methods to facilitate post-estimation tasks such as prediction, credible interval computation, computation of approximate p-value for smooth \& parametric terms alike, and sampling from the normal approximation to $\boldsymbol{\beta}|\mathbf{y},\boldsymbol{\lambda}$. For an overview see the documentation of the \texttt{GAMM} class at \url{https://jokra1.github.io/mssm}.

\subsection{Estimating Mixed Sparse Generalized Additive Models}\label{sec:GAM}

The conventional definition \parencite[e.g.,][]{hastie_generalized_1986,hastie_generalized_1990} of the GAMM stated in the Introduction assumes $Y_i \sim \mathcal{E}(\mu_i,\sigma)$, where $\mathcal{E}$ denotes any member of the exponential family of distributions. In addition, a known function $g(\mu)$ of the mean, rather than $\mu$ directly, is represented by an additive combination of smooth functions and other parametric terms. To ensure that Algorithm (\ref{alg:sparse_gamm}) generalizes to the estimation of GAMMs we start with replacing step 1 in the algorithm with the steps outlined in Algorithm (\ref{alg:pseudo_dat}). The latter completes the so-called Fisher scoring iteration \parencite[e.g.,][]{hastie_generalized_1990,wood_fast_2011,wood_generalized_2017-2}. Under appropriate step-length control, here based on the strategy employed by \Textcite{wood_generalized_2017-1} and implemented in steps 6 and 7 of the algorithm, the latter converges\footnote{Convergence can again be determined based on monitoring the change in $\mathcal{L}_\lambda$ or alternatively the penalized deviance \parencite[e.g.,][]{wood_generalized_2017-1}.} to $\hat{\boldsymbol{\beta}}$, corresponding to the maximizer of the penalized log-likelihood $\mathcal{L}_\lambda$ for any $\boldsymbol{\lambda}$ \parencite[e.g.,][]{wood_fast_2011,wood_generalized_2017-2}. If $g$ is chosen to coincide with the canonical link of the chosen exponential family member $\mathcal{E}$, $\mathbf{H}=\mathbb{E}\{\mathbf{H}\}=\mathbf{X}^\top\mathbf{W}\mathbf{X}/\phi$ at convergence of the aforementioned routine, otherwise $\mathbb{E}\{\mathbf{H}\} = \mathbf{X}^\top\mathbf{W}\mathbf{X}/\phi$, where $\mathbb{E}\{\mathbf{H}\}$ again denotes the expected Hessian of the negative log-likelihood \parencite[e.g.,][]{wood_fast_2011}.

Computing a Newton update for $\boldsymbol{\lambda}$ based on the methods proposed by \Textcite{wood_smoothing_2016} would now require evaluation of $\frac{\partial \mathbf{W}}{\partial \lambda_l}$ and $\frac{\partial^2 \mathbf{W}}{\partial \lambda_l \partial \lambda_j}$ for all $\lambda_l,\lambda_j \in \boldsymbol{\lambda}$ \parencite[see also][]{wood_fast_2011}. \Textcite{wood_generalized_2017} suggest instead to alternate between Algorithm (\ref{alg:pseudo_dat}) and EFS updates to $\boldsymbol{\lambda}$ \parencite[see also][]{wood_generalized_2017-2}. The resulting strategy remains conceptually very similar to the ones proposed by \Textcite{wood_smoothing_2016} and \Textcite{wood_fast_2011}. As mentioned in the Background section, the EFS update by \Textcite{wood_generalized_2017} simply assumes that $\mathbf{H}$, or $\mathbf{W}$ in the GAMM case, is independent of $\boldsymbol{\lambda}$ (i.e., that in general $\forall \lambda_r \in \boldsymbol{\lambda}~ \frac{\partial \mathbf{W}}{\partial \lambda_r}=\mathbf{0}$). If this assumption is met, the derivatives of $\mathcal{V}$ with respect to the regularization parameters $\boldsymbol{\lambda}$ and the EFS update itself can both be computed exactly as for a Gaussian model \parencite[see Appendix \ref{sec:AppendixRemlDeriv} and][]{wood_generalized_2017-1}: all that is necessary when computing Equations (\ref{eq:laplace_reml_grad}) and (\ref{eq:efs}) is to replace $\mathbf{X}$ with $\sqrt{\mathbf{W}}\mathbf{X}$, where $\mathbf{W}$ is the final state of the weight matrix obtained at convergence of Algorithm (\ref{alg:pseudo_dat}) and $\sqrt{\mathbf{W}}$ denotes the matrix obtained by computing the square root of every element on the diagonal of $\mathbf{W}$. Apart from these changes, estimation of GAMMs for which $\mathbf{H}$ (i.e., $\mathbf{W}$) does not depend on $\boldsymbol{\lambda}$ can thus proceed as described in the previous section. 

\begin{algorithm}
\caption{\textit{Pseudo-data transformation}}\label{alg:pseudo_dat}
\begin{algorithmic}[1]
\INPUT Vector of observations $\mathbf{y}$, model matrix $\mathbf{X}$, current coefficient estimate $\hat{\boldsymbol{\beta}}$, link function $g$, variance function $V_{\mathcal{E}}$ of family $\mathcal{E}$, penalized log-likelihood $\mathcal{L}_\lambda(\boldsymbol{\hat{\boldsymbol{\beta}}})$ under current estimate $\boldsymbol{\lambda}$, current estimate $\boldsymbol{\lambda}$

\For{$i = 1, 2, ...$}
    \State Compute $\boldsymbol{\mu}=g^{-1}(\mathbf{X}\hat{\boldsymbol{\beta}})$
    \State Compute $\mathbf{z} = g(\boldsymbol{\mu}) + \frac{\partial g(\boldsymbol{\mu})}{\partial \boldsymbol{\mu}} * (\boldsymbol{y} - \boldsymbol{\mu})$
    \State Compute the diagonal matrix $\mathbf{W}$, where\linebreak\hspace*{1.2em} $\mathbf{W}_{i,i} = 1 / \left[\left(\frac{\partial g(\boldsymbol{\mu})}{\partial \boldsymbol{\mu}}\right)^2 * V_{\mathcal{E}}(\boldsymbol{\mu}))\right]$
    \State Obtain $\boldsymbol{\hat{\beta}}^*$ by solving the penalized normal equations\linebreak\hspace*{1.2em} with $\mathbf{\sqrt{\mathbf{W}}}\mathbf{z}$ instead of $\mathbf{y}$ and $\sqrt{\mathbf{W}}\mathbf{X}$ instead of $\mathbf{X}$
    \State Compute $\mathcal{L}_\lambda(\boldsymbol{\hat{\beta}}^*)$ given $\boldsymbol{\lambda}$ 
    \While{$\mathcal{L}_\lambda(\boldsymbol{\hat{\beta}}) < \mathcal{L}_\lambda(\boldsymbol{\hat{\beta}}^*)$}
    
        \State $\boldsymbol{\hat{\beta}}^* = (\boldsymbol{\hat{\beta}}^* + \boldsymbol{\hat{\beta}})/2$
        \State Repeat steps 6-7
    
    \EndWhile
    \State Set $\boldsymbol{\hat{\beta}}=\boldsymbol{\hat{\beta}}^*$ and $\mathcal{L}_\lambda(\boldsymbol{\hat{\beta}}) =\mathcal{L}_\lambda(\boldsymbol{\hat{\beta}}^*)$
    \State Exit the loop if converged
\EndFor
\end{algorithmic}
\end{algorithm}

\Textcite{wood_generalized_2017} point out that the same strategy could in principle be used to estimate GAMMs for which $\mathbf{W}$ \emph{does} depend on $\boldsymbol{\lambda}$. For these models the EFS update will be approximate, because it will not in general be the case that $\forall~\lambda_r \in \boldsymbol{\lambda}~\frac{\partial \mathbf{W}}{\partial \lambda_r}=\mathbf{0}$ \parencite[e.g.,][]{wood_generalized_2017}. In consequence, alternating between Algorithm (\ref{alg:pseudo_dat}) and EFS updates to $\boldsymbol{\lambda}$ will not necessarily produce a final estimate $\hat{\boldsymbol{\lambda}}$ that maximizes the REML criterion of the GAMM \parencite[e.g.,][]{wood_generalized_2017,wood_generalized_2017-2}. As pointed out by \Textcite{wood_generalized_2017}, this complicates step-length control for $\boldsymbol{\lambda}$, because relying on the REML criterion to decide whether corrections are necessary is no longer justified. In principle, it would be possible to simply drop step-length control for $\boldsymbol{\lambda}$ for GAMMs, relying on the aforementioned theoretical results presented by \Textcite{wood_generalized_2017} which suggest that corrections to proposals $\boldsymbol{\Delta}_{\boldsymbol{\lambda}}$ should rarely be needed in practice.

Alternatively, step-length for $\boldsymbol{\lambda}$ could exploit the fact that the EFS update as defined in Equation (\ref{eq:efs}), again assuming that $\mathbf{X}$ is replaced with $\sqrt{\mathbf{W}}\mathbf{X}$, would remain exact for a Gaussian model of the final transformed data $\mathbf{z} \sim \mathcal{N}(\mathbf{X}\boldsymbol{\beta},\mathbf{W}^{-1}\phi)$ instead \parencite[e.g.,][]{wood_generalized_2015,wood_generalized_2017-1}. This observation forms the basis of the ``Penalized-quasi-likelihood'' \parencite[PQL;][]{breslow_approximate_1993} or ``performance-oriented-iteration'' \parencite[POI;][]{gu_cross-validating_1992} approach to GAMM estimation \parencite[see][for an overview]{wood_generalized_2017-2}. Conventionally, the PQL/POI approach to GAMM estimation proceeds by ``linearizing'' the model: $\mathbf{W}$ and $\mathbf{z}$ are updated once, followed by estimation of $\boldsymbol{\lambda}$ (and thus $\boldsymbol{\beta}$) via Newton's method assuming that $\mathbf{z} \sim \mathcal{N}(\mathbf{X}\boldsymbol{\beta},\mathbf{W}^{-1}\phi)$ \parencite[e.g.,][]{wood_generalized_2015}. Subsequently, $\mathbf{W}$ and $\mathbf{z}$ are updated, given the new estimate for $\boldsymbol{\beta}$, and $\boldsymbol{\lambda}$ is estimated anew for this updated linearized (i.e., Gaussian) model. A more performant variant of this estimation strategy was proposed by \Textcite{wood_generalized_2017-1}, who suggested to instead re-linearize the model (i.e., update $\mathbf{W}$ and $\mathbf{z}$) after every update to $\boldsymbol{\lambda}$ (and $\boldsymbol{\beta}$).

While both of these approaches are considerably more efficient than alternative methods \parencite[e.g.,][]{wood_fast_2011,wood_smoothing_2016}, the PQL/POI assumption about the distribution of the $z_i$ immediately appears misguided. Indeed, it can generally only be justified asymptotically, relying on the central limit theorem \parencite[e.g.,][]{wood_generalized_2015,wood_generalized_2017-1,wood_generalized_2017-2}. Specifically, \Textcite{wood_generalized_2017-1} show that the linearized model $\mathcal{N}(\mathbf{X}\boldsymbol{\beta},\mathbf{W}^{-1}\phi)$ can be replaced with an alternative one, for which the normal assumption becomes justifiable as $N/N_p \to \infty$. Crucially, the authors show that inference for $\boldsymbol{\lambda}$ remains unaffected by the replacement, which warrants reliance on the linearized model $\mathcal{N}(\mathbf{X}\boldsymbol{\beta},\mathbf{W}^{-1}\phi)$ for estimation in practice, as long as $N_p << N$. Conveniently, this also provides justification to perform step-length control for $\boldsymbol{\lambda}$ based on the linearized model with $\mathbf{W}$ and $\mathbf{z}$ fixed \parencite[e.g.,][]{wood_generalized_2015,wood_generalized_2017-1}.

For example, \Textcite{wood_generalized_2017-1} rely on steps 11-14 of Algorithm (\ref{alg:sparse_gamm}) to achieve step-length control for $\boldsymbol{\lambda}$ for GAMMs. Given current $\mathbf{W}$ and $\mathbf{z}$ they again determine the need for a correction to $\boldsymbol{\lambda}$ based on the sign of $\boldsymbol{\nabla}_{\boldsymbol{\lambda} + \boldsymbol{\Delta}_{\boldsymbol{\lambda}}}^\top \boldsymbol{\Delta_{\boldsymbol{\lambda}}}$, where the Gradient of the REML criterion $\boldsymbol{\nabla}_{\boldsymbol{\lambda} + \boldsymbol{\Delta}_{\boldsymbol{\lambda}}}$ defined in Equation (\ref{eq:laplace_reml_grad}) is computed for the current linearized model with $\sqrt{\mathbf{W}}\mathbf{X}$ instead of $\mathbf{X}$ \parencite[e.g.,][]{wood_generalized_2017-1}. Note, that the maximizer of the penalized log-likelihood of the current linearized model required for this purpose is simply $\hat{\boldsymbol{\beta}}^*$, defined in step 6 of Algorithm (\ref{alg:pseudo_dat}) and computed based on the current $\mathbf{W}$ and $\mathbf{z}$. Conceptually, the update $\boldsymbol{\Delta_{\boldsymbol{\lambda}}}$ is thus accepted if the Gradient of the REML criterion for the current linearized/Gaussian model after the update still matches the general direction of the update.

It could be argued that the protection against divergence offered by this check is limited, because the linearized model (i.e., $\mathbf{W}$ and $\mathbf{z}$) changes immediately afterwards. Instead, it could be checked whether the Gradient of the REML criterion for the \emph{updated} linearized model after an update to $\boldsymbol{\lambda}$ still points in the general direction of the update. While this form of step-length control also cannot fully prevent divergence, it has the advantage that it can be performed as part of the estimation strategy described here, alternating between Algorithm (\ref{alg:pseudo_dat}) and EFS updates to $\boldsymbol{\lambda}$. In fact, no further changes to Algorithm (\ref{alg:sparse_gamm}) are necessary to implement this form of step-length control when step 1 is replaced with Algorithm (\ref{alg:pseudo_dat}) and $\mathbf{X}$ is subsequently replaced with $\sqrt{\mathbf{W}}\mathbf{X}$, where $\mathbf{W}$ again reflects the state of the weights at convergence of Algorithm (\ref{alg:pseudo_dat}): for every proposed update $\boldsymbol{\Delta}_{\boldsymbol{\lambda}}$ the maximizer of the penalized log-likelihood $\hat{\boldsymbol{\beta}}$ for the GAMM given $\boldsymbol{\lambda} + \boldsymbol{\Delta}_{\boldsymbol{\lambda}}$ is found in a first step (step 9 in Algorithm (\ref{alg:sparse_gamm})) and the proposed update is accepted only if the Gradient of the REML criterion for the updated linearized model, based on the weights at convergence of Algorithm (\ref{alg:pseudo_dat}), continues to point in the general direction of the update (steps 10-11 in Algorithm (\ref{alg:sparse_gamm})).

Conceptually, this alternative form of step-length control is quite different from the one implemented by \Textcite{wood_generalized_2017-1}. However, assuming a normal distribution for the final linearized model based on the weights at convergence of Algorithm (\ref{alg:pseudo_dat}) can again be justified based on the large sample result from \Textcite{wood_generalized_2017-1}. Similarly, at convergence of Algorithm (\ref{alg:pseudo_dat}), $\hat{\boldsymbol{\beta}}$ coincides with $\hat{\boldsymbol{\beta}}^*$, the maximizer of the penalized log-likelihood of the final linearized model. Thus, the Gradient of the REML criterion of the final linearized model can again be computed with $\sqrt{\mathbf{W}}\mathbf{X}$ instead of $\mathbf{X}$. Alternatively, the EFS update could also be used as a direct replacement for the Newton update to $\boldsymbol{\lambda}$ in the estimation strategies by \Textcite{wood_generalized_2015} and \Textcite{wood_generalized_2017-1}. Utilizing the same form of step-length control relied upon by \Textcite{wood_generalized_2017-1} then requires no additional justification. Relying on the estimation strategy by \Textcite{wood_generalized_2017-1}, modified to rely on the EFS update, also has the advantage that it will often be more efficient than the strategy outlined here for large multi-level models. Fortunately, implementing this strategy only requires a few modifications to Algorithm (\ref{alg:sparse_gamm}).

Specifically, step 1 in Algorithm (\ref{alg:sparse_gamm}) needs to be replaced with step 5 of Algorithm (\ref{alg:pseudo_dat}). This step finds $\hat{\boldsymbol{\beta}}^*$, the solution to the penalized normal equations given $\boldsymbol{\lambda}$ and based on the current weights $\mathbf{W}$ and transformed data $\mathbf{z}$\footnote{Initial values for the weights $\mathbf{W}$ and transformed data $\mathbf{z}$ can be obtained from an initial estimate for $\boldsymbol{\mu}$ \parencite[see][]{wood_generalized_2017-1}.}. When this step is computed for the first time we simply set $\hat{\boldsymbol{\beta}}=\hat{\boldsymbol{\beta}}^*$. In all but the first fitting iterations, two extra steps need to be completed as well. The first extra step needs to enforce step-length control for $\boldsymbol{\beta}$ and should be implemented before checking for convergence in step 6 of Algorithm (\ref{alg:sparse_gamm}). This can be achieved almost exactly as outlined in steps 6-7 of Algorithm (\ref{alg:pseudo_dat}). At this point in the fitting iteration $\hat{\boldsymbol{\beta}}^*$ will correspond to the updated estimate of the coefficients given $\boldsymbol{\lambda} + \boldsymbol{\Delta_{\boldsymbol{\lambda}}}$, obtained at step 9 in Algorithm (\ref{alg:sparse_gamm}). Step-length control for $\boldsymbol{\beta}$ is necessary here since $\hat{\boldsymbol{\beta}}^*$ does not necessarily have to increase the penalized log-likelihood of the actual GAMM given $\boldsymbol{\lambda} + \boldsymbol{\Delta_{\boldsymbol{\lambda}}}$, compared to what is achieved by the current estimate of the coefficients $\hat{\boldsymbol{\beta}}$ \parencite[e.g.,][]{wood_generalized_2017-1}. The second extra step, which only has to happen in case convergence has not yet been achieved, needs to update $\mathbf{W}$ and $\mathbf{z}$ (i.e., complete steps 3-5 in Algorithm (\ref{alg:pseudo_dat})). Finally, $\mathbf{X}$ and $\mathbf{y}$ again have to be replaced with the current versions of $\sqrt{\mathbf{W}}\mathbf{X}$ and $\sqrt{\mathbf{W}}\mathbf{z}$ at any step of Algorithm (\ref{alg:sparse_gamm}).

Before considering how these different strategies are implemented in the MSSM toolbox, it is important to reiterate that independent of the specific estimation strategy and form of step length control, the EFS update will not generally produce a final estimate $\hat{\boldsymbol{\lambda}}$ that maximizes $\mathcal{V}$. In practice, especially when $N_p << N$, it nevertheless often produces estimates for $\boldsymbol{\beta}$ and $\boldsymbol{\lambda}$ that are very close to those obtained from more complex strategies \parencite[e.g.,][]{wood_fast_2011,wood_smoothing_2016}. In the upcoming Section \ref{sec:GAMMLSS} we discuss in more detail why that is the case.

\subsubsection{Implementation in MSSM}

The \texttt{GAMM} class implemented in MSSM also supports estimation of Generalized models, as discussed in this section. For estimation it supports both the strategy alternating between Algorithm (\ref{alg:pseudo_dat}) and EFS updates to $\boldsymbol{\lambda}$ and the more efficient strategy outlined by \Textcite{wood_generalized_2017-1}, modified to rely on the EFS update by \Textcite{wood_generalized_2017}. By default, MSSM adopts the latter strategy and re-linearizes the model (i.e., updates $\mathbf{W}$ and $\mathbf{z}$) after every EFS update to $\boldsymbol{\lambda}$. Step-length control for $\boldsymbol{\lambda}$ is performed for the current linearized model as described in this section and proposed previously by \Textcite{wood_generalized_2017-1}. The \texttt{fit} method of the GAMM class also supports a \texttt{max\_inner} argument however. When this argument is set to a value greater than 1, MSSM will rely on the alternative strategy alternating between Algorithm (\ref{alg:pseudo_dat}) and EFS updates to $\boldsymbol{\lambda}$. The \texttt{max\_inner} argument then determines the maximum number of iterations completed by Algorithm (\ref{alg:pseudo_dat}). As discussed in this section, this results in a different step-length control strategy for $\boldsymbol{\lambda}$. If preferred, step-length control for $\boldsymbol{\lambda}$ can also be disabled entirely (i.e., via the \texttt{control\_lambda} argument). By default, the QR approach discussed in Section \ref{sec:AM} is used to solve for $\boldsymbol{\beta}$ given $\mathbf{W}$, $\mathbf{z}$, and $\boldsymbol{\lambda}$, independent of the value for \texttt{max\_inner}. However, the more efficient Cholesky approach can be selected as well.

The package currently supports GAMMs with a Gaussian, Gamma, Binomial, Inverse Gaussian, or Poisson distribution. Additionally, MSSM currently supports the identity, log, logit, and inverse link functions. Implementation of alternative distributions and link functions not currently supported is facilitated through simple template classes requiring users to provide only a minimum of information. We again refer to the documentation available at \url{https://jokra1.github.io/mssm} for more details and now return to the problem of estimating even more general models.

\subsection{Estimating General Mixed Sparse Smooth Models}\label{sec:GAMMLSS}

As mentioned in the Background section it will often be desirable to extend the model even further, assuming only that $Y_i \sim \mathcal{F}(\mu_i,\phi_i,\ldots,\tau_i)$. Notably, rather than just specifying an additive model of the mean, as would be the case for a GAMM, additive models are now specified for each additional parameter of $\mathcal{F}$, or some known function $g$ of it. As discussed, this encompasses ``Generalized Additive Model of Location Scale and Shape'' \parencite[GAMLSS;][]{rigby_generalized_2005,wood_smoothing_2016} but also ``extended GAMMs'' for which any extra parameters are treated as constant \parencite[i.e., like $\phi$;][]{wood_smoothing_2016}. Even more generally, researchers might want to specify additive models for the parameters (or some known function of them) of a ``non-standard'' but regular log-likelihood, resulting in a ``general smooth model'' \parencite[e.g.,][]{wood_smoothing_2016}. Importantly, the models of different parameters can, but do not have to, differ in their predictor structure, so that some parameters might for example vary as a function of time, while others do not.

Estimating such a generic smooth model with a log-likelihood $\mathcal{L}$ that is a function of $N_m$ parameters (e.g., $N_m=2$ for a Gaussian GAMMLSS model of $\mu$ and $\phi=\sigma^2$) requires estimating $N_m$ sets of coefficients $\boldsymbol{\beta}_m$, each of dimension $N_{pm}$. Together with a parameter-specific model matrix $\mathbf{X}^m$ and link function $g_m$, the coefficient subsets $\boldsymbol{\beta}_m$ all parameterize the additive model of a specific parameter of $\mathcal{L}$. The $\boldsymbol{\beta}_m$ can be concatenated, which results in the overall coefficient vector $\boldsymbol{\beta}$ of dimension $N_p = \sum_{m=1}^M N_{pm}$, including all coefficients to be estimated by the model.

As for any smooth model, suitable $\boldsymbol{\lambda}$ parameters will have to be estimated for any smooth terms present in any of the $\mathbf{X}^m$. To facilitate this, it is again useful to embed the penalty matrix $\mathcal{S}_{\boldsymbol{\lambda}}^{j}$ corresponding to the $j$-th smooth term into a $N_p*N_p$ zero block, so that the overall smoothness penalty can again be computed as $\sum_j\boldsymbol{\beta}^\top\mathbf{S}^j_{\lambda}\boldsymbol{\beta}$. Finally, note that $\phi$ no longer scales $\boldsymbol{\lambda}$, since the former -- if present at all (e.g., as a parameter of $\mathcal{F}$) -- will now be parameterized via a specific subset $\boldsymbol{\beta}_\phi$ of coefficients.

Considering the notational changes necessary to accommodate estimation of $N_m$ distribution parameters and the results discussed in the previous section, the main change necessary to ensure that Algorithm (\ref{alg:sparse_gamm}) accommodates the estimation of general smooth models including GAMMLSS, again regards estimation of $\boldsymbol{\beta}$. In principle, estimation of the latter by means of a conventional Newton update, only requires, in addition to the negative Hessian of the log-likelihood $\mathbf{H}$, the Gradient $\nabla_{\boldsymbol{\beta}}^\mathcal{L} = \partial \mathcal{L} /\partial \boldsymbol{\beta}$ of the log-likelihood \parencite[e.g.,][]{wood_smoothing_2016}.

Notably, the structure and degree of sparsity of $\mathbf{H}$ and thus $\mathcal{H}=\mathbf{H} + \mathbf{S}_\lambda$ will no longer just depend on the specification (i.e., $\mathbf{X}_m$) of the models for the $N_m$ distribution parameters, but also on the specific log-likelihood $\mathcal{L}$ (e.g., on $\mathcal{F}$ for GAMMLSS). For GAMMLSS, efficient accumulation of $\mathbf{H}$ in sparse matrix storage can be achieved by exploiting the transformation strategy discussed by \Textcite{wood_smoothing_2016}. The strategy is designed to obtain $\nabla_{\boldsymbol{\beta}}^\mathcal{L}$ and $\mathbf{H}$ from first and second order (pure \& mixed) partial derivatives of $\mathcal{L}$ with respect to the different parameters of $\mathcal{F}$, which are readily available for a large number of distributions \parencite[e.g.,][]{rigby_generalized_2005}. The strategy is particularly efficient in case the mixed partial derivatives of $\mathcal{L}$ with respect to the parameters of $\mathcal{F}$ are zero. This will be the case for many models that are of interest in the cognitive sciences, including Gaussian and Gamma location \& scale models. For such models, $\mathbf{H}$ will generally be block-diagonal -- with block $m$ holding the partial second derivatives of $\mathcal{L}$ with respect to $\boldsymbol{\beta}_m$. Obtaining the latter only involves multiplying the pure second partial derivative of $\mathcal{L}$ with respect to the $m$-th parameter by $[\mathbf{X}^m]^\top\mathbf{X}^m$ \parencite[see][]{wood_smoothing_2016}. Thus, as long as $[\mathbf{X}^m]^\top\mathbf{X}^m$ is sparse, which will be the case for the multi-level models of interest here, block $m$ of $\mathbf{H}$ will be sparse as well, so that the computations involved in these transformations carry little additional computational overhead compared to those required to enable estimation for GAMMs.

For even more generic smooth models beyond GAMMLSS, researchers will generally have to provide code to directly compute the Gradient $\nabla_{\boldsymbol{\beta}}^\mathcal{L}$ and negative Hessian $\mathbf{H}$ of $\mathcal{L}$ with respect to coefficients $\boldsymbol{\beta}$. As discussed in the Background section, the fact that the structure of the Hessian is problem-specific makes it difficult to provide unified guidelines on how to best accumulate it in sparse matrix storage. In the next section we discuss a possible solution to this problem, but for the remainder of this section we simply assume that $\mathbf{H}$ is available in sparse matrix storage.

Algorithm (\ref{alg:gammlss_beta}) summarizes how $\mathbf{H}$ and $\nabla_{\boldsymbol{\beta}}^\mathcal{L}$ can then be combined in a Newton update to estimate $\boldsymbol{\beta}$ based on the strategy described by \Textcite{wood_smoothing_2016}. As suggested by \Textcite{wood_smoothing_2016} the actual Newton update for $\boldsymbol{\beta}$ (steps 5-10 in Algorithm (\ref{alg:gammlss_beta})) has been modified to accommodate the fact that, depending on the choice of the initial estimate for $\boldsymbol{\beta}$ and $\boldsymbol{\lambda}$, $\mathbf{H}$ and thus $\mathcal{H}$ can actually be \emph{expected} to be rank deficient, that is positive semi-definite or even indefinite, in early fitting iterations. This would prevent the formation of Cholesky factor $\mathbf{L}$\footnote{Note that $\mathbf{L}$ in Algorithm (\ref{alg:gammlss_beta}) is again computed for $\mathbf{P}[\mathcal{H}+\mathbf{I}\epsilon_{\mathcal{H}}]\mathbf{P}^\top$, where $\mathbf{P}$ is again used to increase the sparsity of $\mathbf{L}$.}.

\begin{algorithm}
\caption{\textit{Coefficient Estimation for Mixed Sparse General Smooth Models}}\label{alg:gammlss_beta}
\begin{algorithmic}[1]
\INPUT Vector of observations $\mathbf{y}$, model matrices $(\mathbf{X}^m)_{m=1,...,N_m}$, link functions $(g_m)_{m=1,...,N_m}$, current coefficient estimate $\hat{\boldsymbol{\beta}}$, $\mathcal{L}_\lambda$ for current model, individual penalty matrices $\mathbf{S}^r$ where $r \in \{1,2,...,N_\lambda\}$, penalty matrix $\mathbf{S}_\lambda$ based on current estimate $\boldsymbol{\lambda}$

\For{$i = 1,2,...$}
    \State Check Convergence
    \State Compute Gradient $\nabla_{\boldsymbol{\beta}}^\mathcal{L}$ and negative Hessian $\mathbf{H}$ of the\linebreak\hspace*{1.2em} log-likelihood at $\hat{\boldsymbol{\beta}}$
    \State Set $\mathcal{H}=\mathbf{H} + \mathbf{S}_\lambda$ and $\nabla_{\boldsymbol{\beta}}^{\mathcal{L}_\lambda} = \nabla_{\boldsymbol{\beta}}^\mathcal{L} - \mathbf{S}_\lambda\hat{\boldsymbol{\beta}}$
    \State Set $\epsilon_{\mathcal{H}}=0$. Attempt Cholesky decomposition\linebreak\hspace*{1.2em} $\mathbf{L}\mathbf{L}^\top=\mathbf{P}[\mathcal{H}+\mathbf{I}\epsilon_{\mathcal{H}}]\mathbf{P}^\top$
    \While{step 3 fails}
        \State Increase $\epsilon_{\mathcal{H}}$, repeat step 3
    \EndWhile
    \State Solve $[\mathcal{H}+\mathbf{I}*\epsilon_{\mathcal{H}}]\Delta_{\boldsymbol{\beta}} =-\nabla_{\boldsymbol{\beta}}^{\mathcal{L}_\lambda}$ for $\Delta_{\boldsymbol{\beta}}$ using $\mathbf{L}$ and $\mathbf{P}$
    \State Set $\boldsymbol{\hat{\beta}}^*=\boldsymbol{\hat{\beta}} + \Delta_{\boldsymbol{\beta}}$
    \State Compute penalized likelihood $\mathcal{L}^*_\lambda$, given new\linebreak\hspace*{1.5em}distribution parameter estimates $(g_m(\mathbf{X}^m\boldsymbol{\hat{\beta}}^*_m))_{m=1,...,N_m}$\linebreak\hspace*{1.5em}and $\boldsymbol{\lambda}$ 
    \While{$\mathcal{L}^*_\lambda < \mathcal{L}_\lambda$}
    
        \State $\boldsymbol{\hat{\beta}}^* = (\boldsymbol{\hat{\beta}}^* + \boldsymbol{\hat{\beta}})/2$
        \State Repeat step 9
    
    \EndWhile
    \State Set $\boldsymbol{\hat{\beta}} = \boldsymbol{\hat{\beta}}^*$ and $\mathcal{L}_\lambda= \mathcal{L}^*_\lambda$
\EndFor

\While{$tr\left(\mathbf{S}^{-}_\lambda \mathbf{S}^r\right)<tr\left(\mathcal{H}^{-1}\mathbf{S}^r\right)~\text{for any}~\mathbf{S}^r$}
    \State Increase $\epsilon_{\mathcal{H}}$, set $\mathcal{H} =\mathbf{H} + \mathbf{I}\epsilon_\mathcal{H} + \mathbf{S}_\lambda$ and form $\mathbf{L}\mathbf{L}^\top=\mathbf{P}\mathcal{H}\mathbf{P}^\top$
\EndWhile

\end{algorithmic}
\end{algorithm}

\Textcite{wood_smoothing_2016} suggest to address this problem by repeatedly adding a small value ($\epsilon_\mathcal{H}$ in Algorithm (\ref{alg:gammlss_beta}) to the diagonal of $\mathcal{H}$ until $\mathcal{H}+\mathbf{I}\epsilon_{\mathcal{H}}$ is positive definite (i.e., full rank). At that point, $\mathbf{L}$ can be computed -- at least in theory -- which in turn enables computation of the next set of coefficients. In practice however, $\mathcal{H}$ (and also $\mathcal{H}+\mathbf{I}\epsilon_{\mathcal{H}}$) will often be ill-conditioned to the degree that relying on a Cholesky decomposition utilizing sparsity-oriented pivoting to obtain $\mathbf{L}$ should be avoided \parencite{wood_smoothing_2016}. For GAMMs, we suggested to rely on a sparse QR decomposition of $\begin{bmatrix}
\mathbf{X}\\
\mathbf{E}_\lambda^\top\\
\end{bmatrix}$ to address this problem, avoiding formation of $\mathbf{X}^\top\mathbf{X} + \mathbf{S}_\lambda$ \parencite[see also][]{wood_fast_2011,wood_generalized_2017-2}. For general smooth models this is not possible. Instead it will generally be necessary to re-parameterize the model before using Algorithm (\ref{alg:gammlss_beta}) and to apply a diagonal pre-conditioner to $\mathcal{H}$ \parencite{wood_smoothing_2016}.

This strategy, reviewed here in Appendix \ref{sec:AppendixTransform}, ensures that computation of $\mathbf{L}$ remains stable \parencite[e.g.,][]{wood_smoothing_2016}. The necessary transformation that is applied to the model can simply be reversed after Algorithm (\ref{alg:gammlss_beta}) converges in $\hat{\boldsymbol{\beta}}$, but it is also possible to compute the next update to $\boldsymbol{\lambda}$ based on this transformed model before reversing the transformation \parencite[see Appendix \ref{sec:AppendixTransform} and also Section 6.2.7 in the book by][]{wood_generalized_2017-2}. 

Independent of whether this transformation strategy is utilized or not, it will generally be necessary to account for the fact that some coefficients might generally not be identifiable (i.e., independent of the choice for $\boldsymbol{\lambda}$), in which case $\mathbf{L}$ will never be computable with $\epsilon_\mathcal{H}=0$. \Textcite{wood_smoothing_2016} outline a strategy to test for the presence of such unidentifiable coefficients after convergence of Algorithm (\ref{alg:gammlss_beta}), based on a scaled version $\mathcal{H}^S = \mathbf{H}/||\mathbf{H}||_F + \tilde{\mathbf{S}}/||\tilde{\mathbf{S}}||_F$ of the negative Hessian of the penalized likelihood of the model. $||\mathbf{H}||_F$ here denotes the Frobenius norm of matrix $\mathbf{H}$ and $\tilde{\mathbf{S}}=\sum_{r=1}^{N_\lambda}\mathbf{S}^r/||\mathbf{S}^r||_F$ denotes the ``balanced penalty matrix'' \parencite{wood_smoothing_2016}. While the strategy by \Textcite{wood_smoothing_2016} is very efficient for dense matrices, it does not benefit from sparsity in the structure of $\mathbf{H}$. In Appendix \ref{sec:AppendixUnidentifiable} we describe how the efficiency of the strategy can be improved and also discuss an alternative based on the methods by \Textcite{foster_rank_1986} and \Textcite{gotsman_computation_2008}. As suggested by \Textcite{wood_smoothing_2016}, Algorithm (\ref{alg:gammlss_beta}) should generally be re-iterated on the reduced set of coefficients, in case any of these strategies deem some coefficients unidentifiable.

After unidentifiable parameters have been excluded and Algorithm (\ref{alg:gammlss_beta}) has been repeated, it would again be possible to compute a Newton update for $\boldsymbol{\lambda}$ as described for example by \Textcite{wood_smoothing_2016}. If the EFS update by \Textcite{wood_generalized_2017} is to be used instead, an additional modification is required however, to guarantee that the EFS update proposed by \Textcite{wood_generalized_2017} is defined whenever the algorithm terminates. Specifically, Theorem 1 from \Textcite{wood_generalized_2017} requires $\mathbf{H}$ to be at least \emph{positive semi-definite} to guarantee that $tr(\mathbf{S}_\lambda^-\mathbf{S}^r) \geq tr([\mathbf{H} + \mathbf{S}_\lambda]^{-1}\mathbf{S}^r)$. Clearly, if this inequality does not hold, Equation (\ref{eq:efs}) could produce negative estimates for some $\lambda_r$, which are theoretically impossible \parencite{wood_generalized_2017}. The estimation strategy for GAMMs outlined in the previous section, alternating between Algorithm (\ref{alg:pseudo_dat}) and the EFS update, avoids this problem because $\mathbf{X}^\top\mathbf{W}\mathbf{X}/\phi$ corresponds to the expectation $\mathbb{E}\{\mathbf{H}\}$ at convergence of Algorithm (\ref{alg:pseudo_dat}), which is at least positive semi-definite \parencite[e.g.,][]{wood_generalized_2017,wood_smoothing_2016}.

Unless $\mathbf{H}$ can again be replaced with the expectation $\mathbb{E}\{\mathbf{H}\}$, it will be necessary to explicitly enforce positive semi-definiteness of $\mathbf{H}$ when Algorithm (\ref{alg:gammlss_beta}) terminates before using the EFS update for the generic models discussed in this section. Importantly, completing the steps taken by \Textcite{wood_smoothing_2016} to ensure positive-definiteness of $\mathcal{H}$ is insufficient: when $\mathbf{H}$ is poorly conditioned, $\mathbf{H} + \mathbf{I}\epsilon_\mathcal{H}$ can still be indefinite even when $\mathcal{H} =\mathbf{H} + \mathbf{I}\epsilon_\mathcal{H} + \mathbf{S}_\lambda$ is positive definite \parencite[e.g.,][]{pya_shape_2015,wood_smoothing_2016}. Testing for this explicitly by means of a rank-revealing decomposition would however be expensive. We suggest to rely on the heuristic strategy implemented in steps 18-20 of Algorithm (\ref{alg:gammlss_beta}) instead, which ensures that $\mathbf{H} + \mathbf{I}\epsilon_\mathcal{H}$ is at least close enough to being positive semi-definite for the required inequality $tr(\mathbf{S}_\lambda^-\mathbf{S}^r) \geq tr([\mathbf{H} + \mathbf{S}_\lambda]^{-1}\mathbf{S}^r)$ to hold.

With this modification in place, the EFS update can be used safely to update $\boldsymbol{\lambda}$, whenever Algorithm (\ref{alg:gammlss_beta}) terminates so that estimation of $\boldsymbol{\beta}$ and $\boldsymbol{\lambda}$ for generic smooth models can again be achieved by alternating between Algorithm (\ref{alg:gammlss_beta}) and EFS updates to $\boldsymbol{\lambda}$. As was the case for GAMMs, this estimation strategy can be implemented by replacing step 1 in Algorithm (\ref{alg:sparse_gamm}) with Algorithm (\ref{alg:gammlss_beta}). Additionally, to ensure generalization it is necessary that $\phi$ is dropped from Equations (\ref{eq:laplace_reml}), (\ref{eq:laplace_reml_grad}), and (\ref{eq:efs}). Additionally, $\mathbf{X}^\top\mathbf{X} + \mathbf{S}_{\lambda}$ has to be replaced by $\mathcal{H}$ in step 2 of Algorithm (\ref{alg:sparse_gamm}). The corresponding trace term can again be computed efficiently as outlined in Section \ref{sec:AM} for strictly additive models, i.e., based on the final version of the Cholesky factor $\mathbf{L}$. 

It is worth highlighting that the resulting estimation strategy continues to support estimation of GAMMs since they are just a special case of a general smooth model \parencite[e.g.,][]{wood_smoothing_2016}. Apart from the fact that Algorithm (\ref{alg:gammlss_beta}) relies on the observed negative Hessian for $\mathbf{H}$, the main difference between the strategy outlined in this section and the strategy outlined in Section \ref{sec:GAM} is that the latter optimizes $\mathcal{V}$ for the scale parameter $\phi$. If a GAMM is to be estimated via Algorithm (\ref{alg:gammlss_beta}) instead, $\phi$ would be an element of $\boldsymbol{\beta}$ and thus chosen to maximize $\mathcal{L}_\lambda$. The resulting estimate $\hat{\phi}$ obtained by optimizing $\mathcal{L}_\lambda$ is well known to underestimate the magnitude of $\phi$, which is why the strategy outlined in Section \ref{sec:GAM} is to be preferred when estimating conventional GAMMs \parencite[e.g.,][]{wood_smoothing_2016,wood_generalized_2017-2}.

\begin{figure*}[!h]
    \caption{Illustration of the Repeated Re-linearization Strategy Performed by the EFS Update}
    \includegraphics[width=\textwidth]{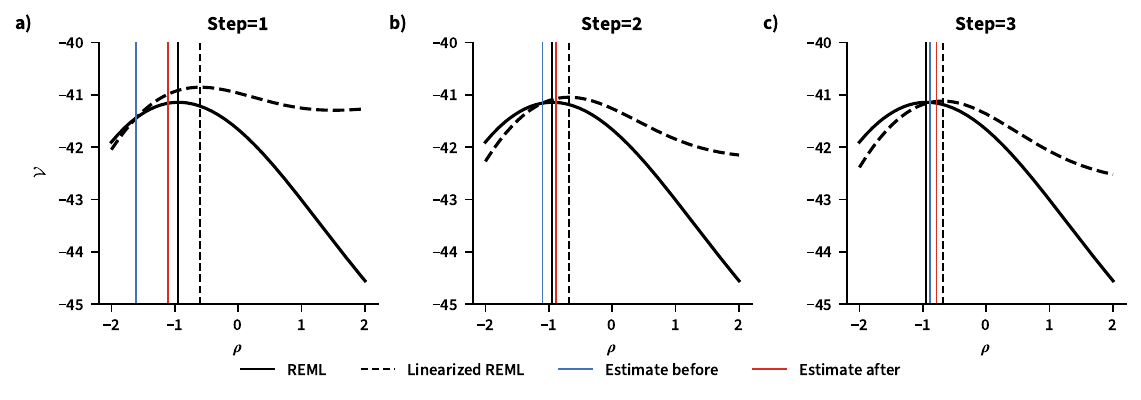}
    \label{fig:Relinearization}
    {\small
         \textit{Note.} Panels a-c of Figure \ref{fig:Relinearization} illustrate the first three EFS updates (i.e., ``steps'') applied to a single $\lambda$ parameter of a proportional hazard model including a single smooth term ($N=25$ observations were simulated). The solid black curve is the same in all three panels and corresponds to the actual REML criterion $\tilde{\mathcal{V}}$ for this model. The solid black vertical line corresponds to the value for $\lambda$ that would maximize the REML criterion. The blue vertical line corresponds to the current estimate of $\boldsymbol{\lambda}$ (i.e., before the update). Related to this is the dashed curve, which corresponds to the approximate (i.e., ``linearized'') REML criterion for which $\mathcal{L}$ has been replaced with $\tilde{\mathcal{L}}|_{\hat{\boldsymbol{\beta}}_{\boldsymbol{\rho}}}$, a second-order Taylor approximation to the log-likelihood around the estimate of the coefficients $\hat{\boldsymbol{\beta}}_{\boldsymbol{\rho}}$ given the current estimate of $\boldsymbol{\lambda}$. The dashed black vertical line again corresponds to the value for $\lambda$ that would maximize the approximate REML criterion. Note that the EFS update takes a step (the vertical red line in each panel corresponds to the updated value for $\lambda$ after the EFS update) towards the maximum of the approximate criterion. Also note, that the approximate criterion and the corresponding maximum are different at every step because the likelihood is re-linearized around the updated coefficient estimate under the new estimate for $\lambda$.
        }
\end{figure*}

Another similarity is that the same caveats discussed in the context of GAMM estimation continue to apply for generic smooth models. Specifically, reliance on the EFS update again assumes that $\mathbf{H}$ is independent of $\boldsymbol{\lambda}$ and will thus only result in a final estimate $\hat{\boldsymbol{\lambda}}$ maximizing $\mathcal{V}$ if the assumption that $\forall~\lambda_r \in \boldsymbol{\lambda}~\frac{\partial \mathbf{H}}{\partial \lambda_r} = \mathbf{0}$ holds in general \parencite[e.g.,][]{wood_generalized_2017}. \Textcite{wood_generalized_2017} point out that at least for GAMMs, dependencies of $\mathbf{H}$ (i.e., $\mathbf{W}$) on $\boldsymbol{\lambda}$ tend to become negligible in the large-sample limit \parencite[see also Section 2.1 in the paper by][]{breslow_approximate_1993}. This observation is, at least in terms of the resulting consequences for estimation, similar to the large sample result proposed by \Textcite{wood_generalized_2017-1}, providing justification for the PQL approach to compute the partial derivatives of $\mathcal{V}$ with respect to the regularization parameters $\boldsymbol{\lambda}$ as would be done for a Gaussian additive model \parencite[e.g.,][]{wood_generalized_2015,wood_generalized_2017-1}. The asymptotic result by \Textcite{wood_generalized_2017-1}, is only really applicable to GAMMs and does not readily generalize to the more generic case \parencite[e.g.,][]{wood_inference_2020}. However, if the importance of dependencies of $\mathbf{H}$ on $\boldsymbol{\lambda}$ tends to diminish eventually, not just for GAMMs but also for more generic models, the EFS update should often produce a final estimate $\hat{\boldsymbol{\lambda}}$ that gets close to the maximizer of the REML criterion in practice, when $N_p << N$. 

Indeed, as mentioned in Section \ref{sec:GAM} the EFS update often produces estimates for $\boldsymbol{\beta}$ and $\boldsymbol{\lambda}$ that are virtually indistinguishable from those obtained from more exact methods \parencite[e.g.,][]{wood_smoothing_2016}. To see why that is the case, we consider a slightly different linearization strategy \parencite[cf.][]{mcgilchrist_estimation_1994}, which extends the basic concept of the PQL approach to generic smooth models (including GAMMs and GAMMLSS).

Specifically, let $\tilde{\mathcal{L}}(\boldsymbol{\beta})|_{\hat{\boldsymbol{\beta}}_{\boldsymbol{\lambda}}}$ denote the second-order Taylor approximation to the log-likelihood $\mathcal{L}$ around $\hat{\boldsymbol{\beta}}_{\boldsymbol{\lambda}}$, the maximizer of the penalized log-likelihood $\mathcal{L}_\lambda$ for a given $\boldsymbol{\lambda}$. Similarly, let $\tilde{\mathcal{L}}_\lambda(\boldsymbol{\beta})|_{\hat{\boldsymbol{\beta}}_{\boldsymbol{\lambda}}}=\tilde{\mathcal{L}}(\boldsymbol{\beta})|_{\hat{\boldsymbol{\beta}}_{\boldsymbol{\lambda}}}  - \frac{1}{2}\sum_j \boldsymbol{\beta}^\top\mathbf{S}^j_{\lambda}\boldsymbol{\beta}$ denote the penalized ``linearized'' log-likelihood based on the second-order Taylor approximation to the log-likelihood $\tilde{\mathcal{L}}|_{\hat{\boldsymbol{\beta}}_{\boldsymbol{\lambda}}}$. Note, that for any given $\boldsymbol{\lambda}$, computation of the EFS update shown in Equation (\ref{eq:efs}) remains unaffected when $\mathcal{L}_\lambda$ is replaced with $\tilde{\mathcal{L}}_\lambda|_{\hat{\boldsymbol{\beta}}_{\boldsymbol{\lambda}}}$ (i.e., when $\mathcal{L}$ is replaced with $\tilde{\mathcal{L}}|_{\hat{\boldsymbol{\beta}}_{\boldsymbol{\lambda}}}$). Thus, without loss of generality we can assume that for a general smooth model the EFS update takes a step $\boldsymbol{\Delta}_\lambda$ towards the maximum of the REML criterion for the current linearized log-likelihood $\tilde{\mathcal{L}}|_{\hat{\boldsymbol{\beta}}_{\boldsymbol{\lambda}}}$ for which $\mathbf{H}$ remains fixed and thus does not change with $\boldsymbol{\lambda}$. However, instead of taking repeated steps towards the maximum of the REML criterion for $\tilde{\mathcal{L}}|_{\hat{\boldsymbol{\beta}}_{\boldsymbol{\lambda}}}$, the log-likelihood is effectively re-linearized around the updated estimate $\hat{\boldsymbol{\beta}}_{\boldsymbol{\lambda}+\boldsymbol{\Delta}_{\boldsymbol{\lambda}}}$ after an EFS update $\boldsymbol{\Delta}_{\boldsymbol{\lambda}}$ to $\boldsymbol{\lambda}$. This is illustrated in Figure \ref{fig:Relinearization}, visualizing the REML criterion for the current linearized log-likelihood before the EFS update is applied to a single $\lambda$ parameter of a simulated proportional hazard model (see the dashed curve). Evidently, the REML criterion for the linearized log-likelihood changes after every update, while the actual REML criterion remains constant (see the solid black curve). Note, how this is comparable to updating $\mathbf{z}$ and $\mathbf{W}$ under the PQL approach to GAMM estimation described in Section \ref{sec:GAM} \parencite[see also][]{hastie_generalized_1990,wood_generalized_2015,wood_generalized_2017-1}. Similarly, if $\mathcal{L}$ truly was a quadratic function it would be approximated perfectly (i.e., without any error) by every linearization $\tilde{\mathcal{L}}|_{\hat{\boldsymbol{\beta}}_{\boldsymbol{\lambda}}}$. Additionally, since $\mathbf{H}$ is stationary for any quadratic log-likelihood and thus independent of $\boldsymbol{\lambda}$, repeated application of the EFS update would produce a final estimate $\hat{\boldsymbol{\lambda}}$ that would maximize the REML criterion. In contrast, the fact that $\tilde{\mathcal{L}}|_{\hat{\boldsymbol{\beta}}_{\boldsymbol{\lambda}}}$ changes with $\boldsymbol{\lambda}$ for any non-quadratic log-likelihood implies that the final estimate $\hat{\boldsymbol{\lambda}}$ is no longer guaranteed to maximize the REML criterion \parencite[e.g.,][see also Figure \ref{fig:Relinearization}c]{wood_generalized_2017}. 

As was the case for GAMMs, optimizing the REML criterion of the linearized log-likelihood $\tilde{\mathcal{L}}|_{\hat{\boldsymbol{\beta}}_{\boldsymbol{\lambda}}}$ when the actual log-likelihood is not truly a quadratic function can to some degree be justified asymptotically. Specifically, \Textcite{wood_smoothing_2016} show that for any given set of regularization parameters $\boldsymbol{\lambda}$ the \emph{penalized log-likelihood} $\mathcal{L}_\lambda$ tends to behave like a quadratic function when $N/N_p \to \infty$, with the third-order derivatives of the log-likelihood being dominated by the elements of $\mathcal{H}$ in the large sample limit \parencite[see Online Supplementary materials B.4 from][]{wood_smoothing_2016}. This not only justifies the normal approximation to $\boldsymbol{\beta}|\mathbf{y},\boldsymbol{\lambda}$ shown in Equation (\ref{eq:posteriorBeta}), but also implies that the error of $\tilde{\mathcal{L}}_\lambda|_{\hat{\boldsymbol{\beta}}_{\boldsymbol{\lambda}}}$, resulting from the second-order Taylor approximation $\tilde{\mathcal{L}}|_{\hat{\boldsymbol{\beta}}_{\boldsymbol{\lambda}}}$ to the log-likelihood, becomes negligible asymptotically \parencite[e.g.,][]{wood_smoothing_2016}. Considering that the error caused by the second-order Taylor approximation $\tilde{\mathcal{L}}|_{\hat{\boldsymbol{\beta}}_{\boldsymbol{\lambda}}}$ to the log-likelihood becomes negligible asymptotically for \emph{any} $\boldsymbol{\lambda}$, it is not unreasonable to simply treat the log-likelihood $\mathcal{L}$ itself like a quadratic function (asymptotically), for which $\mathbf{H}$ is independent of $\boldsymbol{\lambda}$.

This is essentially the assumption of the ``Best Linear Unbiased Predictor'' (BLUP) approach to REML estimation of variance parameters \parencite[e.g.,][]{mcgilchrist_restricted_1991,mcgilchrist_reml_1993,mcgilchrist_estimation_1994}. This approach treats $\mathcal{L}$ as asymptotically quadratic and thus replaces it with the second-order Taylor approximation $\tilde{\mathcal{L}}|_{\hat{\boldsymbol{\beta}}_{\mathcal{L}}}$ of the log-likelihood around the maximizer of the \emph{un-penalized} log-likelihood $\hat{\boldsymbol{\beta}}_{\mathcal{L}}$ for the purpose of estimation. Note, that $\hat{\boldsymbol{\beta}}_{\mathcal{L}}$ is in principle ill-defined for models involving parameters only identifiable under penalization (i.e., random effects). However, if $\mathcal{L}$ truly was a quadratic function, $\tilde{\mathcal{L}}|_{\hat{\boldsymbol{\beta}}_{\boldsymbol{\lambda}}}$ could readily be substituted for $\tilde{\mathcal{L}}|_{\hat{\boldsymbol{\beta}}_{\mathcal{L}}}$ during estimation \parencite[cf.][]{mcgilchrist_estimation_1994}.

These results suggest that even though the EFS estimate $\hat{\boldsymbol{\lambda}}$ is not guaranteed to maximize Equation (\ref{eq:laplace_reml}) in practice, it will get close as long as the log-likelihood is well-approximated by a quadratic function. The large sample result by \Textcite{wood_smoothing_2016} suggests that this will often be the case when $N_p << N$ and $N$ is large. Additionally, these results suggest that, as was the case for GAMMs, (approximate) step-length control for $\boldsymbol{\lambda}$ can again be based on the Gradient of the REML criterion specified in Equation (\ref{eq:laplace_reml_grad}) but computed, as suggested in the previous paragraph and in Section \ref{sec:GAM}, assuming that $\forall~\lambda_r \in \boldsymbol{\lambda}~\frac{\partial \mathbf{H}}{\partial \lambda_r} = \mathbf{0}$.

\subsubsection{Implementation in MSSM}

The \texttt{GAMMLSS} class in MSSM was designed specifically to support estimation of GAMMLSS models and currently supports Gaussian (mean and standard deviation), Multinomial, and Gamma (mean and scale) distributions. The class relies on the derivative transformation strategy, described by \Textcite{wood_smoothing_2016} and mentioned briefly in the previous section, to obtain $\boldsymbol{\nabla}_{\boldsymbol{\beta}}^{\mathcal{L}}$ and $\mathbf{H}$ from the first and (mixed) second-order partial derivatives of a specific distribution with respect to its parameters (e.g., $\mu$ and $\phi$ for the Gamma distribution). To implement an additional distribution for GAMMLSS modeling, users need to implement the \texttt{GAMLSSFamily} class. The latter requires users to implement code to evaluate the aforementioned partial derivatives, which are widely known for a large amount of distributions, and a function to compute the log-likelihood. This is facilitated through robust implementations of many distributions in Python packages like Scipy \parencite{virtanen_scipy_2020} or Numpy \parencite{harris_array_2020}.

The \texttt{GSMM} class was designed to support the most generic kind of smooth model. Generally, users only need to implement the \texttt{GSMMFamily} class. The latter requires users to implement three methods to compute the log-likelihood, $\boldsymbol{\nabla}_{\boldsymbol{\beta}}^{\mathcal{L}}$, and $\mathbf{H}$. By default, MSSM only passes $\hat{\boldsymbol{\beta}}$, $\mathbf{y}$, the link functions $g_m$, and the model matrices $\mathbf{X}^m$ to theses methods whenever they are evaluated. However, the constructor of the \texttt{GSMMFamily} class supports extra arguments, which can readily be accessed from within the methods. Currently, estimation of cox proportional hazard models is facilitated through the \texttt{GSMMFamily} class. Like the \texttt{GAMM} class, the \texttt{GAMMLSS} and \texttt{GSMM} classes both implement methods to facilitate the post-estimation tasks described in Section \ref{sec:AM} for GAMMLSS and even more generic smooth models respectively. Documentation for these classes is available at \url{https://jokra1.github.io/mssm} and contains example code showing how to implement and estimate generic smooth models in MSSM.

To summarize, the theoretical framework supporting estimation and regularization of sparse mixed general smooth models is now in place: Algorithm (\ref{alg:sparse_gamm}) enables this for even the most generic smooth models \parencite[e.g.,][]{wood_smoothing_2016}, with a ``non-standard'' but regular log-likelihood (e.g., for a Cox proportional hazard model) as discussed in the Background section. However, for estimation to remain memory efficient for more general models, $\mathbf{H}$ needs to be sparse and researchers have to produce efficient and stable code so that it can be evaluated at each iteration of Algorithm (\ref{alg:gammlss_beta}). Unfortunately, $\mathbf{H}$ will often not be sparse for more generic models. Even when it is, the requirement to implement efficient code to evaluate it still places a steep burden on the researcher wanting to estimate such models. Therefore we argued that it would be desirable to enable estimation and automatic regularization of generic smooth models without requiring access to the exact Hessian. In the following section, we develop an algorithm that achieves this, by taking a ``quasi-Newton'' approach \parencite[e.g.,][]{nocedal_numerical_2006}. As discussed in the next section, this facilitates estimation of both $\boldsymbol{\beta}$ and $\boldsymbol{\lambda}$ based on the Gradient of the log-likelihood alone. Estimation thus remains memory efficient, independent of whether $\mathbf{H}$ is sparse or not.

\section{A Limited-memory Quasi-EFS Update for $\boldsymbol{\lambda}$}\label{sec:lqefs}

In the context of smooth models, quasi-Newton approaches have previously been used to enable estimation of $\boldsymbol{\beta}$ but also in place of an exact Newton update to $\boldsymbol{\lambda}$ \parencite[e.g.,][]{wood_smoothing_2016,pya_shape_2015,wood_inference_2020}. However, the quasi-Newton update to $\boldsymbol{\lambda}$ still requires computation of the Gradient of the REML criterion (e.g., Equation (\ref{eq:laplace_reml_grad})) and thus up to third-order derivatives of $\mathcal{L}$. In contrast, quasi-Newton estimation of $\boldsymbol{\beta}$ only requires evaluating the Gradient $\nabla_{\boldsymbol{\beta}}^\mathcal{L}$ of the log-likelihood, and thus has the desired theoretical complexity we hope to achieve -- but for an update to $\boldsymbol{\lambda}$. In this section we develop a limited-memory quasi-Newton EFS (L-qEFS) update to $\boldsymbol{\lambda}$ than retains the same low theoretical complexity of the quasi-Newton update to $\boldsymbol{\beta}$ and remains memory-efficient, independent of whether the Hessian of the log-likelihood has a sparse structure or not. Before considering how this update can be derived in Sections \ref{sec:lqefsBFGS} and \ref{sec:lqefsSR1}, the next section briefly reviews the general idea behind quasi-Newton approaches in the context of estimating $\boldsymbol{\beta}$ for smooth models \parencite[for a more detailed introduction we refer to][]{dennis_numerical_1996,nocedal_numerical_2006}. Finally, Section \ref{sec:lqefsInPractice} addresses how the update can be used in practice and as part of typical post-estimation tasks (e.g., credible interval computation, posterior sampling). 

\subsection{Quasi-Newton Estimation of $\boldsymbol{\beta}$}

Quasi-Newton approaches like the popular BFGS update \parencite[e.g.,][]{nocedal_numerical_2006} typically accumulate an approximation matrix $\hat{\mathbf{V}}$ that is substituted for the inverse of the exact negative Hessian (i.e., for $\mathbf{V}=\mathcal{H}^{-1}$ in the context of estimating $\boldsymbol{\beta}$ for a smooth model) to compute the Newton update, which in the literature is commonly denoted by $\mathbf{s}$ (instead of $\Delta_{\boldsymbol{\beta}}$ used in previous sections of this paper), used to update the current coefficient estimate $\hat{\boldsymbol{\beta}}$. The quasi-Newton update is usually paired with a more specific type of line search, which scales the step by a factor $\alpha$, to ensure that sufficient progress is made on the specific optimization problem \parencite[see][]{nocedal_numerical_2006}.

Independent of the specific method, the update $\hat{\mathbf{V}}^{i}$ to the current approximation matrix $\hat{\mathbf{V}}^{i-1}$ is typically chosen to maintain two properties shared with the actual inverse of the negative Hessian. Specifically, the update is usually chosen from the set of \emph{symmetric solutions} to the ``\emph{secant equation}'' $\hat{\mathbf{V}}^{i}\boldsymbol{\nu} = \alpha\mathbf{s}$, where $\boldsymbol{\nu}= -\nabla_{\hat{\boldsymbol{\beta}} + \alpha\mathbf{s}}^{\mathcal{L}_\lambda} + \nabla_{\hat{\boldsymbol{\beta}}}^{\mathcal{L}_\lambda}$ is equal to the difference in the negative Gradient of the penalized log-likelihood after the (scaled) step $\mathbf{s}$ was used to update the current coefficient estimate $\hat{\boldsymbol{\beta}}$ \parencite[e.g.,][]{nocedal_numerical_2006}. The focus on the secant equation can be motivated by considering a second-order Taylor expansion of the negative (penalized) log-likelihood to be optimized (e.g., $-\mathcal{L}_\lambda$) at the current estimate $\hat{\boldsymbol{\beta}}$, which results in $-\mathcal{L}_\lambda(\boldsymbol{\beta}) = $

\begin{equation}\label{eq:TaylorLLK}
    -\mathcal{L}_\lambda(\hat{\boldsymbol{\beta}}) - \left[\nabla_{\hat{\boldsymbol{\beta}}}^{\mathcal{L}_\lambda}\right]^\top( \boldsymbol{\beta}-\hat{\boldsymbol{\beta}}) + \frac{1}{2}( \boldsymbol{\beta}-\hat{\boldsymbol{\beta}})^\top\mathcal{H}( \boldsymbol{\beta}-\hat{\boldsymbol{\beta}}) + \epsilon,
\end{equation}

where $\nabla_{\hat{\boldsymbol{\beta}}}^{\mathcal{L}_\lambda}$ denotes the Gradient of the penalized log-likelihood at estimate $\hat{\boldsymbol{\beta}}$ and $\epsilon$ denotes the error term remaining for non-quadratic log-likelihood functions \parencite[e.g.,][]{nocedal_numerical_2006}. For $\boldsymbol{\beta}  = \hat{\boldsymbol{\beta}} + \alpha\mathbf{s}$, Equation (\ref{eq:TaylorLLK}) can be treated as a function of the scaled step $\alpha\mathbf{s}$ instead. Computing the Gradient of Equation (\ref{eq:TaylorLLK}) with respect to the latter followed by some rearranging of terms then again results approximately in the secant equation $\mathcal{H}^{-1}\left[-\nabla_{\hat{\boldsymbol{\beta}} + \alpha\mathbf{s}}^{\mathcal{L}_\lambda} + \nabla_{\hat{\boldsymbol{\beta}}}^{\mathcal{L}_\lambda}\right] \approx \alpha\mathbf{s}$, which provides the justification for choosing the quasi-Newton approximation matrix from the set of exact solutions to this equation \parencite[e.g.,][]{nocedal_numerical_2006}. Different quasi Newton methods then impose different additional constraints on the solution $\hat{\mathbf{V}}^{i}$. The popular BFGS update for example chooses $\hat{\mathbf{V}}^{i}$ to be the closest candidate to $\hat{\mathbf{V}}^{i-1}$, in terms of a weighted Frobenius norm, from the set of solutions satisfying the secant equation and further ensures that the approximation matrix remains positive definite \parencite[e.g.,][]{nocedal_numerical_2006}.

Once a strategy to compute the update $\hat{\mathbf{V}}^{i}$ has been chosen, the quasi-Newton update can be implemented almost exactly like the Newton routine used to obtain $\hat{\boldsymbol{\beta}}$ shown in Algorithm (\ref{alg:gammlss_beta}). The necessary steps are outlined in Algorithm (\ref{alg:qdNewton}), assuming that a BFGS update is used to approximate the Hessian and assuming that any scale parameter $\phi$ of the log-likelihood to be optimized is parameterized via a coefficient in $\boldsymbol{\beta}$ so that no scaling of $\boldsymbol{\lambda}$ takes place (or alternatively, that $\phi$ already has been used to scale $\boldsymbol{\lambda}$).

\begin{algorithm}
\caption{\textit{Direct Quasi Newton Coefficient update for Smooth Models}}\label{alg:qdNewton}
\begin{algorithmic}[1]
\INPUT Vector of observations $\mathbf{y}$, current coefficient estimate $\hat{\boldsymbol{\beta}}$, penalty matrix $\mathbf{S}_\lambda$ based on current estimate $\boldsymbol{\lambda}$

\State Compute $\nabla_{\boldsymbol{\beta}}^{\mathcal{L}_\lambda 1} = \partial \mathcal{L}_{\lambda} /\partial \boldsymbol{\beta}|_{\hat{\boldsymbol{\beta}}} = \partial \mathcal{L} /\partial \boldsymbol{\beta}|_{\hat{\boldsymbol{\beta}}} - \mathbf{S}_\lambda\hat{\boldsymbol{\beta}}$

\For{$i = 1, 2, ...$}
    \If{i > 1}
        \State Exit the loop if converged
    \EndIf
    
    \State Compute $\mathbf{s}_i =\hat{\mathbf{V}}^{i-1}[-\nabla_{\boldsymbol{\beta}}^{{\mathcal{L}_\lambda i}}]$ 
    \State Set $\mathbf{s}_i = \alpha\mathbf{s}_i$, $\alpha$ meets the Wolfe conditions for $\mathcal{L}_\lambda$
    \State Set $\boldsymbol{\hat{\beta}}=\boldsymbol{\hat{\beta}} + \mathbf{s}_i$
    \State Compute $\nabla_{\boldsymbol{\beta}}^{\mathcal{L}_\lambda i+1}$ by repeating step 1 at new $\boldsymbol{\hat{\beta}}$
    \State Set $\boldsymbol{\nu}_i = -\nabla_{\boldsymbol{\beta}}^{\mathcal{L}_\lambda i+1} + \nabla_{\boldsymbol{\beta}}^{\mathcal{L}_\lambda i}$
    \State Update $\hat{\mathbf{V}}^{i}$ using $(\boldsymbol{\nu}_i,\mathbf{s}_i)$ for example via the BFGS update
\EndFor

\end{algorithmic}
\end{algorithm}

Step 6 in the algorithm constitutes the quasi-newton update to the coefficient vector. As mentioned, a line search is then performed (step 7), finding a scaling factor $\alpha$ that here satisfies the Wolfe conditions \parencite[e.g.,][]{nocedal_numerical_2006}, to finalize the step computation. Apart from enforcing step-length control for $\boldsymbol{\beta}$, choosing $\alpha$ based on the Wolfe conditions also ensures the aforementioned positive-definiteness of the approximate inverse of the negative Hessian of the log-likelihood $\hat{\mathbf{V}}^{i}$, when using the BFGS update \parencite[see][for details]{nocedal_numerical_2006}. Finally, given the step $\mathbf{s}_i$ and $\boldsymbol{\nu}_i$, again denoting the change in the negative Gradient of the penalized log-likelihood after taking the (scaled) step $\mathbf{s}_i$, $\hat{\mathbf{V}}^{i}$ is computed via the BFGS update \parencite[again, we refer to][and Section \ref{sec:lqefsBFGS} in this paper for details on the update]{nocedal_numerical_2006}.

The update outlined in Algorithm (\ref{alg:qdNewton}) is attractive, not just because it only requires information about the Gradient and log-likelihood, but also from a computational perspective: because the BFGS update allows to maintain an approximation of the inverse of the negative penalized Hessian directly, finding the quasi Newton step requires only a simple product between the approximation matrix and Gradient vector. The fact that using the BFGS update enforces positive definiteness of the approximate inverse $\hat{\mathbf{V}}^{i}$ is also appealing, because - assuming that the model is not miss-specified and that no parameters are unidentifiable - the true negative Hessian of the penalized likelihood should itself be positive definite at the converged estimate $\hat{\boldsymbol{\beta}}$ \parencite[e.g.,][]{wood_smoothing_2016}. As discussed in Section \ref{sec:GAMMLSS} and by \Textcite{wood_smoothing_2016}, this does not necessarily hold during early iterations of estimation. However, in those cases the exact Newton update also falls back to perturbing the estimate for $\mathcal{H}$ with a multiple of the identity matrix to ensure positive definiteness at every iteration (e.g., step 7 of Algorithm (\ref{alg:gammlss_beta})). Hence, maintaining a consistently positive definite approximation to $\hat{\mathbf{V}}^{i}$ throughout all iterations of Algorithm (\ref{alg:qdNewton}) is no less appropriate.

Considering that $\hat{\mathbf{V}}^{i}$ is used as a replacement for $\mathbf{V}$ to compute the Newton step, it might be tempting to substitute $\hat{\mathbf{V}}$ -- the final approximation matrix -- for the exact inverse in the EFS update to turn it into the desired quasi-Newton update requiring only the Gradient of the log-likelihood. Unfortunately, this is unlikely to result in reasonable updates to $\boldsymbol{\lambda}$. To see this, consider that there is absolutely no guarantee that the approximation to $\hat{\mathbf{V}}$ can be as neatly decomposed as the exact inverse $\mathbf{V} = \mathcal{H}^{-1} =  [\mathbf{H} + \mathbf{S}_\lambda]^{-1}$: the quasi-Newton update does not include any obvious mechanism that would ensure that $\hat{\mathbf{V}}^{-1} - \mathbf{S}_\lambda$ retains information about the curvature of the log-likelihood, which would be expected from a good approximation to the negative Hessian of a quadratic log-likelihood $\mathbf{H}$ \parencite[e.g.,][]{nocedal_numerical_2006}.

Matters are made worse by the fact that $\hat{\mathbf{V}}^{-1} - \mathbf{S}_\lambda$ can easily become indefinite. As discussed in the previous section, this would clash with Theorem 1 from \Textcite{wood_generalized_2017}, requiring $\mathbf{H}$ to be at least \emph{positive semi-definite} to ensure that $tr(\mathbf{S}_\lambda^-\mathbf{S}^r) > tr([\mathbf{H} + \mathbf{S}_\lambda]^{-1}\mathbf{S}^r)$ and the EFS update to be defined. Indeed, when \Textcite{wood_generalized_2017} speculated about the possibility of using a quasi-Newton approach to the EFS update, they suggested that this would require a quasi-Newton approximation to the negative Hessian of the log-likelihood $\mathbf{H}$, rather than $\mathcal{H}$.

This suggests using an ``indirect'' quasi Newton update, which, in contrast to the ``direct update'' outlined in Algorithm (\ref{alg:qdNewton}), maintains and works with a positive definite approximation to the inverse of the negative Hessian $\hat{\mathbf{H}}$ of the un-penalized log-likelihood instead. Estimation of $\boldsymbol{\beta}$ still remains possible, because it is straightforward to approximate the negative Hessian of the penalized log-likelihood $\mathcal{H}$ from an approximation to $\mathbf{H}$. The next section outlines the steps necessary as part of such an indirect update, and illustrates remaining challenges that have to be addressed if $\hat{\mathbf{H}}$ is to be used as part of a quasi-EFS update.

\subsection{Indirect Quasi-Newton estimation of $\boldsymbol{\beta}$}

Algorithm (\ref{alg:qNewton}) outlines the steps necessary as part of an indirect quasi-Newton update to estimate $\boldsymbol{\beta}$, again assuming that a BFGS update is used. At every iteration of the algorithm, a first step $\mathbf{s}'_{i}$ is taken away from the current estimate of the coefficients $\hat{\boldsymbol{\beta}}$ towards the maximum of the \emph{log-likelihood} $\mathcal{L}$. This step is used to update an approximation to the inverse of the negative Hessian of the log-likelihood $\hat{\mathcal{I}}^{i-1}$ at $\hat{\boldsymbol{\beta}}$, as well as the approximation of the negative Hessian of the log-likelihood $\hat{\mathbf{H}}^{i-1}$ (steps 5-10 in Algorithm (\ref{alg:qNewton})). Note, that it is not necessary to explicitly compute the inverse $[\hat{\mathcal{I}}^{i-1}]^{-1}$ to arrive at $\hat{\mathbf{H}}^{i-1}$. Instead, the BFGS update can be used to simultaneously maintain approximations for both $\hat{\mathcal{I}}^{i}$ and $\hat{\mathbf{H}}^{i}$ \parencite[e.g.,][]{nocedal_numerical_2006}.

\begin{algorithm}
\caption{\textit{Indirect Quasi Newton Coefficient update for Smooth Models}}\label{alg:qNewton}
\begin{algorithmic}[1]
\INPUT Vector of observations $\mathbf{y}$, current coefficient estimate $\hat{\boldsymbol{\beta}}$, penalty matrix $\mathbf{S}_\lambda$ based on current estimate $\boldsymbol{\lambda}$

\For{$i = 1, 2, ...$}
    \If{i > 1}
        \State Exit the loop if converged
    \EndIf
    \State Compute $\nabla_{\boldsymbol{\beta}}^{\mathcal{L} 1} = \partial \mathcal{L} /\partial \boldsymbol{\beta}|_{\hat{\boldsymbol{\beta}}}$
    \State Compute $\mathbf{s}'_{i} =\hat{\mathcal{I}}^{i-1}[-\nabla_{\boldsymbol{\beta}}^{\mathcal{L}1}]$
    \State Set $\mathbf{s}'_i = \alpha'\mathbf{s}'_i$, $\alpha'$ meets the Wolfe conditions for $\mathcal{L}$
    \State Compute $\nabla_{\boldsymbol{\beta}}^{\mathcal{L}2} = \partial \mathcal{L} /\partial \boldsymbol{\beta}|_{\hat{\boldsymbol{\beta}}+\mathbf{s}'_{i}}$
    \State Set $\boldsymbol{\nu}'_{i} = -\nabla_{\boldsymbol{\beta}}^{\mathcal{L}1} + \nabla_{\boldsymbol{\beta}}^{\mathcal{L}2}$
    \State Update $\hat{\mathbf{H}}^{i}$ and $\hat{\mathcal{I}}^{i}$ using $(\boldsymbol{\nu}'_{i},\mathbf{s}'_{i})$ for example via the\linebreak\hspace*{1.5em}BFGS update
    \State Form $\hat{\mathbf{V}}'^{i} = [\hat{\mathcal{H}^i}]^{-1} =  [\hat{\mathbf{H}}^{i} + \mathbf{S}_{\lambda}]^{-1}$
    \State Compute $\nabla_{\boldsymbol{\beta}}^{\mathcal{L}_\lambda i} = \nabla_{\boldsymbol{\beta}}^{\mathcal{L}1}  - \mathbf{S}_\lambda\hat{\boldsymbol{\beta}}$
    \State Compute $\mathbf{s}_i =\hat{\mathbf{V}}'^{i}[-\nabla_{\boldsymbol{\beta}}^{\mathcal{L}_\lambda i}]$ 
    \State Set $\mathbf{s}_i = \alpha\mathbf{s}_i$, $\alpha$ meets the Wolfe conditions for $\mathcal{L}_\lambda$
    \State Set $\boldsymbol{\hat{\beta}}=\boldsymbol{\hat{\beta}} + \mathbf{s}_i$
\EndFor

\end{algorithmic}
\end{algorithm}

Achieving this largely requires repeating the steps outlined in Algorithm (\ref{alg:qdNewton}), now designed as if optimizing $\mathcal{L}$ instead of $\mathcal{L}_\lambda$. Specifically, the negative Gradient $-\nabla_{\boldsymbol{\beta}}^{\mathcal{L}1}$ of the log-likelihood at the current coefficient estimate $\hat{\boldsymbol{\beta}}$ is evaluated in a first step. $\mathbf{s}'_{i}$ is then obtained by computing the quasi-Newton update $\mathbf{s}'_{i} =\hat{\mathcal{I}}^{i-1}[-\nabla_{\boldsymbol{\beta}}^{\mathcal{L}1}]$. After completing a first line-search to scale $\mathbf{s}'_{i}$, the BFGS update to $\hat{\mathbf{H}}^{i}$ and $\hat{\mathcal{I}}^{i}$ can again be performed, based on $\mathbf{s}'_{i}$ and the difference in the negative Gradient of the log-likelihood $\boldsymbol{\nu}'_{i} = -\nabla_{\boldsymbol{\beta}}^{\mathcal{L}1} + \nabla_{\boldsymbol{\beta}}^{\mathcal{L}2}$ after taking it.

Subsequently, the approximation to the negative Hessian of the log-likelihood is used to approximate the inverse of the negative Hessian of the \emph{penalized} log-likelihood. Specifically, the latter is taken to be $\hat{\mathbf{V}}'^{i} = [\hat{\mathcal{H}^i}]^{-1} =  [\hat{\mathbf{H}}^{i} + \mathbf{S}_{\lambda}]^{-1}$. $\hat{\mathbf{H}}^{i}$ is chosen here because it satisfies the secant equation for the step $\mathbf{s}'_{i}$ taken with origin $\hat{\boldsymbol{\beta}}$.

Importantly, the inverse necessary to obtain $\hat{\mathbf{V}}'^{i}$ is guaranteed to be defined, because $\hat{\mathcal{H}}^i$ is positive definite since $\hat{\mathbf{H}}^{i}$ and $\mathbf{S}_\lambda$ are positive definite \parencite[by the BFGS update; ][]{nocedal_numerical_2006} and positive semi-definite respectively. Together with the Gradient of the penalized likelihood $\nabla_{\boldsymbol{\beta}}^{\mathcal{L}_\lambda i}$, computation of which can re-use the first of the two previously computed Gradients of $\mathcal{L}$, this enables computation of a second quasi-Newton update $\mathbf{s}_{i}$, this time aimed at the maximum of the actual objective -- the penalized log-likelihood $\mathcal{L}_\lambda$ (steps 11-15 in Algorithm (\ref{alg:qNewton})). The positive definiteness of $\hat{\mathbf{V}}'^{i}$ ensures that $\mathbf{s}_{i}$ is an ascent direction for $\mathcal{L}_\lambda$, while the scaling by $\alpha$, obtained from the second line search, ensures that $\mathcal{L}_\lambda$ actually increases in practice \parencite[sufficiently;][]{nocedal_numerical_2006}.

The indirect update outlined in Algorithm (\ref{alg:qNewton}), while computationally more costly, ensures that the approximate inverse of the negative Hessian of the penalized log-likelihood remains decomposable into the inverse of a sum of a positive definite and positive semi-definite matrix, i.e., $\hat{\mathbf{V}}'^{i} = [\hat{\mathcal{H}^i}]^{-1} =  [\hat{\mathbf{H}}^{i} + \mathbf{S}_{\lambda}]^{-1}$ where $\mathbf{x}^\top\hat{\mathbf{H}}^{i}\mathbf{x} > 0$ and $\mathbf{x}^\top\mathbf{S}_{\lambda}\mathbf{x} \geq 0$ for any non-zero vector $\mathbf{x} \in \mathbb{R}^{N_p}$. This guarantees that Theorem 1 of \Textcite{wood_generalized_2017} will hold whenever Algorithm (\ref{alg:qNewton}) terminates, which in turn ensures that the EFS update would be defined if the final approximation $\hat{\mathbf{V}}'$ would be substituted for the exact inverse. In principle, due to the way it is chosen, the final approximation $\hat{\mathbf{H}}$ is also more likely to be a reasonable approximation to the Hessian of the log-likelihood at convergence compared to $\hat{\mathbf{V}}^{-1} - \mathbf{S}_\lambda$ from the previous section.

However, while a full rank approximation to (the inverse of) the negative Hessian of the penalized log-likelihood, as utilized in Algorithm (\ref{alg:qdNewton}), is generally justifiable, the true $\mathbf{H}$ will generally only be positive semi-definite if multi-level models are to be estimated involving multiple random effects \parencite[e.g.,][]{wood_generalized_2017}. In those cases, enforcing a positive definite approximation could negatively impact the performance of quasi Newton methods \parencite[e.g.,][]{nocedal_numerical_2006}. If the approximation to the (inverse of the) Hessian remains poor, Algorithm (\ref{alg:qNewton}) could result in poor estimates for the coefficients $\hat{\boldsymbol{\beta}}$ and might take a long time to converge, for example if very small values for either $\alpha$ or $\alpha'$ are consistently necessary to ensure that the steps taken actually result in a sufficient increase of the penalized or un-penalized likelihood respectively \parencite[e.g.,][]{nocedal_numerical_2006}. 

Using a poor approximation to the (inverse of the) Hessian as part of a qEFS update would all but guarantee poor estimates for the regularization parameters $\boldsymbol{\lambda}$ as well. For once, if Algorithm (\ref{alg:qdNewton}) consistently fails to converge to the maximizer $\hat{\boldsymbol{\beta}}$ of $\mathcal{L}_\lambda$, this will introduce bias into the EFS update through the magnitude of the update's denominator $\hat{\boldsymbol{\beta}}^\top\mathbf{S}^r\hat{\boldsymbol{\beta}}$. Similarly, if the final approximation $\hat{\mathbf{H}}$ to the Hessian is poor, in the sense that the log-determinant of $\hat{\mathbf{V}}' =  [\hat{\mathbf{H}} + \mathbf{S}_{\lambda}]^{-1}$ is far from the log-determinant of the inverse of the true penalized Hessian, then there is little hope that $tr([\hat{\mathbf{H}} + \mathbf{S}_\lambda]^{-1}\mathbf{S}^r)$ will be close in magnitude to the exact derivative based on $\mathbf{H}$. Even if a positive definite approximation results in good estimates for $\hat{\boldsymbol{\beta}}$, enforcing this still biases the log-determinant of $\hat{\mathbf{V}}'$ and thus computation of $tr([\hat{\mathbf{H}} + \mathbf{S}_\lambda]^{-1}\mathbf{S}^r)$. All of these can contribute to poor estimates of the direction the next update to $\boldsymbol{\lambda}$ should take. A final problem is that, in the absence of problem-specific knowledge, any quasi-Newton approximation to $\mathbf{H}$ or its inverse $\mathbf{V}$ would result in a $N_p*N_p$ matrix that is all but guaranteed to be dense, reducing the feasibility of both Algorithms (\ref{alg:qdNewton}) and (\ref{alg:qNewton}) and a qEFS update for the generic large-scale multi-level models that are of interest here.

In the next sections we show how Algorithms (\ref{alg:qdNewton}) and (\ref{alg:qNewton}) can be combined and modified to address these limitations. In a first step, Section \ref{sec:lqefsBFGS} shows how $\hat{\mathbf{H}}^i$, $\hat{\mathcal{I}}^i$ , and $\hat{\mathbf{V}}^i$ can be parameterized by limited-memory BFGS (L-BFGS) approximations to avoid the steep memory requirement induced by the density of the conventional quasi-Newton approximations \parencite[e.g.,][]{liu_limited_1989,nocedal_updating_1980,nocedal_numerical_2006}. At every iteration, the latter are represented implicitly, based only on the last $N_{V} < N_p$ pairs of vectors $\boldsymbol{\nu}_i$ and $\mathbf{s}_i$. If $N_{V} << N_p$, the latter are much cheaper to store. We then show how a limited memory representation for $\hat{\mathbf{V}}'^i$ can be computed from the limited memory representation of $\hat{\mathbf{H}}^i$, which also enables computation of $tr([\hat{\mathbf{H}} + \mathbf{S}_\lambda]^{-1}\mathbf{S}^r)$ without ever having to accumulate the explicit $N_p*N_p$ representation of $\hat{\mathbf{H}}$. Combined, these steps enable computation of the desired L-qEFS update, which requires only the Gradient of the log-likelihood and remains memory-efficient independent of whether $\mathbf{H}$ is sparse or not.

In Section \ref{sec:lqefsSR1} we address the remaining problems that can arise when $\hat{\mathcal{I}}^{i}$ and $\hat{\mathbf{H}}^{i}$ are forced to remain positive-definite. We show how a limited-memory variant of the alternative Symmetric Rank 1 (SR1) update can be used instead of the L-BFGS update to approximate $\hat{\mathcal{I}}^{i}$ and $\hat{\mathbf{H}}^{i}$ \parencite[e.g.,][]{byrd_representations_1994,nocedal_numerical_2006}. The SR1 update does not enforce positive definiteness of $\hat{\mathbf{H}}^i$ (or $\hat{\mathcal{I}}^i$) and can often achieve better approximations of the true Hessian \parencite[e.g.,][]{nocedal_numerical_2006}. Because the update does not enforce positive definiteness, it is also no longer necessary to find an $\alpha'$ that meets the Wolfe conditions at every iteration of the indirect update. Instead, the latter can be chosen to meet the simpler Armijo conditions \parencite[e.g.,][]{nocedal_numerical_2006} -- which is computationally less expensive. It does however introduce a new problem: the approximations $\hat{\mathbf{H}}^i$ and $\hat{\mathcal{I}}^i$ can become indefinite. To address this, we show how a minimum of positive semi-definiteness can be enforced efficiently for the limited memory representations of $\hat{\mathbf{H}}^i$ and $\hat{\mathcal{I}}^i$, using the implicit eigen-decomposition approach by \Textcite{burdakov_efficiently_2017}.

\subsection{L-qEFS update based on L-BFGS representations}\label{sec:lqefsBFGS}

In this section we show how limited-memory BFGS (L-BFGS) representations can be obtained for $\hat{\mathbf{H}}^i$, $\hat{\mathcal{I}}^i$, $\hat{\mathcal{H}}^i$, and $\hat{\mathbf{V}}^i$ and how these enable efficient computation of $\hat{\mathbf{V}}'^i$ and the trace term $tr([\hat{\mathbf{H}} + \mathbf{S}_\lambda]^{-1}\mathbf{S}^r)$ required for the L-qEFS update. As mentioned in the previous section, there are some problems with using the BFGS update to approximate $\hat{\mathbf{H}}^i$, $\hat{\mathcal{I}}^i$. We will ignore this for a moment and only get back to these issues in the next sub-section, where the SR1 update will be introduced in more detail.

In many ways, the L-BFGS update is similar to the BFGS update, since both updates utilize the update vectors $\boldsymbol{\nu}$ and $\mathbf{s}$ and maintain a positive semi-definite approximation to the (inverse of the) Hessian. The main difference is that the L-BFGS retains a queue of the latest $N_{V}$ pairs of update vectors and constructs an approximation from this set alone \parencite[e.g.,][]{byrd_representations_1994}. Specifically, at every iteration a new pair ($\boldsymbol{\nu}_i$, $\mathbf{s}_i$) enters the queue and, if the queue is full, the oldest pair is discarded. The approximation matrix is then re-computed based on the updated content in the queue. As mentioned, the approximation matrix is never actually represented as a dense $N_p * N_p$ matrix. Instead, quantities involving the (inverse of) the negative Hessian, such as the product necessary to find the Newton step $\mathbf{s}$, are computed directly from the limited-memory or ``compact representation'' of the approximation matrix \parencite[e.g.,][]{byrd_representations_1994,nocedal_numerical_2006,brust_useful_2025}. 

Theorem 2.2 and Equation 3.1 in \Textcite{byrd_representations_1994} define the compact representation for the L-BFGS approximation of the inverse of the negative Hessian. Based on this, we can define the compact representation for $\hat{\mathbf{V}}^i$ at the end of iteration $i$ as $\hat{\mathbf{V}}^i = $

\begin{equation}\label{eq:LbfgsInvHessian}
\hat{\mathbf{V}}^0 + \left[\mathrm{S} ~ ~ \hat{\mathbf{V}}^0\mathrm{Y}\right] \begin{bmatrix}
\mathbf{R}^{-\top}(\mathbf{D} + \mathrm{Y}^\top\hat{\mathbf{V}}^0\mathrm{Y})\mathbf{R}^{-1} & -\mathbf{R}^{-\top}\\
-\mathbf{R}^{-1} & \mathbf{0}\\
\end{bmatrix} \begin{bmatrix}
\mathrm{S}^\top\\
\mathrm{Y}^\top\hat{\mathbf{V}}^0\\
\end{bmatrix}.
\end{equation}

Note, super-scripts indicating the iteration of Algorithm (\ref{alg:qdNewton}) have been omitted on the right-hand side to avoid clutter. In practice, all of the matrix blocks on the right-hand side can be expected to change between iterations \parencite[e.g.,][]{byrd_representations_1994}, so that in general $\hat{\mathbf{V}}^0=[\hat{\mathbf{V}}^0]^i$, $\mathrm{Y}=\mathrm{Y}^i$, $\mathrm{S}=\mathrm{S}^i$, $\mathbf{R}=\mathbf{R}^i$, and $\mathbf{D}=\mathbf{D}^i$. Intuitively, $\hat{\mathbf{V}}^i$ is thus represented by matrix $[\hat{\mathbf{V}}^0]^i$, which is further refined by an additive adjustment term. For efficiency reasons, $[\hat{\mathbf{V}}^0]^i$ is usually chosen to be a multiple $\gamma_i$ of the identity matrix and a popular choice is to set $\gamma_i=\frac{\mathbf{s}_{i}^\top\boldsymbol{\nu}_i}{\boldsymbol{\nu}_{i}^\top\boldsymbol{\nu}_i}$, which aims to approximate an eigenvalue of $\mathbf{V}$ \parencite[e.g.,][]{nocedal_numerical_2006}. The columns of the $N_p*N_\mathbf{V}$ matrices $\mathrm{Y}^i = [\boldsymbol{\nu}_{i-N_\mathbf{V}+1},\boldsymbol{\nu}_{i-N_\mathbf{V}+2},...,\boldsymbol{\nu}_{i}]$ and $\mathrm{S}^i = [\mathbf{s}_{i-N_\mathbf{V}+1},\mathbf{s}_{i-N_\mathbf{V}+2},...,\mathbf{s}_{i}]$ hold the last $N_V$ update vectors. Note, that they will have fewer columns during the first iterations of Algorithm (\ref{alg:qdNewton}). $\mathbf{R}^i$ and $\mathbf{D}^i$ are both of dimension $N_V*N_V$. From \Textcite{byrd_representations_1994}, we have that $\mathbf{R}^i$ is upper-triangular with elements $\mathbf{R}^i_{lj} = \mathbf{s}_{i - N_V + l}^\top~\boldsymbol{\nu}_{i - N_V + j}$ and $\mathbf{D}^i$ is diagonal with elements $\mathbf{D}^i_{ll}=\mathbf{R}^i_{ll}$. Importantly, except for $\hat{\mathbf{V}}^0$, computation of which is trivial, none of the remaining matrices involved in Equation (\ref{eq:LbfgsInvHessian}) actually have to be re-computed entirely at a new iteration of Algorithm (\ref{alg:qdNewton}). Instead, they are all updated recursively to reflect updates to the queue of update vectors \parencite[see][for details on the recursive computations]{byrd_representations_1994}. The compact representation in Equation (\ref{eq:LbfgsInvHessian}) can now be re-expressed as

\begin{equation}\label{eq:LbfgsInvHessian2}
\hat{\mathbf{V}}^i = \hat{\mathbf{V}}^0 + \mathbf{U}\mathbf{B}\mathbf{U}^\top,
\end{equation}

\noindent
where $\mathbf{U}=\left[\mathrm{S} ~ ~ \hat{\mathbf{V}}^0\mathrm{Y}\right]$ is a $N_p*(2N_\mathbf{V})$ matrix,

\begin{equation}\label{eq:LbfgsInvHessian3}
\mathbf{B} = \begin{bmatrix}
\mathbf{R}^{-\top}(\mathbf{D} + \mathrm{Y}^\top\hat{\mathbf{V}}^0\mathrm{Y})\mathbf{R}^{-1} & -\mathbf{R}^{-\top}\\
-\mathbf{R}^{-1} & \mathbf{0}\\
\end{bmatrix}
\end{equation}

\noindent
is a $(2N_V)*(2N_V)$ matrix, and super-scripts indicating iterations of Algorithm (\ref{alg:qdNewton}) have again been omitted to avoid clutter. Based on this, the product between $\hat{\mathbf{V}}$ and any vector $\mathbf{a}$ (e.g., $-\nabla_{\boldsymbol{\beta}}^{\mathcal{L}_\lambda i}$) is simply

\begin{equation}\label{eq:lbfgsProd}
\hat{\mathbf{V}}^i\mathbf{a} = \hat{\mathbf{V}}^0\mathbf{a} + \mathbf{U}\mathbf{B}\mathbf{U}^\top\mathbf{a},
\end{equation}

\noindent
which does not require forming the $N_p * N_p$ matrix $\hat{\mathbf{V}}^i$ explicitly and benefits from sparsity of $\hat{\mathbf{V}}^0$ \parencite[e.g.,][]{byrd_representations_1994}. This enables efficient computation of step 6 in Algorithm (\ref{alg:qdNewton}).

By a straight-forward generalization of Equation (\ref{eq:LbfgsInvHessian}) the dense approximation matrix $\hat{\mathcal{I}}^i$, required by the indirect update described in Algorithm (\ref{alg:qNewton}), can also be replaced with a compact representation. Specifically, we can define

\begin{equation}\label{eq:LbfgsInvUPHessian}
\hat{\mathcal{I}}^i = \mathbf{I}\gamma'_{i} + \mathbf{U'}\mathbf{B'}\mathbf{U'}^\top,
\end{equation}

\noindent
where $\gamma'_{i}$, $\mathbf{U'}$, and $\mathbf{B'}$ are defined as $\gamma_{i}$, $\mathbf{U}$, and $\mathbf{B}$ for Equation (\ref{eq:LbfgsInvHessian}) but based on $\mathbf{s}'$ and $\boldsymbol{\nu}'$, defined in steps 7 and 9 of Algorithm (\ref{alg:qNewton}), rather than $\mathbf{s}$ and $\boldsymbol{\nu}$. Equation (\ref{eq:lbfgsProd}) then again enables efficient computation of the quasi-Newton step towards the maximum of the log-likelihood required for step 6 in Algorithm (\ref{alg:qdNewton}).

\Textcite{byrd_representations_1994} also show how a compact representation for the inverse of a compact L-BFGS representation, like the ones defined in Equations (\ref{eq:LbfgsInvHessian2}) and (\ref{eq:LbfgsInvUPHessian}), can be obtained efficiently. This permits a definition for the compact representation of the L-BFGS approximation of the negative Hessian of the log-likelihood,

\begin{equation}\label{eq:LbfgsUPHessian}
\hat{\mathbf{H}}^i = \mathbf{I}\frac{1}{\gamma'_{i}} + \mathbf{Q'}\mathbf{C'}\mathbf{Q'}^\top.
\end{equation}

We omit the details on how to compute $\mathbf{Q'}$ and $\mathbf{C'}$ -- which can be found for example in Equation 2.17 of \Textcite{byrd_representations_1994} -- but note that computing them has a complexity comparable to computing $\mathbf{U'}$ and $\mathbf{B'}$ and also relies entirely on the content of the queue holding $\mathbf{s}'$ and $\boldsymbol{\nu}'$.

For the second quasi-Newton step in the indirect update, towards the maximum of the penalized log-likelihood (step 13 in Algorithm (\ref{alg:qNewton})), we also require $\hat{\mathbf{V}}'^{i} = [\hat{\mathcal{H}^i}]^{-1} =  [\hat{\mathbf{H}}^{i} + \mathbf{S}_{\lambda}]^{-1}$. In fact, we need a compact representation for $\hat{\mathbf{V}}'^{i}$ to ensure that the update remains efficient. Based on Equation (\ref{eq:LbfgsUPHessian}), it is clear that

\begin{equation}\label{eq:LbfgsHessian}
\hat{\mathcal{H}}^i = \mathcal{H}^0 + \mathbf{Q'}\mathbf{C'}\mathbf{Q'}^\top,
\end{equation}

\noindent
where $\mathcal{H}^0 = \mathbf{I}\frac{1}{\gamma'_{i}} + \mathbf{S}_{\lambda}$. Using the modified Woodburry identity from \Textcite{henderson_deriving_1981}, the desired compact representation can then be computed as

\begin{equation}\label{eq:LbfgsInvHessian5}
\begin{split}
    \hat{\mathbf{V}}'^{i} &= [\hat{\mathcal{H}^i}]^{-1}\\&=[\mathcal{H}^0]^{-1} - \mathbf{M}\mathbf{A}^{-1}\mathbf{N},
\end{split}
\end{equation}

\noindent
where $\mathbf{M}=[\mathcal{H}^0]^{-1}\mathbf{Q'}$, $\mathbf{A}=\mathbf{I} + \mathbf{C'}\mathbf{Q'}^\top[\mathcal{H}^0]^{-1}\mathbf{Q'}$, and $\mathbf{N}=\mathbf{C'}\mathbf{Q'}^\top[\mathcal{H}^0]^{-1}$. $\mathbf{A}$, while not symmetric, is non-singular, so the inverse $\mathbf{A}^{-1}$ is guaranteed to be defined at every iteration of Algorithm (\ref{alg:qNewton}) \parencite[see][]{henderson_deriving_1981}. The required inverse is also inexpensive to compute, because $\mathbf{A}$ is only of dimension $(2N_\mathbf{V}) * (2N_\mathbf{V})$ \parencite[][]{henderson_deriving_1981}. While the inverse $[\mathcal{H}^0]^{-1}$ is of $N_p *N_p$ matrix $\mathcal{H}^0$, the latter continues to be symmetric and sparse even for large multi-level models and can thus again be computed quickly by means of a Cholesky decomposition that pivots for sparsity. Based on the result defined in Equation (\ref{eq:LbfgsInvHessian5}) we can again use Equation (\ref{eq:lbfgsProd}) to compute the quasi-Newton step towards the maximum of the penalized likelihood efficiently (i.e., in Algorithm (\ref{alg:lqEFS})), avoiding explicit formation of $\hat{\mathbf{V}}'^{i}$.

At this point, the approximation matrices in both the direct and indirect update can be replaced with compact representations and we can address the computation of $tr(\mathbf{\hat{\mathbf{V}}}^{'i}\mathbf{S}^r)$, required to compute the qEFS update after Algorithm (\ref{alg:qNewton}) terminates. Given the final compact representation $\hat{\mathbf{V}}'$ and considering Equation (\ref{eq:lbfgsProd}), this trace term can be computed as

\begin{equation}\label{eq:lbfgsTrace}
tr(\mathbf{\hat{\mathbf{V'}}}\mathbf{S}^r)=\sum_l^{N_p} [\mathcal{H}^0]^{-1}_l[\mathbf{S}^r]_l^\top - \mathbf{M}_l\mathbf{A}^{-1}(\mathbf{N}[\mathbf{S}^r]_l^\top),
\end{equation}

\noindent
where $[\mathcal{H}^0]^{-1}_l$ denotes row $l$ of matrix $[\mathcal{H}^0]^{-1}$, $[\mathbf{S}^r]^\top_l$ denotes column $l$ of matrix $\mathbf{S}^r$, and $\mathbf{M}_l$ denotes row $l$ of matrix $\mathbf{M}$. Note that in the context of mixed sparse models, the sum will never have to be computed over all $N_p$ elements, since matrices like $\mathbf{S}^r$ will only contain a small handful of non-zero columns, the indices of which will generally be known in advance. Additionally, the matrix vector product $\mathbf{N}[\mathbf{S}^r]_l^\top$ will fully benefit from the small number of non-zero rows in non-zero columns $[\mathbf{S}^r]^\top_l$ of $\mathbf{S}^r$, so that each trace -- and ultimately the L-qEFS update -- can be computed very efficiently.

The computations discussed so far are sufficient for an L-qEFS update that relies exclusively on the L-BFGS update. However, before the update can be evaluated, Algorithms (\ref{alg:qdNewton}) and (\ref{alg:qNewton}), both modified to utilize the compact representations derived in this section, have to be combined. In a first step, the direct quasi Newton update outlined in Algorithm (\ref{alg:qdNewton}) has to be iterated until convergence in $\hat{\boldsymbol{\beta}}$ is (nearly) reached. For this purpose Algorithm (\ref{alg:qdNewton}) can be used largely as is, also in combination with the (L) BFGS update, because maintaining a positive definite approximation $\hat{\mathbf{V}}^i$ of the inverse of the negative penalized Hessian is less problematic. Since previous research revealed that the direct Newton update to obtain $\hat{\boldsymbol{\beta}}$ works well in practice, use of Algorithm (\ref{alg:qdNewton}) for this first step is generally justified \parencite[e.g.,][]{pya_shape_2015,wood_smoothing_2016}. Subsequently, at least $N_{V}$ additional iterations of Algorithm (\ref{alg:qdNewton}) have to be completed to obtain limited memory approximations to $\hat{\mathbf{H}}$ and $\hat{\mathcal{I}}$ at $\hat{\boldsymbol{\beta}}$. In combination with Equations (\ref{eq:LbfgsInvHessian5}) and (\ref{eq:lbfgsTrace}) these approximations can then be used to compute the actual L-qEFS update efficiently. The necessary steps have been collected in Algorithm (\ref{alg:lqEFS}).

\begin{algorithm}
\caption{\textit{L-qEFS Update for Smooth Models}}\label{alg:lqEFS}
\begin{algorithmic}[1]
\INPUT Vector of observations $\mathbf{y}$, current coefficient estimate $\hat{\boldsymbol{\beta}}$, individual penalty matrices $\mathbf{S}^r$ where $r \in \{1,2,...,N_\lambda\}$, penalty matrix $\mathbf{S}_\lambda$ based on current estimate $\boldsymbol{\lambda}$,
the number $N_{V}$ of vectors to store for the approximation $\hat{\mathcal{H}}^{-1}$, and an upper limit of iterations $N_i$

\State Compute $\nabla_{\boldsymbol{\beta}}^{\mathcal{L}_\lambda 1} = \partial \mathcal{L}_{\lambda} /\partial \boldsymbol{\beta}|_{\hat{\boldsymbol{\beta}}} = \partial \mathcal{L} /\partial \boldsymbol{\beta}|_{\hat{\boldsymbol{\beta}}} - \mathbf{S}_\lambda\hat{\boldsymbol{\beta}}$

\For{$i = 1, ..., N_i-N_{V}$}
    \If{i > 1}
        \State Exit the loop if converged
    \EndIf
    
    \State Compute $\mathbf{s}_i =\hat{\mathbf{V}}^{i-1}[-\nabla_{\boldsymbol{\beta}}^{\mathcal{L}_\lambda i}]$ 
    \State Set $\mathbf{s}_i = \alpha\mathbf{s}_i$, $\alpha$ meets the Wolfe conditions for $\mathcal{L}_\lambda$
    \State Set $\boldsymbol{\hat{\beta}}=\boldsymbol{\hat{\beta}} + \mathbf{s}_i$
    \State Compute $\nabla_{\boldsymbol{\beta}}^{\mathcal{L}_\lambda i+1}$ by repeating step 1 at new $\boldsymbol{\hat{\beta}}$
    \State Set $\boldsymbol{\nu}_i = -\nabla_{\boldsymbol{\beta}}^{\mathcal{L}_\lambda i+1} + \nabla_{\boldsymbol{\beta}}^{\mathcal{L}_\lambda i}$
    \If{i > $N_{V}$}
        \State Delete $\mathbf{s}_{i - N_{V} + 1}$ and $\boldsymbol{\nu}_{i - N_{V} + 1}$
    \EndIf
    \State Implicitly update $\hat{\mathbf{V}}^{i}$ using $(\boldsymbol{\nu}_i,\mathbf{s}_i)$ via L-BFGS update
\EndFor

\For{$i = 1, ..., N_{V}$ or until convergence is reached}
    \State Compute $\nabla_{\boldsymbol{\beta}}^{\mathcal{L}1} = \partial \mathcal{L} /\partial \boldsymbol{\beta}|_{\hat{\boldsymbol{\beta}}}$
    \State Compute $\mathbf{s}'_{i} =\hat{\mathcal{I}}^{i-1}[-\nabla_{\boldsymbol{\beta}}^{\mathcal{L}1}]$
    \State Set $\mathbf{s}'_{i} = \alpha'\mathbf{s}'_i$, $\alpha'$ meets the Wolfe (for L-BFGS\linebreak\hspace*{1.5em}update) or Armijo (for L-SR1 update) conditions for\linebreak\hspace*{1.5em}$\mathcal{L}$
    \State Compute $\nabla_{\boldsymbol{\beta}}^{\mathcal{L}2} = \partial \mathcal{L} /\partial \boldsymbol{\beta}|_{\hat{\boldsymbol{\beta}}+\mathbf{s}'_{i}}$
    \State Set $\boldsymbol{\nu}'_{i} = -\nabla_{\boldsymbol{\beta}}^{\mathcal{L}1} + \nabla_{\boldsymbol{\beta}}^{\mathcal{L}2}$
    \If{i > $N_{\mathbf{V}}$}
        \State Delete $\mathbf{s}'_{i - N_{V} + 1}$ and $\boldsymbol{\nu}'_{i - N_{V} + 1}$
    \EndIf
    \State Implicitly update $\hat{\mathbf{H}}^{i}$ and $\hat{\mathcal{I}}^{i}$ using $(\boldsymbol{\nu}'_{i},\mathbf{s}'_{i})$ via modified\linebreak\hspace*{1.5em}L-SR1 update or L-BFGS update
    \State Implicitly form $\hat{\mathbf{V}}'^{i} = [\hat{\mathcal{H}^i}]^{-1} =  [\hat{\mathbf{H}}^{i} + \mathbf{S}_{\lambda}]^{-1}$
    \State Compute $\nabla_{\boldsymbol{\beta}}^{\mathcal{L}_\lambda i} = \nabla_{\boldsymbol{\beta}}^{\mathcal{L}1}  - \mathbf{S}_\lambda\hat{\boldsymbol{\beta}}$
    \State Repeat steps 6-8 with $\nabla_{\boldsymbol{\beta}}^{\mathcal{L}_\lambda i}$ and $\hat{\mathbf{V}}'^{i}$ instead of $\hat{\mathbf{V}}^{i-1}$ to\linebreak\hspace*{1.5em}update $\boldsymbol{\hat{\beta}}$, choose $\alpha$ in step 7 to meet the Wolfe (for\linebreak\hspace*{1.5em}L-BFGS update) or Armijo (for L-SR1 update)\linebreak\hspace*{1.5em}conditions for $\mathcal{L}_\lambda$
\EndFor

\State Implicitly compute $tr(\hat{\mathbf{V}}'^{i}\mathbf{S}^r)$ for every $\mathbf{S}^r$ via Equation (\ref{eq:lbfgsTrace}) and use the result to compute the L-qEFS update via Equation (\ref{eq:efs})
\end{algorithmic}
\end{algorithm}

Specifically, steps 1-15 in Algorithm (\ref{alg:lqEFS}) complete the steps implemented in Algorithm (\ref{alg:qdNewton}) but emphasize that $\hat{\mathbf{V}}^i$ is parameterized via a compact representation. Steps 2-15 are iterated until convergence in $\boldsymbol{\beta}$ is (nearly) reached. Conversely, the second phase of the algorithm consists of repeating steps 17-28 in Algorithm (\ref{alg:lqEFS}) which reflect the steps implemented in Algorithm (\ref{alg:qNewton}), again adjusted to emphasize that $\hat{\mathcal{I}}^i$, $\hat{\mathbf{H}}^i$, and $\hat{\mathbf{V}}'^{i}$ are all represented implicitly. Importantly, at least $N_V$ iterations of steps 17-28 are completed to update the approximation of the negative Hessian of the log-likelihood $\hat{\mathbf{H}}^i$ with information collected in the vicinity of $\hat{\boldsymbol{\beta}}$. If convergence is not reached afterwards, more updates can be performed in this second phase theoretically. In that case, a queue needs to be maintained for $(\mathbf{s}'_i,\boldsymbol{\nu}'_i)$ as well, which needs to be updated to reflect discarded and new update vectors (steps 22-24 in Algorithm (\ref{alg:lqEFS})). The final step in Algorithm (\ref{alg:lqEFS}) prepares computation of the actual L-qEFS update by computing $tr(\hat{\mathbf{V}}'^{i}\mathbf{S}^r)$ via Equation (\ref{eq:lbfgsTrace}) for every matrix $\mathbf{S}^r$ required by the model to be estimated.

As discussed earlier, relying on the L-BFGS update to maintain the approximations $\hat{\mathcal{I}}^i$ and $\hat{\mathbf{H}}^i$ has the important benefit that $\hat{\mathbf{H}}^{i}$ remains positive definite, ensuring that the EFS update is defined whenever Algorithm (\ref{alg:lqEFS}) terminates. However, we also discussed potential problems that can result from enforcing $\hat{\mathbf{H}}^i$ to remain positive definite, especially if the true observed Hessian is only positive semi-definite. We suggested that it might thus be more desirable to replace the L-BFGS update in steps 2-15 of Algorithm (\ref{alg:lqEFS}) with a limited-memory variant of the related SR1 update (i.e., the L-SR1 update), which does not enforce positive definiteness. The next section reviews how compact representations can be obtained for the L-SR1 update \parencite[e.g.,][]{byrd_representations_1994} and how these representations can be used as part of a L-qEFS update.

\subsection{L-qEFS update based on L-SR1 Representations}\label{sec:lqefsSR1}

For our review of compact representations for $\hat{\mathcal{I}}^i$ and $\hat{\mathbf{H}}^i$ based on the L-SR1 update, we again follow \Textcite{byrd_representations_1994}. Importantly, while the approximations $\hat{\mathcal{I}}^i$ and $\hat{\mathbf{H}}^{i}$ continue to be symmetric when using the (L) SR1 update, they do not have to remain positive definite but can become indefinite instead \parencite[e.g.,][]{byrd_representations_1994,nocedal_numerical_2006}. We discuss how to address this problem in Section \ref{sec:eigenHI} and for now simply assume that the update results in a positive definite approximation matrix. We rely on Equation 5.9 in the paper of \Textcite{byrd_representations_1994} to define the L-SR1 update for the inverse of the negative Hessian of the log-likelihood as  $\hat{\mathcal{I}}^i = $

\begin{equation}\label{eq:SR1InvHessian}
\mathcal{I}^0 + \left[\mathrm{S} - \mathcal{I}^0\mathrm{Y}\right] \left[\mathbf{R} + \mathbf{R}^\top -  \mathbf{D} -\mathrm{Y}^\top\mathcal{I}^0\mathrm{Y}\right] \left[\mathrm{S} - \mathcal{I}^0\mathrm{Y}\right]^\top,
\end{equation}

\noindent
where $\mathcal{I}^0=\mathbf{I}\gamma'_i$ (or simply $\mathbf{I}$ if $\gamma'_i$ becomes negative), $\mathbf{R}$, $\mathrm{S}$, and $\mathrm{Y}$ are as defined as in Equation (\ref{eq:LbfgsInvHessian}). This enables updated definitions for $\mathbf{U'}=\mathrm{S} - \mathcal{I}^0\mathrm{Y}$ and $\mathbf{B'}=\mathbf{R} + \mathbf{R}^\top - \mathbf{D} -  \mathrm{Y}^\top\mathcal{I}^0\mathrm{Y}$, which can now be substituted for $\mathbf{U'}$ and $\mathbf{B'}$ in Equation (\ref{eq:LbfgsInvUPHessian}) to ensure that the computations outlined therein reflect the current approximation to the inverse of the negative Hessian of the log-likelihood, independent of whether the L-BFGS or the L-SR1 update is used. Notably, under the SR1 update, $\mathbf{B'}$ is only of dimension $N_V * N_V$ instead of $(2N_V) * (2N_V)$.

An intriguing property of the L-SR1 update is that all that would be necessary to obtain the compact representation for the negative Hessian of the log-likelihood $\hat{\mathbf{H}}^i$ instead, would be to swap $\mathcal{I}^0$ for $[\mathcal{I}^0]^{-1}$, $\mathbf{s}'_i$ for $\boldsymbol{\nu}'_i$, and $\boldsymbol{\nu}'_i$ for $\mathbf{s}'_i$ \parencite[see][]{byrd_representations_1994}. This can be shown \parencite[e.g., Equation 5.2 in the paper of][]{byrd_representations_1994,nocedal_numerical_2006} to be equivalent to setting $\mathbf{Q'}=\mathrm{Y} - [\mathcal{I}^0]^{-1}\mathrm{S}$ and $\mathbf{C'}=\mathbf{D} + \mathbf{L} + \mathbf{L}^\top -   \mathrm{S}^\top[\mathcal{I}^0]^{-1}\mathrm{S}$, where $\mathbf{L} = \mathrm{S}^\top\mathrm{Y} - \mathbf{R}$ with the right-hand side again matching the definitions given for Equation (\ref{eq:LbfgsInvHessian}). These can now again be substituted for $\mathbf{Q'}$ and $\mathbf{C'}$ in Equation (\ref{eq:LbfgsUPHessian}). Notably, computation of the compact representation for $\hat{\mathbf{V}}'^{i}$ then also remains agnostic to whether the L-SR1 or L-BFGS update is used, and can proceed as outlined in Equations (\ref{eq:LbfgsHessian}) and (\ref{eq:LbfgsInvHessian5}). The same is true for the computation of $tr(\hat{\mathbf{V}}'\mathbf{S}^r)$, which still can proceed as outlined in Equation (\ref{eq:lbfgsTrace}). Note, that matrix $\mathbf{A}$ in Equations (\ref{eq:LbfgsInvHessian5}) and (\ref{eq:lbfgsTrace}) will again be of dimension $(N_\mathbf{V}) * (N_\mathbf{V})$ when the SR1 update is used, rather than $(2N_\mathbf{V}) * (2N_\mathbf{V})$.

As long as the L-SR1 update results in approximations that are positive definite, computation of the entire L-qEFS update can thus proceed largely along the same lines, independent of whether the L-SR1 or L-BFGS update is used in Equations (\ref{eq:LbfgsInvUPHessian}-\ref{eq:LbfgsUPHessian}) to update $\hat{\mathcal{I}}^i$ and $\hat{\mathbf{H}}^i$. The main difference is that less memory will be required under the L-SR1 update, since the dimensions of the inner matrices $\mathbf{B'}$, $\mathbf{C'}$, and $\mathbf{A}$ will be smaller. Matters become more complicated if $\hat{\mathbf{H}}^i$ (and/or $\hat{\mathcal{I}}^{i-1}$) becomes indefinite. As mentioned before, in that case it will be necessary to replace $\hat{\mathcal{I}}^{i-1}$ in step 18 and/or $\hat{\mathbf{H}}^i$ in step 26 of Algorithm (\ref{alg:lqEFS}) with near-by positive semi-definite matrices $[\hat{\mathcal{I}}^+]^{i-1}$ and $[\hat{\mathbf{H}}^+]^{i}$ respectively. The following sub-section outlines how this can be achieved efficiently, based on implicit Eigendecompositions of $\hat{\mathcal{I}}^{i-1}$ and $\hat{\mathbf{H}}^i$. Section \ref{sec:lqefsInPractice} then discusses practical matters such as step-length control for $\boldsymbol{\lambda}$ updates and how post-estimation tasks such as credible interval computation, model selection, and posterior sampling can be achieved under the L-qEFS update.

\subsubsection{Computing Eigendecompositions for $\hat{\mathcal{I}}$ and $\hat{\mathbf{H}}$}\label{sec:eigenHI}

Many different strategies could be conceived to select the desired near-by positive semi-definite matrices $[\hat{\mathcal{I}}^+]^{i-1}$ and $[\hat{\mathbf{H}}^+]^{i}$. One simple option would be to rely on the strategy employed for the exact Newton update in Algorithm (\ref{alg:gammlss_beta}), that is to add a multiple of the identity to both approximations until a Cholesky succeeds. However, for the limited-memory approximations used here a more targeted adjustment is possible, which is to select the positive semi-definite matrices for $[\hat{\mathcal{I}}^+]^{i-1}$ and $[\hat{\mathbf{H}}^+]^{i}$ that are closest, in terms of the Frobenious norm, to $\hat{\mathcal{I}}^{i-1}$ and $\hat{\mathbf{H}}^i$ respectively. \Textcite{higham_computing_1988} show that the desired matrices are unique and, using the example of $[\hat{\mathbf{H}}^+]^{i}$, are obtained by forming $[\hat{\mathbf{H}}^+]^{i}=\mathrm{U}\Sigma^+\mathrm{U}^\top$. The right-hand side is obtained by first computing the Eigendecomposition $\hat{\mathbf{H}}^i=\mathrm{U}\Sigma\mathrm{U}^\top$. Subsequently, $\Sigma^+$ is simply set to be equal to the diagonal matrix $\Sigma$ with exact zeros in place of negative Eigenvalues. The same approach could be used to obtain $[\hat{\mathcal{I}}^+]^{i-1}$.

While theoretically attractive, computing Eigendecompositions of two $N_p * N_p$ matrices is not feasible in practice when working with large multi-level models. Fortunately, when working with compact representations, the same result can be achieved using the implicit Eigendecomposition approach proposed originally by \citeauthor{burdakov_efficiently_2017} (\citeyear{burdakov_efficiently_2017}; see \citeauthor{brust_useful_2025}, \citeyear{brust_useful_2025} for examples). This only requires computing an Eigendecomposition of two $N_V * N_V$ matrices as well as a ``thin'' QR-decomposition of two $N_p * N_V$ matrices \parencite[e.g.,][]{golub_matrix_2013}. Specifically, following the alternative derivations by \Textcite{erway_efficiently_2015}, this approach yields

\begin{equation}\label{eq:SR1UPHessianPSD}
[\hat{\mathbf{H}}^+]^{i}=\mathbf{I}\frac{1}{\gamma'_i} + \mathrm{P'}\Sigma'^+\mathrm{P'}^\top,
\end{equation}

\noindent
where $\Sigma'^+_{ll} = \Sigma'_{ll} - min(0,\Sigma'_{ll} + \frac{1}{\gamma'_i})$ and $P'=\mathrm{Q}^1\mathrm{U}$ are obtained from the Eigendecomposition $\mathrm{R}^1\mathbf{C'}[\mathrm{R}^1]^\top = \mathrm{U}'\mathrm{\Sigma'}[\mathrm{U'}]^\top$, where $\mathrm{Q}^1\mathrm{R}^1=\mathbf{Q'}$ is the aforementioned thin QR-decomposition so that $\mathrm{Q}^1$ has $N_V$ columns and $\mathrm{R}^1$ is a $N_V*N_V$ upper-triangular matrix \parencite[e.g.,][]{burdakov_efficiently_2017,erway_efficiently_2015,brust_useful_2025}. This result is reviewed in more detail in Appendix \ref{sec:AppendixImplicitEigen}, but by the definition of $\Sigma'^+$  it is clear that addition of $\mathbf{I}\frac{1}{\gamma'_i}$ to $\mathrm{P'}\Sigma'^+\mathrm{P'}^\top$ results in a positive semi-definite matrix, since the Eigenvalues from $\Sigma'^+$ have been shifted up by just enough, to ensure that negative ones become zero when $\frac{1}{\gamma'_i}$ is added.

Importantly, Equation (\ref{eq:SR1UPHessianPSD}) (and Equation (\ref{eq:SR1UPHessianPSDAPD}) in Appendix C) 
is still a compact representation for $[\hat{\mathbf{H}}^+]^{i}$. Thus, when updating the definitions for $\mathbf{Q'}=\mathrm{P'}$ and $\mathbf{C'}=\Sigma'^+$ once more, the compact representation for $[\hat{\mathbf{H}}^+]^{i}$ can directly be substituted for $\hat{\mathbf{H}}^{i}$ when computing $\hat{\mathbf{V'}}^{i}$ based on Equations (\ref{eq:LbfgsHessian}) and (\ref{eq:LbfgsInvHessian5}). Note, that the modified Woodburry identity by \Textcite{henderson_deriving_1981} used in Equation (\ref{eq:LbfgsInvHessian5}) does not require an inverse of matrices $\mathbf{Q'}$ and $\mathbf{C'}$ and thus remains applicable when substituting the positive semi-definite matrix $[\hat{\mathbf{H}}^+]^{i}$. Computation of $tr(\hat{\mathbf{V}}'\mathbf{S}^r)$, can also still proceed just as outlined in Equation (\ref{eq:lbfgsTrace}). Except that the EFS update is technically only defined when $\hat{\mathbf{H}}^+ + \mathbf{S}_\lambda$ is positive-definite \parencite[e.g.,][]{wood_generalized_2017}, which cannot be guaranteed when the final approximation $\hat{\mathbf{H}}^+$ is computed as shown in Equation (\ref{eq:SR1UPHessianPSD}). To avoid this problem, a small multiple of the machine precision can additionally be added to negative Eigenvalues in $\Sigma'^+$, ensuring that $\hat{\mathbf{H}}^+$ is numerically positive definite.

The approach discussed above also generalizes directly to $\hat{\mathcal{I}}^{i-1}$: based on Equation (\ref{eq:LbfgsInvUPHessian}), $[\hat{\mathcal{I}}^+]^{i-1}$ can be obtained from a thin QR-decomposition $\mathrm{Q}^1\mathrm{R}^1=\mathbf{U'}$ and subsequent Eigendecomposition of $\mathrm{R}^1\mathbf{B}'[\mathrm{R}^1]^\top$. $[\hat{\mathcal{I}}^+]^{i-1}$ can then be substituted for $\hat{\mathcal{I}}^{i-1}$ when computing the quasi-Newton direction in step 18 of Algorithm (\ref{alg:lqEFS}). Note, that $[\hat{\mathcal{I}}^+]^{i}$ is not necessarily an actual inverse of $[\hat{\mathbf{H}}^+]^{i}$. While this is unlikely to matter in practice, the issue can be avoided by again using the modified Woodburry identity by \Textcite{henderson_deriving_1981} to compute the inverse of $[\hat{\mathbf{H}}^+]^{i}$ (e.g., Equation (\ref{eq:LbfgsInvHessian5})). The latter will generally exist if the Eigenvalues in $\Sigma'$ are modified as outlined above to ensure that $[\hat{\mathbf{H}}^+]^{i}$ is numerically positive definite.

Finally, should it ever happen that no suitable $\alpha'$ can be found in step 19 of Algorithm (\ref{alg:lqEFS}) for a given $[\hat{\mathcal{I}}^+]^{i-1}$ or that $||\alpha'\mathbf{s}'_{i}|| \approx 0$, the line-search in step 19 of Algorithm (\ref{alg:lqEFS}) can be repeated in the direction of $\nabla'^1_{\boldsymbol{\beta}}$ rather than $\mathbf{s}'_i$. This avoids taking no step if $\hat{\mathcal{I}}^{i-1}$ has (temporarily) become a poor approximation of the inverse of the negative Hessian of the log-likelihood, providing the opportunity that the approximation improves again.

\subsection{Practical Application of the L-qEFS Update}\label{sec:lqefsInPractice}

Independent of whether a L-SR1 or L-BFGS update is utilized in the second phase of Algorithm (\ref{alg:lqEFS}), a few remarks are in order to ensure optimal performance of the update in practical application. The first remark regards initialization of the queue holding $\mathbf{s}'$ and $\boldsymbol{\nu}'$ in the second phase of the Algorithm (see step 22-24 in Algorithm (\ref{alg:lqEFS})). Both the PQL approach and EFS update assume that $\mathbf{H}$ does not depend on $\boldsymbol{\lambda}$ \parencite[e.g.,][]{breslow_approximate_1993,wood_generalized_2015,wood_generalized_2017,wood_generalized_2017-1}, which suggests that the final approximations $\hat{\mathbf{H}}$ and $\hat{\mathcal{I}}$ remain useful -- even after an update to $\boldsymbol{\lambda}$. Hence, the queue holding $\mathbf{s}'$ and $\boldsymbol{\nu}'$ should be initialized to reflect the final state of the queue before the latest update to $\boldsymbol{\lambda}$.

Regarding the topic of initialization, it should also be highlighted that the L-qEFS update is more sensitive than the full Newton update outlined in Algorithm (\ref{alg:gammlss_beta}) to the initial estimate for $\boldsymbol{\beta}$. If the latter is far away from the optima of both $\mathcal{L}$ and $\mathcal{L}_\lambda$, it might happen that neither the direct nor the indirect update in Algorithm (\ref{alg:lqEFS}) succeed in finding an acceptable step-length (i.e., $\alpha$ or $\alpha'$) -- especially if the L-qEFS update is to be based on L-BFGS approximations to the (inverse) of the Hessian matrices. To alleviate this risk, the initial estimate $\hat{\boldsymbol{\beta}}$ should be updated multiple times based on the Gradient of $\mathcal{L}_\lambda$. If step-length control immediately fails for these Gradient-based steps as well, the initial estimate $\hat{\boldsymbol{\beta}}$ can be shrunk towards a random sample from $\mathcal{N}(\mathbf{0},\mathbf{I})$. This can be repeated until Gradient updates repeatedly result in acceptable steps, in which case the quasi-Newton routines of Algorithm (\ref{alg:lqEFS}) are more likely to succeed in finding acceptable steps as well.

A third remark regards the size $N_V$ of the queue holding $\mathbf{s}'$ and $\boldsymbol{\nu}'$. In previous applications of limited-memory quasi-Newton approaches, $N_V$ has typically been set to a small number such as 5 or 10 \parencite[e.g.,][]{nocedal_updating_1980,liu_limited_1989,byrd_representations_1994,nocedal_numerical_2006}. In the context of estimating the regularization parameters $\boldsymbol{\lambda}$ it is however not enough that use of $\hat{\mathbf{V}}'$ results in sufficiently large updates to the estimate of $\boldsymbol{\beta}$. Instead, $\hat{\mathbf{V}}'$ here also needs to approximate important qualities of the true inverse $\mathbf{V}$ so that using it as part of the L-qEFS update results in appropriate updates to $\boldsymbol{\lambda}$. As a consequence, it will generally be desirable to set $N_V$ to a larger number -- with the optimal choice depending on the specific model to be estimated. In simulations (see Section \ref{sec:Simulations}) $N_V=30$ performed well for a range of likelihood functions and situations in which $40 \leq N_p \leq 240$. In practice, it will generally be advisable to consider multiple values for $N_V$ and to pick the model resulting in the highest $\mathcal{L}_\lambda$.

Another, perhaps even more important, choice regards the need for a strategy that attempts to minimize the risk of divergence and ensures that estimation terminates. In principle, this is the same problem also faced when using the full EFS update (i.e., based on an exact Hessian) to estimate $\boldsymbol{\lambda}$: even if $\hat{\mathbf{V'}}$ approximates the true inverse of the negative penalized Hessian well, an update to $\boldsymbol{\lambda}$ is not guaranteed to improve the Laplace approximate REML criterion \parencite[see][]{wood_generalized_2017}. Technically, step-length control for $\boldsymbol{\lambda}$ could again be based on checking the (approximate) Gradient of the REML criterion, neglecting any dependency of $\mathbf{H}$ on $\boldsymbol{\lambda}$, as discussed in Sections \ref{sec:GAM} and \ref{sec:GAMMLSS}. However, because the approximate Gradient would have to be based on the approximation $\hat{\mathbf{V'}}$ as well, this alone is unlikely to guard against divergence.

In Section \ref{sec:GAM}, we discussed the alternative option of simply skipping step length control entirely. The EFS update is unlikely to produce steps that would throw off the estimation routine entirely, providing some justification for this alternative in the context of GAMM estimation \parencite[see][]{wood_generalized_2017}. However, this does not have to be the case when working with approximate Hessians under the L-qEFS update, which is why the issue of step-length control becomes more pressing. Matters are made worse by the fact that when Algorithm (\ref{alg:lqEFS}) terminates after an update to $\boldsymbol{\lambda}$, $tr(\hat{\mathbf{V'}}\mathbf{S}^r)$ might suddenly be of a very different magnitude compared to before the update, for some or all penalty terms. Such a magnitude shift could result from the fact that the new $\hat{\mathbf{V'}}$ is a worse approximation to the inverse of the Hessian. More generally however, since each approximation is based on a relatively small number $N_\mathbf{V}$ of update vectors, changing even just one or two update vectors can result in a drastic effect on the magnitude of $tr(\hat{\mathbf{V'}}\mathbf{S}^r)$. In the absence of tight step-length control for $\boldsymbol{\lambda}$, this again promotes divergence.

To prevent this, we suggest to keep track of the last approximation $\hat{\mathbf{V'}}^*$ that resulted in an ``acceptable'' update to $\boldsymbol{\lambda}$. When computing the next update to $\boldsymbol{\lambda}$ after Algorithm (\ref{alg:lqEFS}) terminates, we can then compute both $tr(\hat{\mathbf{V'}}\mathbf{S}^r)$ based on the most recent state of the approximation matrix $\hat{\mathbf{V'}}$ and $tr(\hat{\mathbf{V'}}^*\mathbf{S}^r)$. Whether the actual L-qEFS update is then based on $\hat{\mathbf{V'}}^*$ or $\hat{\mathbf{V'}}$ is determined by checking if

\begin{equation}\label{eq:lqefsInequality}
\left(\frac{1}{N_\lambda}\sum_r^{N_\lambda} |t_r^{*}|\right) < \left(\frac{1}{N_\lambda}\sum_r^{N_\lambda} |t_r^{}|\right),
\end{equation}

\noindent
where $t_r^{*} = (tr(\mathbf{S}^-_\lambda\mathbf{S}^r) - tr(\hat{\mathbf{V'}}^*\mathbf{S}^r)) - \hat{\boldsymbol{\beta}}^\top\mathbf{S}^r\hat{\boldsymbol{\beta}}$ and computation of $t_r^{}$ is based on $\hat{\mathbf{V'}}$ instead of $\hat{\mathbf{V'}}^*$ but otherwise identical. If the inequality in Equation (\ref{eq:lqefsInequality}) does not hold, we set $\hat{\mathbf{V'}}^*=\hat{\mathbf{V'}}$ and the EFS update based on $\hat{\mathbf{V'}}$ is considered ``acceptable''.

This can be motivated by the realization discussed by \Textcite{wood_generalized_2017}, that the EFS update aims to ``balance'' the inequality

\begin{equation}\label{eq:lqefsInequality2}
(tr(\mathbf{S}^-_\lambda\mathbf{S}^r) - tr(\mathbf{V}\mathbf{S}^r)) \neq \hat{\boldsymbol{\beta}}^\top\mathbf{S}^r\hat{\boldsymbol{\beta}}.
\end{equation}

Clearly, use of the inequality in Equation (\ref{eq:lqefsInequality}) as a heuristic to determine whether the L-qEFS update should be based on $\hat{\mathbf{V'}}^*$ or $\hat{\mathbf{V'}}$ enforces a preference for the option that, given the current estimate of $\boldsymbol{\lambda}$ and on average across penalty terms, moves the second inequality stated in Equation (\ref{eq:lqefsInequality2}) closer to a state of equilibrium. In practice, relying on this strategy drastically reduces the volatility of the L-qEFS update and is usually sufficient to ensure convergence even when no additional step-length control for $\boldsymbol{\lambda}$ based on the approximate Gradient of the REML criterion is performed. One might object that this reduction in volatility is actually \emph{too} drastic, since -- in the most extreme case -- all updates to $\boldsymbol{\lambda}$ might be based on the initial solution for $\hat{\mathbf{V'}}$. However, as mentioned multiple times throughout this paper, using the EFS update and PQL approach anyway imply the expectation that $\frac{\partial\mathbf{H}}{\partial \lambda_r}=\mathbf{0}$ for all $\lambda_r \in \boldsymbol{\lambda}$, which implies that $\mathbf{H}$ should not change with $\lambda_r$, again justifying the reliance on Equation (\ref{eq:lqefsInequality}) to prevent extreme changes to $\hat{\mathbf{V'}}$ between updates to $\boldsymbol{\lambda}$.

Finally, note that the stabilizing transformation strategy described by \Textcite{wood_fast_2011} and \Textcite{wood_smoothing_2016} continues to be applicable when relying on the L-qEFS update. While re-parameterization is not necessary when relying on the L-qEFS update, doing so can still improve the stability of the computations outlined in Algorithm (\ref{alg:lqEFS}). Additionally, applying and reversing the transformation is trivial: if the model is re-parameterized with transformation matrix $\mathbf{T}$ as described in Appendix \ref{sec:AppendixTransform} before using Algorithm (\ref{alg:lqEFS}), then $\mathbf{T}\mathrm{S}$ and $\mathbf{T}\mathrm{Y}$ are the matrices holding the update vectors after reversing the transformation. They can then be used exactly as described in this section to compute $\hat{\mathbf{H}}$, $\hat{\mathbf{V'}}$, and the L-qEFS update (the latter could however also be computed before reversing the transformation; see Appendix \ref{sec:AppendixTransform}).

To summarize, step 1 in Algorithm (\ref{alg:sparse_gamm}) can be replaced by Algorithm (\ref{alg:lqEFS}). $tr(\hat{\mathbf{V'}}\mathbf{S}^r)$ or $tr(\hat{\mathbf{V'}}^*\mathbf{S}^r)$, the choice being determined based on Equation (\ref{eq:lqefsInequality}), can then be used to compute the L-qEFS update (steps 2-3 in Algorithm (\ref{alg:sparse_gamm})). This proposal can either be accepted directly or optionally be applied conditionally, pending verification against the approximate Gradient of the REML criterion (steps 8-12 in Algorithm (\ref{alg:sparse_gamm})). Combined, these steps enable approximate estimation of both $\boldsymbol{\beta}$ and $\boldsymbol{\lambda}$ based on likelihood and Gradient alone, the latter of which could in principle be approximated as well \parencite[e.g.,][]{nocedal_numerical_2006}.

Before concluding this section, we briefly focus on the open issue that $\mathcal{V}$ is not just required during estimation, but also afterwards. For example, computation of credible intervals involves $\mathcal{V}$ \parencite[e.g.,][]{wood_generalized_2017-2,wood_inference_2020}. It might now be tempting to rely on $\hat{\mathbf{V}}'$ for these post-estimation steps as well. From a practical perspective, at least the computation of credible intervals could be implemented efficiently when working with the compact representation for the latter. Equation (\ref{eq:lbfgsProd}) enables relatively efficient computation of products like $\mathbf{X}\hat{\mathbf{V}}'\mathbf{X}^\top$ \parencite[e.g.,][]{byrd_representations_1994} -- which would be required for credible interval computation \parencite[e.g.,][]{wood_generalized_2017-2} -- again without the need to explicitly form $\hat{\mathbf{V}}'$. However, because it cannot be guaranteed that $\hat{\mathbf{V}}'$ results in a good approximation of the exact inverse, there is also no guarantee that relying on $\hat{\mathbf{V}}'$ to compute credible intervals results in good function coverage. In practice, such intervals should thus be considered very cautiously. 

From a practical perspective, post-estimation tasks that require access to the Cholesky factor of $\mathbf{V}$ or $\mathcal{H}$, such as sampling from the conditional normal posterior of the coefficients (see Equation (\ref{eq:posteriorBeta})) or computing model selection criteria, are more complicated -- at least for large models. This is because naive computation of the Cholesky factor of $\hat{\mathbf{V}}'$ or $\hat{\mathcal{H}}$ will require explicitly forming the corresponding $N_p*N_p$ matrices. The Cholesky factor of $\hat{\mathcal{H}}$ at least can be obtained more efficiently based on the compact representation defined in Equation (\ref{eq:LbfgsHessian}). In a first step, we obtain matrix $\mathbf{K}$ so that $\mathbf{K}\mathbf{K}^\top=\hat{\mathcal{H}}^0$. This can be achieved most efficiently by a Cholesky decomposition of $\hat{\mathcal{H}}^0$ that pivots for sparsity, followed by un-pivoting to obtain $\mathbf{K}$. Then we obtain matrix $\mathbf{E}$ so that $\mathbf{E}\mathbf{E}^\top=\mathbf{C}'$. For this a singular value decomposition can be formed, the cost of which is negligible since $\mathbf{C}'$ is only of dimension $N_V*N_V$ \parencite[if the L-SR1 update is used; e.g.,][]{henderson_deriving_1981}. Subsequently, we form the sparsity-preserving QR-decomposition $\begin{bmatrix}
\mathbf{K}^\top\\
\mathbf{E}^\top\mathbf{Q}'^\top\\
\end{bmatrix}\mathbf{P}^\top=\mathbf{Q}\mathbf{R}$ \parencite[see][]{dennis_numerical_1996}. $\mathbf{R}^\top$ is a pivoted version of the Cholesky factor of $\hat{\mathcal{H}}$ and we can thus use it exactly as described in Section \ref{sec:AM} to obtain $\mathbf{L}^{-1}$. Previous research has also shown that it would be possible to maintain an additional BFGS approximation for the Cholesky factor of $\hat{\mathbf{H}}$ \parencite[e.g.,][]{goldfarb_factorized_1976,dennis_numerical_1996} and an interesting direction for future research would be to see how this could be achieved for the compact representations of $\hat{\mathbf{V}}'$ or $\hat{\mathcal{H}}$, as defined in Equations (\ref{eq:LbfgsInvHessian5}) and (\ref{eq:LbfgsHessian}) respectively, to make the process of finding the Cholesky factor even more efficient. 

If the Cholesky factor is cheap to obtain, the L-qEFS update can also be used as part of a Metropolis sampler for the Bayesian posteriors $\boldsymbol{\beta}|\mathbf{y},\boldsymbol{\lambda}$ or $\boldsymbol{\beta}|\mathbf{y}$ \parencite[e.g.,][]{wood_inference_2020}. Specifically, $\mathcal{N}(\hat{\boldsymbol{\beta}},\hat{\mathbf{V}}')$ can readily serve as a stationary proposal distribution to sample from $\boldsymbol{\beta}|\mathbf{y},\boldsymbol{\lambda}$, but ``random walk proposals'' based on $\hat{\mathbf{V}}'$ are also possible \parencite[][has the details]{wood_inference_2020}. To efficiently sample from $\boldsymbol{\beta}|\mathbf{y}$ instead, would require a proposal distribution for $\boldsymbol{\lambda}$ as well \parencite[see next section and also again][for possible options]{wood_inference_2020}. As discussed in this section, $\hat{\mathbf{V}}'$ can readily be updated to reflect a change in $\mathbf{S}_\lambda$ whenever a proposal for $\boldsymbol{\lambda}$ is accepted. The result can then be used in the next iteration when proposing new samples for $\boldsymbol{\beta}$. Use of such samplers allows to further improve the accuracy of the estimates of $\boldsymbol{\beta}$ and $\boldsymbol{\lambda}$ while also providing access to more accurate credible intervals, useful for inference \parencite[e.g.,][]{wood_generalized_2017-2,wood_inference_2020}.

\subsubsection{Implementation in MSSM}

To facilitate access to the L-qEFS update, Algorithm (\ref{alg:lqEFS}) has been implemented in MSSM and can be used to estimate general smooth models as supported by the \texttt{GSMM} class. Various arguments can be passed to the \texttt{fit} method of the \texttt{GSMM} class, allowing to take control over different aspects of the algorithm such as the number of update vectors to accumulate (i.e., $N_V$), initialization, step length control, and whether to rely on the SR1 or BFGS update. Notably, if a model is to be estimated via the L-qEFS update it is technically sufficient to implement only the \texttt{GSMMFamily}'s method to compute the log-likelihood. In that case MSSM approximates the Gradient via finite differencing \parencite[e.g.,][]{dennis_numerical_1996,nocedal_numerical_2006}. When the \texttt{GSMMFamily}'s method to compute the Gradient is implemented as well, MSSM relies on that instead. For more details, we refer to the documentation of the \texttt{GSMM}'s \texttt{fit} method, available at \url{https://jokra1.github.io/mssm}.

This concludes our presentation of a unified framework for efficient estimation and automatic regularization of generic multi-level smooth models. We now return to the final problem formulated in the Introduction and Background sections: correcting for uncertainty in the final estimate $\hat{\boldsymbol{\lambda}}$ and selecting between different smooth models in the absence of higher-order derivatives of the log-likelihood.

\section{Estimating Uncertainty in $\hat{\boldsymbol{\lambda}}$ and Model Selection in the Absence of Higher-order Derivatives}\label{sec:Uncertainty}

Estimation is usually only a means to an end, especially in the context of modeling experimental data. Instead, inference about the effect of experimental manipulations is usually of primary interest. For smooth models, answers to these questions are commonly sought via model comparisons, commonly based on the conditional version of the Akaike Information Criterion \parencite[cAIC][]{akaike_information_1992,greven_behaviour_2010}. For convenience, we re-state the latter in Equation (\ref{eq:aic}). 

\begin{equation}\label{eq:aic}
cAIC = -2\mathcal{L}({\hat{\beta}}) + 2tr(\mathbf{V}\mathbf{H})
\end{equation}

As discussed, the trace $\tau =tr(\mathbf{V}\mathbf{H})$ provides the ``effective degrees of freedom'' of the model \parencite[see][]{hastie_generalized_1990,wood_smoothing_2016} acting as a measure of model complexity. However, we also discussed that computed like this, $\tau$ is essentially conditioned on $\boldsymbol{\lambda}$, which means that the estimate of $\boldsymbol{\lambda}$ is treated as being free of uncertainty \parencite[e.g.,][]{wood_smoothing_2016}. \Textcite{wood_smoothing_2016} proposed two additive correction terms, $\mathbf{V}^J$ and $\mathbf{V}^L$, that transform the ``conditional'' covariance matrix $\mathbf{V}$ into the ``marginal'' covariance matrix $\tilde{\mathbf{V}}$, so that $\boldsymbol{\beta} | \mathbf{y} \sim \mathcal{N}(\hat{\boldsymbol{\beta}},\tilde{\mathbf{V}})$ holds approximately and the problematic conditioning has been eliminated \parencite[see also][]{greven_comment_2016}. However, computation of the two correction terms necessary to arrive at $\tilde{\mathbf{V}}$ generally requires access to third and fourth order derivatives of the log-likelihood, an estimate $\hat{\boldsymbol{\lambda}}$ that maximizes the REML criterion of Equation (\ref{eq:laplace_reml}), and quickly becomes expensive for the multi-level models of interest here. We briefly review these issues below. The remainder of this section is then dedicated towards the development of extensions and alternatives to the correction proposed by \Textcite{wood_smoothing_2016}, which remain applicable for multi-level models estimated via the methods discussed in this paper. Note, that for this section we again assume that any optional $\phi$ parameter already has been used to scale $\boldsymbol{\lambda}$, so that it can be omitted in all equations.

The requirement for third and fourth order derivatives and the need for a maximizer of Equation (\ref{eq:laplace_reml}) result from the fact that, to derive the correction terms, \Textcite{wood_smoothing_2016} rely on the asymptotic distributional result for $\boldsymbol{\rho}=log(\boldsymbol{\lambda})$ which they state as

\begin{equation}\label{eq:posterior_rho}
    \boldsymbol{\rho} | \mathbf{y} \sim \mathcal{N}(\hat{\boldsymbol{\rho}}^*,\mathbf{V}^\rho),
\end{equation}

\noindent
where $\hat{\boldsymbol{\rho}}^*=log(\hat{\boldsymbol{\lambda}}^*)$ denotes the log of the estimate $\hat{\boldsymbol{\lambda}}^*$ obtained by maximizing the Laplace approximate REML criterion -- which is not necessarily the estimate obtained under the EFS update \parencite[e.g.,][]{wood_generalized_2017}. For the covariance matrix, the authors propose the estimate $\mathbf{V}^\rho=[-\frac{\partial^2 \mathcal{V}}{\partial \boldsymbol{\rho} \partial \boldsymbol{\rho}^\top}_{|\hat{\boldsymbol{\rho}}^*}]^{-1}$, which is the inverse\footnote{As pointed out by \Textcite{wood_smoothing_2016}, this inverse might in practice become a pseudo-inverse in case $\partial \mathcal{V}/\partial\rho_r = \partial^2 \mathcal{V}/\partial\rho_r^2=0$ for some log regularization parameters, which is the expected behavior when $\lambda_r \rightarrow \infty$. Due to the wide range of plausible values for the $\lambda_r$ parameters, additional regularization of the (pseudo) inverse will also be prudent \parencite[e.g.,][]{wood_smoothing_2016,wood_generalized_2017-2}} of an estimate of the Fisher information at $\hat{\boldsymbol{\rho}}^*$ \parencite[e.g.,][]{davison_statistical_2003}. Use of the latter as an estimate for the covariance matrix and the normal approximation for the posterior $\boldsymbol{\rho}|\mathbf{y}$ in Equation (\ref{eq:posterior_rho}), are justifiable asymptotically, based on the large sample distribution for $\hat{\boldsymbol{\rho}}^*$ \parencite[e.g.,][]{davison_statistical_2003,wood_core_2015,wood_smoothing_2016}. As discussed in the Background section (see also Appendix \ref{sec:AppendixRemlDeriv}), and more extensively by \Textcite{wood_smoothing_2016}, computation of $\frac{\partial^2 \mathcal{V}}{\partial \boldsymbol{\rho} \partial \boldsymbol{\rho}^\top}$ requires $\frac{\partial \mathbf{H}}{\partial \rho_l}$ and $\frac{\partial^2 \mathbf{H}}{\partial \rho_l \partial \rho_j}$ for all $\rho_l,\rho_j \in \boldsymbol{\rho}$ which, if computed by means of implicit differentiation, in turn require the third and fourth order derivatives of the log-likelihood respectively.

To understand why the complexity of the correction terms becomes problematic for large multi-level models, it is necessary to inspect the derivations of $\mathbf{V}^J$ and $\mathbf{V}^L$. To obtain the latter two, \Textcite{wood_smoothing_2016} rely on the (conditional) distributional results for $\boldsymbol{\beta}$ and $\boldsymbol{\rho}$ -- shown here in Equations (\ref{eq:posteriorBeta}) and (\ref{eq:posterior_rho}). Working under the assumption that these results hold, the authors obtain a third distributional result for

\begin{equation}\label{eq:posterior_beta2}
    \boldsymbol{\beta} | \mathbf{y} \sim \hat{\boldsymbol{\beta}}_{\boldsymbol{\rho}} + [\mathbf{L}^{\boldsymbol{\rho}}]^{-\top}\mathbf{z}~~\text{where}~\mathbf{z} \sim \mathcal{N}(\mathbf{0},\mathbf{I})
\end{equation}

\noindent
has a normal distribution with an identity for the covariance matrix, $\hat{\boldsymbol{\beta}}_{\boldsymbol{\rho}}$ is the maximizer of the penalized likelihood $\mathcal{L}^\lambda$ (i.e., Equation (\ref{eq:pen_llk})) for a given $\boldsymbol{\lambda}=exp(\boldsymbol{\rho})$, and $[\mathbf{L}^{\boldsymbol{\rho}}]$ is the Cholesky factor of the negative Hessian of the penalized log-likelihood $-\frac{\partial^2 \mathcal{L}_\lambda}{\partial \boldsymbol{\beta}\partial\boldsymbol{\beta}^\top}\Big\rvert_{\hat{\boldsymbol{\beta}}_{\boldsymbol{\rho}}}$ for a specific $\boldsymbol{\lambda}=exp(\boldsymbol{\rho})$.

\Textcite{wood_smoothing_2016} then proceed to show that $\tilde{\mathbf{V}}$ is the approximate covariance matrix of Equation (\ref{eq:posterior_beta2}). To do so, they consider a first-order Taylor series expansion of Equation (\ref{eq:posterior_beta2}), treated as a function of $\boldsymbol{\rho}$, around $\hat{\boldsymbol{\rho}}^*$. This results in

\begin{equation}\label{eq:posterior_betaT}
    \boldsymbol{\beta} | \mathbf{y} \sim
    \hat{\boldsymbol{\beta}}_{\hat{\boldsymbol{\rho}}^*} + [\mathbf{L}^{\hat{\boldsymbol{\rho}}^*}]^{-\top}\mathbf{z} + \mathbf{J}(\boldsymbol{\rho} - \hat{\boldsymbol{\rho}}^*) + \sum_k^{N_\lambda}\frac{\partial [\mathbf{L}^{\boldsymbol{\rho}}]^{-\top}}{\partial \rho_k}\Bigg\rvert_{\hat{\boldsymbol{\rho}}^*}(\boldsymbol{\rho} - \hat{\boldsymbol{\rho}}^*) + \epsilon,
\end{equation}

\noindent
where $\mathbf{J}^\top_r = \frac{d \hat{\boldsymbol{\beta}}}{d\rho_r}\Big\rvert_{\hat{\rho}^*}$ and $\epsilon$ denotes the error of the first order expansion, which is subsequently ignored \parencite[][]{wood_smoothing_2016}. The authors show that the three random vectors, separated by square brackets in Equation (\ref{eq:posterior_betaT}), are uncorrelated so that $\tilde{\mathbf{V}}$ is simply the sum of the variances of the aforementioned three vectors (i.e., their individual covariance matrices). 

Upon inspection of Equation (\ref{eq:posteriorBeta}), it becomes evident that $ \hat{\boldsymbol{\beta}}_{\hat{\boldsymbol{\rho}}^*} + [\mathbf{L}^{\hat{\boldsymbol{\rho}}^*}]^{-\top}\mathbf{z}$ is a sampler for the normal approximation to $\boldsymbol{\beta} | \mathbf{y},\hat{\boldsymbol{\lambda}}^*$, which establishes $\left[-\frac{\partial^2 \mathcal{L}_\lambda}{\partial \boldsymbol{\beta}\partial\boldsymbol{\beta}^\top}\Big\rvert_{\hat{\boldsymbol{\beta}}_{\hat{\boldsymbol{\rho}}^*}}\right]^{-1}$ as the covariance matrix for this first term. \Textcite{wood_smoothing_2016} then provide the covariance matrices\footnote{These results follow from the standard result that the covariance matrix of $\mathbf{A}\mathbf{b}$, where $\mathbf{A}$ is a fixed matrix and $\mathbf{b}$ is a random vector, is $\mathbf{A}cov(\mathbf{b})\mathbf{A}^\top$.} $\mathbf{V}^J = \mathbf{J}\mathbf{V}^\rho\mathbf{J}^\top$ for $\mathbf{J}(\boldsymbol{\rho} - \hat{\boldsymbol{\rho}}^*)$ and $\mathbf{V}^L$, with

\begin{equation}\label{eq:V2}
\mathbf{V}^L_{jm}=\sum_i^{N_p}\sum_l^{N_\lambda}\sum_k^{N_\lambda} \frac{\partial {[\mathbf{L}^{\boldsymbol{\rho}}]^{-\top}_{ji}}}{\partial \rho_k}\Bigg\rvert_{\hat{\boldsymbol{\rho}}^*} \mathbf{V}^\rho_{kl} \frac{\partial {[\mathbf{L}^{\boldsymbol{\rho}}]^{-\top}_{mi}}}{\partial \rho_l}\Bigg\rvert_{\hat{\boldsymbol{\rho}}^*},
\end{equation}

\noindent
for $\sum_k^{N_\lambda}\frac{\partial [\mathbf{L}^{\boldsymbol{\rho}}]^{-\top}}{\partial \rho_k}\Big\rvert_{\hat{\boldsymbol{\rho}}^*}(\boldsymbol{\rho} - \hat{\boldsymbol{\rho}}^*)$. Taken together, this yields the final approximate result for

\begin{equation}\label{eq:posterior_betaT2}
\begin{split}
    &\boldsymbol{\beta} | \mathbf{y} \sim \mathcal{N}\left(\boldsymbol{\beta}_{\hat{\rho}},\tilde{\mathbf{V}}\right),~\text{where}\\&\tilde{\mathbf{V}}= \left[-\frac{\partial^2 \mathcal{L}_\lambda}{\partial \boldsymbol{\beta}\partial\boldsymbol{\beta}^\top}\Big\rvert_{\hat{\boldsymbol{\beta}}_{\hat{\boldsymbol{\rho}}^*}}\right]^{-1} + \mathbf{V}^J + \mathbf{V}^L.
\end{split}
\end{equation}

\noindent
The choice for the mean can be verified by taking the expectation over the random vectors in Equation (\ref{eq:posterior_betaT}) \parencite[e.g.,][]{wood_smoothing_2016}.

These results reveal why efficient evaluation of $\tilde{\mathbf{V}}$ quickly becomes difficult for multi-level models: computation of $\mathbf{V}^L$ which, as pointed out by \Textcite{wood_smoothing_2016}, has a complexity of $O(N_\lambda * N_p^3)$, scales very poorly with the number of coefficients implied by the model to be estimated (e.g., $N_p$). While, as mentioned earlier and discussed in more detail by \Textcite{wood_smoothing_2016}, the Laplace approximation anyway dictates a modest number of coefficients for all but strictly additive models, this complexity will still be noticeable for all but the simplest smooth models. Computation of the derivative\footnote{Omitted here for the sake of brevity. The computations can be found for example in the Supplementary materials from \Textcite{wood_smoothing_2016} or in Appendix B.7 of the book by \Textcite{wood_generalized_2017-2}.} of $[\mathbf{L}^{\boldsymbol{\rho}}]^{-\top}$, itself required to compute $\mathbf{V}^L$, also again requires $\forall~\rho_r \in \boldsymbol{\rho}~\frac{\partial \mathbf{H}}{\partial \rho_r}$ and thus generally access to third-order derivatives of the log-likelihood. Considering that computation of $\mathbf{V}^\rho$ anyway requires up to fourth-order derivatives, this last point is however not an immediate problem.

Perhaps more importantly, these results also indicate that computation of $\mathbf{V}^\rho$ and $\tilde{\mathbf{V}}$ will only be justified when relying on the EFS update if $\forall~\rho_r \in \boldsymbol{\rho}~\frac{\partial \mathbf{H}}{\partial \rho_r} = \mathbf{0}$ and $\forall~ \rho_l,\rho_j \in \boldsymbol{\rho}~ \frac{\partial^2 \mathbf{H}}{\partial \rho_l \partial \rho_j} = \mathbf{0}$ hold in general. Only then will the final EFS estimate $\hat{\boldsymbol{\lambda}}$ maximize the REML criterion so that $\hat{\boldsymbol{\rho}}=\hat{\boldsymbol{\rho}}^*$, where $\hat{\boldsymbol{\rho}}=log(\hat{\boldsymbol{\lambda}})$, ensuring that Equations (\ref{eq:posterior_rho}) and (\ref{eq:posterior_beta2}) and thus the correction terms by \Textcite{wood_smoothing_2016} are well grounded. For such models, the elements of the Hessian of the REML criterion with respect to the log-smoothing parameters $\boldsymbol{\rho}_r$ can be computed as $\frac{\partial^2 \mathcal{V}}{\partial \rho_j \partial \rho_l}=$

\begin{equation}\label{eq:pql_hess_rho}
\begin{split}
    &-\gamma^{jl}\frac{\lambda_l}{2}\hat{\boldsymbol{\beta}}^\top\mathbf{S}^l\hat{\boldsymbol{\beta}} - \frac{d\hat{\boldsymbol{\beta}}^\top}{d \rho_j}\mathcal{H}\frac{d\hat{\boldsymbol{\beta}}}{d\rho_l} -\lambda_j \mathbf{S}^j\hat{\boldsymbol{\beta}}\frac{d\hat{\boldsymbol{\beta}}}{d\rho_l} - \lambda_l\mathbf{S}^l\hat{\boldsymbol{\beta}}\frac{d\hat{\boldsymbol{\beta}}}{d\rho_j}\\&
    -\frac{1}{2}\left(\gamma^{jl}\lambda_ltr\left(\mathcal{H}^{-1}\mathbf{S}^l\right) - \lambda_j\lambda_ltr\left(\mathcal{H}^{-1}\mathbf{S}^j\mathcal{H}^{-1}\mathbf{S}^l\right)\right)\\&
    +\frac{1}{2}\left(\gamma^{jl}\lambda_ltr\left(\mathbf{S}^-\mathbf{S}^l\right) - \lambda_j\lambda_ltr\left(\mathbf{S}^-\mathbf{S}^j\mathbf{S}^-\mathbf{S}^l\right)\right),
\end{split}
\end{equation}

\noindent
where $\gamma^{jl}=(1 - min(1,|j-l|)$ and $\hat{\boldsymbol{\beta}}$ again denotes the maximizer of the penalized log-likelihood for the current estimate of $\boldsymbol{\lambda}$. This result is derived in Appendix \ref{sec:AppendixRemlDeriv} of this paper but also follows directly from the results reviewed in Appendix B of the paper by \Textcite{wood_fast_2011} and the derivations provided by \Textcite{wood_smoothing_2016} and also \Textcite{wood_generalized_2017-1}. Note that Equation (\ref{eq:pql_hess_rho}) is exact for the kind of strictly additive (Gaussian) smooth models discussed in Section \ref{sec:AM}. Because Equation (\ref{eq:pql_hess_rho}) can be computed without access to higher-order derivatives, it is tempting to simply commit to the assumption that $\frac{\partial \mathbf{H}}{\partial \rho_l} = \mathbf{0}$ and $\frac{\partial^2 \mathbf{H}}{\partial \rho_l \partial \rho_j} = \mathbf{0}$ hold in general and for all $\rho_l,\rho_j \in \boldsymbol{\rho}$, so that Equation (\ref{eq:pql_hess_rho}) can at least be used to \emph{approximate} $\mathbf{V}^\rho$, and ultimately $\tilde{\mathbf{V}}$ and $\tau'$, for more general smooth models as well. This possibility is explored in the following section.

\begin{figure*}[!t]
    \caption{Illustration of the Behavior of the REML Criterion Based on a Second Order Approximation to $\mathcal{L}$ around $\hat{\boldsymbol{\beta}}_{\hat{\boldsymbol{\rho}}}$}
    \includegraphics[width=\textwidth]{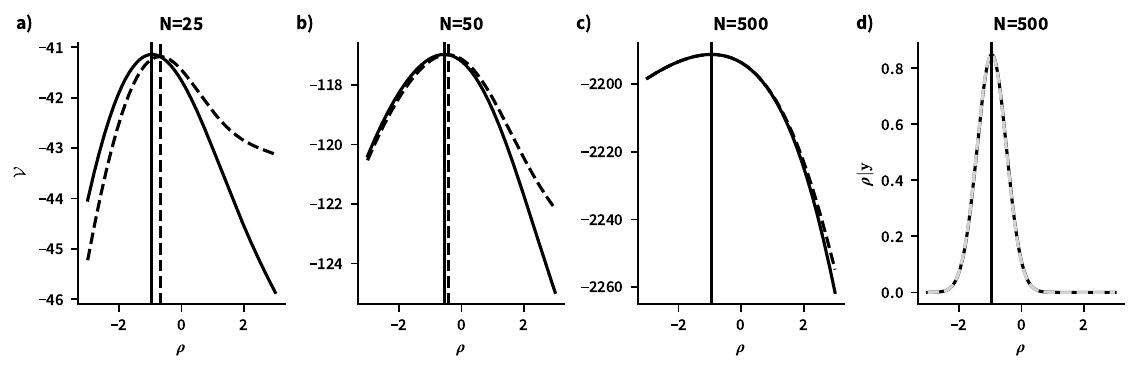}
    \label{fig:TaylorConvergence}
    {\small
         \textit{Note.} Panels a-c of Figure \ref{fig:TaylorConvergence} show an approximate REML criterion $\tilde{\mathcal{V}}$ (dashed line) for which $\mathcal{L}$ has been replaced with $\tilde{\mathcal{L}}|_{\hat{\boldsymbol{\beta}}_{\hat{\boldsymbol{\rho}}}}$, a second-order Taylor approximation to the log-likelihood around the final estimate of the coefficients $\hat{\boldsymbol{\beta}}_{\hat{\boldsymbol{\rho}}}$ obtained under the EFS update, and the ``exact'' (still Laplace-approximate) REML criterion $\mathcal{V}$ (solid line) defined in Equation (\ref{eq:laplace_reml}). Different panels depict both criteria for different sample sizes for the same proportional hazard likelihood. Panel d shows, for a sample size of $N=500$, the densities of the normal approximation to the posterior $\boldsymbol{\rho}|\mathbf{y}$, once (solid black line) parameterized with $\hat{\boldsymbol{\rho}}^*$ and $\mathbf{V}^\rho$ as proposed by \Textcite{wood_generalized_2017-1} and once (dashed gray line) parameterized with the EFS estimates for $\boldsymbol{\rho}$ and $\mathbf{V}^\rho$, where the latter is computed based on Equation (\ref{eq:pql_hess_rho}). Vertical lines in all panels show the estimates $\hat{\boldsymbol{\rho}}=log(\hat{\boldsymbol{\lambda}})$ (dashed line) and $\hat{\boldsymbol{\rho}}^*=log(\hat{\boldsymbol{\lambda}}^*)$. The former is the EFS estimate \parencite[e.g.,][]{wood_generalized_2017} the latter maximizes the Laplace approximate REML criterion. At $N=500$ the two estimates coincide in this example.
        }
\end{figure*}

\subsection{A PQL-like Estimate of $\tau'$}\label{sec:pql_cAIC}

For GAMMs, some justification for this can be derived from the classical PQL approach \parencite[e.g.,][]{wood_generalized_2015}. As discussed in Section \ref{sec:GAM}, adopting the PQL approach essentially means that the smoothing parameters for a GAMM $y_i \sim \mathcal{E}(g^{-1}(\mathbf{X}\boldsymbol{\beta}),\phi)$ are estimated for a sequence of approximately additive (i.e., ``linearized'') models $z_i \sim \mathcal{N}(\mathbf{X}\boldsymbol{\beta},\frac{1}{W_{ii}}\phi)$ of the transformed data $\mathbf{z}$ instead \parencite[see also][]{wood_generalized_2017-1}. Assuming that the normality assumption of this linearized model is correct, $\left[-\frac{\partial^2 \mathcal{V}}{\partial \boldsymbol{\rho} \partial \boldsymbol{\rho}^\top}\Big\rvert_{\hat{\boldsymbol{\rho}}}\right]^{-1}$, computed as shown in Equation (\ref{eq:pql_hess_rho}), will be a reasonable estimate for the uncertainty in $\hat{\boldsymbol{\lambda}}$ obtained from the final linearized additive model \parencite[i.e., for the final transformed data $\mathbf{z}$ and weight matrix $\mathbf{W}$; see section 3.4.3 in the book by][]{wood_generalized_2017-2}. Since the central-limit result established by \citeauthor{wood_generalized_2017-1} (\citeyear{wood_generalized_2017-1}; see also Section \ref{sec:GAM}) provides asymptotic justification for the normality assumption, this estimate can be used in practice to approximate $\tilde{\mathbf{V}}$ as long as $N_p << N$. The \texttt{bam} function, available in mgcv, for example sets $\tilde{\mathbf{V}} = \mathbf{V} + \mathbf{V}^J$ and uses this result to compute $\tau'$ for GAMMs estimated via the PQL-based methods by \Textcite{wood_generalized_2015} and \Textcite{wood_generalized_2017-1}. Here $\mathbf{V}$ again denotes the estimate for the covariance matrix of the normal approximation to $\boldsymbol{\beta}|\mathbf{y},\boldsymbol{\lambda}$ at the final PQL estimate obtained for $\boldsymbol{\lambda}$ and $\mathbf{V}^J$ is based on $\mathbf{V}^\rho$ computed as shown in Equation (\ref{eq:pql_hess_rho}).

In principle, relying on the alternative linearization strategy of Section \ref{sec:GAMMLSS}, the same argument can be made for any generic smooth model. For this purpose, let $\tilde{\mathcal{L}}|_{\hat{\boldsymbol{\beta}}_{\boldsymbol{\rho}}}$ again denote the second-order Taylor approximation to the log-likelihood $\mathcal{L}$ around $\hat{\boldsymbol{\beta}}_{\boldsymbol{\rho}}$, the maximizer of the penalized log-likelihood $\mathcal{L}_\lambda$ for a given $\boldsymbol{\rho}$. As discussed in Section \ref{sec:GAMMLSS}, replacing $\mathcal{L}$ with $\tilde{\mathcal{L}}|_{\hat{\boldsymbol{\beta}}_{\boldsymbol{\rho}}}$ can again be justified asymptotically for any given $\boldsymbol{\lambda}$. In consequence, $\mathbf{V}^\rho$ when computed as shown in Equation (\ref{eq:pql_hess_rho}), will again be a reasonable estimate for the uncertainty in $\hat{\boldsymbol{\lambda}}$ for the final linearization $\tilde{\mathcal{L}}|_{\hat{\boldsymbol{\beta}}_{\hat{\boldsymbol{\rho}}}}$. Again, this suggests that this estimate of $\mathbf{V}^\rho$ can be used in practice, as long as $N_p << N$, to approximately quantify uncertainty in the estimate $\hat{\boldsymbol{\rho}}$.

If we adopt the approach by \Textcite{mcgilchrist_restricted_1991} and \Textcite{mcgilchrist_estimation_1994} and simply assume that the log-likelihood is reasonably approximated by a quadratic function (i.e., that $\mathbf{H}$ is reasonably stationary), use of Equation (\ref{eq:pql_hess_rho}) to compute $\mathbf{V}^\rho=-\left[\frac{\partial^2 \mathcal{V}}{\partial \boldsymbol{\rho} \partial \boldsymbol{\rho}^\top}\right]^{-1}\Big\rvert_{\hat{\boldsymbol{\rho}}}$ will not only be justified but also produce approximately the same estimate of $\mathbf{V}^\rho$ used by \Textcite{wood_smoothing_2016}. As discussed in Section \ref{sec:GAMMLSS}, the final EFS estimate $\hat{\boldsymbol{\rho}}$ should end up to be approximately equal to the REML maximizer $\hat{\boldsymbol{\rho}}^*$ when the log-likelihood is reasonably quadratic. More importantly, in the neighborhood of $\hat{\boldsymbol{\rho}}$ the REML criterion itself will then also be reasonably approximated by a variant for which $\mathcal{L}$ has been replaced with $\tilde{\mathcal{L}}|_{\hat{\boldsymbol{\beta}}_{\hat{\boldsymbol{\rho}}}}$.

As discussed in Section \ref{sec:GAMMLSS}, there is reason to believe that $\mathbf{H}$ can often be treated as reasonably stationary (i.e., the log-likelihood can be reasonably approximated by a quadratic function) in practice, when $N_p << N$ and $N$ is large. This is illustrated in Figure \ref{fig:TaylorConvergence}, visualizing how the ``approximate'' REML criterion $\tilde{\mathcal{V}}$ based on $\tilde{\mathcal{L}}|_{\hat{\boldsymbol{\beta}}_{\hat{\boldsymbol{\rho}}}}$ compares to the ``exact'' REML criterion $\mathcal{V}$ based on $\mathcal{L}$ for a proportional hazard model involving a single smooth term and for different $N$ ($N_p=20$). Note, that the assumption that $\mathbf{H}$ is stationary is clearly unreasonable for small $N$, which is why $\tilde{\mathcal{V}}$ offers a poor approximation to $\mathcal{V}$ in the neighborhood of $\hat{\boldsymbol{\rho}}$ (e.g., $N=25$; Figure \ref{fig:TaylorConvergence}a). However, in this case the log-likelihood $\mathcal{L}$ is better approximated by $\tilde{\mathcal{L}}|_{\hat{\boldsymbol{\beta}}_{\hat{\boldsymbol{\rho}}}}$ as $N$ increases, so that the assumption that $\mathbf{H}$ is stationary becomes increasingly more reasonable. As a consequence, the EFS estimate for $\hat{\boldsymbol{\rho}}$ (dashed vertical line), clearly different from the REML maximizer $\hat{\boldsymbol{\rho}}^*$ (solid vertical line) for small $N$, also becomes virtually indistinguishable from the latter at $N=500$ (see Figure \ref{fig:TaylorConvergence}c). Finally, because $\mathbf{H}$ can reasonably be treated as stationary when $N=500$, Equation (\ref{eq:pql_hess_rho}) also produces a good ``EFS estimate'' of the covariance matrix $\mathbf{V}^\rho$ of the normal approximation to $\boldsymbol{\rho}|\mathbf{y}$ (see Figure \ref{fig:TaylorConvergence}d). As long as $\mathbf{H}$ is reasonably stationary in practice, the EFS estimate for $\mathbf{V}^\rho$ can thus be expected to be a good approximation to the estimate used by \Textcite{wood_smoothing_2016}\footnote{Note, that computation of the derivative of $[\mathbf{L}^{\boldsymbol{\rho}}]^{-\top}$, provided in the Supplementary materials from \Textcite{wood_smoothing_2016}, would also have to be based on the assumption that $\forall~\rho_r \in \boldsymbol{\rho}~\frac{\partial \mathbf{H}}{\partial \rho_r} = \mathbf{0}$} and should thus result in similar estimates of $\tilde{\mathbf{V}}$ and $\tau'$ required for an approximately corrected cAIC.

To summarize, Equation (\ref{eq:pql_hess_rho}) provides one path to approximately estimate (and correct for) uncertainty in $\hat{\boldsymbol{\lambda}}$ for generic models estimated via the methods discussed in this paper. For strictly additive models use of Equation (\ref{eq:pql_hess_rho}) to compute $\mathbf{V}^\rho$ and ultimately $\tilde{\mathbf{V}}$ exactly recovers the proposal by \Textcite{wood_smoothing_2016} and remains perfectly valid when using the EFS update by \Textcite{wood_generalized_2017} to estimate $\boldsymbol{\lambda}$. However, while Equation (\ref{eq:pql_hess_rho}) can be evaluated relatively efficiently (see Appendix \ref{sec:AppendixRemlDeriv}), computation of $\tilde{\mathbf{V}}$ still has a complexity that quickly becomes prohibitive for large multi-level additive models. For such models, (approximately) computing $\tau'=tr(\tilde{\mathbf{V}}\mathbf{H})$ without having to compute $\tilde{\boldsymbol{\mathbf{V}}}$ would be preferable. 

The next section discusses how to obtain an alternative Monte Carlo (MC) estimate for $\tau'$, which does not require computation of $\tilde{\boldsymbol{\mathbf{V}}}$. For more generic models, the assumptions we made in motivating the use of Equation (\ref{eq:pql_hess_rho}) to compute EFS estimates for $\mathbf{V}^\rho$, $\tilde{\mathbf{V}}$, and $\tau'$ are quite strong. While there is reason to believe that, as long as $N_p << N$ and $N$ is large, these assumption will often be met in practice, this will generally have to be validated for individual models (e.g., Figure \ref{fig:TaylorConvergence}). It would thus again be desirable to find a way to compute $\tilde{\mathbf{V}}$ and $\tau'$ that does not rely as much on the assumption that $\mathbf{H}$ is reasonably stationary. In Section \ref{sec:MCGS} we discuss how the MC estimate from Section \ref{sec:MCAM} can be generalized to achieve this, based on an extension of the strategy by \Textcite{greven_comment_2016}.

\subsection{Monte Carlo Estimation of $\tau'$ for Strictly Additive Models}\label{sec:MCAM}

We start with considering that $\tau=tr(\mathbf{V}\mathbf{H}) \approx tr(\mathbb{E}_{\boldsymbol{\beta|\mathbf{y}},\hat{\boldsymbol{\rho}}^*}\{(\hat{\boldsymbol{\beta}}_{\hat{\boldsymbol{\rho}}^*} - \boldsymbol{\beta})(\hat{\boldsymbol{\beta}}_{\hat{\boldsymbol{\rho}}^*} - \boldsymbol{\beta})^\top\}\mathbf{H})$ while $\tau'=tr(\tilde{\mathbf{V}}\mathbf{H}) \approx tr(\mathbb{E}_{\boldsymbol{\beta|\mathbf{y}}}\{(\hat{\boldsymbol{\beta}}_{\hat{\boldsymbol{\rho}}^*} - \boldsymbol{\beta})(\hat{\boldsymbol{\beta}}_{\hat{\boldsymbol{\rho}}^*} - \boldsymbol{\beta})^\top\}\mathbf{H})$, where  $\mathbf{H}$ refers to the negative Hessian of the log-likelihood at $\hat{\boldsymbol{\beta}}_{\hat{\boldsymbol{\rho}}^*}$ in both cases. That is, $\tau$ relies on an estimate (i.e., $\mathbf{V}$ given ${\hat{\boldsymbol{\rho}}^*}$) of the expectation $\mathbb{E}\{(\hat{\boldsymbol{\beta}}_{\hat{\boldsymbol{\rho}}^*} - \boldsymbol{\beta})(\hat{\boldsymbol{\beta}}_{\hat{\boldsymbol{\rho}}^*} - \boldsymbol{\beta})^\top\}$ taken with respect to the normal approximation to the \emph{conditional} posterior $\boldsymbol{\beta}|\mathbf{y},\hat{\boldsymbol{\rho}}^*$, while $\tau'$ relies on an estimate (i.e., $\tilde{\mathbf{V}}$) of the same expectation, but taken with respect to the normal approximation to the \emph{marginal} posterior $\boldsymbol{\beta}|\mathbf{y}$ instead \parencite[e.g.,][]{wood_smoothing_2016}. Following \Textcite{greven_comment_2016} we can rely on the law of total covariance to obtain a different expression for this last expectation and thus also for $\tau'$. Specifically, the latter is approximately equal to

\begin{equation}\label{eq:exptau}
\begin{split}
    &tr(\mathbb{E}_{\boldsymbol{\beta|\mathbf{y}}}\{(\hat{\boldsymbol{\beta}}_{\hat{\boldsymbol{\rho}}^*} - \boldsymbol{\beta})(\hat{\boldsymbol{\beta}}_{\hat{\boldsymbol{\rho}}^*} - \boldsymbol{\beta})^\top\}\mathbf{H})\\&=
    tr\left(\left[\mathbb{E}_{\boldsymbol{\rho}|\mathbf{y}}\{\mathbb{E}_{\boldsymbol{\beta}|\mathbf{y},\boldsymbol{\rho}}\{(\hat{\boldsymbol{\beta}}_{\boldsymbol{\rho}} - \boldsymbol{\beta})(\hat{\boldsymbol{\beta}}_{\boldsymbol{\rho}} - \boldsymbol{\beta})^\top\}\} + cov(\hat{\boldsymbol{\beta}}_{\boldsymbol{\rho}})\right]\mathbf{H}\right)\\&=
    tr\left(\left[\mathbb{E}_{\boldsymbol{\rho}|\mathbf{y}}\{\mathbb{E}_{\boldsymbol{\beta}|\mathbf{y},\boldsymbol{\rho}}\{(\hat{\boldsymbol{\beta}}_{\boldsymbol{\rho}} - \boldsymbol{\beta})(\hat{\boldsymbol{\beta}}_{\boldsymbol{\rho}} - \boldsymbol{\beta})^\top\}\}\mathbf{H}\right] + \left[cov(\hat{\boldsymbol{\beta}}_{\boldsymbol{\rho}})\mathbf{H}\right]\right)\\&=
    tr\left(\left[\mathbb{E}_{\boldsymbol{\rho}|\mathbf{y}}\{\mathbb{E}_{\boldsymbol{\beta}|\mathbf{y},\boldsymbol{\rho}}\{(\hat{\boldsymbol{\beta}}_{\boldsymbol{\rho}} - \boldsymbol{\beta})(\hat{\boldsymbol{\beta}}_{\boldsymbol{\rho}} - \boldsymbol{\beta})^\top\}\}\mathbf{H}\right]\right) + tr\left(\left[cov(\hat{\boldsymbol{\beta}}_{\boldsymbol{\rho}})\mathbf{H}\right]\right),
\end{split}
\end{equation}

\noindent
where expectations $\mathbb{E}_{\boldsymbol{\rho}|\mathbf{y}}$ are taken with respect to the normal approximation to the posterior for $\boldsymbol{\rho}$ \parencite[e.g.,][]{greven_comment_2016} defined in Equation (\ref{eq:posterior_rho}). As was done by \Textcite{wood_smoothing_2016}, we can again approximate the expectation $\mathbb{E}_{\boldsymbol{\beta}|\mathbf{y},\boldsymbol{\rho}}\{(\hat{\boldsymbol{\beta}}_{\boldsymbol{\rho}} - \boldsymbol{\beta})(\hat{\boldsymbol{\beta}}_{\boldsymbol{\rho}} - \boldsymbol{\beta})^\top\}\}$ with $\mathbf{V}$ given $\boldsymbol{\rho}$ (see Equation (\ref{eq:posterior_betaT})). Additionally, the first-order Taylor expansion of the random variable $\hat{\boldsymbol{\beta}}_{\boldsymbol{\rho}}$ is simply $\hat{\boldsymbol{\beta}} + \mathbf{J}(\boldsymbol{\rho} - \hat{\boldsymbol{\rho}}) + \epsilon$ (see again Equation (\ref{eq:posterior_betaT})) which again suggests using $\mathbf{V}^J$ as an estimate for $cov(\hat{\boldsymbol{\beta}}_{\boldsymbol{\rho}})$. Taking this into account, we arrive at

\begin{equation}\label{eq:exptau2}
\begin{split}
    \tau' &\approx tr(\mathbb{E}_{\boldsymbol{\rho}|\mathbf{y}}\{[\mathbf{V]}^{\boldsymbol{\rho}}\}\mathbf{H}) + tr(\mathbf{V}^J\mathbf{H})\\&=
    \mathbb{E}_{\boldsymbol{\rho}|\mathbf{y}}\{tr([\mathbf{V}]^{\boldsymbol{\rho}}\mathbf{H})\} + tr(\mathbf{V}^J\mathbf{H}),
\end{split}
\end{equation}

\noindent
where $[\mathbf{V}]^{\boldsymbol{\rho}}$ denotes the aforementioned estimate for $\mathbf{V}$ given $\boldsymbol{\rho}$ (not to be confused with the covariance matrix $\mathbf{V}^\rho$ of the normal approximation to the posterior $\boldsymbol{\rho}|\mathbf{y}$ defined in Equation (\ref{eq:posterior_rho})) and the second line is justified because $\mathbf{H}$ is being treated as fixed here. The remaining expectation $\mathbb{E}_{\boldsymbol{\rho}|\mathbf{y}}\{tr([\mathbf{V}]^{\boldsymbol{\rho}}\mathbf{H})\}$ has to be approximated through numerical integration \parencite[e.g.,][]{greven_comment_2016}. Substituting a simple MC estimate \parencite[e.g.,][]{robert_monte_2004} for the expectation results in

\begin{equation}\label{eq:exptau3}
\begin{split}
    &\tau' \approx \left(\sum_{r=1}^{N_{r}} \omega_\mathcal{N}(\boldsymbol{\rho}_r)tr([\mathbf{V}]^{\boldsymbol{\rho}_r}\mathbf{H})\right) + tr(\mathbf{V}^J\mathbf{H}),
\end{split}
\end{equation}

\noindent
where $\omega_\mathcal{N}(\boldsymbol{\rho})=\frac{1}{N_r}$ and the $\boldsymbol{\rho}_r$ are sampled from the normal approximation to the posterior given in Equation (\ref{eq:posterior_rho}). Note that $tr(\left[[\mathbf{V}]^{\hat{\boldsymbol{\rho}}} + \mathbf{V}^J\right]\mathbf{H})$ can readily be used as a lower-bound for the result in Equation (\ref{eq:exptau3}). In practice, for models involving a moderate number of $\lambda$ parameters ($\approx$ 5-10), computing $\tau'$ as shown in Equation (\ref{eq:exptau3}) with $150 \leq N_{r}\leq250$ will often result in better performance than what can be achieved by this lower-bound alone. The reason for this is that the normal approximation to the posterior $\boldsymbol{\rho} |\mathbf{y}$ will often have non-negligible support only for candidate vectors $\boldsymbol{\rho}_r$ that are close to $\hat{\boldsymbol{\rho}}$. While the requirement for $150 \leq N_{r}\leq250$ might still appear to be impractical, the cost of evaluating the entire term in square brackets of Equation (\ref{eq:exptau3}) is comparable to completing $N_{r}$ fitting iterations of Algorithm (\ref{alg:sparse_gamm}), because all the terms necessary to evaluate $tr([\mathbf{V}]^{\boldsymbol{\rho}_r}\mathbf{H})$ have to be computed as part of a fitting iteration anyway. In addition, for strictly additive models $tr([\mathbf{V}]^{\boldsymbol{\rho}_r}\mathbf{H})$ itself can be computed with practically no overhead \parencite[e.g.,][]{wood_generalized_2017} from the traces $tr([\mathbf{V}]^{\boldsymbol{\rho}_r}\mathbf{S}^r)$ evaluated for all $\lambda_r\in\boldsymbol{\lambda}$, which again have to be computed for the EFS update (see Equation (\ref{eq:efs})) at every iteration of Algorithm (\ref{alg:sparse_gamm}). For very large but sparse multi-level additive models, computing Equation (\ref{eq:exptau3}) will thus often be much faster than computing $\tau'$ after first computing $\tilde{\mathbf{V}}$ which has a complexity of $O(N_\lambda * N_p^3)$ \parencite[e.g.,][]{wood_smoothing_2016}.

\subsection{Monte Carlo Estimation of $\tau'$ for General Smooth Models}\label{sec:MCGS}

Equation (\ref{eq:exptau3}) first has to be modified to also be of practical relevance for more generic smooth models, because the same problems surrounding the use of Equation (\ref{eq:pql_hess_rho}) to approximate $\mathbf{V}^\rho$ and $\tilde{\mathbf{V}}$ continue to apply. Specifically, samples $\boldsymbol{\rho}_r$ obtained from a Normal distribution parameterized by the EFS estimates for $\boldsymbol{\rho}$ and $\mathbf{V}^{\rho}$ will not necessarily be distributed according to the normal approximation to the posterior $\boldsymbol{\rho}|\mathbf{y}$ defined in Equation (\ref{eq:posterior_rho}). As a consequence, the weighted sum in Equation (\ref{eq:exptau3}) will not necessarily be a good estimate of the required expectation $\mathbb{E}_{\boldsymbol{\rho}|\mathbf{y}}$ -- independent of the value chosen for $N_r$. To alleviate these problems, the expectation $\mathbb{E}_{\boldsymbol{\rho}|\mathbf{y}}$ can be taken with respect to the \emph{Bayesian posterior} $\boldsymbol{\rho}|\mathbf{y}$ rather than the normal approximation. Application of Bayes rule reveals that

\begin{equation}\label{eq:proportionality}
    p(\boldsymbol{\rho} | \mathbf{y}) \propto p(\mathbf{y}|\boldsymbol{\rho})p(\boldsymbol{\lambda}) \approx exp(\mathcal{V}(\boldsymbol{\lambda}))p(\boldsymbol{\lambda}),
\end{equation}

\noindent
i.e., that the density $p(\boldsymbol{\rho} | \mathbf{y})$ of the Bayesian posterior $\boldsymbol{\rho}|\mathbf{y}$, not just the normal approximation from Equation (\ref{eq:posterior_rho}), is proportional to $p(\mathbf{y}|\boldsymbol{\lambda})p(\boldsymbol{\lambda})$ \parencite[see also][]{greven_comment_2016}. Here, $p(\mathbf{y}|\boldsymbol{\lambda})$ again denotes the Bayesian marginal likelihood while $p(\boldsymbol{\lambda})$ denotes the density of a prior placed on $\boldsymbol{\rho}$. This proportionality continues to hold approximately when the Laplace-approximate result $exp(\mathcal{V}(\boldsymbol{\lambda}))$ defined in Equation (\ref{eq:laplace_bml}) is substituted for $p(\mathbf{y}|\boldsymbol{\lambda})$ \parencite[e.g.,][]{greven_comment_2016}. 

\Textcite{greven_comment_2016} outline a grid-based strategy that can be used to numerically approximate expectations $\mathbb{E}_{\boldsymbol{\rho}|\mathbf{y}}$ taken with respect to the Bayesian posterior when finite Uniform priors are placed on each $\rho_r \in \boldsymbol{\rho}$. Specifically, they suggest to cover the prior space implied by the finite limits enforced for each $\rho_r \in\boldsymbol{\rho}$ with a fine equidistant grid of $N_r$ candidate vectors $\boldsymbol{\rho}_r$. Since $p(\boldsymbol{\rho}|\mathbf{y}) \propto p(\mathbf{y}|\boldsymbol{\rho}) \approx exp(\mathcal{V}(\boldsymbol{\rho}))$ holds in the interior of the prior space, expectations $\mathbb{E}_{\boldsymbol{\rho}|\mathbf{y}}$ can then again be evaluated by means of a weighted sum, where the weights are defined \parencite[e.g.,][]{greven_comment_2016} as 

\begin{equation}\label{eq:remlweights}
\begin{split}
    \omega_\mathcal{V}(\boldsymbol{\rho})=\frac{exp(\mathcal{V}(exp(\boldsymbol{\rho})))}{\sum_{r=1}^{N_{r}}exp(\mathcal{V}(exp(\boldsymbol{\rho}_r)))}.
\end{split}
\end{equation}

\noindent
Unfortunately, maintaining a fine grid resolution quickly becomes prohibitively expensive for models involving more than one smoothing penalty, since $N_r$ would grow rapidly. It would thus be desirable to again rely on a MC estimate for expectations $\mathbb{E}_{\boldsymbol{\rho}|\mathbf{y}}$ taken with respect to the Bayesian posterior. For more generic smooth models, we propose to rely on ``importance sampling'' \parencite[e.g.,][]{casella_post-processing_1998,geweke_bayesian_1989,robert_monte_2004} to obtain this MC estimate. The latter again corresponds to a weighted sum as shown in Equation (\ref{eq:exptau3}) but with ``importance weights'' in place of the simple MC weights $\omega(\boldsymbol{\rho})=\frac{1}{N_r}$. The weights defined by \citeauthor{greven_comment_2016} (\citeyear{greven_comment_2016}; see also Equation (\ref{eq:remlweights})) are actually a special case of these importance weights, which are defined \parencite[e.g.,][]{casella_post-processing_1998,geweke_bayesian_1989} as

\begin{equation}\label{eq:importanceweights}
\begin{split}
    \omega(\boldsymbol{\rho})=\frac{exp(\mathcal{V}(exp(\boldsymbol{\rho})) + log(p(\boldsymbol{\rho}))- log(q(\boldsymbol{\rho})))}{\sum_{r=1}^{N_{r}}exp(\mathcal{V}(exp(\boldsymbol{\rho}_r)) + log(p(\boldsymbol{\rho}_r)) - log(p(\boldsymbol{\rho}_r)))},
\end{split}
\end{equation}

\noindent
where $q(\boldsymbol{\rho})$ denotes the density of the ``proposal distribution'' used to generate the $N_r$ candidate vectors $\boldsymbol{\rho}_r$, and $p(\boldsymbol{\rho})$ again corresponds to the density of the specific prior placed on $\boldsymbol{\rho}$. Importantly, while the normal approximation to the posterior $\boldsymbol{\rho}|\mathbf{y}$ would be the natural choice for the proposal distribution, a normal distribution parameterized with the EFS estimates for $\boldsymbol{\rho}$ and $\mathbf{V}^{\rho}$ can also be used instead. Problems can arise mainly in case the resulting proposal distribution underestimates the variance of some $\rho_r \in \boldsymbol{\rho}$. However, this risk can be mitigated by relying on a T-distribution with relatively low degrees of freedom to generate the candidate vectors $\boldsymbol{\rho}_r$ instead \parencite[e.g.,][]{wood_inference_2020}. Therefore, use of $\mathbf{V}^{\rho}$ computed as shown in Equation (\ref{eq:pql_hess_rho}) is far less problematic for this purpose, despite the strong assumptions surrounding use of the latter for generic smooth models. In practice, when the proposal distribution is initialized like this it usually produces samples that are probable under the target posterior.  

To numerically approximate expectations $\mathbb{E}_{\boldsymbol{\rho}|\mathbf{y}}$, taken with respect to the target posterior for a given prior, $\omega(\boldsymbol{\rho})$ can thus be substituted for $\omega(\boldsymbol{\rho})$ in Equation (\ref{eq:exptau3}). Generally this alternative definition for the weights makes Equation (\ref{eq:exptau3}) more expensive to compute, since the REML score has to be evaluated for every candidate vector $\boldsymbol{\rho}_r$. However, if the required log-determinant in Equation (\ref{eq:laplace_reml}) is based on the Cholesky of the pre-conditioned Hessian of the penalized log-likelihood for a given $\boldsymbol{\rho}_r$ the increase in complexity remains moderate \parencite[see][for a strategy to efficiently compute the remaining generalized log-determinant]{wood_fast_2011}.

Importantly, for more generic smooth models we will also want to substitute an alternative estimate for $cov(\hat{\boldsymbol{\beta}}_\rho)$ that does not assume that $\boldsymbol{\rho}$ is distributed according to the normal approximation of the posterior. For this we can again follow \Textcite{greven_comment_2016} and rely on the fact that $cov(\hat{\boldsymbol{\beta}}_\rho) =$

\begin{equation}\label{eq:covBp}
 \mathbb{E}_{\boldsymbol{\rho}|\mathbf{y}}\{\hat{\boldsymbol{\beta}}_{\boldsymbol{\rho}}\hat{\boldsymbol{\beta}}_{\boldsymbol{\rho}}^\top\} - \mathbb{E}_{\boldsymbol{\rho}|\mathbf{y}}\{\hat{\boldsymbol{\beta}}_{\boldsymbol{\rho}}\}\mathbb{E}_{\boldsymbol{\rho}|\mathbf{y}}\{\hat{\boldsymbol{\beta}}_{\boldsymbol{\rho}}^\top\}.
\end{equation}

Clearly, the expectations in Equation (\ref{eq:covBp}) can again be taken with respect to the Bayesian posterior $\boldsymbol{\rho}|\mathbf{y}$ and thus be replaced with MC estimates as well. After substituting the result shown in Equation (\ref{eq:covBp}) for $tr(\mathbf{V}^J\mathbf{H})$ in Equation (\ref{eq:exptau3}), a little re-organization of the resulting trace terms then finally yields the alternative MC estimate  

\begin{equation}\label{eq:exptau4}
\begin{split}
    \tau' \approx &\left(\sum_{r=1}^{N_r} \omega(\boldsymbol{\rho}_r)tr([\mathbf{V}]^{{\boldsymbol{\rho}}_r}\mathbf{H})\right) +
    \left(\sum_{r=1}^{N_{r}} \omega(\boldsymbol{\rho}_r)  \hat{\boldsymbol{\beta}}_{\boldsymbol{\rho}_r}^\top\mathbf{H}\hat{\boldsymbol{\beta}}_{\boldsymbol{\rho}_r}\right)-\\
    &\left[\sum_{r=1}^{N_{r}} \omega(\boldsymbol{\rho}_r)\hat{\boldsymbol{\beta}}_{\boldsymbol{\rho}_r}\right]^\top
    \mathbf{H}
    \left[\sum_{r=1}^{N_{r}} \omega(\boldsymbol{\rho}_r) \hat{\boldsymbol{\beta}}_{\boldsymbol{\rho}_r}\right]. 
\end{split}
\end{equation}

\noindent
Finally, the choice of $\mathbf{H}$ in Equation (\ref{eq:exptau4}) needs to be discussed for general smooth models. As mentioned in the previous section, the latter here denotes the negative Hessian of the log-likelihood at $\hat{\boldsymbol{\beta}}_{\hat{\boldsymbol{\rho}}^*}$ \parencite[or the expected version;][]{wood_smoothing_2016}. However, for more generic models the negative Hessian of the log-likelihood at $\hat{\boldsymbol{\beta}}_{\hat{\boldsymbol{\rho}}^*}$ will generally not be available (because $\hat{\boldsymbol{\rho}}^*$ will not be available). The simplest option would be to substitute the negative Hessian of the log-likelihood at $\hat{\boldsymbol{\beta}}_{\hat{\boldsymbol{\rho}}}$, i.e., the final EFS estimate, instead. As discussed, $\mathbf{H}$ will often be reasonably stationary in practice, in which case this would again be justified. An alternative that avoids this assumption would be to re-compute the negative Hessian of the log-likelihood at an estimate of the mean of the Bayesian posterior $\boldsymbol{\beta}|\mathbf{y}$ \parencite[cf.][]{wood_just_2016}. This is readily motivated by considering that $\hat{\boldsymbol{\beta}}_{\hat{\boldsymbol{\rho}}^*}$ tends to the mean of the normal approximation to the posterior $\boldsymbol{\beta}|\mathbf{y}$ \parencite[e.g.,][]{wood_smoothing_2016}. We can again approximate this mean as $\mathbb{E}_{\boldsymbol{\beta}|\mathbf{y}}\{\boldsymbol{\beta}\}\approx\sum_{r=1}^{N_{r}} \omega(\boldsymbol{\rho}_r) \hat{\boldsymbol{\beta}}_{\boldsymbol{\rho}_r}$ and the negative Hessian of the log-likelihood evaluated at this mean estimate can then be substituted for $\mathbf{H}$ in Equation (\ref{eq:exptau4}).

Use of Equation (\ref{eq:exptau4}) to estimate $\tau'$ in practice again requires specifying the number of samples $N_r$ on which to base the MC estimate. Additionally, a suitable prior distribution needs to be selected, and the choice is likely to affect the required number of samples $N_r$. One option would be to follow \Textcite{greven_comment_2016} and to place finite Uniform priors on each $\rho_r \in \boldsymbol{\rho}$. However, often it will be more efficient to use the chosen proposal distribution (i.e., Normal or T distribution with $\mathbf{V}^\rho$ from Section \ref{sec:pql_cAIC} for covariance matrix) for the prior for $\boldsymbol{\rho}$ as well. In that case, $\forall~\boldsymbol{\rho}_r ~ log(q(\boldsymbol{\rho}_r))=log(p(\boldsymbol{\rho}_r))$ and the importance weights simplify to the weights proposed by \Textcite{greven_comment_2016} shown in Equation (\ref{eq:remlweights}). Using a normal (or T) distribution for the prior, parameterized by the EFS estimate for $\boldsymbol{\rho}$ and $\mathbf{V}^{\rho}$ from Section \ref{sec:pql_cAIC} will generally be justified in practice, for the same reasons outlined in Section \ref{sec:MCAM}: $exp(V(exp(\boldsymbol{\rho}_r)))$ will often be non-negligible\footnote{This will not necessarily be the case in the direction of parameters $\lambda_r \to \infty$. However, as pointed out by \Textcite{greven_comment_2016}, the correction by \Textcite{wood_smoothing_2016} also does not account for this and yet continues to work well in practice.} only for candidate vectors $\boldsymbol{\rho}_r$ close to the EFS estimate $\hat{\boldsymbol{\rho}}$ -- even if the latter does not exactly maximize the REML criterion. This also suggests that, for models involving a moderate number of $\lambda$ parameters, a MC estimate based on the weights in Equation (\ref{eq:remlweights}) will often offer a good approximation for the expectations $\mathbb{E}_{\boldsymbol{\rho}|\mathbf{y}}$ even when $150\leq N_{r}\leq250$. Note also that Equation (\ref{eq:exptau3}), $tr\left(\left[[\mathbf{V}]^{\hat{\boldsymbol{\rho}}} + \mathbf{V}^J\right]\mathbf{H}\right)$, and even the PQL-like estimate of $\tau'$ from Section \ref{sec:pql_cAIC} all continue to be plausible lower-bounds, which can readily be enforced when relying on Equation (\ref{eq:exptau4}) to compute $\tau'$ independent of the chosen prior. 

To summarize the results from this section, we suggested to rely on Equation (\ref{eq:pql_hess_rho}) to approximate the covariance matrix $\mathbf{V}^\rho$ of the normal approximation to the marginal posterior $\boldsymbol{\rho}|\mathbf{y}$, the covariance matrix $\tilde{\mathbf{V}}$ of the normal approximation to the marginal posterior $\boldsymbol{\beta}|\mathbf{y}$, and $\tau'$, required for the uncertainty-corrected version of the cAIC \parencite[e.g.,][]{wood_smoothing_2016}. This was motivated by the desire to ensure that approximately correcting for uncertainty in the estimate $\hat{\boldsymbol{\lambda}}$ remains possible, even when higher-order derivatives are unavailable. Reliance on Equation (\ref{eq:pql_hess_rho}) implies that all three quantities ($\mathbf{V}^\rho$, $\tilde{\mathbf{V}}$, and $\tau'$) will generally be approximate, unless $\mathbf{H}$ will be stationary, which will for example be the case for Gaussian models and some canonical GAMs \parencite[][]{breslow_approximate_1993,wood_generalized_2017}.

We reviewed large-sample results \parencite[e.g.,][]{breslow_approximate_1993,wood_smoothing_2016,wood_generalized_2017-1,wood_generalized_2017}, which suggest that these approximate quantities will often still be useful in practice, also for more generic models, as long as $N_p << N$ (see also Sections \ref{sec:GAM} and \ref{sec:GAMMLSS}). However, in cases where prior knowledge or an investigation of the REML criterion (e.g., Figure \ref{fig:TaylorConvergence}) suggest strong dependencies of $\mathbf{H}$ on $\boldsymbol{\rho}$, the alternative MC strategy from this section should be used to compute the cAIC (see Equation (\ref{eq:exptau4})). This alternative does not require strong assumptions, such as requiring that $\mathbf{H}$ is reasonably stationary, and can readily be used for any smooth model -- independent of whether or not the EFS update has been used during estimation. Generally, this comes at the cost of a potentially expensive sampling step. The advantage exploited here is that the EFS estimates for $\hat{\boldsymbol{\beta}}$ and $\hat{\boldsymbol{\rho}}$ can be used in combination with Equation (\ref{sec:pql_cAIC}) to generate samples $\boldsymbol{\rho}_r$ that will generally be probable under the Bayesian posterior $\boldsymbol{\rho} |\mathbf{y}$. Hence, this strategy will often still produce reasonable approximations to $\tau'$ for models involving a modest number of smoothing parameters even when only $150 \leq N_r \leq 250$ samples are generated -- especially when the estimate of $\tau'$ from Section \ref{sec:pql_cAIC} is used as a lower bound. For more complex models it will generally be advisable to increase $N_r$.

Relying on a MC estimate of $\tau'$ will also be useful when working with very large but sparse strictly additive mixed models. While the $O(N_\lambda*(N_p)^3)$ complexity of directly computing the correction by \Textcite{wood_smoothing_2016} will be prohibitive for these models, relying on the MC estimate defined in Equation (\ref{eq:exptau3}) is more likely to remain feasible.

Before evaluating the performance of the strategies discussed in this section in a simulation study, we would like to point out that the cAIC is by no means the only answer to model comparison or selection questions. For example, approximate confidence intervals and p-values for the null-hypothesis that a smooth terms is zero everywhere over the domain of its covariates are readily available, even for more generic models, and do not require higher-order derivative information nor accounting for uncertainty in $\hat{\boldsymbol{\lambda}}$ to provide well-calibrated results \parencite[e.g.,][]{wahba_bayesian_1983,wood_p-values_2013,marra_coverage_2012}. In some situations it might nevertheless be desirable to obtain credible intervals based on the normal approximation to the marginal posterior $\boldsymbol{\beta}|\mathbf{y}$, which can be achieved using $\tilde{\mathbf{V}}$ \parencite[e.g.,][]{wood_generalized_2017-2}. Note, that to this end Equations (\ref{eq:exptau3}) and (\ref{eq:exptau4}) can both be modified to compute $\tilde{\mathbf{V}}$ instead of $\tau'$, which was the original proposal by \Textcite{greven_comment_2016}.

Yet another alternative form of model selection which remains feasible under the EFS update is penalty-based term selection as proposed by \Textcite{marra_practical_2011}. Penalty-based term selection works by obtaining an additional penalty matrix for a specific smooth term that regularizes any function in the Kernel of the original penalty matrix $\mathcal{S}^r$, allowing to shrink the corresponding smooth function entirely to $\mathbf{0}$ \parencite[][]{marra_practical_2011}.

\subsection{Implementation in MSSM}

The MSSM toolbox supports all of these alternative strategies and also allows to estimate $\mathbf{V}^\rho$, $\tilde{\mathbf{V}}$, and $\tau'$ via the strategies discussed in Sections \ref{sec:pql_cAIC}, \ref{sec:MCAM}, and \ref{sec:MCGS}. By default, it relies on the strategy described in Section \ref{sec:pql_cAIC} for Gaussian models to compute $\tilde{\mathbf{V}}$ and to correct the cAIC (i.e., to compute $\tau'$). For very large Gaussian mixed models it instead opts for the MC strategy of Section \ref{sec:MCAM} by default. For more generic smooth models it falls back to the strategy of Section \ref{sec:MCGS} instead (assuming that the proposal distribution is used as prior as well). For an overview over the different functions and the different arguments they support, we refer to the documentation of the \texttt{compare} and \texttt{utils} modules, available at \url{https://jokra1.github.io/mssm}.

This marks the conclusion of our discussion of the final problem formulated in the Introduction and Background section: with the steps described in this section it is again possible to (approximately) estimate uncertainty in the estimate $\hat{\boldsymbol{\lambda}}$ and to correct for this uncertainty, for example when using the cAIC to select between smooth models estimated based on the methods presented in this paper. In the next section, we will evaluate the performance of the theoretical framework, and our implementation in the MSSM Python toolbox, across multiple simulation studies. Subsequently, we will consider several real data examples.

\section{Results}\label{sec:Results}
\begin{figure*}[!h]
    \caption{Overview of Functions Used Across Simulation Studies}
    \includegraphics[width=\textwidth]{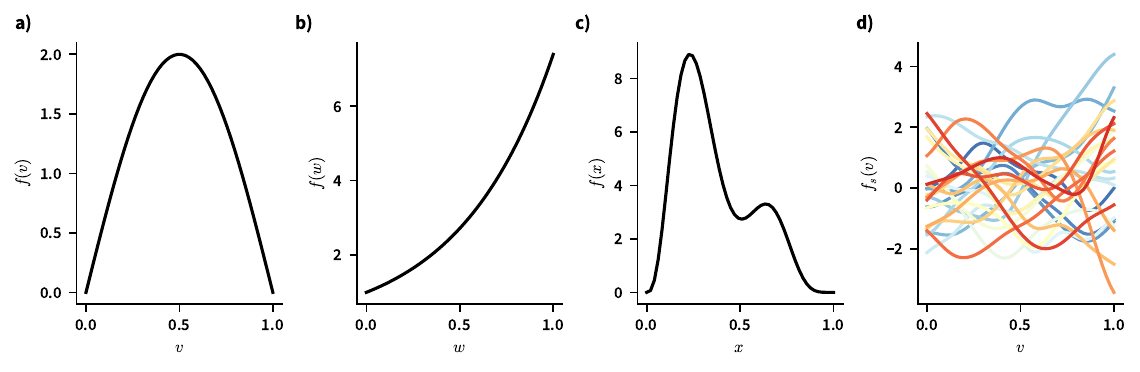}
    \label{fig:sim_func}
    {\small
         \textit{Note.} Figure \ref{fig:sim_func} shows an overview of the functions used to model the linear predictor across simulations studies \parencite[previously used by][]{gu_minimizing_1991,wood_smoothing_2016}. Panels a, b, and c show the functions corresponding of the fixed predictors $v$, $w$, and $x$ used in every simulation study. Since $\forall~ z_i ~ f(z_i) = 0$, it is not visualized here. Panel d shows a sample of the subject-specific ``random smooth'' functions of $v$ used in the second simulation study. The latter were re-sampled for every simulated dataset.
        }
\end{figure*}

\subsection{Simulation Studies}\label{sec:Simulations}

To test the algorithms and extensions reviewed in the previous sections implemented in the Mixed-Sparse-Smooth-Model (MSSM) Python toolbox, a series of five simulation studies was conducted. All studies relied on the same four functions $f(v)$, $f(w)$, $f(x)$, and $f(z)$ as ground-truth to simulate fixed effects. These were proposed originally by \textcite{gu_minimizing_1991} and were used in the simulations by \Textcite{wood_smoothing_2016}. All functions but $f(z)$, which is zero everywhere over the covariate $z$, are depicted in Figure (\ref{fig:sim_func}a-c). 

The first simulation study, a replication of a study performed by \Textcite{wood_smoothing_2016}, was designed to verify that the algorithms reviewed here remain stable in the presence of strongly correlated predictor variables. This scenario is often used to highlight the problems that can arise when relying on algorithms that are based on the PQL approach or optimize (i.e., pivot) for sparsity rather than stability \parencite[e.g.,][]{wood_smoothing_2016}, and was thus considered a good test-case for the robustness of the strategies we presented.

Simulation two was designed to investigate the performance of the algorithms for multi-level models. 250 observations were simulated for each of 20 subjects, all of which showed a different non-linear deviation from the effect of one continuous predictor, to be captured by random smooth terms. A random sample of these subject-specific deviations $f_s(v)$, is shown in Figure \ref{fig:sim_func}d. This simplified setting is sufficient to reveal the performance gains that are possible when using the algorithms discussed in this paper, while also remaining relatively cheap to simulate. Section \ref{sec:DataExamples} features much more complex multi-level smooth models of millions of observations that can only be estimated when using MSSM.

The final two simulation studies, again replications of studies conducted by \Textcite{wood_smoothing_2016}, focused on the problem of correcting for uncertainty in $\hat{\boldsymbol{\lambda}}$ when computing the conditional AIC for the purpose of model selection. In Section \ref{sec:Uncertainty}, we discussed multiple strategies that aim to facilitate this. These strategies were now tested in two selection problems. Specifically, Simulation 4 focused on selecting between a model including a smooth term and a model excluding it. Simulation 5 on the other hand focused on selecting a random effect.

\begin{table*}[!t]
    \centering
    \caption{Model Estimation Algorithms}
    \renewcommand{\arraystretch}{1.25}
    \begin{tabular}{p{.05\textwidth}|p{.625\textwidth}|p{.25\textwidth}}
         \toprule
         \centering \textbf{ID} & \centering \textbf{Estimation Algorithm} &  \centering \textbf{Related Sections \& Software} \tabularnewline
         \midrule
          \centering \emph{a1} & Based on methods discussed by \Textcite{wood_smoothing_2016}. Newton's method is used to estimate $\boldsymbol{\beta}$ and $\boldsymbol{\rho}=log(\boldsymbol{\lambda})$. For GAMMs the method by \Textcite{wood_fast_2011} is used instead. In both cases, the Laplace approximate REML criterion in Equation (\ref{eq:laplace_reml}) is optimized exactly. & Sections \ref{sec:AM}, \ref{sec:GAM}, and \ref{sec:GAMMLSS} of this paper and the \texttt{gam} function available in mgcv \\
          \centering \emph{a2} & Based on methods discussed by \Textcite{wood_generalized_2015}. Fisher scoring is used to estimate $\boldsymbol{\beta}$. Given pseudo-data $\mathbf{z}$ and weight matrix $\mathbf{W}$, $\boldsymbol{\rho}$ is then estimated via Newton's method as if $z_i \sim \mathcal{N}(\mathbf{X}\boldsymbol{\beta},\frac{1}{W_{ii}}\phi)$. The Laplace approximate REML criterion in Equation (\ref{eq:laplace_reml}) is generally \emph{not} optimized exactly. & Sections \ref{sec:AM} and \ref{sec:GAM} of this paper and the \texttt{bam} function available in mgcv\\
          \midrule
          \centering \emph{a3} & Based on Algorithm (\ref{alg:sparse_gamm}). Alternating between single updates to $\hat{\boldsymbol{\beta}}$ and EFS updates to $\boldsymbol{\lambda}$. Essentially this is the PQL strategy proposed by \Textcite{wood_generalized_2017-1} with the EFS update by \Textcite{wood_generalized_2017} in place of the Newton update to $\boldsymbol{\rho}$. The Laplace approximate REML criterion in Equation (\ref{eq:laplace_reml}) is generally \emph{not} optimized exactly. & Sections \ref{sec:AM} and  \ref{sec:GAM} of this paper and the \texttt{GAMM} class available in MSSM\\
          \centering \emph{a4} & Based on Algorithm (\ref{alg:sparse_gamm}) in combination with Algorithm (\ref{alg:pseudo_dat}). Steps in Algorithm (\ref{alg:pseudo_dat}) are iterated until convergence in $\hat{\boldsymbol{\beta}}$ is reached for every proposed update to $\boldsymbol{\lambda}$. Otherwise like $s3$. &  Sections \ref{sec:AM} and  \ref{sec:GAM} of this paper and the \texttt{GAMM} class available in MSSM\\
          \centering \emph{a5} & Based on Algorithm (\ref{alg:sparse_gamm}) in combination with Algorithm (\ref{alg:gammlss_beta}). Newton's method is used to estimate $\boldsymbol{\beta}$. The strategy reviewed in appendix \ref{sec:AppendixUnidentifiable} is used to check for unidentifiable parameters before updating $\boldsymbol{\lambda}$ via the EFS update. Essentially this is the strategy proposed by \Textcite{wood_smoothing_2016} with the EFS update in place of the exact Newton update. The Laplace approximate REML criterion in Equation (\ref{eq:laplace_reml}) is generally \emph{not} optimized exactly. &  Sections \ref{sec:AM}, \ref{sec:GAM}, and \ref{sec:GAMMLSS} of this paper and the \texttt{GAMMLSS} and \texttt{GSMM} classes available in MSSM\\
          \midrule
          \centering \emph{a6} & Based on the L-qEFS update shown in Algorithm (\ref{alg:lqEFS}). Quasi-Newton estimates are obtained of $\boldsymbol{\beta}$ and $\boldsymbol{\lambda}$ based on Gradient of the log-likelihood alone. The Laplace approximate REML criterion in Equation (\ref{eq:laplace_reml}) is generally \emph{not} optimized exactly. &  Sections \ref{sec:GAMMLSS} and \ref{sec:lqefs} of this paper and the \texttt{GSMM} class available in MSSM\\
         \bottomrule
    \end{tabular}

    {\small
    $~$
    
    \textit{Note.} Table \ref{tab:model_overview} provides an overview of the algorithms used to obtain the six model estimates across simulation studies.
    }
    \label{tab:model_overview}
\end{table*}

Six different algorithms were used to estimate the true models from which observations were generated during the simulations. Table \ref{tab:model_overview} provides an overview. We use \emph{a1} to refer to the \emph{baseline algorithm}. This algorithm relied on the methods by \Textcite{wood_smoothing_2016} available via the \texttt{gam} function in mgcv in R \parencite{wood_smoothing_2016,wood_generalized_2017-2}. As a consequence, this algorithm utilizes up to fourth order derivatives of the log-likelihood. If the model to be fitted is a GAMM, the \texttt{gam} function in mgcv actually relies on the method proposed by \Textcite{wood_fast_2011}, which can however be considered a GAM-specific implementation of the general framework by \Textcite{wood_smoothing_2016}. Additionally, and exclusively for simulations in which only a single parameter of the log-likelihood was parameterized by an additive model, we use \emph{a2} to refer to the \emph{PQL baseline algorithm}. This algorithm relied on the methods by \Textcite{wood_generalized_2015}, available via the \texttt{bam} function in mgcv in R, to estimate the model.

Algorithms \emph{a3}, \emph{a4}, and \emph{a5} all relied on Algorithm (\ref{alg:sparse_gamm}) and thus the EFS update by \Textcite{wood_generalized_2017}. Algorithms \emph{a3} and \emph{a4} relied on the pseudo-data transformation described in Section \ref{sec:GAM} and outlined in Algorithm (\ref{alg:pseudo_dat}). For both, the sparse Cholesky strategy described in the section on additive models was used to solve for $\hat{\boldsymbol{\beta}}$ in step four of Algorithm (\ref{alg:pseudo_dat}). As discussed in Section \ref{sec:AM}, the alternative sparse QR strategy is more robust to numerical problems so relying on the Cholesky approach instead allows to get an estimate on the lower bound of performance that can be expected. Algorithm \emph{a5} relied on Algorithm (\ref{alg:gammlss_beta}) to estimate $\hat{\boldsymbol{\beta}}$. Finally, algorithm \emph{a6} relied on the SR-1 variant of the L-qEFS update, outlined in Algorithm (\ref{alg:lqEFS}) and described in Section \ref{sec:lqefsSR1}, to estimate $\boldsymbol{\beta}$ and $\boldsymbol{\lambda}$. Across all simulations we set $N_V=30$.

mgcv version 1.9-3 and MSSM version 1.0.1 were used for all simulations. All smooth terms were parameterized with 10 B-spline basis functions \parencite{eilers_flexible_1996}. All code used as part of the simulations is available on GitHub at \url{https://github.com/JoKra1/mssm-supplementary}.

\subsubsection{Simulation 1}

\begin{figure*}[!t]
    \caption{Simulation 1 Results}
    \includegraphics[width=\textwidth]{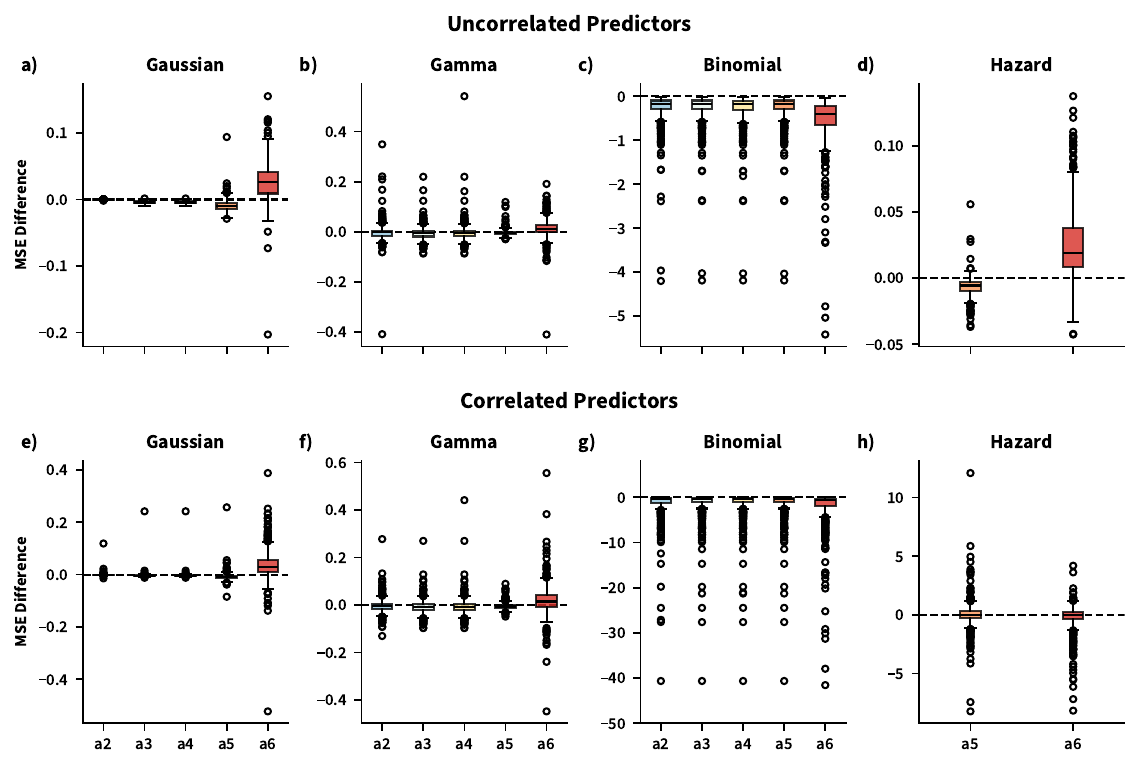}
    \label{fig:sim1_main}
    {\small
         \textit{Note.} Figure \ref{fig:sim1_main} shows boxplots of standardized MSE score differences. Scores were first normalized by the median MSE score computed per panel. MSE scores of algorithm \emph{a1} were then subtracted from the MSE scores of the remaining algorithms. Positive values thus indicate a lower MSE of algorithm \emph{a1}. Negative scores indicate a lower MSE of the alternative algorithm. The top row shows the results from simulations in which the predictor variables were un-correlated. The bottom row shows the results from simulations in which the predictor variables were correlated. For the correlated Binomial case (panel g) extreme MSE scores (MSE > $1e3$ up to $1e6$) were visible in 8 simulations for algorithm \emph{a1}. These cases are not visible, because the y-axis of panel g has been truncated at -50 to preserve interpretability.
        }
\end{figure*}

For the first study, we simulated data from Gaussian, Binomial, Gamma, and Proportional hazard likelihoods using the same linear predictor $\eta_i = f(v_i) + f(w_i) + f(x_i) + f(z_i)$. For the first three, the linear predictor was used to parameterize the true mean $\mu_i=g^{-1}(\eta_i)$. For the Gaussian likelihood, the identity link was used for $g$, i.e., $g(\mu_i)=\mu_i=\eta_i$. The canonical ``logit'' link function was used for $g$ for the Binomial likelihood, i.e., $log(\frac{\mu_i}{1-\mu_i})=\eta_i$. For the Gamma likelihood the non-canonical log link function was used for $g$ i.e., $log(\mu_i)=\eta_i$.

The true scale parameter for the Gaussian and Gamma likelihoods was set to $\phi=2$. To keep the signal-to-noise ratio comparable across models, we followed the strategy by \Textcite{wood_smoothing_2016} and scaled the linear predictor for the Binomial likelihood by the same $\phi=2$. For the Proportional hazard likelihood, observations were simulated from a Weibull distribution, with survival function $S(t) = - exp({-\frac{\eta}{10} t^\frac{1}{\phi}})$, again with $\phi=2$.

For each likelihood, we simulated two types of datasets: 500 data-sets were simulated in which predictors were uncorrelated and 500 data-sets were simulated in which predictors were highly correlated \parencite[see][for details on how this was achieved]{wood_smoothing_2016}. Each dataset consisted of $N=500$ simulated observations and the simulated values for all predictor variables $v$, $w$, $x$, and $z$ covered the range from -1 to 1.

To evaluate the performance of the different algorithms used for estimation, we again followed \Textcite{wood_smoothing_2016} and computed the mean squared error (MSE) between the true linear predictor $\boldsymbol{\eta}$ and the predicted version $\hat{\boldsymbol{\eta}}=\mathbf{X}\hat{\boldsymbol{\beta}}$, where $\mathbf{X}$ is again the model matrix and $\hat{\boldsymbol{\beta}}$ denotes the final set of coefficients obtained from the different algorithms summarized in Table \ref{tab:model_overview}.

Figure \ref{fig:sim1_main} shows boxplots of standardized MSE score differences \parencite[e.g.,][]{wood_smoothing_2016}. To obtain the standardized scores, the MSE scores were first normalized by the median MSE score, computed separately per panel. The standardized MSE scores of algorithm \emph{a1} were then subtracted from the standardized MSE scores of all remaining algorithms to obtain the difference scores visualized. Thus, positive values in Figure \ref{fig:sim1_main} indicate a lower MSE of the baseline algorithm \emph{a1}, while negative values imply a lower MSE score for the corresponding alternative algorithm in a particular simulation.

In the uncorrelated predictor case (top row), algorithms \emph{a3}, \emph{a4}, and \emph{a5} -- based on the methods discussed in Sections \ref{sec:AM}, \ref{sec:GAM}, and \ref{sec:GAMMLSS} of this paper -- on average achieved lower MSE scores than the baseline algorithm \emph{a1}. This was also the case for algorithm \emph{a2}, which relied on the methods by \Textcite{wood_generalized_2015}. While these differences were marginal, except for the Binomial likelihood (Figure \ref{fig:sim1_main}c, top row), it is still encouraging that the algorithms implemented in MSSM were at least as robust as the ones implemented in mgcv in this simulation study. For all but the Binomial likelihood, the MSE scores of algorithm \emph{a6} -- based on the L-qEFS update of Section \ref{sec:lqefs} -- were on average slightly higher than those of algorithm \emph{a1}. In contrast, for the Binomial likelihood algorithm \emph{a6} achieved the lowest MSE on average, outperforming all other algorithms. Generally these differences were marginal as well however, and are unlikely to be of practical significance.

The bottom row of Figure \ref{fig:sim1_main} shows results for the correlated predictor case. For all but the hazard likelihood, algorithms \emph{a2}, \emph{a3}, \emph{a4}, and \emph{a5} again achieved lower MSE scores than the baseline algorithm \emph{a1} on average. The differences generally remained small. However, in a small set of 8 simulations, extreme MSE scores were visible for algorithm \emph{a1} (MSE > $1e3$ up to $1e6$; omitted from the plot for interpretability reasons) for the Binomial likelihood. Upon inspection, the algorithm implemented by the \texttt{gam} function in mgcv clearly diverged\footnote{In these simulations mgcv warned that the model failed to converge. It is quite likely that a higher limit on the maximum number of iterations, a restart based on a randomly perturbed coefficient vector, or a switch to the alternative algorithm developed by \Textcite{wood_smoothing_2016} would be sufficient to address these problems.} in those cases and produced poor estimates of the underlying functions shown in Figure \ref{fig:sim_func}. In contrast, none of the remaining algorithms suffered from such extreme cases of divergence. MSE scores were again higher on average for algorithm \emph{a6}, compared to those of algorithm \emph{a1}. This was not the case for the Binomial likelihood, where algorithm \emph{a6} again achieved the lowest MSE on average. Generally, the differences between algorithm \emph{a6} and the remaining alternative algorithms (i.e., \emph{a2}-\emph{a5}) became even smaller in the correlated predictor case.

\begin{table*}[!t]
    \centering
    \caption{Simulation 2 Results}
    \renewcommand{\arraystretch}{1.25}
    \begin{tabular}{p{.1\textwidth}|p{.07\textwidth}|p{.07\textwidth}|p{.07\textwidth}|p{.11\textwidth}|p{.11\textwidth}?p{.075\textwidth}|p{.085\textwidth}|p{.075\textwidth}}
         \toprule
          \textbf{Likelihood} & \multicolumn{5}{c?}{\textbf{Median MSE Difference}} & \multicolumn{3}{c}{\textbf{Median CPU Time Ratio}}
         \tabularnewline
         \hline
          & \centering \emph{a2}-\emph{a1} & \centering \emph{a3}-\emph{a1} & \centering \emph{a4}-\emph{a1} & \centering \emph{a5}-\emph{a1} & \centering \emph{a6}-\emph{a1} & \centering \emph{a3}/\emph{a2} & \centering (\emph{a4}$|$\emph{a5})/\emph{a1} & \centering \emph{a6}/\emph{a1} 
         \tabularnewline
         \hline
         \centering \textbf{Gaussian} & -0.0 & -0.002 & -0.002 & -0.001 & 0.081 & 1.375 & 0.039 & 11.104 \tabularnewline
         \hline
         \centering \textbf{Gamma} & 0.001 & -0.002 & -0.002 & -0.002 & 0.09 & 0.282 & 0.208 & 7.955 \tabularnewline
         \hline
         \centering \textbf{Binomial} & -0.07 & -0.073 & -0.074 & -0.073 & -0.022 & 0.221 & 0.254 & 13.804 \tabularnewline
         \hline
         \centering \textbf{Hazard} & \centering - & \centering - & \centering - & -0.005 & -0.018 & \centering - & 2.646 & 67.681 \tabularnewline
         \midrule
         \centering \textbf{Gaussian} ($\mu$ \& $\phi$) & \centering - & \centering - & \centering - & 0.0 (-0.006) & 0.08 (0.116) & \centering - & 0.261 & 3.557 \tabularnewline
         \hline
         \centering \textbf{Gamma} \linebreak ($\mu$ \& $\phi$) & \centering - & \centering - & \centering - & -0.0 (0.028) & 0.193 (0.287) & \centering - & 0.728 & 5.116 \tabularnewline
         \bottomrule
    \end{tabular}

    \flushleft{\small
    $~$

    \textit{Note.} Table \ref{tab:sim2_main} provides an overview of the results from the second simulation study. Columns 2-6 depict the median of the standardized MSE score differences for every likelihood and per comparison. MSE scores were again standardized as was done in the first simulation study. Values below zero thus again indicate lower MSE scores compared to the baseline algorithm. Values of 0.0 and -0.0 indicate comparisons for which the absolute value of the median was smaller than $6e{-4}$. For settings in which multiple parameters of the log-likelihood were modeled, the value in parentheses corresponds to the median for the scale parameter (i.e., $\phi$). Columns 7-9 show the median of relative CPU time required by the estimation algorithms. Relative CPU times were obtained for each simulated dataset by dividing the CPU time of algorithms \emph{a2}, \emph{a3}, \emph{a4} (\emph{a5} whenever \emph{a4} was not applicable), and \emph{a6} by the CPU time of the conceptually most similar baseline algorithm (i.e., either \emph{a1} or \emph{a2}). Values below 1 thus indicate lower CPU time compared to the baseline algorithm (i.e., estimating the model took less time in MSSM than in mgcv).
    }
    \label{tab:sim2_main}
\end{table*}

Overall, the results from this simulation suggest that the algorithms discussed in this paper and implemented in the MSSM toolbox are robust to the kind of numerical problems that can result from correlated predictor variables during estimation. The results from the Binomial likelihood are particularly encouraging, because this case is well known to often cause problems for models estimated via some form of the PQL approach \parencite[e.g.,][]{wood_smoothing_2016,wood_generalized_2017-2}. However, neither the MSE scores of algorithm \emph{a2} nor those of algorithm \emph{a3} suggested any case of clear divergence in the correlated predictor case. In fact, the algorithm requiring the least information about the likelihood, algorithm \emph{a6}, achieved the lowest average MSE score (see the online repository containing the simulation code).

In contrast, compared to most of the alternative algorithms, MSE scores were generally slightly higher for algorithm \emph{a1} implemented by the \texttt{gam} function in mgcv. In addition, this was the only algorithm that showed clear signs of divergence in a small set of simulations for the Binomial likelihood. Considering that this was the only algorithm that did not neglect dependencies of the negative Hessian of the log-likelihood $\mathbf{H}$ on the regularization parameters $\boldsymbol{\lambda}$ during estimation, this outcome is surprising. However, we caution against overstating the importance of these results. Specifically, this result should \emph{not} be interpreted as evidence against the usefulness or importance of the high theoretical complexity of the method proposed by \Textcite{wood_fast_2011}. On most likelihoods considered here, the differences between all algorithms were generally small and unlikely to be of practical significance. Additionally, in the original simulations performed by \Textcite{wood_smoothing_2016}, their method, which can be considered an extension of the strategy by \Textcite{wood_fast_2011}, did not show any signs of divergence in the Binomial case. The results here thus merely suggest the obvious: that high theoretical complexity and reliance on algorithms that numerically pivot for stability (see Section \ref{sec:AM} for a discussion) cannot fully prevent the possibility of divergence. Similarly, the results suggest that using algorithms that pivot for sparsity \emph{can} still result in stable estimates in the presence of correlated predictors -- but this is also not guaranteed to hold in every possible scenario.

\subsubsection{Simulation 2}

For the second simulation study, a similar setup was used: we considered Gaussian, Binomial, Gamma, and Proportional hazard likelihoods. For the Gaussian and Gamma likelihood we again considered the constant $\phi=2$ case, but also the case where $\phi$ was parameterized by a second linear predictor. For the constant $\phi$ case, the same linear predictor was again used to parameterize the mean of the Gaussian, Binomial, and Gamma likelihoods and the survival function of the hazard model. In the second study the ground truth of this linear predictor was specified as $\eta_i = f(v_i) + f(w_i) + f(x_i) + f(z_i) +f_{s(i)}(v_i)$, where $s(i)$ indexes the simulated subject from which observation $i$ was obtained. The $f_{s(i)}(v_i)$ were treated as random, i.e. they were re-sampled for every simulated data set and every simulated subject (see Figure \ref{fig:sim_func}d). The same link functions used in the first study were used here again.

As mentioned, for the Gaussian and Gamma likelihoods we also considered the case of two linear predictors requiring estimation, one to parameterize the mean and one for the scale parameter. The true linear predictor for the mean was specified as $\eta^\mu_i = f(v_i) + f(w_i) +f_{s(i)}(v_i)$, while the true linear predictor for the scale parameter was specified as $\eta_i^\phi = f(x_i) + f(z_i)$. The log link was used for the scale parameter for both likelihoods, i.e., $log(\phi_i) = \eta_i^\phi$. For every likelihood, the continuous predictors for each of the 100 simulated datasets were again obtained by sampling from a uniform distribution. No explicit correlation was enforced between them. Every dataset consisted of 5000 simulated observations (based on 20 simulated subjects).

Table \ref{tab:sim2_main} summarizes the results from this second study. Columns 2-6 show the median of the differences in standardized MSE scores between the alternative algorithms and the baseline algorithm (i.e., \emph{a1}) for every likelihood. Note that two median MSE scores are provided for the likelihoods involving two linear predictors -- the one in parentheses corresponds to the scale parameter $\phi$. The results largely match those of the first simulation study. Generally, differences between the MSE score of algorithm \emph{a1} and those of algorithms \emph{a3}, \emph{a4}, and \emph{a5} were small. The latter typically achieved slightly lower MSE scores, which is also reflected in the median MSE scores. MSE score differences between algorithm \emph{a1} and \emph{a6} were more variable across likelihoods. For the Gaussian and Gamma likelihoods, higher median MSE scores were visible for algorithm \emph{a6}. A reversal of that pattern was again visible for the Binomial likelihood, with algorithm \emph{a6} achieving lower MSE scores on average than algorithm \emph{a1}. However, algorithm \emph{a6} no longer clearly outperformed the remaining algorithms for the Binomial likelihood in this second study.

The remaining columns show the median of the relative CPU times required to fit each model for all likelihoods. As a reminder, algorithm \emph{a3} is conceptually most similar to the PQL algorithm proposed by \Textcite{wood_generalized_2015} used by algorithm \emph{a2}. Hence, the more informative comparison here is between algorithms \emph{a3} and \emph{a2}, rather than \emph{a3} and \emph{a1}. This comparison is therefore shown in Table \ref{tab:sim2_main}. Algorithms \emph{a4} (i.e., \emph{a5} for the hazard likelihood and the two-predictor Gaussian and Gamma likelihoods) and \emph{a6} were again compared to algorithm \emph{a1}. Values below 1 in these remaining columns of Table \ref{tab:sim2_main} thus imply that fitting a particular model usually took less time in MSSM than in mgcv.

Because the model matrix $\mathbf{X}$ for the linear predictor of the mean is quite sparse in these studies, which in all but the hazard case can be exploited efficiently by Algorithm (\ref{alg:sparse_gamm}), this simulation study favors the methods discussed in this paper. Indeed, algorithms \emph{a3} and \emph{a4} (\emph{a5}) required less time to fit on average for almost all the likelihoods considered here. Exceptions are the hazard likelihood and the Gaussian likelihood with constant $\phi=2$.

As mentioned, the hazard case is special, in the sense that $\mathbf{H}$ is generally completely dense, despite the sparsity of $\mathbf{X}$ in this study. In these cases, the sparse Cholesky decomposition performed as part of algorithm \emph{a5} (i.e., Algorithm (\ref{alg:gammlss_beta})) is very inefficient. As a consequence, algorithm \emph{a5} generally took longer to estimate the model, compared to algorithm \emph{a1}. In contrast, algorithm \emph{a5} was generally faster whenever the likelihood featured sparsity in $\mathbf{H}$ that could be exploited. For example algorithm \emph{a5} was usually at least five times faster for the Gaussian likelihood with two linear predictors and twice as fast for the Gamma likelihood with two linear predictors. In cases where $\mathbf{H}$ shows no signs of sparsity that can be exploited, and when this knowledge is available in advance, it might thus be desirable to substitute a faster dense Cholesky decomposition for the sparse variant. In that case algorithm \emph{a5} would often require less CPU time than algorithm \emph{a1} (again).

For the Gaussian likelihood with constant $\phi$, algorithm \emph{a3} took generally more time to estimate the model, compared to algorithm \emph{a2}. This can largely be attributed to the fact that up-front decompositions of $\mathbf{X}^\top\mathbf{X}$ and $\mathbf{X}^\top\mathbf{y}$ are obtained as part of algorithm \emph{a2} in the Gaussian case, which pays off during subsequent iterations \parencite[see][for details]{wood_generalized_2015,wood_generalized_2017-2}. In the remaining cases (i.e., Gamma \& Binomial likelihoods) algorithm \emph{a3}, benefiting from the sparsity of $\sqrt{\mathbf{W}}\mathbf{X}$, again achieved a consistently lower time-to-fit compared to algorithm \emph{a2}. For example, MSSM (algorithm \emph{a3}) was usually at least five times faster than mgcv (algorithm \emph{a2}) for the Binomial likelihood. For the Gamma likelihood, MSSM took a bit longer to estimate the models but was usually still around four times faster. 

Generally, these results highlight that considerable computational savings are possible when fitting multi-level GAMMs and GAMMLSS with the algorithms discussed in this paper -- as long as there is sparsity in the structure of $\mathbf{H}$ that can be exploited efficiently. In some simulations, the methods described here were 5-10 times faster and this difference is only going to increase in case of experiments involving more subjects or more complex random effect structures.

Algorithm \emph{a6}, relying on the L-qEFS update, again did not achieve MSE scores as low as those of alternative algorithms. However, in practice it is more important that the produced estimates enable the same conclusions. This was the case in this simulation study: a visual inspection of the estimates produced by algorithm \emph{a6} revealed them to generally still be sufficiently similar to those obtained from more complex algorithms\footnote{The online repository contains code to create these visualizations.}. Estimates produced by algorithm \emph{a6} differed the most for $f(z)$. This function was generally strongly penalized, since the true function $f(z)$ was zero everywhere, and algorithm \emph{a6} showed signs of under-smoothing in some simulations. In these simulations, algorithm \emph{a6} however also generally produced wide credible intervals for $f(z)$, which showed good coverage of the alternative estimate produced by more complex algorithms as well as the true function (i.e., the zero line).

Algorithm \emph{a6} generally also took the longest to estimate the model across all likelihoods. This is likely to result from the fact that each update to the estimate for $\boldsymbol{\beta}$ is less effective than a complete Newton update (step 9 in Algorithm (\ref{alg:gammlss_beta})), so that more updates are required overall. In addition, each update requires a more expensive line search (at steps 7 \& 28 in Algorithm (\ref{alg:lqEFS})), which requires repeated re-evaluation of the (penalized) log-likelihood and the Gradients. On the upside, algorithm \emph{a6} only required information about the Gradient of the log-likelihood. This translates directly into less theoretical work required upfront, in case the higher order derivatives are not available directly. In addition, all estimates here were obtained for $N_V=30$ resulting in a comparatively low memory requirement for algorithm \emph{a6} (especially for the hazard log-likelihood). In practice, the higher CPU times and MSE scores will thus often be acceptable when these considerations matter. In case lower MSE scores are deemed more important than the time it takes to fit and the amount of memory required, $N_V$ could be increased further, which might improve the MSE performance of this algorithm.

To summarize, the estimation algorithms implemented in the MSSM toolbox generally performed competitively in a direct comparison with the algorithms implemented in mgcv in both simulation studies: the algorithms implemented in MSSM achieved comparable MSE scores and required less time-to-fit for many multilevel models (except for the L-qEFS update). In the simulation studies discussed in the upcoming section the performance of the model selection strategies implemented in MSSM was evaluated.

\subsubsection{Simulation 3 \& 4}

\begin{figure*}[!t]
    \caption{Simulation 3 \& 4 Results}
    \includegraphics[width=\textwidth]{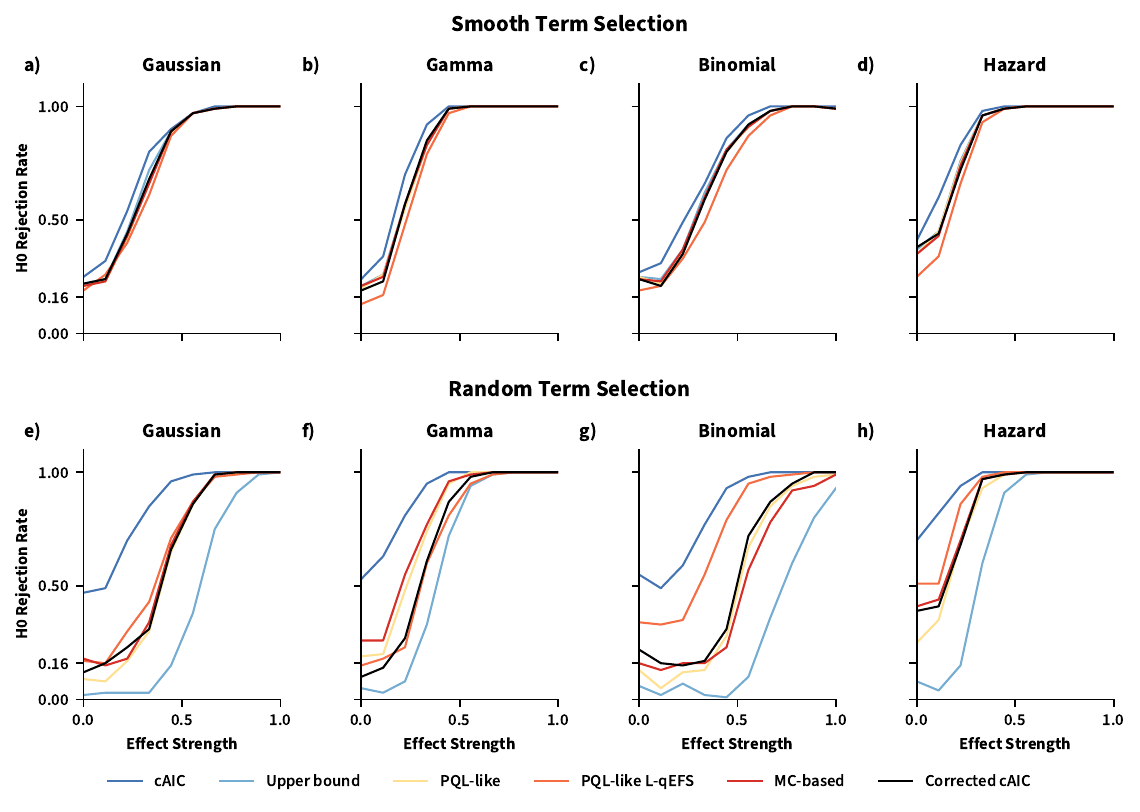}
    \label{fig:sim45_main}
    {\small
         \textit{Note.} Figure \ref{fig:sim45_main} shows the percentage of simulations in which the more complex model was accepted based on different versions of the cAIC and for different simulated effect strengths. The top row shows results for simulations in which a smooth term was subject to selection, while the bottom row shows results for simulations in which a random effect (intercept) was subject to selection. In all panels, the dark blue line corresponds to the conventional cAIC, the light blue line corresponds to the heuristic upper bound for the cAIC proposed by \Textcite{wood_smoothing_2016}, the black line corresponds to the corrected cAIC by \Textcite{wood_smoothing_2016}, the yellow line corresponds to the approximately corrected PQL-like cAIC from Section \ref{sec:pql_cAIC}, the orange line corresponds to the approximately corrected PQL-like cAIC from Section \ref{sec:pql_cAIC} that relies on quasi-Newton approximations $\hat{\mathbf{V}'}$ and $\hat{\mathbf{H}}$, and the red line corresponds to the cAIC based on a Monte-Carlo (MC) estimate for $\tau'$ computed as shown in Equation (\ref{eq:exptau3}) for the Gaussian likelihood and Equation (\ref{eq:exptau4}) for all remaining ones. The .16 tick-mark on the y-axis corresponds approximately to the false-positive rate that could be expected from an optimal cAIC criterion when there is no effect (i.e., $e=0$) \parencite[e.g.,]{greven_behaviour_2010,wood_smoothing_2016}.
        }
\end{figure*}

Following \Textcite{wood_smoothing_2016} this last set of simulation studies involved two types of nested model comparisons: one in which a smooth term was subjected to selection and one in which a random intercept was subjected to selection. For the first simulated comparison, the true linear predictor was set to $\eta_i = e*f(x_i) + f(z_i) + f(w_i) + f(v_i)$ where $e\in [0,1]$, acting as the simulated effect strength, was manipulated to range from 0 to 1. For each simulated dataset a model that estimated this linear predictor was compared to a model that excluded $f(x_i)$.

For the second simulated comparison, the true linear predictor was instead set to $\eta_i = f(x_i) + f(z_i) + f(w_i) + f(v_i) + b_{s(i)}$, where $b_{s(i)}$ denotes the random intercept corresponding to a specific level of a 40-level factor $s$. All the $b_{s(i)}$ were sampled from $\mathcal{N}(0,e)$, so that $e$ again takes on the role of the effect strength -- this time by acting as the standard deviation for the true distribution of the random intercepts \parencite[e.g.,][]{wood_smoothing_2016}. For each simulated dataset a model that estimated this linear predictor was again compared to a model that excluded the random intercepts.

We again considered Gaussian, Gamma, Binomial, and Proportional hazard likelihoods. For $e$, 10 values were considered, spaced equidistantly to cover the interval from 0 to 1. 100 datasets were simulated for each comparison and each level of $e$. Each simulated dataset consisted of 500 observations. The values for $x$, $z$, $w$, and $w$ were again all drawn independently from a Uniform distribution covering the interval $[-1,1]$. $\phi$ was again set to 2 for all simulations. All models were estimated once using algorithms \emph{a1}, \emph{a3} (\emph{a5} for the hazard case), and \emph{a6}.

For every simulation, the outcome of the model comparison was based on different versions of the conditional AIC \parencite[cAIC;][]{akaike_information_1992,greven_behaviour_2010}. Specifically, the complex model was accepted whenever it achieved a lower cAIC score and this acceptance rate was monitored separately for each version of the cAIC considered here \parencite[e.g.,][]{burnham_model_2004}. The first version was simply the conventional cAIC defined in Equation (\ref{eq:aic}), computed from the model estimate obtained from algorithm \emph{a3} (i.e., \emph{a5}). We also considered a heuristic upper-bound for the cAIC \parencite[see][]{wood_smoothing_2016}, again computed from the model estimate obtained from algorithm \emph{a3} (\emph{a5}). The uncertainty corrected cAIC proposed by \Textcite{wood_smoothing_2016} was based on the estimate obtained from algorithm \emph{a1} and obtained via the \texttt{AIC} function in R.

We also considered two variants of the ``PQL-like'' approximate cAIC discussed in Section \ref{sec:pql_cAIC}. The first variant was based on the final estimate $\hat{\boldsymbol{\rho}}=log(\hat{\boldsymbol{\lambda}})$ obtained from algorithm \emph{a3} (i.e., \emph{a5}), while the second variant was based on algorithm \emph{a6}. For the latter, $\hat{\mathbf{H}}$ and $\hat{\mathbf{V}'}$, the final quasi-Newton approximations to the (inverse of the) negative Hessian of the (penalized) log-likelihood were substituted for their exact counterparts for all necessary computations. Finally, we considered a version of the cAIC based on a Monte-Carlo (MC) estimate for $\tau'$. Specifically, for the Gaussian likelihood we considered the MC estimate discussed in Section \ref{sec:MCAM}. The alternative MC estimate discussed in Section \ref{sec:MCGS} was considered for all other likelihoods (based on the weights proposed by \Textcite{greven_comment_2016}; depicted here in Equation (\ref{eq:remlweights})). Note, that for the simulations we did not rely on the PQL-like estimate of $\tau'$ discussed in Section \ref{sec:pql_cAIC} as a lower bound. While this will generally be desirable in practice (see Section \ref{sec:MCGS}), here we were interested in the ``raw'' performance of the MC estimates. Therefore, we only used the weaker lower bound of $tr(\left[[\mathbf{V}]^{\hat{\boldsymbol{\rho}}} + \mathbf{V}^J\right]\mathbf{H})$ for both MC estimates (see Section \ref{sec:MCAM}). 

Figure \ref{fig:sim45_main} shows the percentage of simulations in which the more complex model was selected (i.e., resulted in a lower cAIC) as a function of $e$ for the 6 cAIC versions, both simulated comparisons, and all of the likelihoods considered here. The top row shows the results for the first simulated comparison, in which a smooth term was subjected to selection. The bottom row shows the results for the second simulated comparison, in which a random intercept was subjected to selection.

As was the case in the simulation conducted by \Textcite{wood_smoothing_2016}, model acceptance was largely insensitive to the chosen cAIC version in the first simulated comparison (see top row). Except for the hazard likelihood, the more complex model was generally chosen in about 16\% of the simulations in the absence of an effect (i.e., for $e=0$). For the hazard case, the false positive rate was generally higher but again similar across cAIC versions. Intriguingly, the approximate PQL-like version of the cAIC based on quasi-Newton approximations $\hat{\mathbf{H}}$ and $\hat{\mathbf{V}'}$ (``PQL-like L-qEFS`` in Figure \ref{fig:sim45_main}d), achieved the lowest false positive rate across likelihoods -- the improvement compared to other variants of the cAIC was generally small however.

Model acceptance was generally more sensitive to the choice of cAIC in the second simulated comparison, involving a random effect. As was the case in the simulation conducted by \Textcite{wood_smoothing_2016}, the conventional cAIC can again be observed to be severely biased towards the more complex model, with all likelihoods showing false positive rates of 50\% or higher in the absence of an effect (i.e., when $e=0$; see Figure \ref{fig:sim45_main}e - h). This tendency has previously been observed by \Textcite{greven_behaviour_2010} as well.

As was the case in previous simulations, the false positive rate decreased drastically when relying on the corrected cAIC (``Corrected cAIC`` in Figure \ref{fig:sim45_main}) proposed by \Textcite{wood_smoothing_2016}. A similar trend was observable for the approximately corrected PQL-like cAIC (``PQL-like`` in Figure \ref{fig:sim45_main}) based on algorithm \emph{a3} (\emph{a5} in the hazard case). However, compared to the more exact criterion by \Textcite{wood_smoothing_2016}, the PQL-like criterion was less likely to accept the more complex model for some likelihoods for small $e$ (e.g., hazard), but more likely to do so for others (e.g., Gamma). This increased variability is likely a result of neglecting the dependencies of $\mathbf{H}$ on $\boldsymbol{\rho}$ during estimation and when computing $\tau'$ for the cAIC (see Section \ref{sec:pql_cAIC} and Equation (\ref{eq:pql_hess_rho}) in particular) -- the impact of this will vary from likelihood to likelihood. Generally, the differences between both criteria were small however, and the approximately corrected PQL-like cAIC based on algorithm \emph{a3} (\emph{a5} in the hazard case) still outperformed both the conventional and the heuristic upper bound of the cAIC.

The alternative version of the cAIC, based on an MC estimate of $\tau$ (``Mc-based`` in Figure \ref{fig:sim45_main}), also in general performed similar to the PQL-like version of the approximately corrected cAIC and the corrected cAIC proposed by \Textcite{wood_smoothing_2016}. In the Gaussian case, the performance of all three criteria is particularly similar. Here the MC-based alternative is computationally quite attractive, because it has the potential to remain practically feasible, even for large-scale multi-level additive models for which evaluating the correction by \Textcite{wood_smoothing_2016} would quickly become prohibitive. The competitive performance of the MC-based version on the remaining likelihoods further suggests that this version of the cAIC is a compelling, if often more expensive, alternative to the correction proposed by \Textcite{wood_smoothing_2016} for more generic models as well.

The approximately corrected PQL-like cAIC based on quasi-Newton approximations $\hat{\mathbf{H}}$ and $\hat{\mathbf{V}'}$ still outperformed the conventional cAIC for every likelihood considered in this second set of simulations. However, compared to alternative corrected versions of the cAIC, this quasi-Newton PQL-like version of the cAIC showed greater bias towards the more complex model in this second set of comparisons for some likelihoods, which was particularly pronounced in the Binomial case. 

Considering the strong performance in the first set of simulations and the fact that this version of the cAIC still consistently outperformed the conventional one, we are cautiously optimistic that approximately correcting the cAIC for uncertainty in $\hat{\boldsymbol{\lambda}}$ remains possible, even when only the Gradient of the log-likelihood is readily available. Future research is needed however, which should focus on identifying circumstances in which the quasi-Newton PQL-like cAIC is likely to retain bias towards the more complex model (e.g., the Binomial likelihood) and how this could be addressed. For the moment, it will thus generally be preferable to opt for one of the alternative correction strategies discussed in Section \ref{sec:Uncertainty} when $\mathbf{H}$ can be provided.

Regarding those alternative strategies, the results from these simulation studies are particularly encouraging. The correction by \Textcite{wood_smoothing_2016} produced a corrected cAIC that generally showed less biased towards the more complex model. The approximately corrected PQL-like cAIC of Section \ref{sec:pql_cAIC} generally achieved a similar performance, but was slightly less consistent across different likelihoods. In practice the PQL-like cAIC thus remains a competitive alternative, at least as long as $N_p << N$ (see Section \ref{sec:pql_cAIC} for a discussion), when the higher order derivatives required for the exact correction are unavailable or utilizing them during estimation is too expensive.

Alternatively, the MC-based strategies to compute $\tau'$ discussed in Sections \ref{sec:MCAM} and \ref{sec:MCGS} can be used, at the cost of a potentially expensive sampling step. In contrast to the PQL-like cAIC, these alternatives requires fewer assumptions for more generic models and, in the Gaussian case, have the potential to be more efficient for large additive multi-level models. In this set of simulations the MC-based alternatives generally performed competitively, not just in the Gaussian case, despite the fact that only $N_r=250$ candidate vectors were generated during the sampling step. To ensure that the high level of performance observed here generalizes, it might however be necessary to increase $N_r$ for models involving more regularization parameters. In practice, where only a small number of comparisons will have to be performed, this will usually still be feasible.

\subsection{Real Data Examples}\label{sec:DataExamples}

We now turn towards real data examples that further show the usefulness of the algorithms discussed in this paper. We will begin by estimating a large multi-level additive model of EEG time-course data collected by \Textcite{krause_word_2024}. Subsequently, we consider two hazard models of their reaction time data.

\subsubsection{A Model of The EEG Time Course}

This section focuses on a model of the EEG time-course recorded at a single electrode (Cz) during a simple lexical decision experiment (judging whether a presented string of characters corresponds to a word or not) described previously by \Textcite{krause_word_2024}. The experiment featured three different types of stimuli (words, pseudo-words, and random strings) and a continuous measure of each stimulus' word-likeness, based on Google search result counts. While words and pseudo-words are naturally more word-like than random strings, this continuous measure of word-likeness also captures differences within these categories \parencite[see][for a more detailed introduction to and discussion of this measure]{hendrix_word_2020,krause_word_2024}. After pre-processing, data was available from 10,948 unique trials (5,402 word, 2,668 pseudo-word, and 2,878 random string trials). The EEG data were down-sampled to 256 Hz and base-line corrected by subtracting the mean amplitude in the 100 ms preceding stimulus onset. For every trial, observations from stimulus to response onset were retained (i.e., trials differed in the number of observations they contributed). In total, this resulted in 1,835,270 EEG amplitude recordings that were available for analysis. 

The EEG time course serves as a proxy of the time course of the cognitive processes completed during a lexical decision (e.g., stimulus perception \& encoding, decision, response planning). Of particular interest to the researchers was the question of whether the, potentially non-linear, effect of word-likeness on EEG amplitude would itself vary smoothly as a function of the time-course from stimulus onset to response. A model-based approach to the analysis of EEG data is particularly attractive here because of the presence of continuous predictor variables such as word-likeness. For example, there is no need to discretize the latter for this type of analysis, which would be required by alternative approaches based on permutations. 

To investigate whether the effect of word-likeness indeed varied non-linearly over time, the mean of the EEG amplitude recorded at a single channel could be modeled as

\begin{equation}\label{eq:eegModel}
\begin{split}
&\mu_i = a_{c(i)} + f_{c(i)}(w_i) + f_{c(i)}(t_i) + f_{c(i)}(w_i,t_i) + f_{s(i)}(t_i),\\
&\text{so that } EEG_{i} \sim \mathcal{N}(\mu_i,\sigma).
\end{split}
\end{equation}

\noindent
The $f_{c(i)}(w_i)$ and $f_{c(i)}(t_i)$ terms capture the partial non-linear effects of word-likeness and time for a specific word-type. They were parameterized with 20 and 10 B-spline basis functions respectively. In contrast, the $f_{c(i)}(w_i,t_i)$ term captures the partial non-linear \emph{interaction effect} of time ($t$) and word-likeness ($w$) for a specific word-type, indexed by $c(i)$, and was parameterized via a tensor smooth constructed from marginals parameterized with 10 (for the effect of time) and 5 (for the effect of word-likeness) B-spline basis functions each. The model also includes a random smooth term $f_{s(i)}(t_i)$ for each unique time-series $s$, each parameterized by 20 B-spline basis functions as well. Inclusion of this term is what makes it impossible to estimate the model via the methods proposed by \Textcite{wood_smoothing_2016}. Since the dataset from \Textcite{krause_word_2024} contains EEG time-courses from 10,948 time-series, parameterizing each of the 10,948 $f_{s(i)}(t_i)$ smooths with 20 basis functions would require estimating 218,960 coefficients\footnote{Actually, 229,908 coefficients have to be estimated for all random smooth terms combined, since an extra constant ``basis function'' has to be added per time-series. See Appendix \ref{sec:AppendixSmoothCon}.}. Not even the algorithms for big additive models, proposed by \Textcite{wood_generalized_2015} and \Textcite{wood_generalized_2017-1} could handle such a large number of random coefficients in a reasonable amount of time: completing a single Newton update to $\boldsymbol{\lambda}$, would likely take weeks \parencite[cf.][]{wood_generalized_2017}.

\begin{figure*}[!h]
    \caption{Additive Model of the EEG Time-course}
    \includegraphics[width=\textwidth]{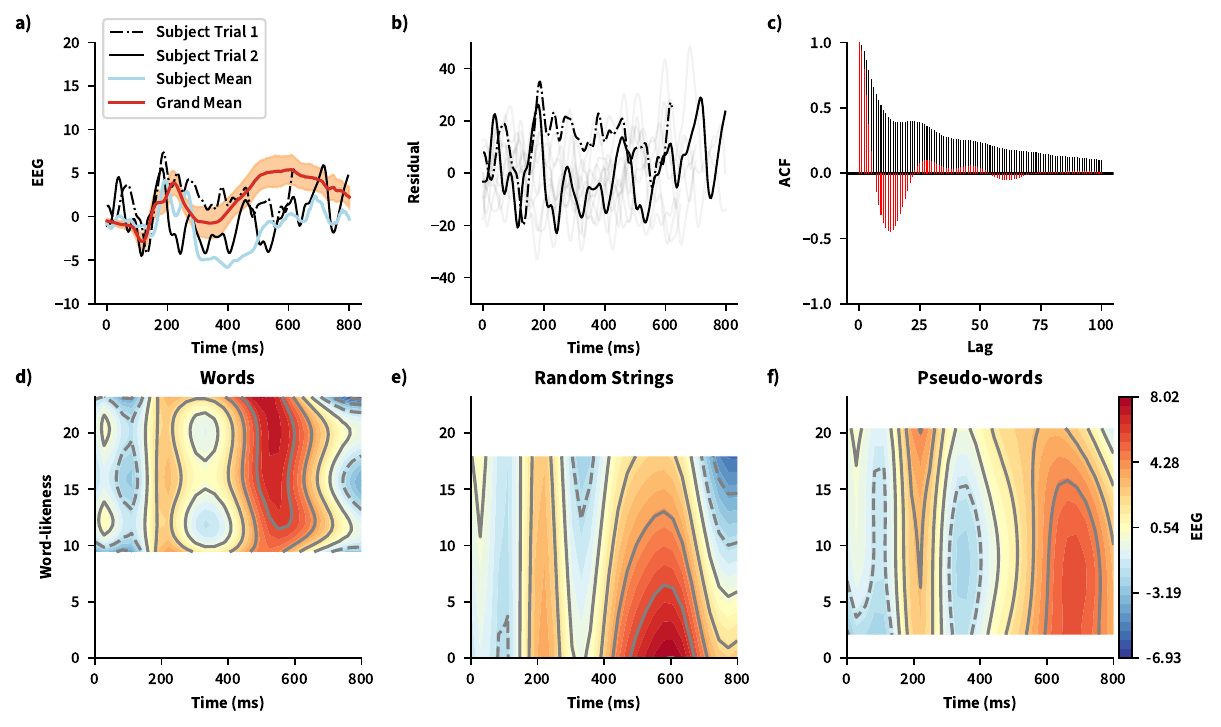}
    \label{fig:eeg_acf}
    {\small
         \textit{Note.} Figure \ref{fig:eeg_acf} shows EEG-time course data from \Textcite{krause_word_2024}, a visualization of the auto-correlation present in the residuals of the data for the models discussed in this section, and the estimates obtained from the final model discussed in this section. Panel a in the top-row shows the average EEG amplitude time-course at channel Cz both on the group level (red) and for an individual subject (light-blue). The thin dark lines reflect EEG time-series from individual trials. Panel b shows the ``residual series'' -- the difference between these EEG time-series from individual trials and the prediction of the first model discussed in this section, excluding random effects. Panel c shows the auto-correlation functions for both models (red corresponds to the second model including random effects). The bottom row (panels d-f) show the predicted EEG based on the ``fixed'' part of the model of the mean, i.e., $a_{c(i)} + f_{c(i)}(w_i) + f_{c(i)}(t_i) + f_{c(i)}(t_i,w_i)$, separately for every word-type.
        }
\end{figure*}

Unfortunately, when working with multi-level time-series data, including series-specific random terms is often mandated to ensure that the model's independence assumption is reasonable. The problem is that neuro-physiological time-series (e.g., EEG recorded on different trials) usually differ substantially in their time-course and temporal dependency structure. Figure \ref{fig:eeg_acf} illustrates the consequences of this for the EEG dataset collected by \Textcite{krause_word_2024}. Figure \ref{fig:eeg_acf}a shows the average EEG time-course evaluated over the first 800 ms after stimulus presentation (red thick line), the average time-course computed for a single subject (blue thick line), and two random trials from the same subject (thin black solid and dash-dot lines). We can now fit the model shown in Equation (\ref{eq:eegModel}) to the data, excluding the $f_{s(i)}(t_i)$ terms. Based on the model specification, the residual-vector $\hat{\boldsymbol{\mu}}-\mathbf{y}$ should match what could be expected when sampling $N$ independent values from $\mathcal{N}(0,\sigma)$. However, the second panel shows that this is clearly not the case: the ``residual-series'', i.e., the difference curves obtained by subtracting from each observed time-series the predicted mean, continue to show very different non-linear trends (see Figure \ref{fig:eeg_acf}b). The temporal dependencies present in the observed time-series thus remain to a large degree visible in the residuals. This is confirmed by considering the auto-correlation function (ACF) of the residuals, which continues to show strong temporal correlation even for large lags (see Figure \ref{fig:eeg_acf}c, black lines). As pointed out in the Introduction, such drastic violations of the independence assumption are well known to result in biased inference \parencite[e.g., confidence intervals that become too narrow, see][]{baayen_cave_2017,wood_generalized_2017-2} which can lead to wrong conclusions about the importance of individual smooth terms. Thus, opting for a simpler model will often not be justifiable.

Including a random smooth term $f_{s(i)}(t_i)$ for each unique time-series $s$  alleviates this problem, since the terms can then capture the trial-level difference between the observed time-course and $f_{c(i)}(w_i) + f_{c(i)}(t_i) + f_{c(i)}(w_i,t_i)$, the ``fixed'' part of the model of the mean (see Section \ref{sec:challenges_mixed}). And, as pointed out by \Textcite{wood_generalized_2017}, under the EFS update it actually becomes possible to estimate the resulting model. Indeed, when Algorithm (\ref{alg:sparse_gamm}) is used together with a Cholesky decomposition that pivots for sparsity as discussed in Section \ref{sec:AM} for additive models, the MSSM Python package takes around 1.5 hours on 10 CPU cores to estimate the 230,151 coefficients of the model specified in Equation (\ref{eq:eegModel}). Importantly, the updated residual vector, taking into account the multi-level nature of $\mu_i$, is much more in line with what could be expected when sampling from $\mathcal{N}(0,\sigma)$ (see Figure \ref{fig:eeg_acf}c for an overlay of the updated auto-correlation function\footnote{Any remaining auto-correlation in a large multi-level model (e.g., as visible in Figure \ref{fig:eeg_acf}c) can usually be accounted for efficiently by means of an auto-regressive model of the residuals \parencite[e.g.,][]{wood_generalized_2015,van_rij_analyzing_2019}.}.).

The bottom row of Figure \ref{fig:eeg_acf} shows how the effect of word-likeness on EEG amplitude changes smoothly as a function of time and for the three different word-types (note that Words never had work-likeness scores less than 9, for example). Especially in the time-frame from 500-700 ms the EEG amplitude appears to change smoothly as a function of word-likeness -- differently so for words compared to the two types of non-words (see Figure \ref{fig:eeg_acf}d-f). An interesting question for future research is whether penalty-based selection \parencite[e.g.,][]{marra_practical_2011} could be used to address questions of model selection when working with EEG data from multiple channels. Given the potentially large number of models that would have to be estimated (one model per electrode), it would still be prudent to restrict the selection approach to a handful of representative channels and to selected effects -- for example only to the word-type specific partial interaction effects of time and word-likeness.

\subsubsection{Hazard Models of Reaction Times}

We now turn to survival models of the reaction time data collected by \Textcite{krause_word_2024}. As mentioned in the Introduction, when working with survival or hazard models we are usually interested in the survival function $S(t) = exp(-H(t))$ where $H(t)$ again denotes the ``cumulative hazard function'' \parencite[see][]{klein_survival_2003}. In the context of lexical decisions, the survival function reflects the probability of still providing a correct response after time-point $t$, when no response has been given so far and $t=0$ was the moment of stimulus onset \parencite[e.g.,][]{hendrix_word_2020}. When smooth functions are incorporated in survival models we can investigate how this probability changes as a function of continuous and categorical predictor variables alike. Because the survival function can again be considered to act as a proxy of the cognitive processes that have to be completed between stimulus onset and response time, this type of analysis can again provide valuable insights into the influence different predictor variables have on processing \parencite[e.g.,][]{hendrix_word_2020}. Additionally, survival models can incorporate trials into estimation on which no response was given before a pre-specified time limit was reached, which are usually excluded in standard models of reaction time data \parencite[e.g.,][]{bender_generalized_2018,hendrix_word_2020}.

\begin{figure*}[!h]
    \caption{Cox Proportional Hazard Model of Reaction Times}
    \includegraphics[width=\textwidth]{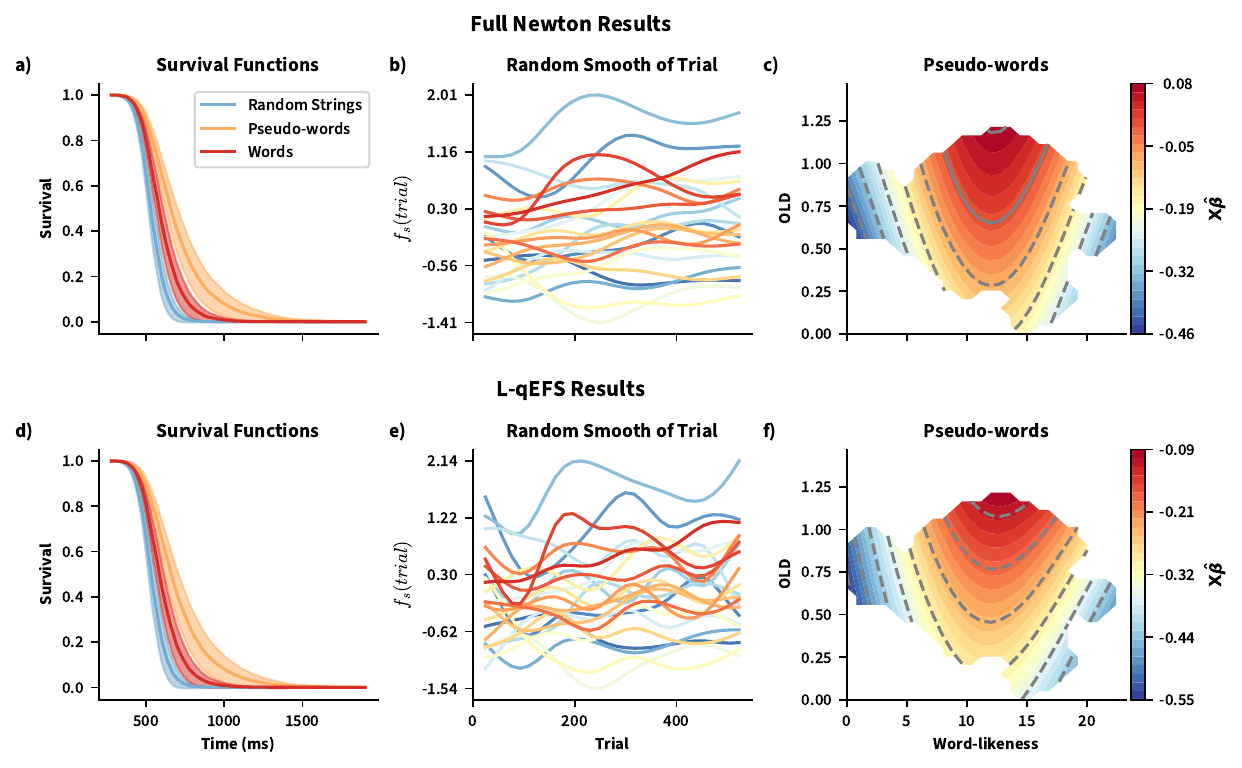}
    \label{fig:coxph}
    {\small
         \textit{Note.} Figure \ref{fig:coxph} shows the estimates obtained from a Cox proportional hazard model, fitted to the reaction time data from \Textcite{krause_word_2024}. The top row shows the estimates obtained when relying on Algorithms (\ref{alg:gammlss_beta}) and (\ref{alg:sparse_gamm}), while the bottom row shows the estimates obtained from the L-qEFS update defined in Algorithm (\ref{alg:lqEFS}). The left-most panels show the estimated survival functions for all word types, evaluated at their average word-likeness and OLD20 scores \parencite[shaded areas correspond to approximate 95\% credible interval;][]{wood_generalized_2017-2}. The central panels show an estimate of the random smooth terms of trial, and the right-most panels show the estimated function of word-likeness and the OLD20 measure for pseudo-words.
        }
\end{figure*}

The Cox-proportional hazard model \parencite[][]{cox_regression_1972}, discussed briefly in the Background section of this paper, is an example of such a survival model. As mentioned in the Introduction, this model represents the log-hazard function, given model matrix row $\mathbf{X}_i$, as $log(h(t|\mathbf{X}_i)) = log(h_0(t)) +\eta_i$, where $h_0(t)$ again denotes the ``baseline hazard'' and $\eta_i=\mathbf{X}_i\boldsymbol{\beta}$ \parencite[see also][]{hastie_generalized_1990}. For the reaction time data collected by \Textcite{krause_word_2024} a suitable linear predictor would for example be $\eta_i= a_{c(i) \neq W} +f_{c(i)}(w_i,o_i) + f_{s(i)}(tr_i)$, where $s(i)$ now indexes the subject from which observation $i$ was collected and $a_{c(i): c(i) \neq W}$ represents offset terms, which are only added for random strings and pseudo-words, but not for words, because the proportional hazard model cannot have an intercept. $o_i$ here represents the OLD20 score for the stimulus present on trial $i$, which is another measure of (orthographic) word-likeness of the stimuli that correlates with $w$ \parencite[see][]{yarkoni_moving_2008,krause_word_2024}. The OLD20 variable was included here in the model to make estimation more complicated, to highlight that the robustness of the L-qEFS update to such correlations, established by the simulation studies, also translates to real data examples. The $f_{c(i)}(w_i,o_i)$ terms capture the potentially non-linear interaction effect of word-likeness and OLD20 scores, which is again allowed to differ in shape between word types. The $f_{s(i)}(tr_i)$ terms are subject-specific random smooth functions of the experimental time-course (i.e., trial) and thus allow $h(t|\mathbf{X}_i)$ to change over the time-course of the experiment -- differently so for every subject.

An estimate for $\boldsymbol{\beta}$ can be obtained by using Algorithm (\ref{alg:gammlss_beta}) together with Algorithm (\ref{alg:sparse_gamm}). However, since the negative Hessian of the partial log-likelihood of the Cox-proportional hazard model is completely dense, an alternative is to rely on the L-qEFS update detailed in Section \ref{sec:lqefs} instead. Figure \ref{fig:coxph} shows the estimated survival functions as well as approximate 95\% credible intervals for the three types of words, as well as the estimated $f_{c(i)}(w_i,o_i)$ terms for words, and the estimated $f_{s(i)}(tr_i)$ terms. The top row shows the estimates obtained when Algorithm (\ref{alg:gammlss_beta}) is combined with Algorithm (\ref{alg:sparse_gamm}) to fit the model. The bottom row shows the results obtained when the L-qEFS update is used instead. Both models were estimated via the MSSM Python toolbox. As was the case in the simulation studies considered in the previous section, the estimates obtained from both models are quite similar. Thus, even for real data, the L-qEFS update, which only relies on the Gradient of the log-likelihood for estimation, produces estimates comparable to those obtained from theoretically more complex algorithms.

As discussed at the end of Section \ref{sec:lqefs}, the approximations to the negative Hessian matrix maintained by the L-qEFS update can also be used to compute credible intervals for the estimated effects. In this example, and also in the second simulation study discussed in Section \ref{sec:Simulations}, these approximate intervals were similar to the ones based on the exact Hessian (see Figure \ref{fig:coxph}a \& d). However, there is no guarantee that this will be the case in all scenarios, so these approximate intervals should only be considered as a rough indicator of the uncertainty in the obtained estimates. As discussed in Section \ref{sec:lqefs}, the L-qEFS estimates and approximation to the Hessian can however be used to initialize a more accurate Metropolis sampler of the posterior distribution $\boldsymbol{\beta} |\boldsymbol{\lambda}, \mathbf{y}$ \parencite[or even $\boldsymbol{\beta} |\mathbf{y}$, see also][]{wood_inference_2020}, from which more accurate interval estimates can readily be obtained. We thus believe that the L-qEFS update will be very useful for dealing with more generic smooth models in practice.

By definition of the hazard function, the Cox model presented here assumes that the effect of predictor variables remains stable throughout the time-course of lexical decisions: only the ``baseline hazard'' $h_0(t)$ changes over time \parencite[][]{klein_survival_2003}. In contrast, considering the model of EEG data discussed in the previous section, it would be more interesting to consider a survival model in which the effect of continuous predictor variables (e.g., word-likeness) on the hazard function can change over the time-course of lexical decisions. As discussed by \Textcite{bender_generalized_2018}, this can be achieved by first breaking up the time-course between stimulus onset and the maximum response time into $N_t$ time-intervals $t_0,t_1,...t_{N_t}$ and then modeling the (log) hazard function as $log(h(t | \mathbf{X}_i)) = f(t_j) + \mathbf{X}_i\boldsymbol{\beta}$ where $f(t_j)$ takes on the role of the baseline hazard -- now represented as a smooth function of time intervals $t_j$. Similarly, $\mathbf{X}\boldsymbol{\beta}$ can now again be used to directly incorporate non-linear interactions between time and continuous predictors like word-likeness (e.g., $f(t_j,w_i)$) into the model of the hazard function.

It turns out, that the resulting model of the hazard function can be estimated approximately by means of a Poisson GAMM, after transforming the original reaction time data \parencite[e.g.,][]{bender_generalized_2018,wood_generalized_2017-2}. The details of this transformation and the use of the Poisson GAMM as a vehicle for estimation have been described extensively by \Textcite{bender_generalized_2018}, so we do not revisit these issues here \parencite[see also][]{hastie_generalized_1990,wood_generalized_2017-2}. In practice, the R package pammtools \parencite[][]{bender_generalized_2018}, which has been developed specifically to facilitate estimation of these models, can be used to apply the transformation to existing reaction time data. One important caveat should be highlighted however: applying this transformation can quickly result in models that are expensive to estimate, especially if random effects are to be involved in the model. The reason is that, as part of the transformation, every trial is essentially expanded into a time-series, with a maximum length of $N_t$ \parencite[][]{bender_generalized_2018}. The latter is typically chosen to be the unique number of response times (in seconds, rounded) in the original response time data \parencite[][]{bender_generalized_2018}. While trials on which the observed reaction time was below the maximum will result in shorter time-series, this will quickly result in a large number of transformed observations. For the modest reaction time dataset from \Textcite{krause_word_2024} for example, which contained 1,031 unique (rounded) reaction times from 10,948 trials completed by 24 subjects, the transformed data obtained via the pammtools package consisted of 3,459,947 observations.

Depending on the specific model of the log hazard function, such a large number of observations can make estimation difficult. For the reaction time data by \Textcite{krause_word_2024}, a relatively complex model would for example be 

\begin{equation}\label{eq:hazModel}
\begin{split}
log(h(t|\mathbf{X}_i))=&a_{c(i)} + f_{c(i)}(t_j) + f_{c(i)}(w_i) + f_{c(i)}(t_j,w_i) ~+\\& f_{s(i)c(i)}(t_j) + f_{s(i)}(tr_i).
\end{split}
\end{equation}

$f_{c(i)}(w_i) + f(t_j,w_i)$ here captures the, potentially time-varying and word-type specific, effect of word-likeness on the log-hazard rate. $f_{s(i)c(i)}(t_j)$ is a random smooth of the lexical decision time-course. Note, that while the random smooth functions are estimated separately for each subject and word type combination they still all share the same smoothness penalty parameters (see Appendix \ref{sec:AppendixSmoothCon}). Similarly, $f_{s(i)}(tr_i)$ is a subject-specific random smooth of the experimental time-course (i.e., trial). Note, that $f_{s(i)c(i)}(t_j)$ essentially captures the subject-specific non-linear deviation from each of the word-type specific baseline hazard functions ($f_{c(i)}(t_j)$). If the marginal smooths and random smooths are all parameterized by 10 basis functions and each of the $f_{c(i)}(t_j,w_i)$ is again represented as a tensor smooth with 5 basis functions used to parameterize the marginals, this model results in $N^p=1785$ coefficients.

While the resulting model could be estimated based on the methods discussed by \Textcite{wood_generalized_2015}, the method by \Textcite{wood_generalized_2017-1} is not just faster but also more efficient here. Using the \texttt{bam} function in mgcv in R, estimation of the model based on the method by \Textcite{wood_generalized_2017-1} takes a bit more than 2 hours on a single CPU core. As discussed in Section \ref{sec:GAM}, both rely on slightly different versions of the PQL approach. However, the advantage, in terms of time to fit, of the method by \Textcite{wood_generalized_2017-1} can largely be attributed to an additional discretization of the continuous predictor variables performed before estimation.

\begin{figure*}[!h]
    \caption{Penalized Additive Hazard Model of Reaction Times}
    \includegraphics[width=\textwidth]{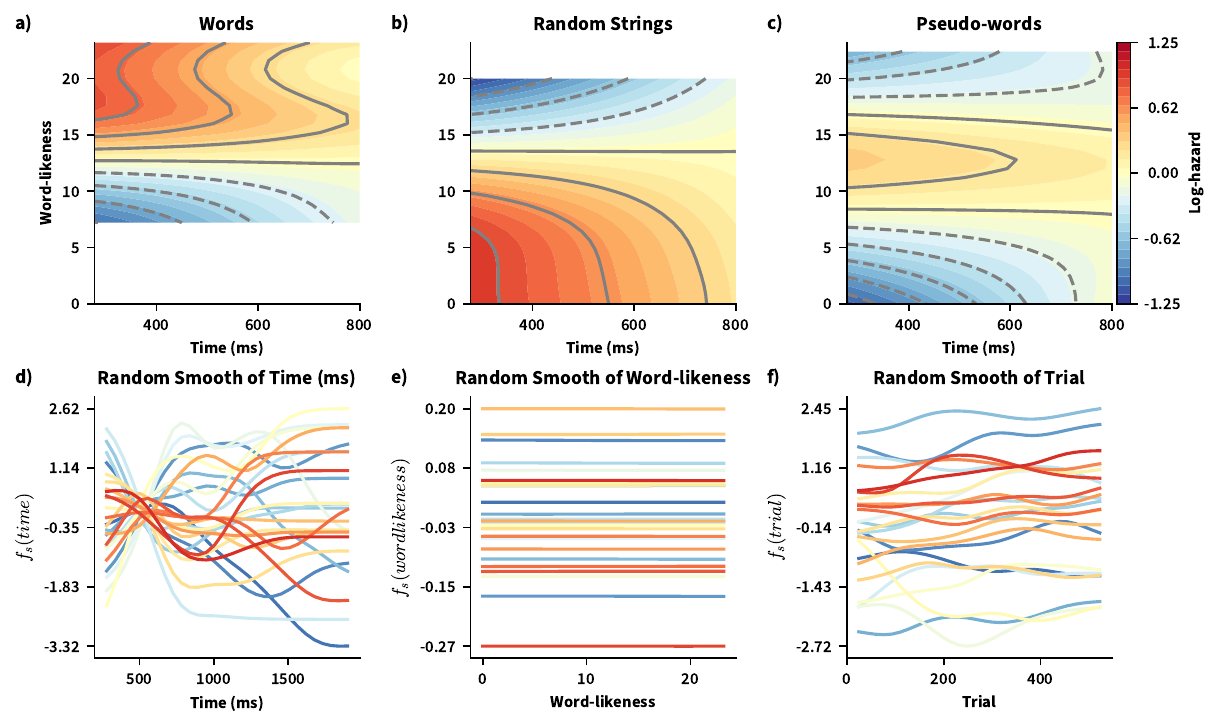}
    \label{fig:coxph2}
    {\small
         \textit{Note.} Figure \ref{fig:coxph2} shows the estimates obtained from the second penalized additive hazard model discussed in this section, fitted again to the reaction time data from \Textcite{krause_word_2024}. The top row shows the estimates for the $f_{c(i)}(w_i) + f_{c(i)}(t_j,w_i)$ smooth terms separately for every word-type. The bottom row shows estimates of the three random smooth terms included in the model.
        }
\end{figure*}

Alternatively, the model can again be estimated via the MSSM Python toolbox, using Algorithm (\ref{alg:sparse_gamm}) and the methods discussed in this paper. If a Cholesky decomposition that pivots for sparsity is used to solve for the coefficients, estimation takes only around 39 minutes on a single CPU core. Notably, MSSM does not require discretization of continuous predictor variables\footnote{In this example, as is often the case in practice, the discretization performed by the method of \Textcite{wood_generalized_2017-1} has little effect on the estimated smooth functions, which are very similar to the ones shown in Figure \ref{fig:coxph2}.}. Naturally, if the complexity of the random effect structure of the model is reduced, the difference in time-to-fit between the different methods becomes smaller. Eventually MSSM will take longer to complete estimation, as the overhead associated with sparse matrix storage starts to dominate the time it takes to fit ``thin'' models of a large number of observations. However, when working with multi-level data, it will generally be desirable to keep the random effect structure on the complex side, within the limits imposed by imperfect data and assumptions of the Laplace approximation, to ensure that inference for the fixed effect estimates remains conservative. Such models will also often require many updates to the regularization parameters $\boldsymbol{\lambda}$, in which case the greater efficiency of the sparse EFS update performed by MSSM will again result in a noticeable reduction of the required time to fit.

Figure \ref{fig:coxph2} depicts the estimated time-varying effect of word-likeness on the log-hazard rate for the three word-types. The estimated smooth functions for words and random strings show essentially inverted effects of word-likeness over time, which aligns with observations made in previous experiments \parencite[e.g.,][]{hendrix_word_2020}. The effect of word-likeness for pseudo-words is generally more subtle and, intriguingly, shows similarities to the effect visible for both words and random strings, with lower log-hazard rates early during the lexical decision time-course at both ends of the word-likeness spectrum.

\section{Discussion}\label{sec:Discussion}
In this paper we discussed the challenges faced when working with multi-level smooth models. The first problem identified was that estimation, automatic regularization, and model selection of more generic smooth models beyond the case of generalized additive models requires researchers to provide efficient code to evaluate 3rd and 4th order derivatives of the log-likelihood \parencite[e.g.,][]{wood_smoothing_2016}. This requirement places a substantial burden on researchers wanting to estimate more general smooth models. The second, related problem, is that automatic regularization of multi-level models based on methods utilizing these higher-order derivatives tends to slow down drastically as the number of random effects included in the model grows. As mentioned in the Introduction, strong dependencies are a common presence in experimental data because data is usually collected over time or from multiple subjects. Models of such data thus routinely have to include many random effects to account for these dependencies so that any conclusions derived from the estimated model are likely to generalize. Thus, there is ample need for fast estimation of multi-level or mixed smooth models and for the possibility to estimate general smooth models without requiring code for 3rd and 4th order derivatives.

To address these challenges we relied in large parts on the algorithm for automatic regularization proposed by \Textcite{wood_generalized_2017}, the Extended Fellner Schall (EFS) update. This update was designed to allow approximate but very efficient estimation and regularization of smooth models based on first and second order derivatives (i.e., Gradient and Hessian) of the log-likelihood alone. We reviewed how the EFS update can be embedded into a unified sparse estimation framework for multi-level sparse smooth models, which permits efficient estimation and regularization of multi-level GAMMs, GAMMLSS, and generic smooth models alike. Subsequently, we extended this framework in two important ways. We devised a limited-memory quasi-Newton variant (L-qEFS) of the EFS update, which enables estimation and automatic regularization of generic smooth models based only on the Gradient of the log-likelihood. This further lowers the burden placed on researchers wanting to estimate generic smooth models and ensures that estimation remains memory-efficient -- even when the Hessian of the log-likelihood is not sparse at all. Additionally, we presented multiple strategies, based on previous work by \Textcite{wood_smoothing_2016} and \Textcite{greven_comment_2016}, that allow to quantify the uncertainty in the estimated degree of regularization. This allows to correct for this source of uncertainty, for example when computing the conditional AIC \parencite[cAIC;][]{akaike_information_1992,greven_behaviour_2010} to select between smooth models.

To facilitate estimation, automatic regularization, and selection of the models discussed throughout this paper, we have implemented all algorithms and proposed extensions in the Mixed-Sparse-Smooth-Model (MSSM) Python toolbox, which is available openly on GitHub and distributed via the Python package Index (pypi). This implementation will hopefully make these models accessible to a larger audience. Documentation is available at \url{https://jokra1.github.io/mssm} for the entire public API of MSSM and we provide a set of examples in a tutorial repository, which also includes functions to visualize and extract model predictions for GAMMs, GAMMLSS, and general (multi-level) smooth models. To verify the performance of this toolbox we conducted multiple simulation studies and considered real data examples, which confirmed that the strategies presented here perform competitively when compared to existing methods \parencite[e.g.,][]{wood_fast_2011,wood_generalized_2015,wood_smoothing_2016,wood_generalized_2017-1}. 

In practice, when working with massive additive mixed models, the MSSM toolbox enables estimation of models that are impossible to fit via alternative software (see Section \ref{sec:DataExamples}). However, already for small optimization problems involving multi-level models (e.g., $N=5000$ observations collected from 20 subjects) the algorithms discussed here can be far more efficient. In situations where higher-order derivatives are unavailable or difficult to implement efficiently the toolbox still enables estimation of more generic models, producing estimates that in practice are often virtually indistinguishable from the ones produced when relying on more complex methods. If exact results are crucial, the estimates obtained from the strategies discussed here can be used to efficiently initialize samplers of the posteriors $\boldsymbol{\beta}|\mathbf{y},\boldsymbol{\lambda}$ and $\boldsymbol{\beta}|\mathbf{y}$ (see the discussion at the end of Section \ref{sec:lqefs}).

For the goal of model selection, the strategies presented here generally came close to the performance achieved by the method proposed by \Textcite{wood_smoothing_2016}, and the corresponding cAIC variants are likely to be a substantial improvement over the conventional cAIC in practice. Already the simplest strategy proposed here, essentially computing the uncertainty correction by \Textcite{wood_smoothing_2016} as if the log-likelihood would match a quadratic function (see Equation (\ref{eq:pql_hess_rho}) and the discussion thereafter in Section \ref{sec:pql_cAIC}), performed well for all the likelihoods considered in the simulations conducted here. This strategy will likely work well for many other general smooth models, including GAMMs of other distributions beyond the ones considered here, as long as $N_p << N$ holds in practice.

For Gaussian additive models, the alternative MC strategy of Section \ref{sec:MCAM} can be used to to approximate the correction proposed by \Textcite{wood_smoothing_2016} as well. This strategy will be useful in practice when working with large multi-level models, for which the method proposed by \Textcite{wood_smoothing_2016} and even the simple approximation based on Equation (\ref{eq:pql_hess_rho}) can quickly become expensive. For other members of the broad class of generic smooth models for which the simple strategy from Section \ref{sec:pql_cAIC} might not be appropriate (see the discussion at the end of Sections \ref{sec:GAMMLSS} and \ref{sec:Uncertainty}), the second MC strategy presented in Section \ref{sec:MCGS} can be used as an alternative. This MC strategy does not require strong assumptions and is generally applicable to any smooth model. In the simulation studies conducted here, both MC strategies achieved competitive performance, even for a modest number of $N_r=250$ MC samples. However, for more complex models it will often be necessary to increase $N_r$ to achieve comparable performance. More research is thus required to establish computationally cheap cut-off criteria that can be used to determine whether sampling should be continued (i.e., whether $N_r$ needs to be increased further) or not. Nevertheless, the results from the simulation studies are encouraging and provide a good starting point for future investigations to build on.

\printbibliography

\appendix
\onecolumn
\newpage
\section{Unified Basis Construction \& Random Smooth Terms}\label{sec:AppendixSmoothCon}
\setlength{\parindent}{0pt}

This section reviews the unified approach to univariate smooth construction discussed previously by \citeauthor{wood_generalized_2017-2} (\citeyear{wood_generalized_2017-2}; see Section 5.4.2 ), \Textcite{wood_smoothing_2016}, and by \Textcite{wood_inference_2020}. We refer to these publications for a more detailed review and here only provide a basic overview which is sufficient to formalize ``random'' smooth terms. 

Consider a simple additive model for which $g(\mu_i) = f(x_i) = \mathbf{X}_i\boldsymbol{\beta}$, where we have omitted an intercept term $\alpha$ for convenience so that $f$ does not need to be subjected to identifiability constraints \parencite[see][for more detailed discussions of identifiability issues]{wood_core_2015,wood_generalized_2017-2}. As discussed in the Background section of the main text, $f$ is represented via
$f(x_{j}) = \sum_k b_{j,k}(x_{j})\beta_{j}$ (see Equation (\ref{eq:smooth_term})) and is penalized by $\mathcal{S}_\lambda=\lambda_1\mathcal{S}$. 

\begin{figure}[h!]
\caption{Demmler \& Reinsch Transformation}
\begin{center}
\includegraphics[width=0.5\linewidth]{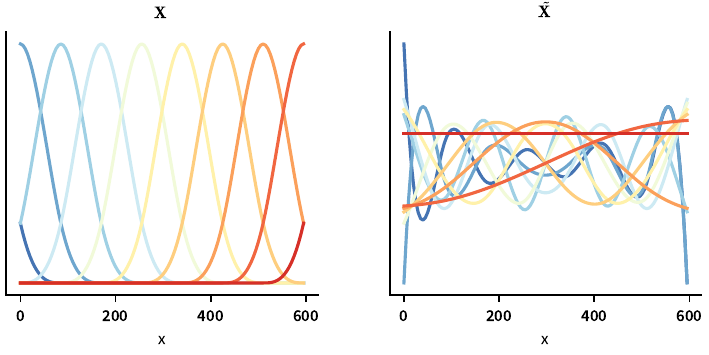}
\end{center}

        {\small
         \textit{Note.} Figure \ref{fig:reparam} shows how the content of the columns of $\mathbf{X}$, holding the 10 B-spline basis functions used to parameterize $f$, changes as a result of transforming $f$ to be approximately in Demmler \& Reinsch parameterization \parencite{demmler_oscillation_1975,wood_smoothing_2016,wood_generalized_2017-2}. Panel a) shows the content of the model matrix (i.e., $\mathbf{X}$) before applying the latter, while panel b) shows the state (i.e., $\tilde{\mathbf{X}}$) after applying the transformation.
        }

\label{fig:reparam}
\end{figure}

As discussed in the Background section, there are many options for the basis functions $b$ and depending on the choice the structure of the penalty matrix $\mathcal{S}$ will have to differ in order to capture the complexity of an estimate of $f$ \parencite[see][for an overview over different basis function sets]{wood_generalized_2017-2}. However, as discussed by \Textcite{wood_smoothing_2016} univariate smooth terms can generally be transformed to approximately be in Demmler \& Reinsch parameterization \parencite{demmler_oscillation_1975,wood_smoothing_2016}, independent of the specific basis chosen. We now review this transformation, based on the simple example outlined above. We use a B-spline basis of $k=10$ to parameterize $f$ and a $k$-dimensional difference matrix of order 1 for $\mathcal{S}$ (so that the size of the Kernel of $\mathcal{S}$ is $N_0=1$), which is a typical choice when working with B-spline basis functions \parencite[e.g.,][]{eilers_flexible_1996}. The first panel in Figure \ref{fig:reparam} shows the 10 B-spline basis functions, i.e., the content of the columns of $\mathbf{X}$, evaluated over the range of covariate $x$.

Following the steps outlined in section 5.4.2 of \Textcite{wood_generalized_2017-2}, transforming the basis functions to be approximately in Demmler \& Reinsch parameterization can then be achieved by first computing $\mathbf{L}$, so that $\mathbf{L}\mathbf{L}^\top = \mathbf{X}^\top\mathbf{X}$. This can be achieved via a Cholesky of QR decomposition as described in Section \ref{sec:AM} of the main text \parencite[see also][]{wood_generalized_2017-2}. Afterwards, the Eigendecomposition $\mathbf{L}^{-T}\mathcal{S}\mathbf{L}^{-1} = \mathbf{U}^\top\mathbf{\Sigma}\mathbf{U}$ is formed. It is assumed that the Eigenvalues in $\Sigma$ are ordered to be of decreasing magnitude. We then solve $\mathbf{L}^\top\mathbf{P} = \mathbf{U}$ for $\mathbf{P}$. The transformation is completed by setting $\tilde{\mathbf{X}}=\mathbf{X}\mathbf{P}$ and $\tilde{\mathcal{S}}=\Sigma$. The second panel in Figure \ref{fig:reparam} shows the resulting approximation to the first 10 Demmler \& Reinsch basis functions, i.e., the content of the columns of $\tilde{\mathbf{X}}$.

As pointed out by \Textcite{wood_generalized_2017-2}, $\tilde{\mathbf{X}}$ and $\tilde{\mathcal{S}}$ have some important and general qualities. For once, the columns in $\tilde{\mathbf{X}}$ will correspond to decreasingly flexible functions, orthogonal to another. The flexibility \parencite[or ``roughness''][]{kimeldorf_correspondence_1970} of the first $N_p - N_0$ functions will be directly related to the magnitude of the Eigenvalues now making up the elements on the diagonal of $\tilde{\mathcal{S}}$: the Eigenvalues will be larger for the most complex (i.e., rough) functions in the first columns and diminish as functions become simpler. This immediately highlights how $\tilde{\mathcal{S}}$ enforces regularization of $\tilde{f}$: For a given $\lambda$ more flexible functions (specifically their coefficients) are drawn towards zero more strongly than less flexible ones \parencite{wood_generalized_2017-2}. Notably, because of the rank-deficiency of $\mathcal{S}$, the last $N_0$ elements on the diagonal of $\tilde{\mathcal{S}}$ will be zero. Clearly, the transformed basis functions in the last $N_0$ columns of $\tilde{\mathbf{X}}$ are thus left un-penalized -- and it should be unsurprising that they are sufficient to construct any function in the Kernel of $\mathcal{S}$ \parencite{wood_generalized_2017-2}. In this example, $N_0=1$, so the only ``basis function'' left un-penalized by $\tilde{\mathcal{S}}$ is the constant (dark red line in panel 2 of Figure \ref{fig:reparam}).

The re-parameterization discussed here also forms the basis of how ``random smooth terms'' are computed in the R-package ``mgcv'' \parencite{wood_generalized_2017-2} and in the MSSM toolbox presented here. As discussed in the main text, random smooth terms often \emph{have} to be incorporated into models of experimental data, to account for dependencies present in the latter. This necessity has previously been discussed by \Textcite{van_rij_analyzing_2019} and \Textcite{baayen_cave_2017} (see also Section \ref{sec:DataExamples} of the main text). Before we discuss how to turn $\tilde{f}$ into a random smooth term it helps to consider that, as discussed in the Background section of the main text, the transformed penalty matrix $\tilde{\mathcal{S}}$ can already be treated as the precision matrix of an improper Bayesian prior $\mathcal{N}(\mathbf{0},\tilde{\mathcal{S}}_\lambda^-)$ placed on the coefficients $\boldsymbol{\beta}$ parameterizing the transformed smooth $\tilde{f}$ \parencite[e.g.,][]{kimeldorf_correspondence_1970,wood_fast_2011,wood_just_2016,wood_generalized_2017-2}. $\tilde{\mathcal{S}}_\lambda^-$ is again used to denote a generalized inverse of $\tilde{\mathcal{S}}_\lambda=\lambda_1\tilde{\mathcal{S}}$ -- note that in the  Demmler \& Reinsch transformation $\tilde{\mathcal{S}}_\lambda^-$ can be computed as an actual inverse of the top-left $(N_p-N_0) * (N_p-N_0)$ block of $\tilde{\mathcal{S}}_\lambda$ \parencite[e.g.,][]{wood_fast_2011,wood_generalized_2017-2}.

Clearly, all that is necessary for $\boldsymbol{\beta}$ to become a completely random vector -- as would be expected from a classical ``random effect'' -- is to leave no column of $\tilde{\mathbf{X}}$ un-penalized \parencite{wood_generalized_2017-2}. In other words, it is necessary to further transform $\tilde{\mathcal{S}}$ to a full rank matrix. The Kernel of the transformed $\tilde{\mathcal{S}}_\lambda^{-1}$ will then be empty and the matrix itself will again correspond to a full rank covariance matrix of a Bayesian prior \parencite[or equivalently a classical Gaussian random effect;][]{wood_generalized_2017-2}. Because of the diagonal structure of $\tilde{\mathcal{S}}$, it is trivial to define $N_0$ additional penalties that act only on one of the columns of $\tilde{\mathcal{X}}$ that are currently not penalized by $\tilde{\mathcal{S}}$. These can then be added to $\tilde{\mathcal{S}}$, so that $\tilde{\mathcal{S}}_\lambda = \lambda_1\tilde{\mathcal{S}} + \sum_{n=1}^{N_0} \lambda_{n+1}\Psi^n$. $\Psi^n_{k-N_0+n,k-N_0+n} = 1$ and otherwise contains only zeroes, so that it exclusively penalizes column $k-N_0+n$ of $\tilde{\mathbf{X}}$. Note that this requires estimation of $N_0$ additional $\lambda$ parameters. After these steps, $\tilde{f}$ is a fully penalized random smooth. As discussed in the Background section and Section \ref{sec:challenges_mixed}, random smooth terms are usually \parencite[i.e., in mgcv,][and also in MSSM]{wood_generalized_2017-2} set up per level of a factor variable (e.g., subjects, trials, etc.). For example, if $[y_i,x_i]$ are collected from multiple subjects, a more plausible model would be $g(\mu_i) = \alpha + f(x_i) + \tilde{f}_{s(i)}(x_i)$, where the $\tilde{f}$ terms are subject-specific random smooth terms. Let $\mathbf{X}$ and $\tilde{\mathbf{X}}$ again correspond to the model matrix and transformed model matrix of the shared $f$ -- term \emph{before} absorbing identifiability constraints. Subsequently, identifiability constraints have to be applied to $f$, due to the presence of an intercept term, and we use $\bar{\mathbf{X}}$ to denote the version of $\mathbf{X}$ after identifiability constraints have been absorbed \parencite{wood_generalized_2017-2}. The setup of the overall model matrix then proceeds as outlined in Section \ref{sec:challenges_mixed}, resulting in the matrix

$$
\begin{bmatrix}
\mathbf{1} & \bar{\mathbf{X}}_{\mathbf{1}} & \tilde{\mathbf{X}}_{\mathbf{1}} & \mathbf{0} & \mathbf{0} & \ldots & \mathbf{0}\\
\vdots & \bar{\mathbf{X}}_{\mathbf{2}} & \mathbf{0} & \tilde{\mathbf{X}}_{\mathbf{2}} & \mathbf{0} & \ldots & \mathbf{0}\\
\vdots & \bar{\mathbf{X}}_{\mathbf{3}} & \mathbf{0} & \ddots & \ddots & \ddots & \mathbf{0}\\
\vdots & \vdots & \vdots & \ddots & \ddots & \ddots & \mathbf{0}\\
\mathbf{1} & \bar{\mathbf{X}}_{\mathbf{N_s}} & \mathbf{0} &\ldots & \mathbf{0} & \mathbf{0} & \tilde{\mathbf{X}}_{\mathbf{N_s}}
\end{bmatrix}.
$$

Bold sub-scripts here again identify sub-blocks of $\bar{\mathbf{X}}$ (i.e., $\tilde{\mathbf{X}}$) belonging to individual subjects. Note, that while separate coefficients are to be estimated per subject-specific random smooth term $\tilde{f}$, all $\tilde{f}$ usually share the same regularization parameters \parencite[i.e., in mgcv,][and also in MSSM]{wood_generalized_2017-2}. Specifically, let $\mathcal{S}$ and $\tilde{\mathcal{S}}$ again denote the penalty matrix and transformed penalty matrix belonging to $f$. The total penalty matrix $\mathbf{S}_\lambda$ belonging to this model is then equal to

$$
\begin{bmatrix}
0 & \mathbf{0}  & \ldots  & \ldots  & \ldots & \mathbf{0}\\
\mathbf{0} & \mathcal{S}_\lambda & \mathbf{0} & \ldots  & \vdots & \vdots\\
\vdots & \mathbf{0} & \tilde{\mathcal{S}}_\lambda & \ddots & \vdots & \vdots\\
\vdots & \vdots & \mathbf{0} & \ddots & \vdots & \vdots\\
\vdots & \vdots & \vdots & \ddots & \tilde{\mathcal{S}}_\lambda & \mathbf{0}\\
\mathbf{0} & \mathbf{0}& \mathbf{0} &\ldots & \mathbf{0} & \tilde{\mathcal{S}}_\lambda
\end{bmatrix},
$$

where $\mathcal{S}_\lambda=\lambda_1\mathcal{S}$ and $\tilde{\mathcal{S}}_\lambda = \lambda_2\tilde{\mathcal{S}} + \sum_{n=1}^{N_0} \lambda_{n+2}\Psi^n$. In total, estimating this specific model thus only requires estimating $2 + N_0$ regularization parameters.
\newpage
\section{Stabilizing Newton's method via the Re-parameterization strategy by \textcite{wood_fast_2011}}\label{sec:AppendixTransform}
\setlength{\parindent}{0pt}

This section reviews the core concept of the stabilizing transformation strategies described previously (in slightly different variations) by \Textcite{wood_fast_2011}, \Textcite{wood_smoothing_2016}, and \Textcite{wood_generalized_2017-1} \parencite[see also Section 6.2.7 in the book by][for an overview]{wood_generalized_2017-2}. 

As discussed in the main text, the negative Hessian of the penalized log-likelihood $\mathcal{H}$ (and also $\mathcal{H}+\mathbf{I}\epsilon_{\mathcal{H}}$) will often be ill-conditioned to the degree that relying on a Cholesky decomposition utilizing sparsity-oriented pivoting to obtain $\mathbf{L}$ where $\mathbf{L}\mathbf{L}^\top=\mathcal{H}+\mathbf{I}\epsilon_{\mathcal{H}}$ -- required for the Newton update to $\boldsymbol{\beta}$ -- should be avoided \parencite{wood_smoothing_2016}. To stabilize the Newton update, \Textcite{wood_smoothing_2016} suggest to re-parameterize the model and to apply a diagonal pre-conditioner to $\mathcal{H}$. Specifically, \Textcite{wood_smoothing_2016} suggest that, after every update to $\boldsymbol{\lambda}$, the transformation strategy described in Appendix B of the paper by \Textcite{wood_fast_2011} should be applied separately to all penalized terms (i.e., smooth or random effect) included in the $N_m$ additive models of the parameters of the log-likelihood \parencite[see also Section 6.2.7 in the book by][]{wood_generalized_2017-2}. As pointed out by \Textcite{wood_smoothing_2016}, it is actually sufficient to complete the re-parameterization once for terms with a single penalty applied to them and for terms with multiple penalties applied to them that target separate sets of coefficients. For these terms \Textcite{wood_smoothing_2016} also suggest to rely on a different transformation strategy than the one described by \Textcite{wood_fast_2011}, which is more efficient while also stabilizing the computations involved in the Newton update. For convenience we here assume that the original strategy outlined by \Textcite{wood_fast_2011} is applied to every term. The result is an orthogonal block-diagonal transformation matrix

$$
\mathbf{T}=\begin{bmatrix}
\mathrm{T}^1 & \mathbf{0} & & \mathbf{0}\\
\mathbf{0} & \ddots & & \mathbf{0}\\
\mathbf{0} & \mathbf{0} & & \mathrm{T}^{N_m}\\
\end{bmatrix}.
$$

$\mathrm{T}^{m}$ itself is a block-diagonal matrix associated with the $m$-th parameter of the log-likelihood \parencite[individual sub-blocks of $\mathrm{T}^{m}$ are associated with individual terms included in the additive model of the $m$-th parameter;][]{wood_smoothing_2016}. The transformation matrix $\mathbf{T}$ and sub-blocks $\mathrm{T}^{m}$ can then be used to re-parameterize the entire model \parencite{wood_smoothing_2016}. Specifically, the total penalty matrix as well as individual (embedded) penalty matrices are replaced by their transformed variants $\mathbf{T}^\top\mathbf{S}_\lambda\mathbf{T}$ and $\mathbf{T}^\top\mathbf{S}^r\mathbf{T}$ respectively. Similarly, $\mathbf{X}^m\mathrm{T}^m$ is substituted for $\mathbf{X}^m$.

Subsequently, Algorithm (\ref{alg:gammlss_beta}) can be used to obtain the estimate $\hat{\boldsymbol{\beta}}$ from the transformed model \parencite{wood_smoothing_2016}. To stabilize the computation of the Cholesky decomposition of $\mathcal{H}$, now the negative Hessian of the penalized log-likelihood of the \emph{transformed model}, \Textcite{wood_smoothing_2016} suggest to substitute $\mathcal{H}'=\mathcal{D}\mathcal{H}\mathcal{D}$ for $\mathcal{H}$ before attempting the first Cholesky decomposition (e.g., after step 4 of Algorithm (\ref{alg:gammlss_beta})). $\mathcal{D}$ is a diagonal pre-conditioner matrix with elements $\mathcal{D}_{ii}=|\mathcal{H}_{ii}|^{-0.5}$ \parencite{wood_smoothing_2016}. Once the Cholesky decomposition completes successfully, application of the pre-conditioner $\mathcal{D}$ can be reversed \parencite[see][for details]{wood_smoothing_2016}. For example, assuming that $\mathbf{L}'$ is the Cholesky factor of $\mathbf{P}\left[\mathcal{H}' +\mathbf{I}\epsilon_{\mathcal{H}}\right]\mathbf{P}^\top$, $\mathbf{L} = \mathcal{D}^{-1}\mathbf{P}^\top\mathbf{L}'$. Similarly, once convergence in $\hat{\boldsymbol{\beta}}$ is reached, the transformations applied to the penalty matrices, model matrices, and all estimated quantities (e.g., $\hat{\boldsymbol{\beta}}$,$\mathbf{L}$) could be reversed as well \parencite{wood_smoothing_2016}. For example, the estimate of the coefficient vector after reversing the transformation would simply be equal to $\mathbf{T}\hat{\boldsymbol{\beta}}$ \parencite[e.g., ][]{wood_fast_2011,wood_generalized_2017-2}.

It is however not necessary to reverse the transformation at this point -- in fact it is useful to keep the model in the transformed state to further stabilize the computations performed as part of any strategy designed to check for unidentifiable coefficients and those required by the EFS update \parencite[e.g.,][]{wood_smoothing_2016,wood_generalized_2017-2}. As discussed in the main text, it will be necessary to account for the fact that some coefficients might generally not be identifiable (i.e., independent of the choice for $\boldsymbol{\lambda}$), in which case $\mathbf{L}$ (i.e., $\mathbf{L}'$) will never be computable with $\epsilon_\mathcal{H}=0$. This check has to be completed after Algorithm (\ref{alg:gammlss_beta}) converges and we described potential strategies to implement this check in Appendix \ref{sec:AppendixUnidentifiable} \parencite[see also][]{wood_smoothing_2016}. Notably, all of these strategies can be performed based on the transformed model.

Similarly, the heuristic strategy designed to ensure that the EFS update is defined whenever Algorithm (\ref{alg:gammlss_beta}) terminates, described in Section \ref{sec:GAMMLSS} and implemented in steps 18-20 of Algorithm (\ref{alg:gammlss_beta}), can be completed before reversing the transformation applied to the model: it can readily be shown that $tr(\mathbf{S}_\lambda^-\mathbf{S}^r)$ and $tr([\mathbf{H} + \mathbf{S}_\lambda]^{-1}\mathbf{S}^r)$ have the same result, independent of whether they are computed for the transformed model or after reversing the transformation. In fact, the entire EFS update by \Textcite{wood_generalized_2017} can be computed before reversing the transformation \parencite[e.g.,][]{wood_generalized_2017-2}. To implement this, only minor modifications to Algorithm (\ref{alg:sparse_gamm}) are necessary (Assuming that the changes discussed in Section \ref{sec:GAMMLSS} have already been implemented). Specifically, all that is necessary is to make sure that the model (i.e., $\mathbf{S}_\lambda$, the $\mathbf{S}^r$, and $\mathbf{X}^m$) is transformed anew after every proposed update to $\boldsymbol{\lambda}$ (i.e., after step 8 and re-forming $\mathbf{S}_\lambda$ in Algorithm (\ref{alg:sparse_gamm})). This transformation needs to be reversed (or rather renewed) in case the update $\boldsymbol{\Delta}_\lambda$ to $\boldsymbol{\lambda}$ needs to be corrected (i.e., as part of step 12 in Algorithm (\ref{alg:sparse_gamm})). Additionally, the transformation will generally have to be reversed as well after proposing the next update $\boldsymbol{\Delta}_\lambda$ (i.e., after step 15 in Algorithm (\ref{alg:sparse_gamm})).

\newpage
\section{Testing for Unidentifiable Coefficients when Working with Sparse Smooth Models}\label{sec:AppendixUnidentifiable}
\setlength{\parindent}{0pt}

This section reviews strategies to check for unidentifiable parameters for general smooth models estimated via Algorithm (\ref{alg:gammlss_beta}). As mentioned in the main text, \Textcite{wood_smoothing_2016} suggest to identify unidentifiable coefficients based on a rank revealing decomposition of $\mathcal{H}^S = \mathbf{H}/||\mathbf{H}||_F + \tilde{\mathbf{S}}/||\tilde{\mathbf{S}}||_F$, where $\mathbf{H}$ is the negative Hessian of the log-likelihood of the (transformed, if the strategy by \Textcite{wood_fast_2011} reviewed in Appendix \ref{sec:AppendixTransform} is used to stabilize computation of the Newton update) model, $\tilde{\mathbf{S}}=\sum_{r=1}^{N_\lambda}\mathbf{S}^r/||\mathbf{S}^r||_F$ denotes the ``balanced penalty matrix'' \parencite{wood_smoothing_2016}, and $||\mathbf{H}||_F$ denotes the Frobenius norm. While Heath's method \parencite[][Section \ref{sec:AM}]{heath_extensions_1982} could be used to get an indication of the specific coefficients causing rank deficiency of $\mathcal{H}^S$, it is unlikely that a QR decomposition of $\mathcal{H}^S$ would benefit from sparsity-oriented pivoting of the columns of $\mathcal{H}^S$. This is less of a problem for GAMMs, because here we can decompose $\mathbf{X}$ rather than $\mathbf{H}$. For models involving only a moderate number of coefficients it would thus usually be just as efficient to follow the approach taken by \Textcite{wood_smoothing_2016}: cast $\mathcal{H}^S$ to a dense matrix and then rely on a QR \parencite[or Cholesky]{wood_smoothing_2016} decomposition utilizing stability-oriented pivoting, often in combination with Cline's method \parencite{cline_estimate_1979}, to identify the rank of $\mathcal{H}^S$ and which elements of $\boldsymbol{\beta}$ have to be dropped. The next update to $\boldsymbol{\lambda}$ is then proposed for the reduced model, excluding the problematic coefficients \parencite[as mentioned in the main text and by][it will be necessary to re-run Algorithm (\ref{alg:gammlss_beta}) on the reduced model]{wood_smoothing_2016}. For large multi-level models, this step will however quickly become extensive.

There are some options to reduce this cost. For once, we could decide to simply accept the set of dropped coefficients once identified: instead of re-starting Algorithm (\ref{alg:gammlss_beta}) based on \emph{all} coefficients for every new proposal of $\boldsymbol{\lambda}$ it could be re-started with the same \emph{reduced} set of coefficients previously deemed unidentifiable. If we only try to identify rank deficiency issues in case Algorithm (\ref{alg:gammlss_beta}) terminates with $\epsilon_\mathcal{H} > 0$, or the final Cholesky factor $\mathbf{L}$ results in a poor condition number estimate \parencite[obtained for example via Cline's method;][]{cline_estimate_1979}, the time-to-fit will often be prolonged only marginally. Considering that, as pointed out by \Textcite{wood_smoothing_2016}, the rank deficiency issues we aim to detect here should be independent of the state of $\boldsymbol{\lambda}$, this approach is also not unreasonable from a theoretical perspective. To further reduce the cost of this step for large multi-level models we can perform the QR decomposition on a sub-block of $\mathcal{H}^S$, obtained by extracting only the rows and columns corresponding to parametric (i.e., linear) terms and smooth functions for which the corresponding penalty matrix as a non-trivial Kernel. This is less exact, but can be motivated by considering the fact that the random effects (and in particular the random smooth terms) that make this step so expensive to compute for large multi-level models are anyway fully penalized -- $\boldsymbol{\lambda}$-independent rank deficiencies in $\mathcal{H}^S$ thus cannot be a result of these terms.

An alternative to a (dense) QR decomposition utilizes stability-oriented pivoting is offered by a generalization of Foster's method \parencite{foster_rank_1986} -- which also aims to check for unidentifiable coefficients after convergence of Algorithm (\ref{alg:gammlss_beta}) based on $\mathcal{H}^S$. Specifically, when Algorithm (\ref{alg:gammlss_beta}) terminates with $\epsilon_\mathcal{H} > 0$, or the final Cholesky factor $\mathbf{L}$ results in a poor condition number estimate, Foster's method \parencite{foster_rank_1986} would proceed by first  approximating the smallest singular value and corresponding singular vector $\mathbf{v}$ for $\mathcal{H}^S$. In case $\mathcal{H}^S$ is rank-deficient the exact version of $\mathbf{v}$ would fall in the Kernel of $\mathcal{H}^S$ in which case an element in $\boldsymbol{\beta}$ that is unidentifiable could be identified (and removed from both $\mathcal{H}^S$ and $\boldsymbol{\beta}$) by finding the index in $\mathbf{v}$ with the maximum absolute value \parencite{foster_rank_1986}. This can then be repeated (each time based on the updated $\mathcal{H}^S$, reflecting the coefficients dropped so far) until a Cholesky decomposition of the reduced $\mathcal{H}^S$ succeeds and produces a Cholesky factor with an acceptable condition number. Originally, \Textcite{foster_rank_1986} suggested to approximate \parencite[e.g., via Cline's method;][]{cline_estimate_1979} the smallest singular value and corresponding singular vector $\mathbf{v}$ for $\mathcal{H}^S$ based on $\mathbf{R}$ -- again obtained from a QR decomposition of $\mathcal{H}^S$. For efficiency reasons, we suggest to instead form the decomposition $\mathcal{H}^S=[\mathbf{P}^r]^T\mathbf{L}\mathbf{U}[\mathbf{P}^c]^T$, where $\mathbf{L}$ is a unit-diagonal lower triangular matrix, $\mathbf{U}$ is upper-triangular, and $\mathbf{P}^r$ and $\mathbf{P}^c$ are permutation matrices designed to balance sparsity in the triangular factors and numerical stability of the decomposition \parencite[e.g.,][]{gotsman_computation_2008}. As discussed by \Textcite{gotsman_computation_2008}, this LU decomposition will often be more efficient than computing a QR-decomposition. Moreover, $\mathbf{U}$ contains information about the Kernel of $\mathcal{H}^S$ and can thus be substituted for $\mathbf{R}$. As pointed out by \Textcite{gotsman_computation_2008}, any $\mathbf{v}$ that falls in the Kernel of $\mathbf{U}$ will also fall in the Kernel of $\mathcal{H}^S$ -- but not necessarily the other way around. Hence, no suitable vector $\mathbf{v}$ might be identifiable based on $\mathbf{U}$, despite $\mathcal{H}^S$ still being rank deficient. \Textcite{gotsman_computation_2008} discuss how this can be addressed in principle, but in practice it will often be advisable to fall back to one of the QR-based strategies outlined at the beginning of this section in case Foster's method based on $\mathbf{U}$ fails to produce a full-rank (reduced) $\mathcal{H}^S$ after multiple iterations.

The MSSM toolbox by default relies on the QR approach outlined above (based on the full version of $\mathcal{H}^S$) and, in case coefficients are deemed unidentifiable, retains the set of dropped coefficients for any subsequent iteration of Algorithm (\ref{alg:gammlss_beta}). It is however also possible to prevent this retention. Additionally, the generalization of Foster's method \parencite{foster_rank_1986} based on the work by \textcite{gotsman_computation_2008} has been implemented as well. For more details we refer to the documentation of the \texttt{GSMM} class, available at \url{https://jokra1.github.io/mssm}.
\newpage
\section{Review of the Implicit Eigendecomposition Approach proposed by \Textcite{burdakov_efficiently_2017}}\label{sec:AppendixImplicitEigen}

\setlength{\parindent}{0pt}
This section reviews the implicit Eigendecomposition approach proposed originally by \Textcite{burdakov_efficiently_2017} to compute the positive semi-definite matrices $[\hat{\mathcal{I}}^+]^{i-1}$ and $[\hat{\mathbf{H}}^+]^{i}$ that are closest, in terms of the Frobenious norm, to $\hat{\mathcal{I}}^{i-1}$ and $\hat{\mathbf{H}}^i$ respectively without forming the explicit Eigendecomposition $\hat{\mathbf{H}}^i=\mathrm{U}\Sigma\mathrm{U}^T$ as discussed in the main text. While the approach was originally proposed by \Textcite{burdakov_efficiently_2017}, \Textcite{erway_efficiently_2015} provide an alternative derivation that better highlights the relation between the implicit decomposition of $\hat{\mathbf{H}}^i$ and the explicit Eigen-decomposition $\mathrm{U}\Sigma\mathrm{U}^T$. We thus follow their approach to show how $[\hat{\mathbf{H}}^+]^{i}$ can be obtained by means of an implicit decomposition. They start by considering the full QR-decomposition $\mathrm{Q}\mathrm{R}=$

\begin{equation}\label{eq:qr}
\left[\mathrm{Q^1~~\mathrm{Q}^2}\right]  \begin{bmatrix}
\mathrm{R}^1\\
\mathbf{0}\\
\end{bmatrix} = \mathbf{Q'},
\end{equation}

where $\mathrm{Q}$ is a $N_p*N_p$ orthogonal matrix of which $\mathrm{Q}^1$ makes up the first $N_V$ columns and $\mathrm{R}^1$ is a $N_V*N_V$ upper-triangular matrix \parencite[]{erway_efficiently_2015}. As pointed out by \Textcite{erway_efficiently_2015}, this decomposition has a purely theoretical purpose and will not have to be formed in practice. We now again follow \Textcite{erway_efficiently_2015} and define the Eigendecomposition

\begin{equation}\label{eq:qr2}
\begin{split}
\mathrm{R}\mathbf{C'}\mathrm{R}^T &=
\begin{bmatrix}
\mathrm{R}^1\mathbf{C'}[\mathrm{R}^1]^T & \mathbf{0}\\
\mathbf{0} & \mathbf{0}\\
\end{bmatrix}=\begin{bmatrix}
\mathrm{U'}\Sigma'\mathrm{U'}^T & \mathbf{0}\\
\mathbf{0} & \mathbf{0}\\
\end{bmatrix}
\\&=\mathrm{U''}\Sigma''\mathrm{U''}^T~\text{with}\\&~\mathrm{U''}=\begin{bmatrix}
\mathrm{U'} & \mathbf{0}\\
\mathbf{0} & \mathbf{I}\\
\end{bmatrix}~\text{and}~
\Sigma''=\begin{bmatrix}
\mathrm{\Sigma'} & \mathbf{0}\\
\mathbf{0} & \mathbf{0}\\
\end{bmatrix}.
\end{split}
\end{equation}

Then, $\mathrm{P}\Sigma''\mathrm{P}^T$, with $\mathrm{P} = \mathrm{Q}\mathrm{U''}$, is the Eigendecomposition of $\mathbf{Q'}\mathbf{C'}\mathbf{Q'}$ \parencite[since $\mathbf{Q}$ is orthogonal;][]{erway_efficiently_2015}. Hence, $\hat{\mathbf{H}}^{i}=\mathbf{I} \frac{1}{\gamma'_i} + \mathrm{P}\Sigma'\mathrm{P}^T$ and $\hat{\mathbf{H}}^{i} = \mathrm{U}\Sigma\mathrm{U}^T = \mathrm{P}(\mathbf{I} \frac{1}{\gamma'_i} + \Sigma'')\mathrm{P}^T$ is the Eigendecomposition of $\hat{\mathbf{H}}^{i}$ \parencite{erway_efficiently_2015}. In combination with Equation (\ref{eq:qr2}), this result reveals that negative Eigenvalues of $\hat{\mathbf{H}}^{i}$ originate from negative Eigenvalues in $\Sigma'$ with a magnitude greater than $\frac{1}{\gamma'_i}$. This makes it straightforward to compute the nearest positive semi-definite matrix $[\hat{\mathbf{H}}^+]^{i}= \mathrm{U}\Sigma^+\mathrm{U}^T = \mathrm{P}(\mathbf{I} \frac{1}{\gamma_i} + \Sigma''^+)\mathrm{P}^T$, where

\begin{equation}\label{eq:evShift}
\Sigma''^+=\begin{bmatrix}
\Sigma'^+ & \mathbf{0}\\
\mathbf{0} & \mathbf{0}\\
\end{bmatrix}~\text{with}~\Sigma'^+_{ll} = \Sigma'_{ll} - min(0,\Sigma'_{ll} + \frac{1}{\gamma'_i}),
\end{equation}

so that all negative Eigenvalues are shifted just enough that they become zero when adding $\frac{1}{\gamma'_i}$ \parencite[see][]{higham_computing_1988}. \Textcite{erway_efficiently_2015} show that in practice $[\hat{\mathbf{H}}^+]^{i}$ can be computed without forming $\mathrm{Q}$, $\mathrm{U''}$, and $\Sigma''$. They point out, that $\frac{1}{\gamma'_i}$ and the decomposition $\mathrm{R}^1\mathbf{C'}[\mathrm{R}^1]^T$ already ``implicitly'' represent the entire range of Eigenvalues of $\hat{\mathbf{H}}^i$. Thus, computing the thin QR-decomposition $\mathrm{Q}^1\mathrm{R}^1=\mathbf{Q'}$ \parencite[e.g.,][]{golub_matrix_2013}, where $\mathrm{Q}^1$ is a $N_p * N_\mathbf{V}$ matrix with orthogonal columns, is sufficient to obtain $[\hat{\mathbf{H}}^+]^{i}$ \parencite[see also][]{burdakov_efficiently_2017}. Specifically, from this decomposition and Equation (\ref{eq:qr2}), which requires only $\mathrm{R}^1$ but not $\mathrm{R}$, it follows that

\begin{equation}\label{eq:evEquality}
\begin{split}
\hat{\mathbf{H}}^i &= \mathbf{I}\frac{1}{\gamma'_i} + \mathrm{P}\Sigma''\mathrm{P}^T\\&=\mathbf{I}\frac{1}{\gamma'_i} + \mathrm{P'}\Sigma'\mathrm{P'}^T,
\end{split}
\end{equation}

where $\mathrm{P'}=\mathrm{Q}^1\mathrm{U'}$. Substituting $\Sigma'^+$ for $\Sigma'$ again in equation (\ref{eq:evEquality}) then finally provides

\begin{equation}\label{eq:SR1UPHessianPSDAPD}
\begin{split}
[\hat{\mathbf{H}}^+]^{i} &= \mathbf{I}\frac{1}{\gamma'_i} + \mathrm{P} \Sigma''^+\mathrm{P}^T\\&= \mathbf{I}\frac{1}{\gamma'_i} + \mathrm{P'}\Sigma'^+\mathrm{P'}^T.
\end{split}
\end{equation}

The second line of equation (\ref{eq:SR1UPHessianPSDAPD}) is the result stated in the main text.
\newpage
\section{Derivation of $\frac{\partial^2 \mathcal{V}}{\partial \rho_j \partial \rho_l}$}\label{sec:AppendixRemlDeriv}
\setlength{\parindent}{0pt}

Equation (\ref{eq:laplace_reml}) in the main text states the log-REML criterion, as $\mathcal{V}=\mathcal{L}_\lambda(\hat{\boldsymbol{\beta}}) + \frac{1}{2}log|\mathbf{S}_\lambda|_+ - \frac{1}{2}log|\mathcal{H}|$ \parencite[see also][]{wood_smoothing_2016}. For convenience, we again assume that if the model to be estimated is a Generalized additive model and thus features a scale parameter $\phi$, $\boldsymbol{\lambda} = \boldsymbol{\lambda}/\phi$. We now seek $\frac{\partial^2\mathcal{V}}{\partial \rho_j \partial \rho_l}$ with $\rho_j=log(\lambda_j)$. From the definition of the REML criterion we have that

\begin{equation}\label{eq:targetDeriv_apdx}
    \frac{\partial^2\mathcal{V}}{\partial \rho_j \partial \rho_l}=\frac{d^2\mathcal{L}_\lambda}{d\rho_jd\rho_l} + \frac{1}{2}\frac{\partial^2log|\mathbf{S}_\lambda|_+}{\partial \rho_j \partial \rho_l} - \frac{1}{2}\frac{\partial^2log|\mathcal{H}|}{\partial \rho_j \partial \rho_l},
\end{equation}

where we use $\mathcal{L}_\lambda$ as a shorthand for $\mathcal{L}_\lambda(\hat{\boldsymbol{\beta}})$. We first obtain the result for the total second derivative $\frac{d^2\mathcal{L}_\lambda}{d\rho_jd\rho_l}$, which requires finding $\frac{d\mathcal{L}_\lambda}{d\rho_j}$ in a first step. We denote the required total derivative as

\begin{equation}\label{eq:total_deriv_apdx}
f'=\frac{d \mathcal{L}_\lambda}{d\rho_j} = \frac{\partial \mathcal{L_\lambda}}{\partial \rho_j} + \frac{\partial \mathcal{L_\lambda}}{\partial \hat{\boldsymbol{\beta}}}\frac{d \hat{\boldsymbol{\beta}}}{d \rho_j},
\end{equation}

where we use $\frac{\partial \mathcal{L_\lambda}}{\partial \hat{\boldsymbol{\beta}}}$ as a shorthand for $\frac{\partial \mathcal{L_\lambda}}{\partial \boldsymbol{\beta}}\Big\rvert_{\hat{\boldsymbol{\beta}}}$. This result follows from considering that $\mathcal{L}_\lambda$ depends on $\boldsymbol{\lambda}$ directly, via the penalty $\frac{1}{2}\hat{\boldsymbol{\beta}}^\top\mathbf{S}_\lambda\hat{\boldsymbol{\beta}}$ subtracted from the log-likelihood $\mathcal{L}$, and indirectly through $\hat{\boldsymbol{\beta}}$ \parencite[e.g.,][]{wood_generalized_2017-2}. Differentiating again, we get the second total derivative $\frac{d^2\mathcal{L}_\lambda}{d\rho_jd\rho_l}$, which, based on Equation (\ref{eq:total_deriv_apdx}), can be defined as

\begin{equation}
f'' = \frac{d^2\mathcal{L}_\lambda}{d\rho_jd\rho_l} = \frac{d f'}{d\rho_l} =\frac{d \frac{\partial \mathcal{L}_\lambda}{\partial \rho_j}}{d \rho_l} + \frac{d \frac{\partial \mathcal{L_\lambda}}{\partial \hat{\boldsymbol{\beta}}}\frac{d \hat{\boldsymbol{\beta}}}{d \rho_j}}{d \rho_l}.
\end{equation}

Computation of $\frac{d \frac{\partial \mathcal{L}_\lambda}{\partial \rho_j}}{d \rho_l}$ now again follows from the total derivative computations outlined in Equation (\ref{eq:total_deriv_apdx}). Specifically,

\begin{equation}
\frac{d \frac{\partial \mathcal{L}_\lambda}{\partial \rho_j}}{d \rho_l} =\frac{\partial^2 \mathcal{L}_\lambda}{\partial \rho_j \partial \rho_l} + \frac{\partial^2 \mathcal{L}_\lambda}{\partial \rho_j \partial \hat{\boldsymbol{\beta}}}\frac{d\hat{\boldsymbol{\beta}}}{d\rho_l}.
\end{equation}

Additionally, $\frac{d \frac{\partial \mathcal{L_\lambda}}{\partial \hat{\boldsymbol{\beta}}}\frac{d \hat{\boldsymbol{\beta}}}{d \rho_j}}{d \rho_l}$ can itself be separated into two terms via the product rule, after which we have

\begin{equation}
\frac{d \frac{\partial \mathcal{L_\lambda}}{\partial \hat{\boldsymbol{\beta}}}\frac{d \hat{\boldsymbol{\beta}}}{d \rho_j}}{d \rho_l}=\frac{d \frac{\partial \mathcal{L_\lambda}}{\partial \hat{\boldsymbol{\beta}}}}{d \rho_l}\frac{d \hat{\boldsymbol{\beta}}}{d \rho_j} + \frac{\partial \mathcal{L_\lambda}}{\partial \hat{\boldsymbol{\beta}}}\frac{d^2 \hat{\boldsymbol{\beta}}}{d \rho_j d \rho_l}.
\end{equation}

By definition, $\frac{\partial \mathcal{L_\lambda}}{\partial \boldsymbol{\beta}}\Big\rvert_{\hat{\boldsymbol{\beta}}}$ evaluates to zero \parencite[e.g.,][]{wood_core_2015,wood_generalized_2017-2}. Hence, we only need to focus on the first term of the sum, computation of which again follows from the total derivative computations in Equation (\ref{eq:total_deriv_apdx}). Specifically,

\begin{equation}
\frac{d \frac{\partial \mathcal{L_\lambda}}{\partial \hat{\boldsymbol{\beta}}}}{d \rho_l}=\frac{\partial^2 \mathcal{L}_\lambda}{\partial \hat{\boldsymbol{\beta}} \partial \rho_l} + \frac{\partial^2 \mathcal{L}_\lambda}{\partial \hat{\boldsymbol{\beta}} \partial \hat{\boldsymbol{\beta}}^\top}\frac{d\hat{\boldsymbol{\beta}}}{d\rho_l}.
\end{equation}

Substituting the total derivative results for the terms in $f''$ we get

\begin{equation}
\begin{split}\label{eq:total_deriv2_apdx}
f''&=\frac{\partial^2 \mathcal{L}_\lambda}{\partial \rho_j \partial \rho_l} + \frac{\partial^2 \mathcal{L}_\lambda}{\partial \rho_j \partial \hat{\boldsymbol{\beta}}}\frac{d\hat{\boldsymbol{\beta}}}{d\rho_l} + \frac{\partial^2 \mathcal{L}_\lambda}{\partial \hat{\boldsymbol{\beta}} \partial \rho_l}\frac{d \hat{\boldsymbol{\beta}}}{d \rho_j} + \frac{d \hat{\boldsymbol{\beta}}^\top}{d \rho_j}\frac{\partial^2 \mathcal{L}_\lambda}{\partial \hat{\boldsymbol{\beta}} \partial \hat{\boldsymbol{\beta}}^\top}\frac{d\hat{\boldsymbol{\beta}}}{d\rho_l} + \frac{\partial \mathcal{L_\lambda}}{\partial \hat{\boldsymbol{\beta}}}\frac{d^2 \hat{\boldsymbol{\beta}}}{d \rho_j d \rho_l}\\
&=\frac{\partial^2 \mathcal{L}_\lambda}{\partial \rho_j \partial \rho_l} + \frac{d \hat{\boldsymbol{\beta}}^\top}{d \rho_j}\frac{\partial^2 \mathcal{L}_\lambda}{\partial \hat{\boldsymbol{\beta}} \partial \hat{\boldsymbol{\beta}}^\top}\frac{d\hat{\boldsymbol{\beta}}}{d\rho_l} + \frac{\partial^2 \mathcal{L}_\lambda}{\partial \rho_j \partial \hat{\boldsymbol{\beta}}}\frac{d\hat{\boldsymbol{\beta}}}{d\rho_l} + \frac{\partial^2 \mathcal{L}_\lambda}{\partial \hat{\boldsymbol{\beta}} \partial \rho_l}\frac{d \hat{\boldsymbol{\beta}}}{d \rho_j}.
\end{split}
\end{equation}

Where the second line again follows from the fact that the final term in the first line cancels by definition of $\hat{\boldsymbol{\beta}}$ \parencite[e.g.,][]{wood_generalized_2017-2}. Now we can address the partial (mixed) derivatives $\frac{\partial^2 \mathcal{L}_\lambda}{\partial \rho_j \partial \rho_l}$ and $\frac{\partial^2 \mathcal{L}_\lambda}{\partial \rho_j \partial \hat{\boldsymbol{\beta}}}$. By definition of the penalized log-likelihood, the former is equal to

\begin{equation}
\frac{\partial^2 \mathcal{L}_\lambda}{\partial \rho_j \partial \rho_l}=-\frac{\partial^2\frac{1}{2}\hat{\boldsymbol{\beta}}^\top\mathbf{S}_\lambda\hat{\boldsymbol{\beta}}}{\partial \rho_j \partial \rho_l} = -\gamma^{jl}\frac{\lambda_l}{2}\hat{\boldsymbol{\beta}}^\top\mathbf{S}^l\hat{\boldsymbol{\beta}},
\end{equation}

where $\gamma^{jl}\in\{0,1\}$ is only non-zero if $j=l$. The mixed partial derivative 

\begin{equation}
\frac{\partial^2 \mathcal{L}_\lambda}{\partial \rho_j \partial \hat{\boldsymbol{\beta}}}=-\frac{\partial\frac{1}{2}\hat{\boldsymbol{\beta}}^\top\mathbf{S}_\lambda\hat{\boldsymbol{\beta}}}{\partial \rho_j \partial \hat{\boldsymbol{\beta}}} = -\frac{\partial\frac{\lambda_j}{2}\hat{\boldsymbol{\beta}}^\top\mathbf{S}^j\hat{\boldsymbol{\beta}}}{\partial \hat{\boldsymbol{\beta}}}= 
-\lambda_j\mathbf{S}^j\hat{\boldsymbol{\beta}}
\end{equation}

follows from the standard result $\frac{\partial\frac{c}{2}\mathbf{b}^\top\mathbf{A}\mathbf{b}}{\partial \mathbf{b}}=c\mathbf{A}\mathbf{b}$ for symmetric matrix $\mathbf{A}=\mathbf{S}^j$. Additionally, because of the symmetry of mixed partial derivatives, computation of $\frac{\partial^2 \mathcal{L}_\lambda}{\partial \hat{\boldsymbol{\beta}} \partial \rho_l} = \frac{\partial^2 \mathcal{L}_\lambda}{\partial \rho_l \partial \hat{\boldsymbol{\beta}}}$, proceeds exactly the same. Substituting the partial derivative results for the terms in Equation (\ref{eq:total_deriv2_apdx}) and using $\mathcal{H}$ to denote $-\frac{\partial^2 \mathcal{L}_\lambda}{\partial \boldsymbol{\beta} \partial \boldsymbol{\beta}^\top}\Big\rvert_{\hat{\boldsymbol{\beta}}}$ then yields

\begin{equation}\label{eq:total_deriv2_final_apdx}
f''=-\gamma^{jl}\frac{\lambda_l}{2}\hat{\boldsymbol{\beta}}^\top\mathbf{S}^l\hat{\boldsymbol{\beta}} - \frac{d\hat{\boldsymbol{\beta}}^\top}{d \rho_j}\mathcal{H}\frac{d\hat{\boldsymbol{\beta}}}{d\rho_l} - \lambda_j\mathbf{S}^j\hat{\boldsymbol{\beta}}\frac{d\hat{\boldsymbol{\beta}}}{d\rho_l} - \lambda_l\mathbf{S}^l\hat{\boldsymbol{\beta}}\frac{d\hat{\boldsymbol{\beta}}}{d\rho_j}.
\end{equation}

Now, all that remains is to address the second partial derivatives for $\frac{\partial^2log|\mathbf{S}_\lambda|_+}{\partial \rho_j \partial \rho_l}$ and $\frac{\partial^2log|\mathcal{H}|}{\partial \rho_j \partial \rho_l}$. Their expressions follow from the same standard result, stated for example in Appendix B of \textcite{wood_fast_2011}, that $\frac{\partial^2log|\mathbf{B}|}{\partial a_j \partial a_l} = tr(\mathbf{B}^{-1}\frac{\partial^2\mathbf{B}}{\partial a_j \partial a_l}) - tr(\mathbf{B}^{-1}\frac{\partial\mathbf{B}}{\partial a_j}\mathbf{B}^{-1}\frac{\partial\mathbf{B}}{\partial a_l})$. Considering that $\mathcal{H}=\mathbf{H}+\mathbf{S}_\lambda$, where $\mathbf{H}$ again denotes $-\frac{\partial^2 \mathcal{L}}{\partial \boldsymbol{\beta} \partial \boldsymbol{\beta}^\top}\Big\rvert_{\hat{\boldsymbol{\beta}}}$, it then follows that

\begin{equation}\label{eq:DerivH_apdx}
    \begin{split}
        \frac{\partial^2log|\mathcal{H}|}{\partial \rho_j \partial \rho_l}&=tr(\mathcal{H}^{-1}\frac{\partial^2\mathcal{H}}{\partial \rho_j \partial \rho_l}) - tr(\mathcal{H}^{-1}\frac{\partial\mathcal{H}}{\partial \rho_j}\mathcal{H}^{-1}\frac{\partial\mathcal{H}}{\partial \rho_l})\\&
        =\gamma^{jl}\lambda_ltr\left(\mathcal{H}^{-1}\left(\frac{\partial^2\mathbf{H}}{\partial \rho_j \partial \rho_l} + \mathbf{S}^l\right)\right) - \lambda_j\lambda_ltr\left(\mathcal{H}^{-1}\left(\frac{\partial\mathbf{H}}{\partial \rho_j} + \mathbf{S}^j\right)\mathcal{H}^{-1}\left(\frac{\partial\mathbf{H}}{\partial \rho_l} + \mathbf{S}^l\right)\right).
    \end{split}
\end{equation}

The equation above matches the result stated in slightly different terms by \textcite{wood_smoothing_2016}. Similarly,

\begin{equation}\label{eq:DerivS_apdx}
        \frac{\partial^2log|\mathbf{S}_\lambda|_+}{\partial \rho_j \partial \rho_l}=\gamma^{jl}\lambda_ltr\left(\mathbf{S}^-\mathbf{S}^l\right) - \lambda_j\lambda_ltr\left(\mathbf{S}^-\mathbf{S}^j\mathbf{S}^-\mathbf{S}^l\right),
\end{equation}

which again makes use of the generalized inverse $\mathbf{S}^-$ \parencite[e.g.,][]{wood_generalized_2017}. As mentioned briefly in Section \ref{sec:GAMMLSS} and described in detail by \Textcite{wood_fast_2011}, the generalized inverse can in practice be replaced by an actual inverse, when relying on the re-parameterization strategy from Appendix B in the paper by \textcite{wood_fast_2011}.

As discussed in the main text, adopting the PQL assumption implies assuming that $\frac{\partial \mathbf{H}}{\partial \rho_l} = \mathbf{0}$ and $\frac{\partial^2 \mathbf{H}}{\partial \rho_l \partial \rho_j} = \mathbf{0}$ hold in general and for all $\rho_l,\rho_j \in \boldsymbol{\rho}$. In that case, after substituting the results defined in Equations (\ref{eq:total_deriv2_final_apdx}), (\ref{eq:DerivH_apdx}), and (\ref{eq:DerivS_apdx}) for the terms in Equation (\ref{eq:targetDeriv_apdx}), we arrive at 

\begin{equation}
\begin{split}
    \frac{\partial^2\mathcal{V}}{\partial \rho_j \partial \rho_l}=~&-\gamma^{jl}\frac{\lambda_l}{2}\hat{\boldsymbol{\beta}}^\top\mathbf{S}^l\hat{\boldsymbol{\beta}} - \frac{d\hat{\boldsymbol{\beta}}^\top}{d \rho_j}\mathcal{H}\frac{d\hat{\boldsymbol{\beta}}}{d\rho_l} -\lambda_j \mathbf{S}^j\hat{\boldsymbol{\beta}}\frac{d\hat{\boldsymbol{\beta}}}{d\rho_l} - \lambda_l\mathbf{S}^l\hat{\boldsymbol{\beta}}\frac{d\hat{\boldsymbol{\beta}}}{d\rho_j}\\&
    -\frac{1}{2}\left(\gamma^{jl}\lambda_ltr\left(\mathcal{H}^{-1}\mathbf{S}^l\right) - \lambda_j\lambda_ltr\left(\mathcal{H}^{-1}\mathbf{S}^j\mathcal{H}^{-1}\mathbf{S}^l\right)\right)\\&
    +\frac{1}{2}\left(\gamma^{jl}\lambda_ltr\left(\mathbf{S}^-\mathbf{S}^l\right) - \lambda_j\lambda_ltr\left(\mathbf{S}^-\mathbf{S}^j\mathbf{S}^-\mathbf{S}^l\right)\right),
\end{split}
\end{equation}

which is the result shown in Equation (\ref{eq:pql_hess_rho}) of the main text and, if the estimated model is a GAM so that $\mathbf{H}=\mathbf{X}^\top\mathbf{W}\mathbf{X}$, equivalent (except for a sign flip) to the PQL derivative result stated below Equation (8) in the paper by \textcite{wood_generalized_2017-1}. In Section \ref{sec:AM} of the main text we already discussed how the trace $tr(\mathcal{H}^{-1}\mathbf{S}^l)$ can be computed efficiently when working with sparse matrix storage and $\mathcal{H}^{-1}$ is obtained by means of a Cholesky decomposition of $\mathcal{H}=\mathbf{P}^\top\mathbf{L}\mathbf{L}^\top\mathbf{P}$, where $\mathbf{P}$ is again a permutation matrix chosen to increase the sparsity of $\mathbf{L}$. This is useful if $\mathbf{H}$ has a structure that can be exploited, which will for example be the case for multi-level GAMs where $\mathbf{H}=\mathbf{X}^\top\mathbf{W}\mathbf{X}$. The trace  $tr(\mathbf{V}\mathbf{S}^j\mathbf{V}\mathbf{S}^l)$ will then again be equal to the sum of squares of the non-zero elements in matrix $[\mathbf{D}^l]^\top\mathbf{P}^\top\mathbf{L}^{-T}\mathbf{L}^{-1}\mathbf{P}\mathbf{D}^j$ where $\mathbf{D}^j[\mathbf{D}^j]^\top = \mathbf{S}^j$ and $\mathbf{D}^l[\mathbf{D}^l]^\top = \mathbf{S}^l$. While this second trace is generally more costly to compute, it only has to be computed once since the result is only required to correct for uncertainty in $\hat{\boldsymbol{\lambda}}$ which is a post-estimation task.
\end{document}